\documentstyle[12pt,epsfig]{article}

\begin{document}
\setlength{\baselineskip}{0.30in}
\newcommand{\be}{\begin{eqnarray}}
\newcommand{\ee}{\end{eqnarray}}
\newcommand{\bi}{\bibitem}
\newcommand{\nue}{\nu_e}
\newcommand{\num}{\nu_\mu}
\newcommand{\nut}{\nu_\tau}
\newcommand{\nus}{\nu_s}
\newcommand{\mne}{m_{\nu_e}}
\newcommand{\mnm}{m_{\nu_\mu}}
\newcommand{\mnt}{m_{\nu_\tau}}
\newcommand{\munu}{\mu_{\nu}}
\newcommand{\lar}{\leftarrow}
\newcommand{\rar}{\rightarrow}
\newcommand{\lrar}{\leftrightarrow}
\newcommand{\nuh}{\nu_h}
\newcommand{\mnut}{m_{\nu_\tau}}
\newcommand{\mnh}{m_{\nu_h}}
\newcommand{\taut}{\tau_{\nut}}
\newcommand{\fg}{f_{\gamma}}
\newcommand{\mnu}{m_{\nu}}
\newcommand{\taunu}{\tau_{\nu}}
\newcommand{\nc}{\newcommand}
\newcommand{\dm}{\delta m^2}

\newcommand{\so}{\, \mbox{sin}\Omega}
\newcommand{\co}{\, \mbox{cos}\Omega}
\newcommand{\mnus}{m_{\nu_s}}
\newcommand{\taus}{\tau_{\nu_s}}
\newcommand{\cost}{\cos \theta}
\newcommand{\sint}{\sin \theta}
\newcommand{\stw}{\sin 2 \theta}
\newcommand{\ctw}{\cos 2 \theta}
\newcommand{\sv}{\sin 2 \theta}
\newcommand{\nua}{\nu_a}

\newcommand{\om}{\omega}
\newcommand{\ds}{\partial \!  \! \! /}
\newcommand{\Zs}{Z \! \! \! \! /}
\newcommand{\Ws}{W \! \! \! \! \! /}

\newcommand{\raa}{\rho_{aa}}
\newcommand{\rss}{\rho_{ss}}
\newcommand{\rsa}{\rho_{sa}}
\newcommand{\ras}{\rho_{as}}

{\hbox to\hsize{April, 2002  \hfill INFN-2002}
\begin{center}
\vglue .06in
{\Large \bf { Neutrinos in cosmology 
  }
}
\bigskip
\\{\bf A.D. Dolgov}
 \\[.05in]
{\it{INFN, sezzione di Ferrara\\
via Paradiso, 12, 44100 - Ferrara, Italy
\footnote{Also: ITEP, Bol. Cheremushkinskaya 25, Moscow 113259, Russia.}
}}\\
\end{center}
\begin{abstract}
Cosmological implications of neutrinos are reviewed. The following 
subjects are discussed at a different level of scrutiny: cosmological 
limits on neutrino mass, neutrinos and primordial nucleosynthesis, 
cosmological constraints on unstable neutrinos, lepton asymmetry of 
the universe, impact of neutrinos on cosmic microwave radiation, 
neutrinos and the large scale structure of the universe, neutrino
oscillations in the early universe, baryo/lepto-genesis and neutrinos,
neutrinos and high energy cosmic rays, and briefly some more exotic 
subjects: neutrino balls, mirror neutrinos, and neutrinos from large 
extra dimensions.
\end{abstract}


{\bf Content}
\begin{enumerate}

\item{}
Introduction
\item{}
Neutrino properties.
\item{}
Basics of cosmology.\\
3.1. Basic equations and cosmological parameters.\\
3.2. Thermodynamics of the early universe.\\
3.3. Kinetic equations.\\
3.4. Primordial nucleosynthesis.  
\item{}
Massless or light neutrinos\\
4.1. Gerstein-Zeldovich limit.\\
4.2. Spectral distortion of massless neutrinos.
\item{}
Heavy neutrinos.\\
5.1. Stable neutrinos, $m_{\nu_h} < 45$ GeV.\\
5.2. Stable neutrinos, $m_{\nu_h} > 45$ GeV.
\item{}
Neutrinos and primordial nucleosynthesis.\\
6.1. Bound on the number of relativistic species.\\
6.2. Massive stable neutrinos. Bounds on $m_{\nut}$.\\
6.3. Massive unstable neutrinos.\\
6.4. Right-handed neutrinos.\\
6.5. Magnetic moment of neutrinos.\\ 
6.6. Neutrinos, light scalars, and BBN.\\
6.7. Heavy sterile neutrinos: cosmological bounds and direct experiment.
\item{}
Variation of primordial abundances and lepton asymmetry of the universe.
\item{}
Decaying neutrinos.\\
8.1. Introduction.\\
8.2. Cosmic density constraints.\\
8.3. Constraints on radiative decays from the spectrum of cosmic 
microwave background radiation.\\
8.4. Cosmic electromagnetic radiation, other than CMB.
\item{}
Angular anisotropy of CMB and neutrinos.
\item{}
Cosmological lepton asymmetry.  \\
10.1. Introduction.\\
10.2. Cosmological evolution of strongly degenerate neutrinos.\\
10.3. Degenerate neutrinos and primordial nucleosynthesis.\\
10.4. Degenerate neutrinos and large scale structure.\\
11.4. Neutrino degeneracy and CMBR.
\item{}
Neutrinos, dark matter and large scale structure of the universe.\\
11.1. Normal neutrinos.\\
11.2. Lepton asymmetry and large scale structure.\\
11.3. Sterile neutrinos.\\
11.4. Anomalous neutrino interactions and dark matter; unstable neutrinos.
\item{}
Neutrino oscillations in the early universe.\\
12.1. Neutrino oscillations in vacuum. Basic concepts.\\
12.2. Matter effects. General description.\\
12.3. Neutrino oscillations in cosmological plasma.\\
$\,\,$ 12.3.1. A brief (and non-complete) review.\\
$\,\,$ 12.3.2. Refraction index.\\
$\,\,$ 12.3.3. Loss of coherence and density matrix.\\
$\,\,$ 12.3.4. Kinetic equation for density matrix.\\
12.4. Non-resonant oscillations.\\
12.5. Resonant oscillations and generation of lepton asymmetry.\\
$\,\,$ 12.5.1. Notations and equations.\\
$\,\,$ 12.5.2. Solution without back-reaction.\\
$\,\,$ 12.5.3. Back-reaction.\\
$\,\,$ 12.5.4. Chaoticity.\\
12.6. Active-active neutrino oscillations.\\
12.7. Spatial fluctuations of lepton asymmetry.\\
12.8. Neutrino oscillations and big bang nucleosynthesis.\\
12.9. Summary.
\item{}
Neutrino balls.
\item{}
Mirror neutrinos.
\item{}
Neutrinos and large extra dimensions.
\item{}
Neutrinos and lepto/baryogenesis.
\item{}
Cosmological neutrino background and ultra-high energy cosmic rays.
\item{}
Conclusion.
\item{}
References.

\end{enumerate}

\section{Introduction}

The existence of neutrino was first proposed by Pauli in 
1930 {\cite{paul1} as an
attempt to explain the continuous energy spectrum observed in
beta-decay~\cite{beta} under the assumption of energy conservation. 
Pauli himself did not
consider his solution to be a very probable one, though today such 
observation would be considered unambiguous proof of the existence
of a new particle. That particle was named ``neutrino'' in 1933,
by Fermi. A good, though brief description of historical events
leading to $\nu$-discovery can be found in ref.~\cite{ramond}.

The method of neutrino detection was suggested by
Pontecorvo~\cite{pontecorvo46}. To this end he proposed
the chlorine-argon reaction and discussed the possibility of
registering solar neutrinos. This very difficult experiment was
performed by Davies et al~\cite{davies68} in 1968, and marked the discovery
neutrinos from the sky (solar neutrinos). 
The experimental discovery of neutrino was carried out
by Reines and Cowan~\cite{reco} in 1956, a quarter of a century
after the existence of that particle was predicted.

In 1943 Sakata and Inou{\"e}~\cite{insa} suggested
that there might be more than one species of neutrino.
Pontecorvo~\cite{pont} in 1959 made a similar conjecture that
neutrinos emitted in beta-decay and in muon decay might be different.
This hypothesis was confirmed in 1962 by Danby et al~\cite{danb},
who found that neutrinos produced in muon decays could create in
secondary interactions only muons but not electrons.
It is established now that there are at least three different
types (or flavors) of neutrinos: electronic ($\nue$), muonic ($\num$),
and tauonic ($\nut$) and their antiparticles.
The combined LEP result~\cite{pdg} based on the measurement of the decay 
width of $Z^0$-boson gives the following number of different 
neutrino species:
$N_\nu = 2.993\pm 0.011$, including all neutral fermions with the normal
weak coupling to $Z^0$ and mass below $m_Z/2 \approx 45$ GeV.

It was proposed by Pontecorvo~\cite{pontos1,pontos2} in 1957 that, in direct
analogy with $(K^0-\bar K^0)$-oscillations, neutrinos may also
oscillate due to $(\bar \nu -\nu)$-transformation. After it was confirmed that
$\nue$ and $\num$ are different particles~\cite{danb}, Maki, Nakagawa, and
Sakata~\cite{mns} suggested the possibility of neutrino flavor oscillations,
$\nue\lrar \num$. A further extension of the oscillation space 
what would permit the violation of the total leptonic charge 
as well as violation of separate lepton flavor charges, 
$\nue \lrar \num$ and  $\nue \lrar \bar\num$, or flavor 
oscillations of Majorana neutrinos was proposed by Pontecorvo and 
Gribov~\cite{pontos3, bpvg}. Nowadays the phenomenon of neutrino 
oscillations attracts great attention in experimental particle 
physics as well as in astrophysics and cosmology. A historical 
review on neutrino oscillations can be found 
in refs.~\cite{nakagawa98,bilenky99-rev}.

Cosmological implications of neutrino physics were first considered 
in a paper by
Alpher et al~\cite{alpher53} who mentioned that neutrinos would be in
thermal equilibrium in the early universe. The possibility that the
cosmological energy density of neutrinos may be larger than the energy
density of baryonic matter and the cosmological implications of this
hypothesis were discussed by Pontecorvo and Smorodinskii~\cite{pontecorvo62}.
A little later Zeldovich and Smorodinskii~\cite{zeldovich62} derived
the upper limit on the density of neutrinos from their gravitational
action. In a seminal paper in 1966, Gerstein and
Zeldovich~\cite{gz} derived the cosmological upper limit on neutrino
mass, see below sec.~\ref{gerzel}. This was done already in the
frameworks of modern cosmology. Since then the interplay between neutrino
physics and cosmology has been discussed in hundreds of papers, where
limits on neutrino properties and the use of neutrinos in solving 
some cosmological problems were considered. Neutrinos could have been 
important in the formation of the large-scale structure (LSS) of the 
universe, in big bang nucleosynthesis (BBN), in anisotropies of cosmic 
microwave background radiation
(CMBR), and some others cosmological phenomena. This is the subject 
of the present review. The field is so vast and the number of published papers
is so large that I had to confine the material strictly to cosmological
issues. Practically no astrophysical material is presented, though in many
cases it is difficult to draw a strict border between the two.
For the astrophysical implications of neutrino physics one can address
the book~\cite{raffelt96} and a more recent review~\cite{raffelt99a}. The
number of publications rises so quickly (it seems, with increasing speed)
that I had to rewrite already written sections
several times to include recent
developments. Many important papers could be and possibly
are omitted involuntary but their absence in the literature list does not
indicate any author's preference. They are just ``large number errors''.
I tried to find old pioneering papers where essential physical mechanisms
were discovered and the most recent ones, where the most accurate treatment
was performed; the latter was much easier because of available astro-ph
and hep-ph archives.


\section{Neutrino properties.\label{2prop} }


It is well established now that neutrinos have standard weak
interactions mediated by $W^{\pm}$- and $Z^0$-bosons in which only
left-handed neutrinos participate. No other interactions of neutrinos
have been registered yet. The masses of neutrinos are either small or zero.
In contrast to photons and gravitons, whose vanishing masses are ensured by
the principles of gauge invariance and general covariance respectively, no similar theoretical principle is known for neutrinos. They may have non-zero
masses and their smallness presents a serious theoretical challenge. For
reviews on physics of (possibly massive) neutrinos
see e.g. the papers~\cite{jv}-\cite{langacker01}.
Direct observational bounds on neutrino masses, found kinematically, are:
\be
\mne < \left\{ \begin{array}{ll}
2.8-2.5 \,\, {\rm eV} & \mbox{\cite{weinheimer99,mnue}},  \\
         10 \,\,{\rm eV} & \mbox{(other groups, see \cite{pdg}) },
 \end{array}\right.
\label{mne}
\ee
\be
\begin{array}{ll}
\mnm < 170 {\rm keV} & \mbox{ \cite{mnum}},
\end{array}
\label{mnm}
\ee
\be
\begin{array}{ll}
\mnt < 18 {\rm MeV} & \mbox{\cite{mnut}},
\end{array}
\label{mnt}
\ee
while cosmological upper limit on masses of light stable neutrinos is about
10 eV (see below, Sec. \ref{gerzel}).

Even if neutrinos are massive, it is unknown if they have Dirac or Majorana
mass. In the latter case processes with leptonic charge non-conservation
are possible and from their absence on experiment, in particular, from the
lower limits on the nucleus life-time with respect to
neutrinoless double beta decay one can deduce an upper limit on the Majorana
mass. The most stringent bound was obtained in Heidelberg-Moscow
experiment \cite{hm}: $\mne < 0.47 $ eV; for the results of other groups
see~\cite{kz}.

There are some experimentally observed anomalies (reviewed e.g.
in refs.~\cite{jv,kz}) in neutrino physics, which
possibly indicate new phenomena and most naturally can be explained by neutrino
oscillations. The existence of oscillations implies a non-zero mass difference
between oscillating neutrino species, which in turn means that at least
some of the neutrinos should be massive. Among these anomalies is the well
known deficit of solar neutrinos, which has been registered by 
several installations: the pioneering
Homestake, GALLEX, SAGE, GNO, Kamiokande and its mighty successor,
Super-Kamiokande. One should also mention the first data recently
announced by SNO~\cite{CNO} where evidence for the presence of
$\num$ or $\nut$ in the flux of solar neutrinos was given. This
observation strongly supports the idea that $\nue$ is mixed with
another active neutrino, though some mixing with sterile ones is not
excluded. An analysis of the solar neutrino data can be
found e.g. in refs.~\cite{bks}-\cite{bahcall01-SNO}.
In the last two of these papers the data from SNO was also used.

The other two anomalies in neutrino physics are, first, the
$\bar \nue$-appearance seen in LSND experiment~\cite{lsnd} in the
flux of $\bar\num$ from $\mu^+$ decay at rest and $\nue$ appearance
in the flux of $\num$ from the $\pi^+$ decay in flight. In a recent
publication~\cite{lsnd01} LSND-group reconfirmed their original results.
The second anomaly is registered in energetic cosmic ray air showers. 
The ratio of $(\num/\nue)$-fluxes is suppressed by factor two
in comparison with
theoretical predictions (discussion and the list of the
references can be found in~\cite{jv,kz}). This effect
of anomalous behavior of atmospheric neutrinos recently received 
very strong support from the Super-Kamiokande observations~\cite{sk98}
which not only confirmed $\num$-deficit but also discovered that the latter
depends upon the zenith angle. This latest result is a very strong argument in
favor of neutrino oscillations. The best fit to the oscillation parameters
found in this paper for $\num \lrar\nut$-oscillations are
\be
\sin^2 2\theta &=&1 \nonumber \\
\Delta m^2 &=& 2.2\,\times\, 10^{-3} \,{\rm eV}^2
\label{sksin}
\ee
The earlier data did not permit distinguishing between the oscillations
$\num \lrar\nut$ and the oscillations of $\num$ into a non-interacting
sterile neutrino, $\nus$, but more detailed investigation gives a strong
evidence against explanation of atmospheric neutrino anomaly by
mixing between $\num$ and $\nus$~\cite{sk00}.

After the SNO data~\cite{CNO} the explanation of the solar neutrino anomaly
also disfavors dominant mixing of $\nue$ with a sterile neutrino and the
mixing with $\num$ or $\nut$ is the most probable case.
The best fit to the solar neutrino anomaly~\cite{bahcall01-SNO}   is provided by MSW-resonance solutions (MSW means
Mikheev-Smirnov~\cite{mikheev} and
Wolfenstein~\cite{wolfenstein}, see sec.~\ref{nuosceu}) 
- either LMA (large mixing angle
solution):
\be
\tan^2 \theta &=& 4.1 \,\times\, 10^{-1}\nonumber \\
\Delta m^2 &=& 4.5\,\times\, 10^{-5} \,{\rm eV}^2
\label{solact}
\ee
or LOW (low mass solution):
\be
\tan^2 \theta &=& 7.1 \,\times\, 10^{-1}\nonumber \\
\Delta m^2 &=& 1.0\,\times\, 10^{-7} \,{\rm eV}^2
\label{solact1}
\ee
Vacuum solution is almost equally good:
\be
\tan^2 \theta &=& 2.4 \,\times\, 10^{0}\nonumber \\
\Delta m^2 &=& 4.6\,\times\, 10^{-10} \,{\rm eV}^2
\label{solvac}
\ee
Similar results are obtained in a slightly earlier paper~\cite{fogli01}.

The hypothesis that there may exist an (almost) new non-interacting sterile
neutrino looks quite substantial but if all the reported neutrino anomalies
indeed exist, it is impossible to describe them all, together with the
limits on oscillation parameters found in plethora of other experiments,
without invoking a sterile neutrino. The proposal to invoke a sterile
neutrino for explanation of the total set of the observed neutrino
anomalies was raised in the papers~\cite{peltoniemi93,caldwell93}.
An analysis of the more recent data and a list of references can be
found e.g. in the paper~\cite{jv}. Still with the exclusion
of some pieces of the data, which may be unreliable, 
an interpretation in terms of three known neutrinos remains
possible~\cite{bhssw,bhs}. For an earlier attempt to ``satisfy everything''
based on three-generation neutrino mixing scheme see e.g.
ref.~\cite{cardall96a}. If, however, one admits that a sterile neutrino
exists, it is quite natural to expect that there exist even three sterile
ones corresponding to the known active species: $\nue$, $\num$, and $\nut$.
A simple phenomenological model for that can be realized with the
neutrino mass matrix containing both Dirac and Majorana mass 
terms~\cite{ad1}. Moreover, the analysis performed in the
paper~\cite{barger01} shows that the combined solar neutrino data are
unable to determine the sterile neutrino admixture.

If neutrinos are massive, they may be unstable.
Direct bounds on their life-times are very loose~\cite{pdg}:
$\tau_{\nue} /\mne > 300$ sec/eV, $\tau_{\num} /\mnm > 15.4$ sec/eV, and
no bound is known for $\nut$. Possible decay channels of a heavier 
neutrino,
$\nu_a$ permitted by quantum numbers are: $\nu_a \rar \nu_b\, \gamma$,
$\nu_a \rar \nu_b\, \nu_c \bar\nu_c$, and $\nu_a \rar \nu_b\,e^-\,e^+$. If
there exists a yet-undiscovered light (or massless) (pseudo)scalar 
boson $J$,
for instance majoron~\cite{maj} or familon~\cite{fam}, another decay 
channel
is possible: $\nu_a \rar \nu_b\,J$. Quite restrictive limits on different
decay channels of massive neutrinos can be derived from cosmological 
data as
discussed below.

In the standard theory neutrinos possess neither electric charge nor magnetic
moment, but have an electric form-factor and their charge radius is 
non-zero, though
negligibly small. The magnetic moment may be non-zero if right-handed neutrinos
exist, for instance if they have a Dirac mass. In this case the magnetic
moment should be proportional to neutrino mass and quite
small~\cite{lee77,marciano77}:
\be
\mu_\nu = {3e G_F  m_\nu\over 8\sqrt{2}\pi^2} \approx 3.2\times 10^{-19}\mu_B
\left(m_\nu / {\rm eV}\right)
\label{munu}
\ee
where $G_F = 1.1664\cdot 10^{-5} {\rm GeV}^{-2}$ is the Fermi coupling
constant, $e=\sqrt {4\pi \alpha} = 0.303$ is the magnitude of electric charge
of electron, and $\mu_B = e/2m_e$ is the Bohr magneton. In terms of the
magnetic field units G=Gauss the Born magneton is equal to
$\mu_B = 5.788\cdot 10^{-15}{\rm MeV /G} $.
The experimental upper limits on magnetic moments of different neutrino
flavors are~\cite{pdg}:
\be
\mu_{\nue} < 1.8\times 10^{-10} \mu_B,\,\,\,
\mu_{\num} < 7.4\times 10^{-10} \mu_B,\,\,\,
\mu_{\nut} < 5.4\times 10^{-7} \mu_B.
\label{muexsp}
\ee
These limits are very far from simple theoretical expectations. However
in more complicated theoretical models much larger values for neutrino
magnetic moment are predicted, see sec.~\ref{ssec:magnmom}.

Right-handed neutrinos may appear not only because of the left-right transformation
induced by a Dirac mass term but also if there exist direct right-handed
currents. These are possible in some extensions of the standard electro-weak
model. The lower limits on the mass of possible right-handed intermediate
bosons are summarized in ref.~\cite{pdg} (page 251). They are typically around
a few hundred GeV. As we will see below, cosmology gives similar or even
stronger bounds.


Neutrino properties are well described by the standard electroweak theory
that was finally formulated in the late 60th in the works of S. Glashow,
A. Salam, and S. Weinberg. Together with quantum chromodynamics (QCD), this
theory forms the so called Minimal Standard Model (MSM) of particle physics.
All the existing experimental data are in good agreement with MSM, except
for observed anomalies in neutrino processes. Today neutrino
is the only open window to new physics in the sense that only in neutrino
physics some anomalies are observed that disagree with MSM.
Cosmological constraints on neutrino
properties, as we see in below, are often more restrictive than
direct laboratory measurements. Correspondingly, cosmology may be more
sensitive to new physics than particle physics experiments.

\section{Basics of cosmology. \label{sec:cosmology} }

\subsection{Basic equations and cosmological parameters.\label{sec:cosmeq}}

We will present here some essential cosmological facts and equations so
that the paper would be self-contained. One can find details e.g. in the
textbooks~\cite{zn}-\cite{peacock99}. Throughout this review we will use
the natural system of units, with $c$, $k$, and $\hbar$ each equaling 1.
For conversion factors
for these units see table~\ref{t:units} which is borrowed from
ref.~\cite{raffelt99t}.

\begin{table}[hbt]
\caption{\label{t:units}Conversion factors for natural units.}
\bigskip
\hbox to\hsize{\hss
\small
\begin{tabular}{|c||c|c|c|c|c|c|c|} \hline
 & $\rm s^{-1}$ & $\rm cm^{-1}$ & K & eV & amu & erg & g \\ \hline \hline
$\rm s^{-1}$ & 1 & $0.334{\times}10^{-10} $ & $0.764{\times}10^{-11} $
& $0.658{\times}10^{-15} $ &  $0.707{\times}10^{-24} $ &
$1.055{\times}10^{-27} $ & $1.173{\times}10^{-48}$ \\ \hline
$\rm cm^{-1}$ &  $2.998 {\times} 10^{10}$ & 1 & $0.229                $
& $1.973 {\times} 10^{- 5} $ &  $2.118 {\times} 10^{-14} $
& $3.161 {\times} 10^{-17} $
& $0.352 {\times} 10^{-37}$ \\ \hline
K             &  $1.310 {\times} 10^{11}$ & 4.369 & $1                    $
& $0.862 {\times} 10^{- 4} $ &  $0.962 {\times} 10^{-13} $ &
$1.381 {\times} 10^{-16} $
& $1.537 {\times} 10^{-37}$ \\ \hline
eV           &  $1.519 {\times} 10^{15}$ & $0.507 {\times} 10^{5}$ &
$1.160 {\times} 10^4$ & $1                    $ &  $1.074 {\times} 10^{-9 } $
& $1.602 {\times} 10^{-12} $ & $1.783 {\times} 10^{-33}$ \\ \hline
amu        &  $1.415 {\times} 10^{24}$ & $0.472 {\times} 10^{14}$ &
$1.081 {\times} 10^{13}$ & $0.931 {\times} 10^9     $ &  $1    $
& $1.492 {\times} 10^{-3 } $ & $1.661 {\times} 10^{-24}$ \\ \hline
erg     &  $0.948 {\times} 10^{27}$ & $0.316 {\times} 10^{17}$ &
$0.724 {\times} 10^{16}$ & $0.624 {\times} 10^{12}  $ &  $0.670 {\times} 10^3$
& 1                       & $1.113 {\times} 10^{-21}$ \\ \hline
  g     &  $0.852 {\times} 10^{48}$ & $2.843 {\times} 10^{37}$ &
$0.651 {\times} 10^{37}$ & $0.561 {\times} 10^{33}  $
&  $0.602 {\times} 10^{24}  $
& $0.899 {\times} 10^{21}$  &  1 \\ \hline
\end{tabular}\hss}
\end{table}

In the approximation of a homogeneous and
isotropic universe, its expansion is described by the
Friedman-Robertson-Walker metric:
\be
ds^2 = dt^2 - a^2 (t)\, { d{\vec r}\,^2 \over 1 + k {\vec r}\,^2 /4}
\label{ds2}
\ee
For the homogeneous and  isotropic  distribution  of  matter  the
energy-momentum tensor has the form
\be
T_0^0 &=& \rho, \nonumber \\
T_i^j &=& -p\delta_i^j ,\,\, (i,j=1,2,3)
\label{t00}
\ee
where $\rho$ and $p$ are respectively energy and pressure densities.
In  this  case  the  Einstein  equations  are   reduced   to   the
following two equations:
\be
\ddot a=-(4\pi G/3)(\rho +3p)a
\label{ddota} \\
{\dot a^2 \over 2} -{4\pi \over 3} G\rho a^2 =-{k \over 2}
\label{dota2}
\ee
where $G$ is the gravitational  coupling  constant, $G\equiv m^{-2}_{Pl}$,
with the Planck mass equal to $m _{Pl} = 1.221\cdot 10^{19}$ GeV. From
equations (\ref{ddota}}) and (\ref{dota2}) follows the covariant law of
energy conservation, or better to say, variation:
\be
\dot \rho =-3H(\rho +p)
\label{dotrho}
\ee
where $H=\dot a /a$ is the Hubble parameter. The critical
or closure energy density is expressed through the latter as:
\be
\rho_c =3H^2/8\pi G \equiv 3H^2m_{Pl}^2/8\pi
\label{rhoc}
\ee
$\rho=\rho_c$
corresponds to eq.~(\ref{dota2}) in the flat case, i.e. for $k=0$.
The present-day value of the critical density is
\be
\rho_c^{(0)} = 3H_0^2 m_{Pl}^2 /8\pi =
1.879 \cdot 10^{-29}\,h^2 \,{\rm g/cm}^3 =
10.54\, h^2\, {\rm keV/cm}^3\,,
\label{rhoc0}
\ee
where $h $ is the  dimensionless value of the present  day
Hubble parameter $H_0$ measured in 100 km/sec/Mpc. The value of the Hubble
parameter is rather poorly known, but it would be possibly safe to say that
$h  = 0.5-1.0$ with the preferred value $0.72\pm 0.08$~\cite{freedman00}.

The magnitude of mass or energy density in the universe,
$\rho$, is usually presented in terms of the dimensionless ratio
\be
\Omega = \rho / \rho_c
\label{omega}
\ee
Inflationary theory predicts $\Omega = 1$ with the accuracy $\pm 10^{-4}$
or somewhat better. Observations are most likely in agreement with 
this prediction, or at least do not contradict it.
There are several different
contributions to $\Omega$ coming from different forms of matter. The cosmic
baryon budget was analyzed in refs.~\cite{persic97s,fhp}. The amount
of visible baryons was estimated 
as $\Omega_b^{vis} \approx 0.003$~\cite{persic97s}, while for the 
total baryonic mass fraction the following range was 
presented~\cite{fhp}:
\be
\Omega_B = 0.007\, - \, 0.041
\label{omb}
\ee
with the best guess $\Omega_B \sim 0.021$ (for $h=0.7$). The recent data
on the angular distribution of cosmic microwave background radiation (relative
heights of the first and second acoustic peaks) add up to the result 
presented, e.g., in ref.~\cite{bernardis01}:
\be
\Omega_B h^2 = 0.022^{+0.004}_{-0.003}
\label{omegabh2}
\ee
Similar results are quoted in the works~\cite{2ndpeak}. 

There is a significant contribution to $\Omega$ from an unknown dark or
invisible matter. Most probably there are several different forms of this
mysterious matter in the universe, as follows from the observations of
large scale structure. The matter concentrated on galaxy cluster scales,
according to classical astronomical estimates, gives:
\be
\Omega_{DM} = \left\{ \begin{array}{ll}
(0.2-0.4)\pm 0.1  & \mbox{\cite{wlf}},  \\
  0.25\pm 0.2  & \mbox{ \cite{rbt} },
 \end{array}\right.
\label{omdm}
\ee
A recent review on the different ways of determining $\Omega_m$ can be
found in~\cite{schindler01}; though most of measurements converge
at $\Omega_m =0.3$, there are some indications for larger or smaller
values.

It was observed in 1998~\cite{perl} through observations of high 
red-sift supernovae that vacuum energy density, or cosmological 
constant, is non-zero
and contributes:
\be
\Omega_{vac} = 0.5-0.7
\label{omvac}
\ee
This result was confirmed by measurements of the position of the first
acoustic peak in angular fluctuations of CMBR~\cite{jaffe00} which is
sensitive to the total cosmological energy density, $\Omega_{tot}$.
A combined analysis of available astronomical data can be found in recent
works~\cite{wang01,tonry01,turner01}, where considerably more accurate
values of basic cosmological parameters are presented.

The discovery of non-zero lambda-term
deepened the mystery of vacuum energy, which is one of the
most striking puzzles in contemporary physics - the fact that any estimated
contribution to $\rho_{vac}$ is 50-100 orders of magnitude larger than the
upper bound permitted by cosmology (for reviews see~\cite{sw2,ad2,ad3}). The
possibility that vacuum energy is not precisely zero speaks in favor of 
adjustment mechanism\cite{ad4}.
Such mechanism would, indeed, predict that 
vacuum energy is compensated only with the accuracy of the order of
the critical energy density, $\rho_c \sim m_{pl}^2 /t^2$ at any epoch of the
universe evolution. Moreover, the non-compensated remnant may be subject to
a quite unusual equation of state or even may not be described by any
equation of state at all. There are many phenomenological models with a
variable cosmological "constant" described in the literature, a list
of references can be found in the review~\cite{aas}. A
special class of matter with the equation of state $p= w\rho$ with
$-1<w<0$ has been named "quintessence"~\cite{cds}. An analysis of
observational data~\cite{ptw} indicates
that $w<-0.6$ which is compatible with simple vacuum energy, $w=-1$. Despite
all the uncertainties, it seems quite probable that about half the matter
in the universe is not in the form of normal elementary
particles, possibly yet unknown, but in some other unusual state of matter.

To determine the expansion regime at different periods cosmological
evolution one has to know the equation of state
$p = p(\rho)$. Such a relation normally holds in some simple and physically
interesting cases, but generally equation of state does not exist. For
a gas of nonrelativistic particles the equation of state is $p=0$ (to be
more precise, the pressure density is not exactly zero but
$p\sim (T/m)\rho \ll \rho$). For the universe dominated by nonrelativistic
matter the expansion law is quite simple if $\Omega =1$:
$a(t)=a_0\cdot (t/t_0)^{2/3}$. It was once believed that nonrelativistic matter
dominates in the universe at sufficiently late stages, but possibly this
is not true today because of a non-zero cosmological constant. Still
at an earlier epoch ($z>1$) the universe was presumably dominated by
non-relativistic matter.

In standard cosmology the bulk of matter was relativistic at much
earlier stages. The equation of state was $p=\rho /3$ and the scale
factor evolved as $a(t)\sim t^{1/2}$. Since at that time $\Omega$ was
extremely close to unity, the energy density was equal to
\be
\rho = \rho_c = {3m_{Pl}^2 \over 32 \pi t^2 }
\label{rhorelc}
\ee
For vacuum dominated energy-momentum tensor, $p=-\rho$, $ \rho =const$,
and the universe expands exponentially, $a(t) \sim \exp (H_v t)$.

Integrating equation (\ref{dota2}) one can express the age of the 
universe through
the current values of the cosmological parameters $H_0$ and $\Omega_j$,
where sub-$j$ refers to different forms of matter with different equations
of state:
\be
t_0 = {1 \over H_0} \int_0^1 {dx \over \sqrt{ 1 - \Omega_{tot} +
\Omega_{m} x^{-1} + \Omega_{rel} x^{-2} + \Omega_{vac}x^2 }}
\label{tu}
\ee
where $\Omega_{m}$, $\Omega_{rel}$, and $\Omega_{vac}$
correspond respectively to the energy density of nonrelativistic
matter, relativistic matter, and to the vacuum energy density
(or, what is the same, to the cosmological constant);
$\Omega_{tot} = \Omega_{m} + \Omega_{rel} + \Omega_{vac}$,
and $H_0^{-1} = 9.778 \cdot 10^9h^{-1}$yr. This expression can be evidently
modified if there is an  additional contribution of matter with the equation
of state $p=w\rho$. Normally $\Omega_{rel}\ll \Omega_m$ because
$\rho_{rel} \sim a^{-4}$ and $\rho_m \sim a^{-3}$. On the other hand
$\rho_{vac} = const$ and it is quite a weird coincidence that
$\rho_{vac} \sim \rho_m$ just today. If $\Omega_{rel}$ and $\Omega_{vac}$
both vanishes, then there is a convenient expression for $t_0$ valid with
accuracy better than 4\% for $0<\Omega <2$:
\be
t_0^{m} = {9.788\cdot 10^9h^{-1} {\rm yr} \over 1 + \sqrt{\Omega} }
\label{tum}
\ee
Most probably, however, $\Omega_{tot} =1$, as predicted by inflationary
cosmology and $\Omega_{vac} \neq 0$. In that case the universe age is
\be
t_0^{lam} =  {6.525\cdot 10^9h^{-1} {\rm yr} \over \sqrt{\Omega_{vac}}}
\ln \left[ {1+\sqrt{\Omega_{vac}}
\over \sqrt{ 1- \Omega_{vac}} }\right]
\label{tulam}
\ee
It is clear that if $\Omega_{vac} >0$, then the universe may be considerably
older with the same value of $h$.
These expressions for $t_0$ will be helpful in what follows for the derivation
of cosmological bounds on neutrino mass.

The age of old globular clusters and nuclear chronology both give
close values for the age of the universe~\cite{wlf}:
\be
t_0 = (14-15)\pm 2 \,{\rm Gyr}
\label{t0}
\ee

\subsection{Thermodynamics of the early universe. \label{thermdyn}}

At early stages of cosmological evolution, particle number densities,
$n$, were so large that the rates of reactions, $\Gamma \sim \sigma n$, were
much higher than the rate of expansion, $H=\dot a /a$ (here $\sigma$ is
cross-section of the relevant reactions). In that period thermodynamic
equilibrium was established with a very high degree of accuracy. For a
sufficiently weak and short-range interactions between particles,
their distribution is represented by the well known Fermi or Bose-Einstein
formulae for the ideal homogeneous gas (see e.g. the book~\cite{ll}):
\be
f^{(eq)}_{f,b}(p) ={1\over \exp \left[ (E-\mu )/T \right] \pm  1 }
\label{ffb}
\ee
Here signs '+' and '$-$' refer to fermions and bosons respectively,
$E=\sqrt {p^2 +m^2} $ is the particle energy,  and  $\mu$  is  their
chemical potential. As is well known, particles and antiparticles in
equilibrium have equal in magnitude but opposite in sign chemical
potentials:
\be
\mu + \bar \mu =0
\label{mumubar}
\ee
This follows from the  equilibrium  condition  for chemical
potentials which for an arbitrary reaction $a_1 +a_2  +a_3  \ldots
\leftrightarrow b_1 +b_2 +\ldots $ has the form
\be
\sum_i \mu_{a_i} =\sum_j \mu_{b_j}
\label{summu}
\ee
and from the fact that particles and antiparticles can annihilate into
different
numbers of photons or into other neutral channels,
$a+\bar a  \rar  2\gamma,\,3\gamma \,,\ldots$. In particular,
the chemical potential of photons vanishes in equilibrium.

If certain particles possess a conserved  charge, their chemical
potential in equilibrium may be non-vanishing. It corresponds  to
nonzero  density  of  this  charge  in  plasma. Thus, plasma  in
equilibrium is completely defined by temperature and by a set of
chemical potentials corresponding to all conserved charges.
Astronomical observations indicate that the cosmological densities - of
all charges - that can be measured, are very small or even
zero. So in what follows we will usually assume that in equilibrium
$\mu_j =0$, except for Sections~\ref{sec:asym}, \ref{leptaslss},
\ref{sec:leptas}, and \ref{sec:spatfluc},
where lepton asymmetry is discussed. In
out-of-equilibrium conditions some effective chemical potentials - not
necessarily just those that satisfy condition (\ref{mumubar}) -  may be generated
if the corresponding charge is not conserved.

The number density of bosons corresponding to distribution (\ref{ffb})
with $\mu =0$ is
\be
n_b \equiv \sum_s \int {f_b (p)  \over  (2\pi)^3}  \,d^3p  =
\left\{ \begin{array}{ll}
\zeta  (3)  g_sT^3/  \pi^2  \approx  0.12gT^3,\,\,  {\rm if}\,\,
 T>m; \\
(2\pi)^{-3/2}g_s (mT)^{3/2} \exp(-m/T),\,\,{\rm if}\,\, T<m.
\end{array}\right.
\label{nb}
\ee
Here summation is made over all spin states of the boson, $g_s$ is the number
of this states, $\zeta (3)\approx 1.202$. In particular the number density
of equilibrium photons is
\be
n_\gamma =0.2404T^3=411.87 (T/2.728 {\rm K})^3\,{\rm cm}^{-3}
\label{ngamma}
\ee
where 2.728 K is the present day temperature of the cosmic microwave
background radiation (CMB).

For fermions the equilibrium number density is
\be
n_f=\cases {{3\over 4}n_b  \approx  0.09g_sT^3, & if  $T>m$;\cr  n_b
\approx (2\pi )^{-3/2}g_s(mT)^{3/2}\exp(-m/T), & if $T<m$. \cr}
\label{nf}
\ee

The equilibrium energy density is given by:
\be
\rho =\sum {1\over 2\pi^2} \int {\,dpp^2E \over \exp(E/T) \pm  1}
\label{rhoeq}
\ee
Here the summation is done  over  all  particle  species  in 
plasma and their spin states. In the relativistic case
\be
\rho_{rel} =(\pi^2/30)g_*T^4
\label{rhorel}
\ee
where $g_*$ is the effective number of relativistic species,
$g_*= \sum [g_b +(7/8)g_f]$, the summation is done over all species and their
spin states. In particular, for photons we obtain
\be
\rho_\gamma  ={\pi^2  \over  15} T^4   \approx   0.2615\left( {T \over
2.728\, {\rm K}}\right)^4 {{\rm eV \over cm^3}}
\approx 4.662\cdot 10^{-34}
\left( {T\over 2.728K}\right)^4  {{\rm g \over cm^3}}
\label{rhogamma}
\ee
The contribution of heavy particles, i.e. with $m>T$, into $\rho$ is
exponentially  small if  the   particles   are   in   thermodynamic
equilibrium:
\be
\rho_{nr} =g_s m\left( {mT  \over  2\pi}\right)^{3/2}
\exp\left( -{m\over   T}\right) \left(1+
{27T\over 8m} +\ldots \right)
\label{rhonr}
\ee
Sometimes the total energy density is described by expression (\ref{rhorel})
with the effective $g_* (T)$ including contributions of all relativistic as
well as non-relativistic species.

As we will see below, the equilibrium for stable particles sooner  or
later breaks down because their number density  becomes  too  small  to
maintain the proper annihilation rate. Hence their  number  density
drops as $a^{-3}$ and not exponentially. This  ultimately  leads to
a dominance of massive  particles in  the  universe.  Their  number
and energy densities could be even higher if they possess a conserved charge
and if the corresponding  chemical  potential is non-vanishing.

Since $\Omega_m$ was very close to unity at early cosmological stages,
the energy density at that time was almost  equal to the  critical
density (\ref{rhorelc}). Taking this into account, it is easy to 
determine the
dependence of temperature on time during RD-stage when  $H=1/2t$ and
$\rho$ is given simultaneously by eqs. (\ref{rhorel}) 
and (\ref{rhorelc}):
\be
t T^2=\left({90\over 32\pi^3} \right)^{1/2} {m_{Pl} \over \sqrt  g_* }=
{2.42\over \sqrt g_*} \,({\rm MeV})^2 \,{\rm sec}
\label{tT2}
\ee
For example, in equilibrium plasma consisting of photons, $e^{\pm}$, and
three types of neutrinos with temperatures above the electron mass but
below the muon mass, $0.5 < T < 100$ MeV,
the effective number of relativistic species is
\be
g_* = 10.75
\label{g1075}
\ee
In the course of expansion and cooling down, $g_*$ decreases as the particle species with $m>T$ disappear from the plasma.
For example, at $T\ll m_e$ when the only relativistic particles are
photons and three types of neutrinos with the temperature
$T_\nu \approx 0.71 \, T_\gamma$ the effective number of species is
\be
g_* = 3.36
\label{g336}
\ee

If all chemical potentials vanish and thermal equilibrium is
maintained, the entropy of the primeval plasma is conserved:
\be
{d\over dt}\left( a^3\, {p+\rho \over T}\right)=0
\label{dtent}
\ee
In fact this equation is valid under somewhat weaker conditions, namely if
particle occupation numbers $f_j$ are arbitrary functions of the
ratio $E/T$ and the quantity $T$ (which coincides with temperature only in
equilibrium) is a function of time subject to the condition~(\ref{dotrho}).

\subsection{Kinetic equations.}

The universe is not stationary, it expands and cools down, and as a result
thermal equilibrium is violated or even destroyed. The evolution of the
particle occupation numbers $f_j$ is usually described by the kinetic
equation in the ideal gas approximation. The latter is valid because the
primeval plasma is not too dense, particle mean free path is much larger
than the interaction radius so that individual distribution functions
$f(E,t)$, describing particle energy spectrum, are physically meaningful.
We assume that $f(E,t)$ depends neither on space point ${\vec x}$
nor on the direction of the particle momentum. It is fulfilled because
of cosmological homogeneity and isotropy.
The universe expansion is taken
into account as a red-shifting of particle momenta, $\dot p=-Hp$. It gives:
\be
{df_i\over dt}={\partial  f_i\over  \partial  t}  +{\partial  f_i
\over \partial  p_i}\dot  p_i  =  {\partial  f_i\over  \partial  t}
-Hp_i{\partial f_i\over \partial p_i}
\label{dndt}
\ee
As a result the kinetic equation takes the form
\be
\left( {\partial \over \partial t} -Hp_i{\partial \over \partial
p_i}\right) f_i(p_i,t)=I^{coll}_i
\label{kin1}
\ee
where $I^{coll}_i$ is the collision integral for the process
$i+Y \lrar Z$:
\be
I^{coll}_i&=&-{(2\pi)^4 \over 2E_i}  \sum_{Z,Y}  \int  \,d\nu_Z  \,d\nu_Y
\delta^4 (p_i +p_Y -p_Z)
 \lbrack \mid A(i+Y\rightarrow  Z)  \mid^2  \nonumber \\
&&f_i  \prod_Y  f\prod_Z  (1\pm  f)-\mid  A(Z\rightarrow  i+Y)\mid^2
\prod_Z f \prod_{i+Y} (1\pm f)\rbrack
\label{si}
\ee
Here $Y$ and $Z$ are  arbitrary, generally  multi-particle  states,
$\prod_Y f$ is  the  product  of  phase space densities  of  particles
forming the state $Y$, and
\be
d\nu_Y = \prod_Y {\overline {dp}} \equiv \prod_Y {d^3p\over (2\pi )^3 2E}
\label{dnuy}
\ee
The signs '+' or '$-$' in $\prod (1\pm f)$ are chosen for  bosons  and
fermions respectively.

It can be easily verified that in the stationary  case  ($H=0$), the
distributions (\ref{ffb}) are indeed solutions of the kinetic 
equation (\ref{kin1}), if one takes into account  the conservation of energy
$E_i +\sum_Y E=\sum_Z E$, and the condition (\ref{summu}).
This follows from the validity of the relation
\be
\prod_{i+Y} f^{(eq)}\prod_Z  (1\pm  f^{(eq)})=
\prod_Z  f^{(eq)}\prod_{i+Y}  (1\pm  f^{(eq)})
\label{prodiy}
\ee
and from the detailed balance  condition,  $\mid  A(i+Y\rightarrow
Z)\mid =\mid A(Z\rightarrow i+Y)\mid$ (with a trivial transformation of
kinematical variables). The last condition is only true  if
the theory is invariant  with respect  to time  reversion.  We know, 
however, that CP-invariance is broken  and, because of the CPT-theorem,
T-invariance is also broken. Thus
T-invariance is only approximate.  Still even if  the  detailed  balance
condition is violated, the form of equilibrium distribution functions
remain the same. This is ensured by the weaker condition~\cite{adcyc}:
\be
\sum_k \int \,d\nu_{Z_k} \delta^4 \left(\sum_{Z_k} p -p_f\right)
\left( \mid A(Z_k \rightarrow f)\mid ^2 -\mid  A(f\rightarrow  Z_k
\mid^2 \right) =0
\label{cycbal}
\ee
where summation is made over all possible states $Z_k$. This condition
can be termed the cyclic balance condition, because it demonstrates
that thermal equilibrium is achieved not by a simple equality of probabilities
of direct and inverse reactions but through a more complicated cycle of
reactions. Equation (\ref{cycbal}) follows  from
the unitarity of $S$-matrix, $S^+ S=SS^+ =1$. In fact, a  weaker
condition is sufficient  for  saving  the  standard  form  of  the
equilibrium distribution functions, namely the  diagonal  part  of
the  unitarity  relation,  $\sum_f  W_{if}  =1$, and  the  inverse
relation $\sum_i W_{if}=1$, where $W_{if}$ is  the  probability  of
transition from the state $i$ to the  state  $f$.  The  premise that the sum of probabilities of all possible events is  unity  is
of course evident. Slightly less evident is the  inverse  relation,
which can be obtained from the first one by the CPT-theorem.

For the solution of kinetic equations, which will be considered below,
it is convenient to introduce the following dimensionless variables:
\be
x= m_0 a\,\,\, {\rm and}\,\, y_j = p_j a
\label{xyi}
\ee
where $a(t)$ is the scale factor and $m_0$ is some fixed parameter with
dimension of mass (or energy). Below 
we will take $m_0 = 1 $ MeV.  The scale factor $a$ is normalized
so that in the early thermal equilibrium relativistic stage $a=1/T$. In
terms of these variables the l.h.s. of kinetic equation (\ref{kin1}) takes
a very simple form:
\be
Hx{\partial f_i \over \partial x } = I^{coll}_i
\label{hxdfdx}
\ee
When the universe was dominated by relativistic matter and when the 
temperature dropped  as $T\sim 1/a$, the Hubble parameter could be taken as
\be
H = 5.44 \sqrt{{g_* \over 10.75}}\, {m_0^2 \over x^2 m_{Pl}}
\label{hofx}
\ee
In many interesting cases the evolution of temperature differs from the simple
law specified above but still the expression (\ref{hofx}) is 
sufficiently accurate.

\subsection{Primordial nucleosynthesis \label{prns} }

Primordial or big bang nucleosynthesis (BBN) is one of the cornerstones
of standard
big bang cosmology. Its theoretical predictions agree beautifully with 
observations of the abundances of the light elements, $^2H$, $^3He$,
$^4He$ and $^7Li$, which span 9 orders of magnitude. Neutrinos play
a significant role in BBN, and the preservation of successful predictions 
of BBN allows one to work our restrictive limits on neutrino properties. 

Below we will present a simple pedagogical
introduction to the theory of BBN and briefly discuss observational data.
The content of this subsection will be used in sec.~\ref{s:bbn}
for the analysis of neutrino physics at 
the nucleosynthesis epoch. A good reference where these
issues are discussed in detail is the book~\cite{bp};
see also the review papers~\cite{schtur,sarkar} and the
paper~\cite{esposito00b} where BBN with degenerate neutrinos is included.

The relevant temperature interval for BBN is approximately from 1 MeV  
to 50 keV. In accordance with eq. (\ref{tT2}) the corresponding time
interval is from 1 sec to 300 sec. When the universe cooled down
below MeV the weak reactions
\be
n+\nu_e &\lrar& p + e^-,
\label{nnue} \\
n+e^+ &\lrar& p +  \bar \nu
\label{ne}
\ee
became slow in comparison with the universe expansion rate, so the
neutron-to-proton ratio, $n/p$, froze at a constant value
$(n/p)_f = \exp \left( -\Delta m /T_f \right)$, where $\Delta m = 1.293$ MeV
is the neutron-proton mass difference and $T_f=0.6-0.7$ MeV is the
freezing
temperature. At higher temperatures the neutron-to-proton ratio was equal 
to its equilibrium value, $(n/p)_{eq} = \exp (-\Delta m /T)$. Below $T_f$ the
reactions (\ref{nnue}) and (\ref{ne}) stopped and the evolution of
$n/p$ is determined only by the neutron decay:
\be
n\rar p+e+\bar \nue
\label{ndecay}
\ee
with the life-time $\tau_n = 887\pm 2$ sec.

In fact the freezing is not an
instant process and this ratio can be determined from numerical 
solution of kinetic equation.
The latter looks simpler
for the neutron to baryon ratio, $r=n/(n+p)$:
\be
\dot r = {(1+3g^2_A) G_F^2 \over 2\pi^3} \left[ A - (A+B)\,r  \right]
\label{dotr}
\ee
where $g_A = -1.267$ is the axial coupling constant and the
coefficient functions are given by the expressions
\be
A = \int^\infty_0 dE_\nu E_\nu^2 E_e p_e f_e(E_e)
\left[1-f_\nu(E_\nu) \right] |_{E_e = E_\nu + \Delta m} +\nonumber \\
 \int_{m_e}^\infty dE_e E_\nu^2 E_e p_e f_{\bar \nu} (E_\nu)
\left[1-f_{\bar e}(E_e \right] \mid_{E_\nu = E_e + \Delta m} +\nonumber \\
\int_{m_e}^{\Delta m} dE_e E_\nu^2 E_e p_e f_{\bar \nu} (E_\nu)
f_e(E_e)\mid_{E_\nu + E_e = \Delta m}\, ,
\label{a}
\ee
\be
B = \int^\infty_0 dE_\nu E_\nu^2 E_e p_e
 f_\nu(E_\nu)\left[1-f_e(E_e) \right]|_{E_e = E_\nu + \Delta m} +\nonumber \\
\int_{m_e}^\infty dE_e E_\nu^2 E_e p_e
f_{\bar e}(E_e) \left[1-f_{\bar \nu}(E_\nu) \right]
\mid_{E_\nu = E_e + \Delta m} +\nonumber \\
\int_{m_e}^{\Delta m} dE_e E_\nu^2 E_e p_e
 \left[1-f_{\bar \nu}(E_\nu) \right]
\left[1-f_e(E_e) \right] \mid_{E_\nu + E_e = \Delta m}
\label{b}
\ee
These rather long expressions are presented here because they explicitly
demonstrate the impact of neutrino energy spectrum and of a possible
charge asymmetry on the $n/p$-ratio. It can be easily verified that for
the equilibrium distributions of electrons and neutrinos the following
relation holds, $A=B\exp\left( -\Delta m/T\right)$. In the high temperature
limit, when one may neglect $m_e$, the function $B(T)$ can be easily
calculated:
\be
B = 48T^5 + 24(\Delta m) T^4 + 4(\Delta m)^2 T^3
\label{bt}
\ee
Comparing the reaction rate, $\Gamma = (1+3g^2_A) G_F^2 B /2\pi^3$ with
the Hubble parameter taken from eq. (\ref{tT2}),
$H = T^2 \sqrt {g_*} /0.6 m_{Pl}$, we find that neutrons-proton ratio
remains close to its equilibrium value for temperatures above
\be
T_{np} = 0.7 \left( {g_*\over 10.75} \right)^{1/6} {\rm MeV}
\label{tnp}
\ee
Note that the freezing temperature, $T_{np}$, depends upon $g_*$, i.e.
upon the effective
number of particle species contributing to the cosmic energy density.

The ordinary differential equation (\ref{dotr}) can be derived from the master
equation (\ref{kin1}) either in nonrelativistic limit or, for more precise
calculations, under the assumption that neutrons and protons are in kinetic
equilibrium with photons and electron-positron pairs with a
common temperature $T$, so that $f_{n,p}\sim \exp (-E/T)$. As we will see
in what follows, this is not true for neutrinos below $T=2-3$ MeV. Due
to $e^+e^-$-annihilation the temperature of neutrinos became different
from the common temperature of photons, electrons, positrons, and baryons.
Moreover, the energy distributions of neutrinos noticeably (at per
cent level) deviate from equilibrium, but the impact of that on
light element abundances is very weak (see sec. 4.2).

The matrix elements of $(n-p)$-transitions as well as phase space integrals
used for the derivation of expressions (\ref{a}) and  (\ref{b}) were
taken in non-relativistic limit. One may be better off taking the exact
matrix elements with finite temperature and radiative corrections to
calculate the $n/p$ ratio with very good precision
(see refs.~\cite{lt,emmp} for details). Since reactions (\ref{nnue}) and
(\ref{ne}) as well as neutron decay are linear with respect to baryons,
their rates $\dot n/n$ do not depend upon the cosmic baryonic number density,
$n_B = n_p+n_n$, which is rather poorly known. The latter is usually expressed
in terms of dimensionless baryon-to-photon ratio:
\be
\eta_{10} \equiv 10^{10} \eta = 10^{10} n_B /n_\gamma
\label{eta}
\ee
Until recently, the most precise way of determining the magnitude 
of $\eta$ was through the abundances of light elements, especially 
deuterium and
$^3 He$, which are very sensitive to it. Recent accurate determination
of the position and height of the second acoustic peak in the
angular spectrum of CMBR~\cite{bernardis01,2ndpeak} allows us to find 
baryonic mass 
fraction independently. The conclusions of both ways seem to converge 
around $\eta_{10} = 5$.

The light element production goes through the chain of reactions:
$p\,(n,\gamma)\,d$, $d\,(p\gamma)\,^3He$, $d\,(d,n)\,^3He$, $d\,(d,p)\,t$,
$t\,(d,n)\,^4He$, etc. One might expect naively that the light nuclei became
abundant at $T=O({\rm MeV})$ because a typical nuclear binding energy is
several MeV or even tens MeV. However, since $\eta =n_B/n_\gamma$ is very
small, the amount of produced nuclei is tiny even at temperatures
much lower than their binding energy. For example, the number
density of deuterium is determined in equilibrium by the equality
of chemical potentials, $\mu_d = \mu_p + \mu_n$. From that and the
expression (\ref{nb}) we obtain: 
\be
n_d =3 e^{   \left(\mu_d -m_d\right) /T  }
\left({m_dT \over 2\pi} \right)^{3/2} =
{3\over 4}n_n n_p e^{ B/T } \left( 2\pi m_d \over m_pm_n T \right)^{3/2}
\label{nd}
\ee
where $B_D = 2.224 $ MeV is the deuterium binding energy and the coefficient
3/4 comes from spin counting factors.  One can see
that $n_d$ becomes comparable to $n_n$ only at the temperature:
\be
T_d = { 0.064 \,{\rm MeV} \over 1 - 0.029 \ln \eta_{10}}
\label{td}
\ee
At higher temperatures deuterium number density in cosmic
plasma is negligible. Correspondingly, the formation of other nuclei,
which stems from collisions with deuterium is suppressed. Only
deuterium could reach thermal equilibrium with protons and neutrons. This
is the so called "deuterium bottleneck". But as soon as $T_d$ is reached,
nucleosynthesis proceeds almost instantly. In fact, deuterium never 
approaches equilibrium abundance because of quick formation of heavier 
elements. The latter are created through two-body nuclear collisions 
and hence the probability of production of heavier elements increases 
with an increase of the baryonic number density. Correspondingly, 
less deuterium survives with
larger $\eta$. Practically all neutrons that had existed in the cosmic
plasma at $T\approx T_d$ were quickly captured into
$^4 He$. The latter has the largest binding energy,
$B_{^4 He} = 28.3$ MeV, and in equilibrium
its abundance should be the largest. Its mass fraction, $Y(^4 He )$,
is determined predominantly by the $(n/p)$-ratio at the moment when
$T\approx T_d$ and is approximately equal to
$2(n/p)/[1+(n/p)] \approx  25\%$.
There is also some production of $^7Li$ at the level
(a few)$\times 10^{-10}$. Heavier elements in the standard model are
not produced because the baryon number density is very small and
three-body collisions are practically absent.

Theoretical calculations of light elements abundances are quite accurate,
given the values of the relevant parameters: neutron life-time, which is
pretty well known now, the number of massless neutrino species, which 
equals 3 in the standard model and the ratio of baryon and photon 
number densities during nucleosynthesis, 
$\eta_{10} = 10^{10} (n_B /n_\gamma)$ (\ref{eta}).
The last parameter brings the largest uncertainty
into theoretical results. There are also some uncertainties
in the values of the nuclear reaction rates which were never
measured at such low energies in plasma
environment. According to the analysis of ref.~\cite{flsv}
these uncertainties could change the mass fraction of $^4 He$
at the level of a fraction of per cent, but for deuterium the
``nuclear uncertainty'' is about 10\% and for $^7 Li$ it is could be 
as much as 25\%. An extensive discussion of
possible theoretical uncertainties and a list of relevant references can be
found in recent works~\cite{lt,emmp}. Typical curves for primordial
abundances of light elements as functions of $\eta_{10}$, calculated with
the nucleosynthesis code of ref.~\cite{kaw}, are presented in
fig. \ref{primabund}.
\begin{figure}[htb]
\begin{center}
  \leavevmode
  \hbox{
    \epsfysize=4.0in
 \epsffile{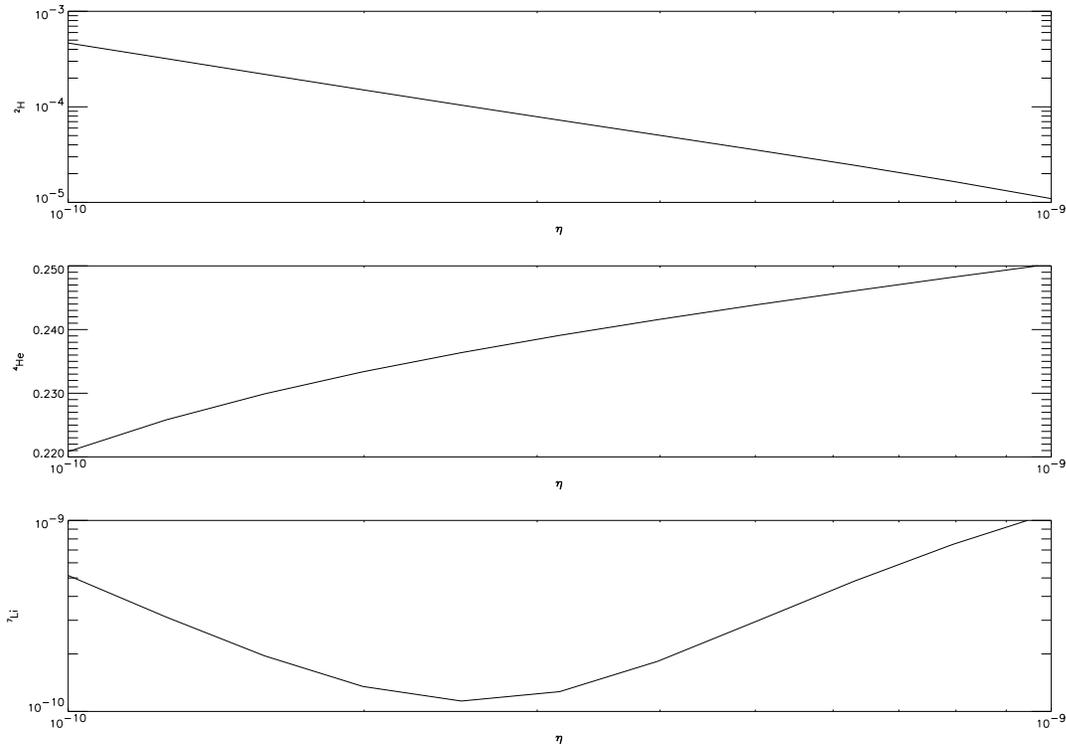}}
\end{center}
\caption{Abundances of light elements $^2H$ (by number) $^4He$ (by mass),
and $^7 Li$ (by number) as functions of baryon-to-photon ratio
$\eta_{10} \equiv 10^{10}n_B/n_\gamma$.
\label{primabund}}
\end{figure}
Another, and a very serious source of uncertainties, concerns
the comparison of theory with
observations. Theory quite precisely predicts {\it primordial} abundances of
light elements, while observations deals with the {\it present day}
abundances. The situation is rather safe for $^4 He$ because this
element is very strongly
bound and is not destroyed in the course of evolution. It can only be created
in stars. Thus any observation of the present-day mass fraction of $^4 He$
gives an upper limit to its primordial value. To infer its primordial value
$Y_p$, the abundance of $^4 He$ is measured together with other heavier
elements, like oxygen, carbon, nitrogen, etc 
(all they are called "metals") and
the data is extrapolated to zero metallicity (see the book~\cite{bp} for
details). The primordial abundance of deuterium is very sensitive to the
baryon density and could be in principle a very accurate indicator of
baryons~\cite{rafs}. However deuterium is fragile and can be easily destroyed.
Thus it is very difficult to infer its primordial abundance based on 
observations at relatively close quarters in the media where a 
large part of matter had been processed by the stars. Recently, 
however, it became possible to observe deuterium
in metal-poor gas clouds at high red-shifts. In these clouds
practically no matter was contaminated by stellar processes so these
measurements are believed to yield the
primordial value of $D /H$. Surprisingly, the results of these measurements
are grouped around two very different values, normal deuterium,
$(D/H)_p \approx 3\cdot 10^{-5} $~\cite{tfb}-\cite{bt2}, which is reasonably
close to what is observed in the Galaxy, and high deuterium,
$(D/H)_p \approx (1-2)\cdot 10^{-4} $~\cite{crw}-\cite{wcl}. The observed
variation may not be real; for example, uncertainties in the velocity
field allow the D/H ratio in the system at $z=0.7$~\cite{wcl} to be as low
as in the two high-z systems \cite{lev}-\cite{ltb}. An interpretation of
the observations in the system at $z=0.7$ under the assumption of a simple
single $(H+D)$-component~\cite{tbl} gives
$8\cdot 10^{-5}< D/H <57 \cdot 10^{-5}$. With the possibility of
a complicated velocity distribution or of a second component in this system
a rather weak limit was obtained~\cite{tbl}, $D/H <50 \cdot 10^{-5}$.
However, it was argued in the recent work~\cite{kirkman01}
that the observed absorption features most probably are not induced by
deuterium and thus the conclusion of anomalously high deuterium in
this system might be incorrect.
On the other hand, there are systems where anomalously low fraction of
deuterium is observed~\cite{lowD}, $D/H \sim (1-2)\cdot 10^{-5}$.
An analysis of the data on D and $^4He$ and recent references can be
found in~\cite{dolgov01tp}.
It seems premature to extract very accurate statements about baryon
density from these observations. The accuracy of the determination of
light element abundances is often characterized in terms of permitted
additional neutrino species, $\Delta N_\nu$.
The safe upper limit, roughly speaking, is that
one extra neutrino is permitted in addition to the known three
(see sec. \ref{knu}). On the other hand, if all observed anomalous
deuterium (high or low) is not real and could be explained by some
systematic errors or misinterpretation of the data
and only ``normal'' data are correct, then BBN would provide quite
restrictive upper bound on the number of additional neutrino species,
$\Delta N_\nu < 0.2$~\cite{tytler00}. For more detail and recent
references see sec.~\ref{knu}.

\section{Massless or light neutrinos \label{massless}}

\subsection{Gerstein-Zeldovich limit \label{gerzel}}

Here we will consider neutrinos that are either massless or so light 
that they had decoupled from the primordial $e^\pm\,\gamma$-plasma
at $T>m_\nu$. A crude estimate of the decoupling
temperature can be obtained as follows. The rate of neutrino interactions with
the plasma is given by:
\be
\Gamma_\nu \equiv \dot n_\nu / n_\nu = \langle \sigma_{\nu e} n_e
\rangle
\label{gammanu}
\ee
where $\sigma_{\nu e}$ is the cross section of neutrino-electron scattering
or annihilation and $\langle...\rangle$ means thermal averaging.
Decoupling occurs when the interaction rate falls below
the expansion rate, $\Gamma_\nu < H$. One should substitute for the
the cross-section
$\sigma_{\nu e}$ the sum of the cross-sections of neutrino elastic scattering
on electrons and positrons and of the inverse annihilation $e^+e^-
\rar \bar \nu \nu$ in the relativistic limit. Using expressions presented in
Table~\ref{table:amplitudes-nu-e0}  we find:
\be
\sigma_{\nu,e} = {5G^2_F\, s\over 3\pi} \left( g^2_L + g^2_R\right)
\label{sigmanue}
\ee
where $s=(p_1 +p_2)^2$, $p_{1,2}$ are the 4-momenta of the initial
particles, and $g_{L,R}$ are the coupling to the left-handed and right-handed
currents respectively, $g_L = \pm 1/2 + \sin^2 \theta_W$ and $g_R =
\sin^2 \theta_W$, plus or minus in $g_L$ stand respectively for $\nue$ or
$\nu_{\mu,\tau}$. The weak mixing angle $\theta_W$ is experimentally
determined as $\sin^2 \theta_W =0.23$.

\begin{table}
\begin{center}
\begin{tabular}{cccc|cc}
& Process &&&& $2^{-5} G^{-2}_{F} S \left| A \right| ^{2}$
\\ \hline\hline
$\nu _{e} + \bar{\nu}_{e} $ & $\rightarrow$ & $  \nu _{e} + \bar{\nu}_{e} $&&&
$4 (p_{1} \cdot p_{4}) (p_{2} \cdot p_{3})$ \\

$\nu _{e} + \nu_{e} $ & $\rightarrow$ & $  \nu _{e} + \nu_{e}$ &&&
$2  (p_{1} \cdot p_{2}) (p_{3} \cdot p_{4})$\\

$\nu _{e}+\bar{\nu}_{e} $ & $\rightarrow$ &
$ \nu _{\mu (\tau)}+\bar{\nu}_{\mu (\tau)} $&&&
$ (p_{1} \cdot p_{4}) (p_{2} \cdot p_{3}) $\\

$\nu_{e} + \bar{\nu}_{\mu (\tau)} $ & $\rightarrow$&
$\nu _{e}+\bar{\nu}_{\mu (\tau)} $&&&
$ (p_{1} \cdot p_{4}) (p_{2} \cdot p_{3})$\\

$\nu _{e} + \nu_{\mu (\tau)} $ & $\rightarrow$ &
$ \nu _{e}+\nu_{\mu (\tau)} $&&&
$ (p_{1} \cdot p_{2}) (p_{3} \cdot p_{4})$\\

$\nu _{e} + \bar{\nu}_{e} $ & $\rightarrow$ & $  e^{+} + e^{-}$ &&&
$4[ ( g_{L}^{2} (p_{1} \cdot p_{4}) (p_{2} \cdot p_{3}) $\\
&&&&&$+ g_{R}^{2}  (p_{1} \cdot p_{3}) (p_{2} \cdot p_{4})$ \\
&&&&&$+ g_{L} g_{R} m _{e}^{2} (p_{1} \cdot p_{2})]$\\

$\nu _{e} +  e^{-} $ & $\rightarrow$ & $ \nu _{e} +  e^{-} $&&&
$4 [ g_{L}^{2} (p_{1} \cdot p_{2}) (p_{3} \cdot p_{4})$ \\
&&&&&$+g_{R}^{2}  (p_{1} \cdot p_{4}) (p_{2} \cdot p_{3})$ \\
&&&&&$-g_{L} g_{R} m _{e}^{2} (p_{1} \cdot p_{3}) ] $\\

$\nu _{e} +  e^{+} $ & $\rightarrow$ & $ \nu _{e} +  e^{+} $&&&

$4 [ g_{R} ^{2} (p_{1} \cdot p_{2}) (p_{3} \cdot p_{4})$ \\
&&&&&$+g_{L}^{2} (p_{1} \cdot p_{4}) (p_{2} \cdot p_{3}) $ \\
&&&&&$-g_{L} g_{R} m _{e}^{2} (p_{1} \cdot p_{3}) ] $
\end{tabular}
\caption{Matrix elements squared for reactions with electron neutrino;
$S$ is the symmetrization factor related to identical particles in
the initial or final state,
$g_{L} = \frac{1}{2} + \sin^{2}\theta_{W}$ and
$g_{R} = \sin^{2} \theta_{W}$. Matrix elements for muon or tau
neutrino processes are obtained by the substitutions
$\nu_e \rightarrow \nu_{\mu,\tau}$ and
$g_{L} \rightarrow \tilde{g}_{L} =g_{L} - 1$.}
\label{table:amplitudes-nu-e0}
\end{center}
\end{table}

We would not proceed along these lines because one can do better by 
using kinetic equation~(\ref{hxdfdx}). We will keep only direct reaction
term in the collision integral and use the matrix elements taken from
the Table~\ref{table:amplitudes-nu-e0}. We estimate the collision
integral in the
Boltzmann approximation. According to calculations of ref.~\cite{dolkai}
this approximation is good with an accuracy of about 10\%.
We also assume that particles with which neutrinos interact, are in
thermal equilibrium with temperature $T$. After straightforward
calculations we obtain:
\be
Hx\,{\partial f_\nu \over f_\nu\,\partial x} =
-{80G_F^2\left( g_L^2 + g_R^2 \right) y \over 3\pi^3 x^5 }
\label{fnue0}
\ee
Using the expression~({\ref{hofx}) and integrating over $x$ we find
for the decoupling temperature of electronic neutrinos
$T_{\nue} = 2.7 y^{-1/3}$ MeV and $T_{\num,\nut} = 4.5 y^{-1/3}$ MeV.
This can be compared with the results of refs.~\cite{dz,eks}.
On the average one can take $\langle y \rangle = 3$ and
$T_{\nue} = 1.87 $ MeV and $T_{\num,\nut} = 3.12 $ MeV. These
results are applicable to the decoupling of neutrinos from the
electromagnetic component of the plasma, i.e. from $e^{\pm}$ and photons.
If we take into account all possible reactions of neutrinos in the
the plasma, including their scattering on themselves, the coefficient
$(g_L^2 + g_R^2)$ should be changed into $(1+g_L^2 + g_R^2)$. These results
are in agreement with refs.~\cite{eks,enqvist92b} (see discussion in
sec.~\ref{coher}). Correspondingly
the decoupling temperature determined with respect to the total reaction
rate would be $T_{\nue} =1.34 $ MeV and $T_{\num,\nut} = 1.5 $ MeV.
Somewhat more accurate calculations of the reaction rates with
Fermi exclusion taken into account were performed in
ref.~\cite{dolgov00}, see eq.~(\ref{gammaj1}) and discussion after it.
The finite temperature corrections to the reaction rates have been
studied in ref.~\cite{fornengo97}. As a result of these corrections
the interaction rate becomes weaker and the decoupling temperature
rises by 4.4\%.

The decoupling temperature depends upon neutrino momentum, so that 
more energetic neutrinos decouple later. In fact the decoupling
temperature is somewhat higher because inverse reactions neglected in
this estimate diminish the reaction rate approximately by half
if the distribution is close to the equilibrium one.
Anyway, it is safe to say that below 2 MeV neutrinos practically became 
non-interacting and their number density remains constant in a comoving
volume, $n_\nu \sim 1/a^3$. At the moment of decoupling the relative
number density of neutrinos was determined by thermal equilibrium and in
the absence of charge asymmetry was given by:
\be
{n_{\nu_j} \over n_\gamma } = {n_{\bar \nu_j} \over n_\gamma } =
{3\over 8}
\label{nnungamma0}
\ee
Later $e^+e^-$-annihilation enlarges the number density of
photons in the comoving volume. This increase can be easily calculated
using the entropy conservation law (\ref{dtent}). The photon number density
increases proportionally to the ratio of the number of species before and
after annihilation. In the case under consideration, it is $(2+7/2)/ 2 = 11/4$.
If no more photons were created during the subsequent expansion, then the
present day neutrino-to-photon ratio should be 
\be
{n_{\nu_j}+n_{\bar \nu_j}\over n_\gamma } = {3\over 11}
\label{nnungamma}
\ee

The number density of photons in CMB radiation is now known with a 
great degree of precision,
see (\ref{ngamma}). From that we find
$n_{\nu_j}+n_{\bar \nu_j }=112/{\rm cm^3}$ for any kind of light
($m< O({\rm MeV})$) neutrino. If neutrinos are massless, they
preserve their initial Fermi
distribution with the present day temperature $T_\nu = 1.95$ K 
(although there are some deviations, which will be  discussed in 
the next subsection). If they
are massive they are much colder. Energy density of massive
neutrinos is
 $\rho_{\nu} =112 \sum_j m_{\nu_j} /{\rm cm}^3 =
\rho_c h^{-2} \sum_j \left(m_{\nu_j}/94 {\rm eV}\right)$
Assuming that $\Omega_\nu =\rho_{\nu}/\rho_c \leq 1$ we obtain the
following upper limit on neutrino masses:
\be
\sum m_{\nu_j} < 94 {\rm eV}\,\, \Omega h^2
\label{mnu1}
\ee
In particular for $h= 0.7$ and $\Omega_{matter} < 0.3$ the mass of
neutrino should be smaller than 14 eV.
This bound was first found by Gerstein and Zeldovich \cite{gz}
and with different modifications was considered in many
subsequent papers. A good account of historical developments that 
led to the discovery of this bound can be found in  ref.~\cite{sarkar1}. 
That account  has been  marred, however, by a serious misquotation of 
the Gerstein and Zeldovich paper. Namely it was claimed~\cite{sarkar1}
that the GZ calculations of the relic neutrino abundance was erroneous
because they assumed that massive neutrinos are
Dirac particles with fully populated right-handed states and that they (GZ)
"did not allow for the decrease in the neutrino temperature relative to
photons due to $e^+e^-$-annihilation". Both accusations are incorrect. It
is explicitly written in GZ paper: "In considering the question of the
possible mass of the neutrino we have, naturally, used statistical formulas for
four-component $m\neq 0$ particles. We know, however, that in accordance with
$(V-A)$-theory, neutrinos having a definite polarization participate
predominantly in weak interactions. Equilibrium for neutrinos for opposite
polarization  is established only at a higher temperature. This, incidentally,
can change the limit on the mass by not more than a factor of 2." It was
also correctly stated there that in equilibrium
$n_\nu /n_\gamma = (3/4)(g_\nu/g_\gamma)$, where $g_a$ is the number of spin
states: "However during the course of cooling... these relations change,
since the annihilation of $e^+e^-$ increases the number of quanta without the
changing the number of neutrinos". Gerstein and Zeldovich used the
result Peebles~\cite{peebheat} to obtain the perfectly correct number 
accepted today: $n_\nu /n_\gamma = 3g_\nu /11$.

The numerical magnitude of the bound
obtained in the original (and perfectly correct!) paper by GZ was relatively
weak, $m_\nu < 400$ eV because they used a very small value for the universe
age, $t_U > 5$ Gyr and a very loose upper limit for the cosmological energy
density, $\rho < 2\cdot 10^{-28} {\rm g/cm}^3$. A somewhat better bound
$m_\nu<  130$ eV  was obtained in subsequent
papers~\cite{masz,szma}. A much
stronger bound $m_\nu <8 $ eV was obtained in paper~\cite{cm} but
this paper is at fault for unnecessarily counting right-handed neutrino 
spin states and of not accounting for extra heating of photons 
by $e^+e^-$-annihilation. With these two effects the limit should be 
bigger by factor 22/3.

Alternatively one can express the cosmological upper bound on neutrino
mass through the limit on the universe age~\cite{dolgov84}:
\be
\sum m_{\nu_j} < 380\,{\rm eV} \left( 2.7\,{\rm K} \over T_\gamma\right)^3
\left( {0.98\cdot 10^{10}\,{\rm years} \over t_U } - h_{100}\right)^2
\label{mnu2}
\ee
The result is valid for cosmology with vanishing lambda-term and is quite
restrictive for the old universe. In the case of non-zero $\Omega_\Lambda$
the universe age limit is not especially useful for neutrino mass. Assuming
a flat universe, $\Omega_m +\Omega_\Lambda =1$, we find (see
eq.~(\ref{tulam})):
\be
t_U = {2\over 3 H\sqrt{\Omega_\Lambda} }
\ln { 1+\sqrt{\Omega_\Lambda} \over \sqrt{\Omega_m}}
\label{tulambda}
\ee
If $h = 0.7$ and $\Omega_m = 0.3$ the universe age is quite large,
$t_U \approx 13.5$ Gyr. However if the universe is considerably older
than that, see e.g. ref.~\cite{rengel01} where the age above 16 Gyr is
advocated, then we need $\Omega_\Lambda > 0.8$, and correspondingly
$\Omega_m < 0.2$. In this case $\sum m_{\nu_j} < 9$ eV.
A similar constraint on neutrino mass by the universe age was
derived in ref.~\cite{bludman92} both for the cases of vanishing and
nonvanishing cosmological constant.

The basic assumptions leading to GZ-bound (\ref{mnu1}) or
(\ref{mnu2}) are quite simple and solid:
\begin{enumerate}
\item{}
Thermal equilibrium in the early universe between neutrinos, electrons,
and photons. It can be verified that this is precisely true down to
temperatures 2-3 MeV.
\item{}
Negligible lepton asymmetry, or in other words vanishing (or near-vanishing)
leptonic chemical potentials. The validity of this assumption has not 
been completely verified observationally. The only reason for that is 
the small value of baryonic 
chemical potential and the belief that lepton asymmetry is generated 
essentially by the same mechanism as the baryonic one. The strongest upper
bound for leptonic chemical potentials comes from primordial nucleosynthesis,
which permits $\xi_{\num,\nut} \equiv \mu_{\num,\nut}/T = O(1)$
and $\xi_{\nue} \equiv |\mu_{\nue}/T| < 0.1$ (see 
secs.~\ref{degnubbn},\ref{ssec:active}).
In derivation of eqs. (\ref{nnungamma0})-(\ref{mnu2}) it was assumed
that the chemical potentials of all neutrinos were zero. Otherwise, 
the upper bound on the mass would be stronger 
by the factor $(1+\Delta k_\nu)$, where $\Delta k_\nu$ is given by
eq.~(\ref{deltanuxi}).
\item{}
No other sources of extra heating for the cosmic photons at $T\leq $ MeV,
except for the above mentioned $e^+e^-$-annihilation.
If the photons of CMBR had been heated at some point between the
neutrino decoupling and
the present day, then the bound on neutrino mass would be correspondingly
weaker. Possible sources of this heating could be decays or annihilation
of new particles, but that could only have taken place sufficiently early,
so that the Planck spectrum of CMBR was not destroyed.
\item{}
Stability of neutrinos on cosmological time scale,
$\tau_\nu \geq 10^{10} {\rm years} $. For example, in the case of
neutrino-majoron coupling the bound on the neutrino mass can be much
less restrictive or completely avoided if the symmetry
breaking scale is below $10^6$ GeV~\cite{chikashige80} and life-time
of even very light neutrinos is very short. A similar weakening of the
bound is found in the familon model~\cite{fam}.
\item{}
No new interactions of neutrinos which could diminish their number 
density, for instance by annihilation, into new lighter particles, 
such as Majorons; and no annihilation of heavier neutrinos into 
lighter ones due to a stronger
interaction than the normal weak one. On the other hand, a new stronger
coupling of neutrinos to electrons or photons could keep neutrinos longer
in equilibrium with photons, so that their number density would not be
diluted by 4/11.
\item{}
The absence of right-handed neutrinos. If neutrinos possess a Majorana mass,
then right-handed neutrinos do not necessarily exist, but if they have a
Dirac mass, both left-handed and right-handed particles must be present.
In this case, one could naively expect that the GZ-bound should be twice
as strong. However, even though right-handed states could exist in 
principle, their number density in the cosmic plasma at $T$ around 
and below MeV would be suppressed. The probability of production of 
right-handed neutrinos by the  normal
weak interaction is $(m_\nu/E)^2$ times smaller than the probability of
production of left-handed ones. It is easy to estimate the number
density of the produced right-handed neutrinos through this
mechanism~\cite{shws,ad1} and to see that they are always far below
equilibrium. Even if there are right-handed currents, one
can see that the interaction with right-handed $W_R$ and/or
$Z_R$ should drop from equilibrium at $T$ above the QCD phase transition
(see sec. \ref{ssec:nur}).
So even if $\nu_R$ were abundant at $T > 100$ MeV their number density
would be diluted by the factor $\sim 1/5$ with respect to $\nu_L$.
\end{enumerate}

A very strong modification of the standard cosmological thermal history
was proposed in ref.~\cite{giudice00a}. It was assumed that the universe
never heated above a few MeV. In such scenario neutrinos would never
be produced in equilibrium amount and therefore, their relative number 
density, compared to photons in CMBR, would be much smaller then the standard
number 3/11. From the condition of preserving big bang nucleosynthesis
the lower limit, $T_{min}$,
on the universe temperature was derived. If the universe
was never heated noticeably above $T_{min}$ neutrinos would
never be abundant in the primeval plasma and  
the upper limit on neutrino mass would become much weaker than~(\ref{mnu1}):
$m_\nu < 210$ keV (or 120 keV for Majorana neutrinos). Such scarce 
neutrinos could form cosmological warm dark matter~\cite{giudice00b}
(see sec.~\ref{11dm}).

\subsection{Spectral distortion of massless neutrinos. \label{masslessdistr}}

It is commonly assumed that thermal relics with $m=0$ are in perfect
equilibrium state even after decoupling. For photons in cosmic microwave
background (CMB) this has been established with a very high degree of 
accuracy. The same assumption has been made about neutrinos, so that 
their distribution is given as
eq.~(\ref{ffb}). Indeed, when the interaction rate is high in comparison
with the expansion rate, $\Gamma_{int} \gg H$, equilibrium is evidently
established. When interactions can be neglected the distribution function
may have an arbitrary form, but for massless particles, equilibrium
distribution is preserved, as long as it had been established earlier 
at a dense and hot stage when the interaction was fast. One can see from 
kinetic equation in the expanding universe (\ref{kin1}) that this is 
indeed true.
The collision integral in the r.h.s.
vanishes for equilibrium functions (\ref{ffb}), where temperature $T$
and chemical potential $\mu$ may be functions of time.
The l.h.s. is annihilated by $f = f^{(eq)}$ if the following condition is
fulfilled for arbitrary values of particle energy $E$ and
momentum $p=\sqrt{E^2 -m^2}$:
\be
{\dot T \over T} + H {p\over E} {\partial E \over \partial p}
- {\mu \over E}\left( {\dot \mu \over \mu} -{\dot T \over T}\right) =0
\label{dott}
\ee
This can only be true if $p=E$ (i.e. $m=0$), $\dot T/T = -H$, and
$\mu \sim T$. One can demonstrate that for massless particles, which initially
possessed equilibrium distribution, temperature and
chemical potential indeed satisfy these requirements and that 
the equilibrium distribution is not destroyed even when the interaction
is switched off.

The same would be true for neutrinos if they decoupled from the electronic 
component of the plasma (electrons, positrons and photons) instantly and 
at the moment when neutrino interactions were strong enough to maintain 
thermal equilibrium with photons and $e^{\pm}$.
According to simple estimates
made in sec. 4.1,
the decoupling temperature, $T_{dec}$, for $\nue$ is about 2 MeV and that
for $\num$ and $\nut$ is about 3 MeV.
In reality, the decoupling is not instantaneous, and even 
below $T_{dec}$ there are some residual interactions between 
$e^{\pm}$ and neutrinos. An important
point is that after neutrino decoupling the temperature of the electromagnetic
component of the plasma became somewhat higher than the neutrino temperature.
The electromagnetic part of the plasma is heated by the annihilation of
{\it massive} electrons and positrons. This is a well-known effect
which ultimately results in the present day ratio of temperatures,
$T_\gamma /T_\nu = (11/4)^{1/3}=1.4$.
During primordial nucleosynthesis the temperature difference between
electromagnetic and neutrino components of the plasma was small but still
non-vanishing. Due to this temperature difference the annihilation of the
hotter electrons/positrons, $e^+ e^- \rightarrow \bar \nu \nu$, heats up
the neutrino component of the plasma and distorts the
neutrino spectrum. The
average neutrino heating under the assumption that their spectrum 
maintains equilibrium was estimated in refs.~\cite{dic}-\cite{rm}. 
However, the approximation
of the equilibrium spectrum is significantly violated and this assumption
was abolished in refs.~\cite{df}-\cite{gg}. In the earlier 
papers \cite{df,dt} the effect was considered in the Boltzmann
Approximation, which very much simplifies calculations. Another simplifying
assumption, used previously, is the neglect of the electron mass in
collision integrals for $\nu e $-scattering and for annihilation
$\bar \nu \nu  \rightarrow e^+e^-$. In ref. \cite{dt} the effect
was calculated numerically, while in ref. \cite{df} an approximate analytical
expression was derived. However in ref. \cite{df} the influence of
the back-reaction that smooths the spectral distortion was 
underestimated due to a numerical error in the integral. When this error 
is corrected, the effect should shrink by half (under the approximations 
of that paper) and the corrected result would be:
\be
{\delta f_{\nue} \over f_{\nue} }\approx 3\cdot 10^{-4} \,\,{E\over T}
\left( {11 E \over 4T } - 3\right)
\label{dff}
\ee
Here $\delta f = f - f^{(eq)}$.
The distortion of the spectra of $\num$ and $\nut$ is approximately twice
weaker. 
Subsequent accurate numerical calculations~\cite{hanmad,dhs0} are in 
reasonable agreement with this expression and with the calculations of
paper~\cite{dt}.

An exact numerical treatment of the problem was conducted in
papers~\cite{hanmad}-\cite{gg}. There is some disagreement 
among them, so we
will discuss the calculations in some detail. The coupled system of
integro-differential kinetic equations (\ref{hxdfdx}) was solved numerically
for three unknown distribution functions, $f_{\nu_j}(x,y)$, where
$j=e,\mu,\tau$. The dimensional variables "time" $x$ and momentum $y$ are
defined in eqs. (\ref{xyi}). The collision integral $I^{coll}$ is
dominated by two-body reactions between different leptons
$1+2 \rightarrow 3+4$, and is given by the expression:
\be
I^{coll} = {1\over 2E_1}\sum \int {d^3 p_2 \over 2E_2 (2\pi)^3}
{d^3 p_3 \over 2E_3 (2\pi)^3}{d^3 p_4 \over 2E_4 (2\pi)^3}
\nonumber \\
(2\pi)^4\delta^{(4)} (p_1+p_2-p_3-p_4) F(f_1,f_2,f_3,f_4)
S\, |A|^2_{12\rightarrow 34}
\label{icoll}
\ee
where $F = f_3 f_4 (1-f_1)(1-f_2)-f_1 f_2 (1-f_3)(1-f_4)$,
$|A|^2$ is weak interaction amplitude squared summed over spins of
all particles except the first one, and $S$
is the symmetrization factor which includes $1/2!$ for each pair of identical
particles in initial and final states and the factor 2 if there are 2 identical
particles in the initial state. The summation is done over all possible
sets of leptons 2, 3, and 4.
The amplitude squared of the relevant processes are presented in
Table~\ref{table:amplitudes-nu-e0}. The expressions in the
Tables are taken from ref.~\cite{dhs0}, while those used in
ref.~\cite{hanmad} and repeated in ref.~\cite{gg} do not take into
account identity of initial particles in the reactions
$\nu_a\nu_a\rightarrow \nu_a \nu_a$  (or with
anti-neutrinos) and hence are erroneously twice smaller than presented here.

It would be natural to assume that 
distribution functions for $\nu_\mu$ and $\nu_\tau$ are equal, 
while the one for
$\nu_e$ is different because the former have only neutral
current interactions at relevant temperatures, while $\nue$ has both
neutral and charged current interactions. One can also assume that the lepton
asymmetry is negligible, so that $f_{\nu} = f_{\bar{\nu}}$.
Therefore there are two unknown functions of two variables, $x$ and $y$:
$f_{\nu_e}$ and $f_{\nu_\mu} = f_{\nu_\tau}$. Since the distributions of
photons and $e^\pm$ are very precisely equilibrium ones, they can be described
by a single unknown function of one variable, namely the temperature,
$T_\gamma(x)$. The chemical potentials are assumed to be vanishingly small.
The third necessary equation is the covariant energy conservation:
\be{
x \; {d\rho (x) \over dx } = -3(\rho + P)
\label{drhodx}
}\ee
where $\rho$ is the total energy density:
\be{
 \rho = {\pi^2 T^4_\gamma\over 15}  + {2\over \pi^2} \int {dq q^2
\sqrt{q^2 + m^2_e} \over \exp {(E/T_\gamma)} +1 } +
{1\over \pi^2} \int dq q^3 \left( f_{\nu_e} + 2 f_{\nu_\mu} \right)
\label{rho}
}\ee
and $P$ is the pressure:
\be{
P = {\pi^2 T^4_\gamma\over 45}  + {2\over 3\pi^2} \int {dq q^4 \over
  [\exp (E/T_\gamma) +1 ]\sqrt{q^2 + m^2_e} } +
{1\over 3\pi^2} \int dq q^3 \left( f_{\nu_e} +2 f_{\nu_\mu} \right)
\label{p}
}\ee

The Hubble parameter, $H=\dot a /a$, which enters the kinetic
equation (\ref{hxdfdx}) is expressed through $\rho$ in the usual way,
$3H^2 m^2_{Pl} =8\pi \rho$, ignoring the curvature term and the
cosmological constant, which are negligible in the essential
temperature range.

The collision integral in eq. (\ref{icoll}}) can be reduced from nine to
two dimensions as described in ref.~\cite{dhs0}. After that, the system of
equations (\ref{hxdfdx},\ref{icoll}-\ref{p}) for three unknown functions
$f_{\nue}, f_{\num,\nut}\,\, {\rm and}\,\,  T_\gamma$ was solved numerically
using the integration method developed in ref.~\cite{st}.

There are three phenomena that play an essential role in the
evolution of neutrino distribution functions. The first
is the temperature difference between photons and $e^\pm$ on one hand 
and neutrinos on the other, which arises due to the
heating of the electromagnetic plasma by $e^+ e^- $-annihilation.
Through interactions between neutrinos
and electrons, this temperature difference
leads to non-equilibrium distortions of the neutrino spectra.
The temperature difference is essential in the interval
$1<x<30$. The second effect is the freezing of the neutrino interactions
because the collision integrals drop as
$1/x^2$. At small $x \ll 1$ collisions are fast but at
$x>1$ they are strongly suppressed.
The third important phenomenon is the elastic $\nu \nu$-scattering
which smooths  down the non-equilibrium corrections to the neutrino
spectrum. It is especially important at small $x<1$.

The numerical calculations of ref.~\cite{dhs0}, which are possibly the most
accurate, have been done in two different but equivalent ways. First, the
system was solved directly, as it is, for the full distribution
functions $f_{\nu_j}(x,y)$
and, second, for the small deviations $\delta _{j}$ from equilibrium
$f_{\nu_j} (x,y) = f_{\nu_j}^{(eq)} (y) \left[1 +\delta_j (x,y))\right]$,
where $f_{\nu_j}^{(eq)} = [\exp( E/T_\nu  ) +1 ]^{-1}$ with $T_\nu = 1/a$.
In both cases the
numerical solution was exact, not perturbative.
So with infinitely good numerical precision the results must be the same.
However since precision is finite, different methods may produce
different results, and their consistency is a good indicator of 
the accuracy of the calculations.
It is convenient to introduce $\delta (x,y)$ because the dominant terms
in the collision integrals, which contain only neutrinos, cancel out; and
sub-dominant terms are proportional to $\delta$. In the parts of the
collision integrals that contain electron distribution functions, there 
is a driving term
proportional to the difference in temperatures $(T_\gamma - T_\nu)$. However
in calculations with complete distribution functions the numerical
value for the Planck mass was taken as $m_{Pl} =10^{19}$ GeV, i.e. without
the factor 1.22. It explains some discrepancies between the results of the
calculations with $f_{\nu_a}$ and with $ \delta f_{\nu_a}$ in ref.~\cite{dhs0}.
This error was corrected in the addendum~\cite{dhsad} and the results of
two different ways of calculations are in perfect agreement, as one 
can see from
Table \ref{tab:twoways}. The first entry in
this Table shows the number of integration points and thus provides a 
measure of the stability of the calculations. The second one, 
$a T_\gamma$, demonstrates how much the photon gas has cooled 
down by sharing part of its energy with
neutrinos. In standard calculations this number is
$T_\gamma /T_\nu = (11/4)^{1/3} = 1.401$ (see discussion below
eq.~(\ref{nnungamma0})).  The relative energy gain of neutrinos,
$\delta \rho_{\nu_a} /\rho_{\nu_a}$ for $\nue$ and $\nu_{\mu,\tau}$ are
presented respectively in the third and fourth columns. They can be
compared with the results of ref.~\cite{hanmad}:
$\delta \rho_{\nu_e} /\rho_{\nu_e} = 0.83\%$ and
$\delta \rho_{\nu_{\mu,\tau}} /\rho_{\nu_{\mu,\tau}} = 0.41\%$. The difference
between the two results may be prescribed to the different accuracies of
ref.~\cite{hanmad} where 35 integration points were taken and of
ref.~\cite{dhs0,dhsad} where 100-400 points were taken.
The last
column presents the effective number of neutrinos at asymptotically
large time. The latter is defined as:
\be
N_{\mbox{eff}} =  \frac{\rho_{\nu_e}+2 \rho_{\nu_\mu}}{\rho_{\nu}^{eq}}
\frac{\rho_\gamma^{eq}}{\rho_\gamma} ~,
\label{N_eff}
\ee
where the photon energy density is
$\rho_\gamma=(\pi^2/15) (a T_{\gamma})^4$ and the
equilibrium quantities are
$\rho_{\nu}^{eq}=(7/8)(\pi^2/15)$ and $\rho_\gamma^{eq}=
(\pi^2/15)(a T_{\gamma}^{eq})^4$.

\begin{table}
\begin{center}
\begin{tabular}{|c||c|c|c|c|c|}
\hline

&&&&&\\
Program&points&
~~$a T_\gamma$~~&~~$\delta\rho_{\nu_e}/\rho_{\nu_e}
 $~~&~~$\delta\rho_{\nu_{\mu}}/\rho_{\nu_{\mu}} $~~&$N_{\mbox{eff}}$\\
&&&&&\\
\hline
\hline
             &100&1.399130& 0.9435\%&0.3948\%&3.03392\\
$\delta(x,y)$&200&1.399135& 0.9458\%&0.3971\%&3.03395 \\
             &400&1.399135& 0.9459\%&0.3972\%&3.03396 \\
\hline
             &100&1.399079& 0.9452\%&0.3978\%&3.03398 \\
$f(x,y)$     &200&1.399077& 0.9459\%&0.3986\%&3.03401 \\
             &400&1.399077& 0.9461\%&0.3990\%&3.03402 \\
\hline
\hline
\end{tabular}
\end{center}
\caption{Two ways of calculation.}
\label{tab:twoways}
\end{table}

There is some disagreement between the calculations of the
papers~\cite{dhs0} and \cite{gg}, though both groups claim high accuracy
of their procedure. The authors of ref.~\cite{gg} have 289 integration
points logarithmically distributed in the
momentum interval $10^{-5.5} \leq q/T \leq 10^{1.7}$ or, in other words, 40
points per decade. 
It seems that there are too many points in the region
of low momenta, where interaction is weak and not essential. Meanwhile, 
the number of points in the important interval of high momenta is
considerably smaller than in refs.~\cite{dhs0,dhsad}, where integration 
points are distributed linearly in momentum interval $0\leq y \leq 20$.
In particular, with the choice of ref.~\cite{gg}, more than half the
points lie in the region
$y<0.1$, which gives only $0.0002 \%$ contribution to the
neutrino energy density~\cite{dhsad}.
In the most important decade, $1<y<10$, there are only 40 points in the
method of ref.~\cite{gg}. This is definitely too little to achieve the
desired accuracy.

Recently calculations of the distortion of neutrino spectrum
were done in ref.~\cite{esposito00} through a radically different method:  
using expansion in interpolating polynomials in momentum. The results
of this work perfectly agree with those of refs.~\cite{dhs0,dhsad}.

In Fig. \ref{deltaofx} the deviations
from the equilibrium distributions, $\delta_{\nu_e}$ and
$\delta_{\nu_{\mu(\tau)}}$ for FD and MB statistics are shown;
$\delta_i$ are plotted for the fixed value
of the  momentum $y=5$ as functions of $x$. The results
for the case of Boltzmann statistics are larger than those for the
Fermi statistics by approximately $25\%$.  For both FD and
MB statistics, the spectral distortion for $\nu_e$ is more than twice 
the size of that for $\nu_\mu$ or $\nu_\tau$. This is due to a stronger 
coupling of
$\nu_e$ to $e^\pm$.

\begin{figure}[htb]
\begin{center}
  \leavevmode
  \hbox{
    \epsfysize=3.0in
    \epsffile{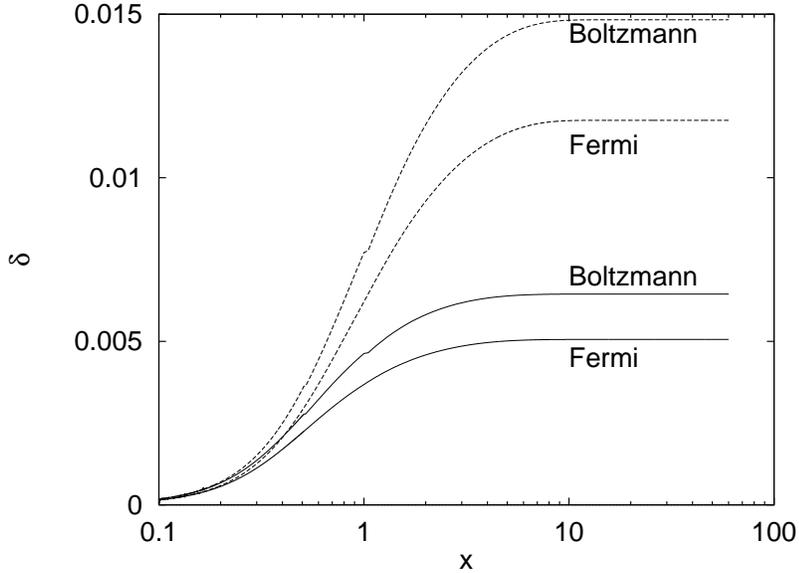}}
\end{center}
\caption{Evolution of non-equilibrium corrections to the
distribution functions $\delta_j = (f_{\nu_j} - f_\nu^{eq})/f_\nu^{eq} $
for running inverse temperature $x$ and fixed dimensionless momentum
$y=5$ for electronic (dotted curves) and muonic (tau) (solid curves)
neutrinos in the cases of  FD and MB statistics.
\label{deltaofx}}
\end{figure}

In Fig. \ref{deltaofy} the asymptotic, when $x\rar \infty$, values of the
corrections to the neutrino distributions
$\delta_j = (f_{\nu_j} - f_\nu^{eq})/f_\nu^{eq} $ are plotted
as functions of the dimensionless momentum $y$.
The dashed lines $a$ and $c$ correspond to Maxwell-Boltzmann statistics
and the solid lines $b$ and $d$ correspond to Fermi-Dirac statistics.
The upper curves $a$ and $b$ are for electronic neutrinos and the lower
curves $c$ and $d$ are for muonic (tau) neutrinos. All the curves can be well
approximated by a second order polynomial in $y$,
$\delta = A y (y -B)$, in agreement with eq. (\ref{dff}) \cite{df}.
\begin{figure}[htb]
\begin{center}
  \leavevmode
  \hbox{
    \epsfysize=3.0in
    \epsffile{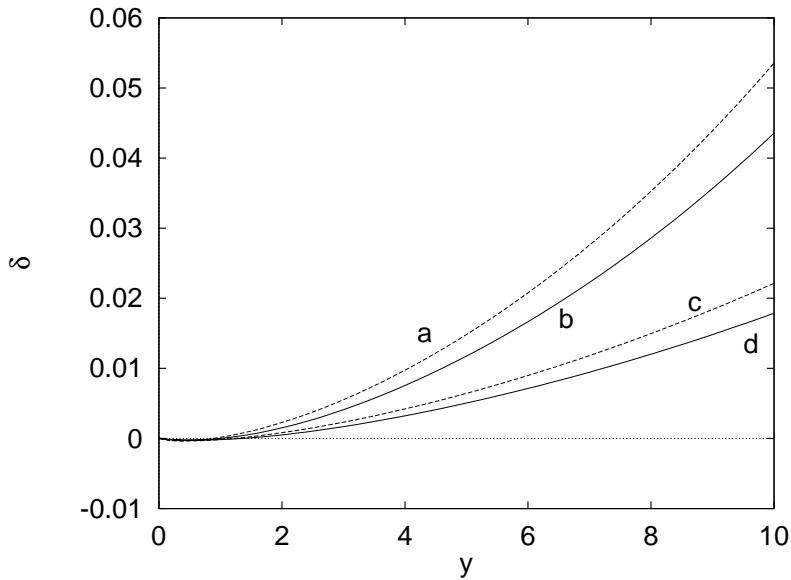}}
\end{center}
\caption{The distortion of the neutrino spectra
$\delta_j = (f_{\nu_j}-f_\nu^{eq})/f_\nu^{eq} $
as functions of the dimensionless momentum $y$ at the final "time" $x=60$.
The dashed lines $a$ and $c$ correspond to Maxwell-Boltzmann statistics,
while the solid lines $b$ and $d$ correspond to Fermi-Dirac statistics.
The upper curves $a$ and $b$ are for electronic neutrinos, while the lower
curves $c$ and $d$ are for muonic (tau) neutrinos. All the curves can be
well approximated by a second order polynomial
in $y$, $\delta = A y (y -B)$.
\label{deltaofy}}
\end{figure}

A simplified hydrodynamic approach to non-equilibrium neutrinos in the
early universe was recently proposed in ref.~\cite{matr}. Though
significantly less accurate, it gives a simple intuitive description 
and qualitatively similar results.

Naively one would expect that the distortion of neutrino spectrum at a
per cent level would result in a similar distortion in the primordial
abundances of light elements. However, this does not occur for the following
reason: An excess of neutrinos at the high energy tail of the spectrum 
results in excessive destruction of neutrons in reaction (\ref{nnue})
and excessive production in reaction (\ref{ne}).
This nonequilibrium contribution into the second process is more efficient
because the number density of protons at nucleosynthesis (when $T\approx 0.7$
MeV) is 6-7 times larger than that of neutrons. So
an excess of high energy neutrinos results in an increase of
the frozen neutron-to-proton ratio, $r=n_n/n_p$, and in a
corresponding increase of $^4 He$. On the other hand, an excess of
neutrinos at low energies results in the decrease of $r$ because reaction
(\ref{ne}) is suppressed due to threshold effects. Moreover, an overall
increase of neutrino energy density leads to a lower freezing
temperature, $T_{np}$, of the reactions~(\ref{nnue},\ref{ne})
and also leads to the decrease of
$r$. 
It happened that the nonequilibrium spectrum distortion discussed above, 
together with the decrease of $T_{np}$, took place between the two 
extremes and that the net influence of this distortion
on e.g. $^4 He$ is minor. The change of the mass fraction of $^4 He$ is
$\sim 10^{-4}$. All the papers~\cite{df}-\cite{dhs0},\cite{fdt} where this
effect was considered are in agreement here.

Thus the present day energy density of relativistic matter, i.e. of 
massless photons and massless neutrinos, with the account of late neutrino
heating, should be a little larger than predicted by the standard instant
freezing approximation. As was mentioned above, the increase of energy
density due to this effect is equivalent to adding 0.03 extra massless
neutrino species into the plasma. There is
another effect of the similar magnitude and sign~\cite{afh,ldht}, namely
finite-temperature electromagnetic corrections to the energy density
of $\gamma e^+ e^-$-plasma. As any second order effect, it diminishes
the energy of the electromagnetic part of the plasma, so that neutrino energy
normalized to the photon energy becomes a little larger. In accordance
with ref.~\cite{ldht} this effect gives 0.01 effective number of extra
neutrino species.
Though quite small, such extra heating of neutrinos may be in principle
registered~\cite{gg,ldht} in high precision measurements of CMB
anisotropies by future MAP or PLANCK satellite missions.
A change in neutrino energy compared to the standard case would result 
in the shift of equilibrium epoch between
matter and radiation, which is imprinted on the form of the angular spectrum
of fluctuations of CMB. If the canonical model can be tested with the
accuracy of about 1\% or better, the minute effects discussed here could be
observed (see however the discussion in sec.~\ref{9cmb}). The total energy
density of relativistic matter in the standard model is given by
\be
\Omega_{rel} = \Omega_\gamma \left[ 1 + 0.68\, {N_\nu \over 3}\,
\left( {1.401\, T_\nu \over T_\gamma}\right)^4 \right]
\label{omegarel}
\ee
where $\Omega_\gamma$ is the relative energy density of cosmic
electromagnetic background radiation (CMBR) and $T_\gamma$ is photon
temperature. The corrections
found in this section and electromagnetic corrections of ref.~\cite{ldht}
could be interpreted as a change of  $N_\nu$ from 3 to 3.04. A detailed
investigation of the effective number of neutrinos 
has been recently done in the paper~\cite{mangano01nnu}. As is summarized 
by the authors the non-equilibrium
heating of neutrino gas and finite temperature QCD corrections lead to
$N_\nu = 3.0395$ in a good agreement with the presented above results.
A similar conclusion is reached in the paper~\cite{steigman01}
where account was taken for possible additional to neutrinos
relativistic degrees of freedom.

\section{Heavy stable neutrinos. \label{sec:heavynu}}

\subsection{Stable neutrinos, $\mnh < 45$ GeV. \label{stabhvnu}}

If neutrino mass is below the neutrino decoupling temperature, 
$T\sim 2-3$ MeV,
the number density of neutrinos at decoupling is not Boltzmann 
suppressed. Within a factor of order unity, it is equal to the number 
density of photons, see
eq. (\ref{nnungamma}). For heavier neutrinos this is not true - the
cross-section of their annihilation is proportional to mass squared
and their number density should be significantly smaller than that of
light ones. Thus, either very light (in accordance with Gerstein-Zeldovich
bound) or sufficiently heavy neutrinos may be compatible with cosmology.
As we will see below, the {\it lower} limit on heavy neutrino mass is a few
GeV. Evidently the bound should be valid for a stable or a long
lived neutrino with
the life-time roughly larger than the universe age. Direct laboratory
measurements (\ref{mne})-(\ref{mnt}) show that none of the three known
neutrinos can be that heavy, so this bound may only refer to a new neutrino
from the possible fourth lepton generation. Below it will be
denoted as $\nu_h$. It is known from the LEP measurements~\cite{pdg}
of the Z-boson width that there are only 3 normal neutrinos with masses
below $m_Z /2$, so if a heavy neutrino exists, it
must be heavier than 45 GeV. It would be natural to expect that such a 
heavy neutral
lepton should be unstable and rather short-lived. Still, we cannot
exclude that there exists the fourth family of leptons which possesses a
strictly conserved charge so the neutral member of this family, if it is
lighter than the charged one, must be absolutely stable.
The experimental data on three families of observed leptons confirm the 
hypothesis of separate leptonic charge conservation, though
it is not excluded that lepton families are mixed by the
mass matrix of neutrinos and hence leptonic charges are non-conserved,
as suggested by the existing indications to neutrino oscillations.

Although direct experimental data for $\mnh$ in a large range of values
are much more restrictive than
the cosmological bound, still we will derive the latter here. The 
reasons for that are partly  historical, and partly related to the fact 
that these arguments, with slight  modifications, can be applied to 
any other particle with a weaker than normal weak interaction, for 
which the LEP bound does not work.
The number density of $\nuh$ in the early universe is depleted through
$\bar \nuh \nuh$-annihilation into lighter leptons
and possibly into hadrons if $\mnh > 100$ MeV. The annihilation rate is
\be
\Gamma_{ann} = \dot n_h /n_h = \sigma_{ann} n_h
\label{dotnl}
\ee
where for a simple estimate, that we will describe below,
the annihilation cross-section can be approximately taken as
$\sigma_{ann} \approx G^2_F \mnh^2$ if $\nuh$ are Dirac neutrinos
(for Majorana neutrinos annihilation proceeds in $p$-wave and the
cross-section is proportional to velocity, see below).
This estimate for the cross-section
is valid if $m_h < m_Z/2 \approx 45$ GeV. If the annihilation
rate is faster than the universe expansion rate, $\Gamma_{ann} >H$,
the distribution of $\nuh$ would be very close to the equilibrium one.
The annihilation effectively stops, freezes, when
\be
\Gamma_{ann} =H,
\label{gammah}
\ee
and, if at that moment $T=T_f<\mnh$, the number and energy densities of 
$\nuh$
would be Boltzmann suppressed. The freezing temperature can be estimated
from the above condition with $H$ taken from eq.~(\ref{hofx}),
$ H= 5.44 (T^2 /m_{Pl})\sqrt{g_*/10.75} $,
and $n_{\nuh}$ taken from the second of eqs. (\ref{nf}).
Substituting these expressions
into condition (\ref{gammah}), we find for the freezing
temperature $x_f\equiv m_h/T_f \approx 20 +3\ln m_h$. Correspondingly
we obtain $n_h/n_\gamma \approx 0.2 x_f^{3/2}\exp\,(-x_f)$.

After the freezing of annihilation, the number density of heavy 
neutrinos would remain constant in comoving volume and it is easy to 
calculate their contemporary energy density, $\rho_h=m_h n_h$.
From the condition
$\rho_h < \rho_c$  (see eq.~(\ref{rhoc0})) we obtain
\be
m_h > 2\, {\rm GeV}
\label{mh}
\ee
This is very close to the more precise, though still not exact, 
results obtained by the standard, more lengthy, method. Those 
calculations are done in the following way. It is assumed that: 
\begin{enumerate}
\item{}
Boltzmann statistics is valid.
\item{}
Heavy particles are in kinetic but not in chemical equilibrium,
i.e their distribution function is given by $f_\nu = \exp [-E/T + \xi(t) ]$.
\item{}
The  products of annihilation are in complete thermal equilibrium state.
\item{}
Charge asymmetry in heavy neutrino sector is negligible, so the effective
chemical potentials are the same for particles and antiparticles, $\xi=
\bar \xi$.
\end{enumerate}
Under these three assumptions a complicated system of integro-differential
kinetic equations can be reduced to an ordinary differential equation
for the number density of heavy particles $n_h (t)$:
\be
\dot n_h +3H n_h = \langle \sigma_{ann} v \rangle
(n^{(eq)2}_{h} -n^2_h)
\label{dotn}
\ee
Here $n^{(eq)}$ is the equilibrium number density, $v$ is the velocity of
annihilating particles, and angular brackets mean thermal averaging:
\be
\langle \sigma_{ann} v \rangle =
\frac{(2\pi)^4}{(n_{\nuh}^{eq})^2}
\int{\overline {dp}}_{\nuh}  {\overline {dp}}_{\nuh'}
 \int{\overline {dp}}_{f}  {\overline {dp}}_{f'}~
 \delta^4 (p+p'-k-k')
 \mid A_{ann} \mid ^2 e^{-\left( E_p  + E_{p'}\right) /T}
\label{sigmav}
\ee
where ${\overline {dp}}= d^3p /(2E\,(2\pi)^3)$ and
$f$ and $f'$ are fermions in the final state (products of annihilation).
Following ref.~\cite{goge} one can reduce integration down to one dimension:
\be
\langle \sigma_{ann} v \rangle = \frac{x}{8 \mnh^5 K_2^2(x)}
\int_{4 \mnh^2}^{\infty}ds~(s-4 \mnh^2)\sigma_{ann}(s)
\sqrt{s} K_1\left( \frac{x\sqrt{s}}{\mnh}\right)
\label{sigmak}
\ee
where $x = \mnh/T$, $K_i(x)$ are the modified Bessel functions of order $i$
(see for instance \cite{gradshtein94}) 
and $s=(p+p')^2$ is the invariant center-of-mass
energy squared of the process $\nuh\, \bar\nuh \leftrightarrow f \,f'$.
Corrections to eq.~(\ref{dotn}) in cases when the particles in question
freeze out semi-relativistically or annihilate
into non-equilibrium background were considered in the
papers~\cite{ad9}-\cite{hannestad99-na}, see also
sec.~\ref{massstabnu}.

Equation (\ref{dotn}) is the basic equation for calculations of frozen
number densities of cosmic relics. It was first used (to the best of my
knowledge) in ref.~\cite{zop} (see also the book~\cite{zn})
to calculate the number density of relic quarks if they existed
as free particles. Almost 15 years later this equation was simultaneously
applied in two papers~\cite{lw,vdz} to the
calculation of the frozen number density of possible heavy neutrinos.
At around the same time there appeared two more papers~\cite{hut,sako}
dedicated to the same subject. In ref.~\cite{hut} essentially the same
simplified arguments as at the beginning of this section were used and the
result~(\ref{mh}) was obtained. In ref.~\cite{sako} it was assumed that
heavy neutrinos were unstable and the bound obtained there is contingent upon
specific model-dependent relations between mass and life-time.
In the papers~\cite{lw,vdz} eq.~(\ref{dotn})
was solved numerically with the result $\mnh > 2.5$ GeV. An approximate,
but quite accurate, solution of this equation is described in the
books~\cite{zn,kt} and in the review paper~\cite{dz}. Another possible
way of approximate analytic solution of this equation, which is a Riccatti
equation, is to transform it into a Schroedinger equation by a standard
method and to solve the latter in quasi-classical approximation. There is
a very convenient and quite precise formula for the present day number
density of heavy cosmic relics derived in the book~\cite{kt}:
\be
{ n_{\nuh} \over s} \approx { 4 x_f \left(g_*^{1/2} / g_{*S} \right)
\over \langle \sigma_{ann} v \rangle m_{Pl} \mnh }
\label{nuhnug}
\ee
where $s \approx 3000/ {\rm cm}^3$ is the present day entropy density,
including photons of CMB with $T_\gamma=2.7$ K and three types of massless
neutrinos with $T_\nu \approx 1.9 $ K; $g_*$ is the effective number of
particle species contributing into energy density, defined in accordance
with eq. (\ref{rhorel});
$g_{*S}$ is the similar quantity for the entropy, $s=g_{*S} (2\pi^2/45) T^3$.
All the quantities are defined at the moment of the freezing of
annihilation, at $T=T_f$;
$x_f = m/T_f \approx \ln (\langle \sigma_{ann} v \rangle m_{Pl} \mnh) $.
Typically $x_f=10-50$.

The results presented above are valid for s-wave annihilation, when the
product $\sigma_{ann} v$ tends to a non-vanishing constant as
$ v\rar 0$. This can be applied to massive Dirac neutrinos. In the case 
of  Majorana neutrinos, for which particles and antiparticles are 
identical, annihilation at low energy can  proceed only in p-wave, so 
$\sigma_{ann} v \sim v^2$. If $\sigma_{ann} v \sim v^{2n}$, the result
(\ref{nuhnug}) is corrected by an extra factor $(n+1)$ in the numerator
and by the factor $1/x_f^n$ due to the cross-section
suppression. A smaller cross-section results in a stronger
bound~\cite{lkr}, $\mnh > 5$ GeV.

As was noticed in ref.~\cite{vdz}, if all dark matter in the 
universe is formed by heavy neutrinos, then their number 
density would increase
in the process of structure formation. This in turn would lead to an
increased rate of annihilation. Since about half of entire energy
release would ultimately go into electromagnetic radiation, which is
directly observable, the lower limit on heavy neutrino mass could be
improved at least up to 12 GeV. Cosmological consequences of existence
of a heavy stable neutral lepton were discussed in ref.~\cite{gllss}.
It was noted, in particular, that these leptons could form galactic
halos and that their annihilation could produce a detectable
electromagnetic radiation. This conclusion was questioned in
ref.~\cite{stecker78} where detailed investigation of the gamma-ray
background from the annihilation of primordial heavy neutrinos was
performed. It was argued that the annihilation radiation from the halo
of our Galaxy could make at most one third of the observed intensity.
The halos of other galaxies could contribute not more than a per cent
of the observed gamma ray background.

On the contrary, in ref.~\cite{zeldovich80kk} very restrictive limits
were advocated: $m_{\nu_h} >15$ GeV from the $\gamma$-ray background 
and $m_{\nu_h} >100$ GeV from the $e$ and $p$ components of cosmic 
rays. The
authors argued that heavy neutrinos would be entrained by baryons to
galactic center in the process of galaxy formation and their number
density would rise in the 
same proportion as the number density of baryons.
However, this result was obtained under assumption of baryonic dominance,
$\rho_b > \rho_{nu_h}$. This is not true in realistic cosmology when
the total density of cold dark matter is much larger than the baryonic 
one. Accordingly the bounds should be noticeably relaxed.

We will not go into more detail because
precise positions of these bounds are not of much interest now.
Indeed, a heavy neutrino, if it exists, must be heavier than 
45 GeV. Still
we will discuss the validity of four assumptions used for the derivation
of eq.~(\ref{dotn}) keeping in mind that this equation is of general
interest. It can be applied to some 
other cases and, in particular, to the 
derivation of the nucleosynthesis bounds on the mass of $\nut$ (see 
sec.~\ref{massstabnu}).

The first assumption of Boltzmann statistics
is quite accurate if $T_f\ll m_h$. The assumption
of kinetic equilibrium is generically fulfilled near annihilation freezing
because kinetic equilibrium is maintained by the scattering of heavy particles
on the light ones with the scattering rate $\sigma_{el} n_0$, while
the rate of annihilation is proportional to the number density of
heavy particles, $\sigma_{ann} n_h$ and the latter
is suppressed as $n_h\sim \exp (-m_h /T)$. In reality
heavy particle spectrum is always somewhat colder than the equilibrium
one. If annihilation does not vanish in the limit of zero momentum, one
may obtain reasonable upper and lower bounds on the frozen number density
of heavy particles making calculations in two extreme cases of all heavy
particles being at rest and in kinetic equilibrium.
The assumption of equilibrium distribution of annihilation products 
may be slightly violated because annihilation of non-equilibrium parents
would created a non-equilibrium final state.
The validity of this
assumption depends upon the rate of thermalization of the annihilation
products. The deviation from equilibrium is a second order effect 
and is normally rather weak.
All three assumptions are well fulfilled for heavy neutrinos with
$m_h \gg 100$ MeV.
Usually eq. (\ref{dotn}) gives a rather good
 approximation to exact results but
e.g. for the case of neutrinos with masses 3-20 MeV, calculations based on
this equation underestimate the result by approximately factor 2. The point is
that for neutrinos in this mass range kinetic equilibrium is broken
simultaneously with the chemical one and deviations from both are quite
significant \cite{ad9,dhsm}.

The fourth hypothesis of vanishingly small lepton asymmetry stands separately
from the above three. While these three have been adopted to simplify the
calculations, the fourth assumption does not serve this purpose. If asymmetry
is non-vanishing kinetic equations can still be reduced to ordinary
differential ones under the same three assumptions presented above.
Lepton asymmetry is an essential unknown parameter and it
is assumed to be small because the baryon asymmetry of the universe is small,
$n_B/n_\gamma \sim (3-5)\cdot 10^{-10}$, though strictly speaking they are
not related. If $n_L/n_\gamma \sim n_B/n_\gamma $ then the quoted here bounds
do not noticeably change. However if the asymmetry is larger by an
order of magnitude or more, then
the number density of heavy leptons, which survived annihilation, would be
determined by the (conserved) leptonic charge density. In particular, if
the lepton asymmetry is close to unity the mass of the corresponding leptons
should be smaller than $\sim 25\,h^2$ eV with $h$ determined in
eq.~(\ref{rhoc0}). In the case of arbitrary chemical potential the 
above limit
is modified by the factor~(\ref{etal}) (see sec.~\ref{degennu}).

If the universe is reheated only up to MeV temperatures, as described in
refs.~\cite{giudice00a,giudice00b}, the lower limit on the neutrino mass
is drastically relaxed, $m_\nu > 4 (3)$ MeV for Dirac (Majorana) particles.

\subsection{Stable neutrinos, $\mnh > 45$ GeV.\label{stabheavy}}

Such heavy neutrinos are not excluded by the measurements of the total decay
width of $Z^0$ and, if they are stable, the cosmological limit on their mass
may be of interest. It has been shown in ref.~\cite{maltoni99} that 
very heavy neutrinos do not decouple from the lower energy sector and their
presence could be observed through radiative corrections in the
precision LEP experiments. According to the results of this paper, 
a relatively light
extra generation, $m<m_Z$, is disfavored by the data and the only
open possibility is a neutral lepton with the mass near 50 GeV.
The minimum of $\chi^2$ for such hypothesis lies between one and two
extra generations~\cite{novikov01}. 
If all 4 particles of a generation are heavier than
$Z$-boson and if new generations are not mixed with the three light 
ones then additional chiral generations are 
not excluded by the precision electroweak data~\cite{novikov01}. 
Moreover, for very heavy neutrinos the Yukawa coupling to
the Higgs boson would be so strong that perturbative calculations
become non-reliable.

The cross-section of $\bar \nuh \nuh$-annihilation
in a renormalizable gauge theory with a weak coupling should behave as
$\sigma_{ann} \sim \alpha^2/s \sim \alpha^2 /\mnh^2$ and in accordance
with eq.~(\ref{nuhnug}) the cosmic energy density of these neutrinos would
behave as $\rho_{\nuh} \sim \mnh^2$. Hence, with an increasing mass,
$\rho_{\nuh}$ would overcome $\rho_c$. The corresponding upper limit found
in ref.~\cite{dz} is $\mnh < 3$ TeV. A somewhat stronger bound,
$\mnh < 5$ TeV, is obtained in ref.~\cite{ot}. However, as was argued in
ref.~\cite{ekm} (see also \cite{ek}), both papers overlook an important
contribution into cross-section. For $\mnh > m_W$ a new channel
 of annihilation becomes open,
$\bar\nuh \nuh \rar W^+W^-$ with the cross-section proportional to
$\alpha^2 (\mnh/m_W)^4 /s$. Near the threshold $s \approx 4\mnh^2$
and $\sigma_{ann} \sim \mnh^2$.
Though the singularity, $1/m_W$, as
$m_W\rar 0$, should not be present in the renormalizable electro-weak
theory, the terms $\sim \mnh^2 /m_W^2$ are possible because both
denominator and numerator
proportionally disappear when symmetry is restored. These terms
come from the strong Yukawa coupling of heavy neutrinos to the Higgs field.
The coupling constant of this interaction is $g= \mnh / \langle H \rangle$,
where $\langle H \rangle \approx 250$ GeV is the vacuum expectation value
of the Higgs field. Taking into account that
$m_W^2 \sim \alpha \langle H \rangle^2$ we obtain the above presented
estimate for the cross-section. According to the calculations of
ref.~\cite{ek} the accurate threshold value is:
\be
\langle v \sigma (\bar\nuh \nuh \rar W^+W^-)\rangle = 
{G_F^2\mnh^2 \over 8\pi}
\label{sigmaww}
\ee

With the account of the rising with $\mnh$ cross-section of the
process $\bar\nuh \nuh \rar W^+W^-$, the energy density
of relic $\nuh$ behaves as $\rho_{\nuh} \sim \mnh^{-2}$ and would never
contradict astronomical upper limit. So it appears at first sight that all
neutrinos heavier than 45 GeV would be cosmologically allowed.
However, this result is obtained in the lowest order of perturbation theory.
With rising $\mnh$ the Yukawa coupling of the Higgs field to $\nuh$
becomes large, $g= \mnh / \langle H \rangle >1$ and we arrive in the
regime of strong interactions where one should not trust perturbative
calculations. There is an absolute upper limit on the partial wave amplitudes
imposed by the $S$-matrix unitarity. According to it, the partial wave
cross-section with angular momentum $J$ cannot exceed
\be
\sigma_J^{max} = {\pi (2J+1) \over p^2}
\label{sigmaj}
\ee
where $p = \sqrt{s-4\mnh^2} /2$ is the momentum of the annihilating particles
in the center of mass frame. The existence of this limit in connection with
cosmic heavy lepton relics was first noted in ref.~\cite{gk} and studied in some
detail in ref.~\cite{ek}. Of course, if all partial waves are saturated, the
total cross-section $\sigma_{tot}^{max} = \sum_J \sigma^{max}_J$ would be
infinitely large. Evidently it never happens. Moreover, partial wave amplitudes
are known to vanish near threshold as $\sim p^J$. Correspondingly annihilation
in S-wave ($J=0$) behaves as $\sigma_0 \sim 1/p$, while annihilation in P-wave
($J=1$) behaves as $\sigma_1 \sim p$, etc. Thus, near threshold only lowest
partial waves are essential.
Using eq. (\ref{sigmak}), one can find for S-wave annihilation in
non-relativistic limit ($x\gg 1$):
\be
\langle \sigma_0^{max} v \rangle ={4\sqrt{\pi x} \over \mnh^2}
\label{sigma0}
\ee
Comparing this with expression (\ref{sigmaww}) we find that the latter
overshoots the unitarity limit when $\mnh > 1.6$ TeV (for $x=\mnh/T_f = 30$).
Analogous boundary for Majorana leptons, which annihilate in P-wave, is
$\mnh > 3.2$ TeV~\cite{ek}.

If the unitarity bound is adopted for the cross-section when 
$\mnh$ is larger then presented above values
the energy density of relic heavy $\nuh$ would be larger than
$\rho_c$ at least for $\mnh > 100-200$ TeV. Thus heavy stable neutrinos
with masses above
these values are excluded. However the limit may be considerably
stronger than that. The point is that strong interaction effects become
significant much below unitarity saturation. It is analogous to
electromagnetic form-factor of nucleons. Though electromagnetic interaction
is quite weak so that unitarity in electromagnetic process
$e^+e^- \rar ({\rm virtual} \,\,\gamma)\rar \bar p p$,
is far from being saturated, the electromagnetic
vertex $\bar pp\gamma$ for photons with a large virtuality is strongly
suppressed due to strong interaction of protons. Similar effects may
significantly suppress $\bar \nuh \nuh$-annihilation into $W^+W^-$. Such
effects would become important at the onset of strong interaction regime, i.e.
for $\mnh > O({\rm TeV})$. So incidentally, the old limit $\mnh < 3-5$ TeV
may come back. Resolving this problem demands more accurate and 
quite difficult calculations of heavy $\bar \nuh \nuh$-annihilation 
in strong interaction regime.

To summarize this discussion, the cosmic energy density, $\rho_{\nuh}$,
of heavy neutrinos with the usual weak interaction
is sketched in fig. (\ref{rholfig}). In the region of very small masses the
ratio of number densities $n_{\nuh}/n_\gamma$ does not depend upon the
neutrino mass and $\rho_{\nuh}$ linearly rises with mass. For larger masses
$\sigma_{ann} \sim \mnh^2$ and $\rho_{\nuh}\sim 1/\mnh^2$. This formally opens
a window for $\mnh$ above 2.5 GeV. A very deep minimum in $\rho_{\nuh}$ near
$\mnh = m_Z /2$ is related to the resonance enhanced cross-section around
$Z$-pole. Above $Z$-pole the cross-section of $\bar \nuh \nuh$-annihilation
into light fermions goes down with mass as $\alpha^2/\mnh^2$ (as in any normal
weakly coupled gauge theory). The corresponding rise in $\rho_{\nuh}$ is
shown by a dashed line. However for $\mnh > m_W$ the contribution of the
channel $\bar \nuh \nuh \rar W^+W^-$ leads to the rise of the cross-section
with increasing neutrino mass as $\sigma_{ann} \sim \alpha^2 \mnh^2 /m_W^4$.
This would allow keeping $\rho_{\nuh}$ well below $\rho_c$
for all masses above
2.5 GeV. The behavior of $\rho_{\nuh}$, with this effect of rising
cross-section included, is shown by the solid line up to
$\mnh =1.5 $ TeV. Above that value it continues as a dashed line.
This rise with mass would break unitarity limit for partial wave
amplitude when $\mnh$ reaches 1.5 TeV (or 3 TeV for Majorana neutrino).
If one takes the maximum value of the S-wave cross-section permitted by
unitarity (\ref{sigma0}), which scales as $1/\mnh^2$, this would give rise
to $\rho_{\nuh} \sim \mnh^2$ and it crosses $\rho_c$ at $\mnh \approx 200$ TeV.
This behavior is continued by the solid line above 1.5 TeV.
However for $\mnh \geq {\rm a\,\, few}\,\, {\rm TeV}$ the Yukawa coupling of
$\nuh$ to the Higgs field becomes strong and no reliable calculations
of the annihilation cross-section has been done in this limit.
Presumably the cross-section is much
smaller than the perturbative result and the cosmological bound
for $\mnh$ is close to several TeV. This possible, though not certain,
behavior is presented by the dashed-dotted line. One should keep in mind,
however, that the presented results for the energy density could only be true
if the temperature of the universe at an early stage was higher than
the heavy lepton mass.

\begin{figure}[htb]
\begin{center}
  \leavevmode
  \hbox{
    \epsfysize=3.0in
    \epsffile{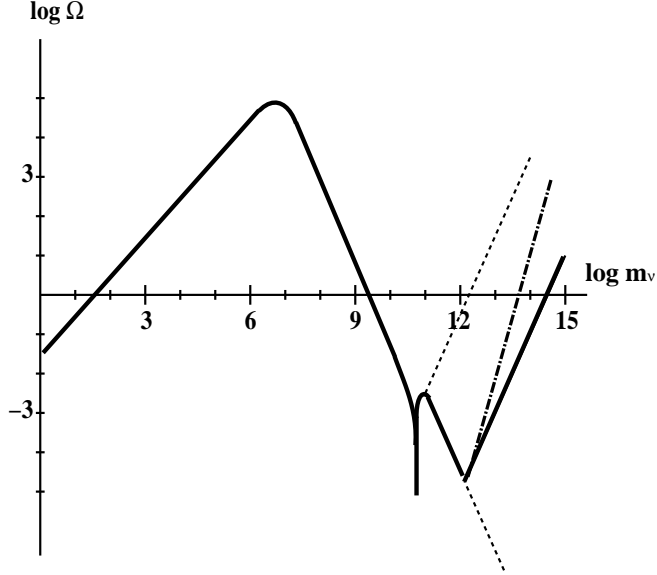}}
\end{center}
\caption{Cosmological energy density of massive neutrinos
$\Omega = \rho_{\nuh} /\rho_c$ as a function of their mass
measured in eV. The meaning of different lines is explained 
in the text.
\label{rholfig}}
\end{figure}

Heavy neutral leptons, compatible with the cosmological constraint
$\rho_{\nuh} < \rho_c$, could be accumulated in the Galaxy and observed
through the flux of cosmic rays created by their annihilation~\cite{tw,fkkm}.
Using the cosmic ray data one could exclude a certain range of neutrino masses.
However the analysis of ref.~\cite{golkon} shows that existing data is not
sufficient to derive any interesting bound.

Another possible way of registering or constraining cosmic heavy
neutrinos is to use the bounds or possible signals of their detection in
terrestrial low background experiments. The excluded mass regions quoted in
ref.~\cite{pdg} are typically from $\sim 10 $ up to hundreds GeV or even
1-2 TeV. All these results are based on the assumption that heavy neutrinos
constitute the bulk of dark matter in the Galaxy. In particular, in our
neighborhood the mass density of dark matter is:
\be
\rho_{\nuh}^{(gal)} \approx \rho_{DM}^{(gal)} \approx 0.3\, {\rm GeV/cm}^3
\label{rhonuhgal}
\ee
However the cosmological energy density of $\nuh$, is much smaller than
$\rho_c$ practically in all interesting parameter range.
So one would expect that their energy density in the Galaxy would
also be smaller than the observed density of dark matter. The
depletion of the galactic energy density due to a smaller original cosmic
energy density of $\nuh$ was not taken properly into account in
refs.~\cite{far1,far2}. The
authors claimed that the cosmological number density of heavy neutrinos
was enhanced in the Galaxy by the factor $3.3\times 10^6$ at least.
For example, for $\mnh = 100 $ GeV the cosmological energy density in
accordance with the results of refs.~\cite{ek,far2} is about $5\times 10^{-3}$
of the critical energy density (one can check that using eq. (\ref{nuhnug})
for the cosmological number density of $\nuh$ and cross-section
(\ref{sigmaww})). With the amplification factor quoted above the galactic
mass density of heavy neutrinos would be approximately 1/3 of the total
mass density of dark matter in our neighborhood. Using the calculated
values of the cosmic energy density of relic heavy $\nuh$ and the amplification
factor $3.3\times 10^6$ the authors of refs.~\cite{far1,far2} were able to
exclude the mass interval 60-290 GeV based
on the data of the underground
experiments on search of WIMPs. However, it seems that the amplification factor
of ref.~\cite{far2} is too large. It is possibly overestimated by
one-two orders of magnitude.
Indeed a reasonable coefficient of enhancement of the galactic mass density
of heavy dark matter particles could be found from the following considerations. If
such particles give a contribution of order unity into $\Omega$, they would
give the observed mass density of dark matter in galaxies. On the other
hand, if some heavy particles contribute only a minor fraction to the total
mass density of cold dark matter, their fraction in galaxies should be also
minor. This argument
invalidates the exclusion of the region $60<\mnh<290$ GeV and, with
the present day data, no mass of heavy lepton is excluded above 45 GeV up to
at least a few (tens) TeV. As we have already argued above,
the concrete position of this upper bound is very difficult to calculate.

In a later paper by the same group~\cite{fargion99}
the range of the galactic amplification factor was taken somewhat
smaller, about $10^5$ at the lower end. The authors concluded that
the annihilation of heavy leptons with the mass between $m_Z/2\,-\,m_Z$
could explain the diffused gamma ray radiation around galactic
plane~\cite{dixon98}.

\section{Neutrinos and primordial nucleosynthesis \label{s:bbn}}

Primordial or big bang nucleosynthesis (BBN) is very sensitive to neutrino
number density and neutrino energy spectrum in the primeval plasma. As we 
have mentioned above, this influence is especially strong for electronic
neutrinos. Any deviation
from standard neutrino physics would have an impact on nucleosynthesis
and may be observed through present-day abundances of light elements.
We will discuss below some possible manifestations of non-standard neutrino
properties in primordial nucleosynthesis and the bounds on neutrino
masses, life-times, and oscillation parameters
that can be deduced from observational data on light element abundances.
A condensed review of neutrino effects in BBN (including inhomogeneous
case) is given in~\cite{kurki99}.

\subsection{Bound on the number of relativistic species \label{knu}}

One of the most impressive results that can be derived from primordial
nucleosynthesis is a bound on the total number of light neutrino flavors,
$N_\nu$. "Light" here means $m_\nu < 1$ MeV, so that these neutrinos are
not Boltzmann suppressed at the nucleosynthesis epoch.
Before LEP data became available, nucleosynthesis was the only
source of information about the value of $N_\nu$.  The first observation
that "if there were
more than two kinds of neutrino the expansion would have to be faster in order
to overcome the gravitational attraction of the extra neutrinos and...
the larger the ratio $He/H$ turns out to be" was made by Hoyle and Tayler
in 1964~\cite{hota}. A similar statement was made by Peebles~\cite{peebheat}, 
that the introduction of a new kind of (two-component) neutrino field would
increase helium abundance by mass from 0.30 to 0.32. Detailed calculations
of the effect were performed by Shvartsman~\cite{vfs} who presented results
for helium mass fractions for different number of neutrino species and
different values of the baryonic mass density. Further development of the idea
was carried out by Steigman, Schramm, and Gunn~\cite{ssg}, who concluded
that the existing data permitted to exclude 5 extra neutrinos, $N_\nu <8$.
As we will see below, accuracy at the present day is considerably better.

Additional particles in the primeval plasma during nucleosynthesis influenced
light element abundances essentially through the following two effects.
First, they shift the frozen neutron-to-proton ratio because the freezing
temperature depends upon the number of particle species in accordance
with eq.~(\ref{tnp}). Second, though the temperature $T_d$ (\ref{td})
when light element formation begins
practically does not depend upon the number of species, the moment of time 
when this temperature is reached, $t(T_d)$, depends upon $g_*$ as seen from
eq.~(\ref{tT2}). Correspondingly, the number of surviving neutrons, which
decay with life-time 887 sec, depends upon $g_*$.

To show the sensitivity of light element abundances to the number of
massless neutrino species we calculated (using code~\cite{kaw})
the mass fraction of $^4 He$ and
the relative number density of deuterium $D/H$ as functions of $N_\nu$
for different values of baryon number density, expressed in terms of
the present day number density of CMB photons,
$\eta_{10} = 10^{10}n_B /n_\gamma$. The results are presented in
figs. (\ref{helknu}) and (\ref{deuterknu}). To avoid possible confusion
let us mention that the results are valid for any relativistic particle species
contributing the same amount of energy
into the total energy density as one two-component neutrino.

\begin{figure}[htb]
\begin{center}
  \leavevmode
  \hbox{
    \epsfysize=3.0in
    \epsffile{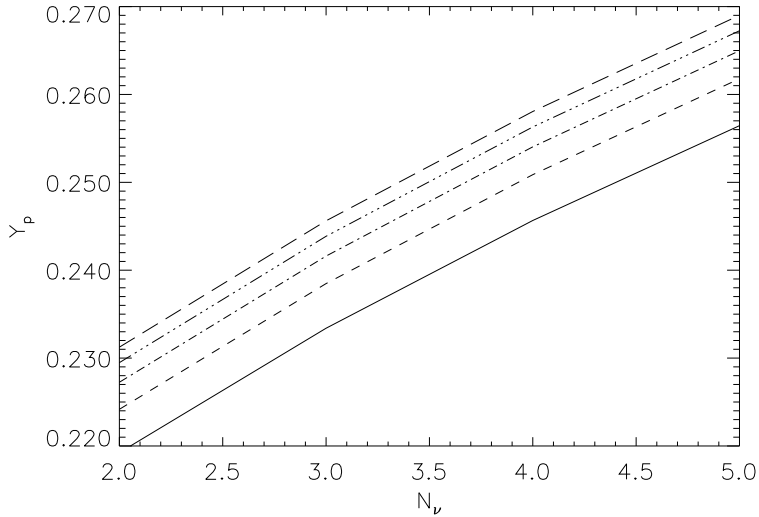}}
\end{center}
\caption{Mass fraction of $^4 He$ as a function of the number of massless
neutrino species. Different curves correspond to different values of the
baryon-to-photon ratio $\eta_{10} \equiv 10^{10}n_B/n_\gamma = 
2,3,4,5,6$ in order of increasing helium abundance.
\label{helknu}}
\end{figure}

\begin{figure}[htb]
\begin{center}
  \leavevmode  
\hbox{
    \epsfysize=3.0in
    \epsffile{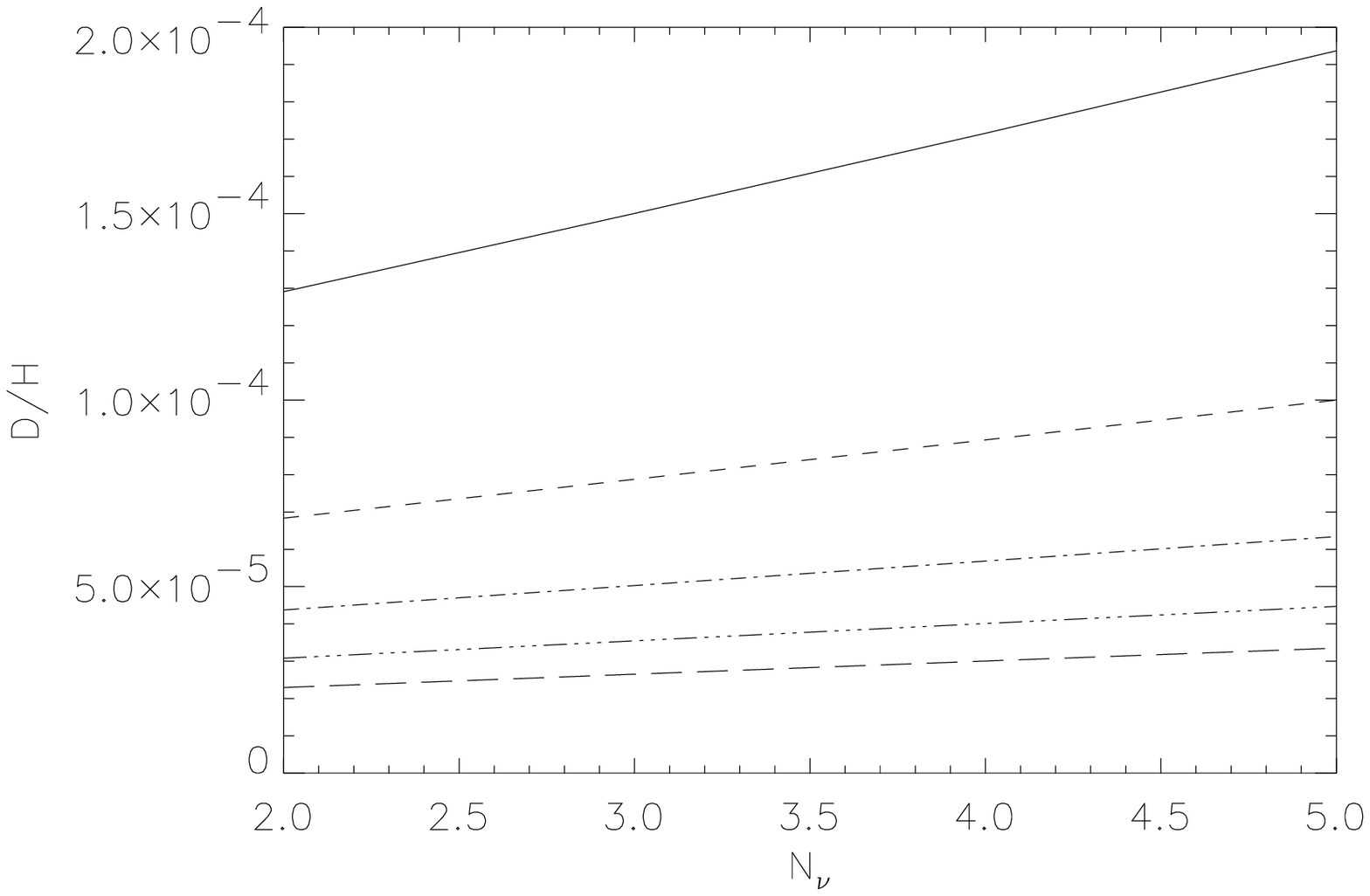}}
\end{center}
\caption{Deuterium-to-hydrogen by number as a function of the number of
massless
neutrino species. Notations are the same as in fig. (\ref{helknu}).
\label{deuterknu}}
\end{figure}

Quite often the impact of nonrelativistic particles on BBN is also described
in terms of the effective number of relativistic particles, which give the same
variation of primordial abundances. One should keep in mind, however, that
the result depends upon the chosen light element. For example, a possibly
massive $\nut$ with $m =10$ MeV shifts $^4 He$ as 2 extra massless neutrinos,
while its impact on $^2 H$ is equivalent to 20 additional massless neutrinos
(see the following subsection). Massive particles, if they are
sufficiently long-lived, play an especially important role in shifting $t(T_d)$ and changing the number density of surviving neutrons.

There are several conflicting papers in the literature presenting different
upper bounds on the allowed value of $N_\nu$. The most restrictive limit
is advocated in refs.~\cite{bntt,tytler00},
$\Delta N_\nu < 0.20$ (at 95\% C.L.). To obtain
such a restrictive result the authors used the measurements of deuterium
in high $z$-clouds~\cite{bt1,bt2} which give $(D/H)_p = (3.4\pm 0.25)
\cdot 10^{-5}$. However, uncertain velocity corrections and the possibility of
a two-component system may invalidate this conclusion (see discussion at the
end of sec. \ref{prns}). Much weaker statements are made in
refs.~\cite{olth,lsv}. According to ref.~\cite{olth} the limit is $N_\nu<4.3$
if $Y_p = 0.238$ and varies  from $N_\nu<3.3$ if $Y_p = 0.225$ to
$N_\nu< 5.3$ if $Y_p = 0.250$; all at  95\% C.L. These results depend upon
the abundance of primordial $^7 Li$ and could be somewhat relaxed. Analysis of
ref.~\cite{lsv} give $N_\nu$ ranging from 2 to 4. A small value, $N_\nu<3$,
would lead to revival of "nucleosynthesis crisis"~\cite{hss}. In particular,
according to ref.~\cite{hata96}, low deuterium observations require
$\Omega_b h^2 = 0.02-0.03$ and $N_\nu = 1.9\pm 0.3$, while high deuterium
data need $\Omega_b h^2 = 0.005-0.01$ and $N_\nu$ compatible with 3.
It is probably too early to worry about these discrepancies, though several
particle physics solutions can be easily found that give $N_\nu <3$
(see the following subsections). The analysis presented
in the review paper~\cite{olive99st} gives $\Delta N_\nu < 0.3$ for low
deuterium and $\Delta N_\nu < 1.8$ for high deuterium.
The same conclusion, $\Delta N_\nu < 0.3$, was reached in the recent
work~\cite{dibari01n} based on the measurements of $\eta_{10}$ in
angular fluctuations of CMBR (see sec.~\ref{prns}).

The latest data on light element abundances,
as discussed above, seem to converge to low (or, better to
say, to normal) deuterium abundance and 
to $N_\nu =3$. Similar results were
obtained earlier in ref.~\cite{cardall96b}. However one still has to be
cautious in making conclusions about the accuracy of determination
of $N_\nu$ from BBN. This limit demands the simultaneous knowledge of
$^4 He$ and $^2 H$, which are necessary to fix two unknown parameters
$\Omega_b h^2$ and $N_\nu$. In ref.~\cite{lsv} an error of determination
of mass fraction of $^4 He$ was taken to be 0.004, while in other works
it was assumed to be twice smaller. Possibly with an independent
measurement of $\Omega_b h^2$ from CMBR, better accuracy in
determination of $N_\nu$ could be achieved.

The new data and new analysis seem to give a convergent mass fraction
of primordial deuterium near $3\cdot 10^{-5}$. Together with the data
on other light elements, this result permits fixing the baryon number
density at BBN with very high precision~\cite{burles01}:
\be
\Omega_b h^2 = 0.020 \pm 0.002\,\,\,(95\% \,\,\,
{\rm confidence\,\,\, level})
\label{omegab-bbn}
\ee
This precision corresponds roughly to 0.2 allowed extra neutrino species
during BBN. However, as is argued in ref.~\cite{dolgov01tp} and in 
the papers quoted above such accuracy at the present time
seems to be overestimated and the safe bound is closer to 1.

\subsection{Massive stable neutrinos. Bound on $\mnt$.\label{massstabnu}}

If neutrinos are stable or have lived longer than the age of the universe,
$t_U = 12-14 $ Gyr, their mass is strongly bounded from above by
Gerstein-Zeldovich limit (see section \ref{gerzel}). However if
$\tau_{\nut} \ll t_U$, tau-neutrinos could be quite heavy, their mass is
only restricted by direct measurements (\ref{mnt}). If the life-time of $\nut$
is larger than the characteristic time scale of
primordial nucleosynthesis, $t_{NS} \sim 300$ sec, they can be considered
effectively stable during BBN and their energy density would be much larger
than the energy density of massless neutrinos. The equilibrium energy density
of massless particles is larger than that of massive ones. But at some stage
$\nut$-annihilation into lighter fermions was frozen down and the actual 
number and energy densities of $\nut$ became much larger than the 
equilibrium values. As a result
a massive $\nut$ would have quite a strong influence on nucleosynthesis.
A large mass of $\nut$, which can be essential for
BBN, is now most probably excluded by the Super-Kamiokande data on atmospheric
neutrino anomaly~\cite{sk98,sk00}. The latter is explained by the
$\num-\nut$ oscillations with a small mass difference. Hence the
$\nut$-mass cannot be noticeably different from the $\num$-mass. Even if
the atmospheric neutrino anomaly is created by the oscillations between
$\num$ and $\nus$, the sterile state with the large mixing angle demanded
by the anomaly, would bring sterile neutrinos into thermal equilibrium
in the early universe (see sec.~\ref{nuosceu}) and that would create
serious problems for nucleosynthesis~\cite{kainulainen98}.
Still even the above is true, there are several physical effects,
as we can see in this section and in sec.~\ref{nuosceu}, that could
diminish the effective number of neutrino species compensating the effect
of additional sterile neutrino. On the
other hand, we cannot absolutely exclude a different interpretation of the data. If an alternative interpretation exists, although it seems unlikely now, $\mnt$ could well be in MeV range. Moreover, even if the results
presented below are not applicable directly to $\nut $, the physics
is still worth discussing and it can be of interest for some
other, yet unknown, possible light particles.

In the first paper~\cite{sako} where the influence of possibly massive
neutrinos on nucleosynthesis was considered, the following two effects were
taken into account. First, a change in the total energy density of the
primeval plasma at BBN due to the presence of massive neutrinos, $\nu_m$.
It was estimated in ref.~\cite{sako} as:
\be
\delta \rho_{\nu_m} \sim {m_\nu \over T_\gamma}\, {\rm Min} \left[ 1,
(10\,{\rm MeV} /m_\nu)^3 \right]
\label{deltarhonum}
\ee
Two terms in the brackets correspond respectively to relativistic
decoupling, when $n_\nu + \bar n_\nu = 3\,n_\gamma /4$, and to non-relativistic
decoupling, when $\rho_\nu \sim m_\nu^{-2}$ in accordance with
eq.~(\ref{nuhnug}). From the limit existing at that time on the mass
fraction of $^4 He$, $Y_p<0.29$, the 
authors concluded that neutrino should
be heavier than 23 MeV or lighter than 70 eV in accordance with GZ bound.
The second effect mentioned in the paper is more model dependent and is
operative only if heavy neutrino could decay into photons. These photons
would alter theoretical predictions for the primordial abundances because
the parameter $\eta = n_B/n_\gamma$ at BBN and at the present time would be
different.

In a subsequent paper~\cite{dktw} a different conclusion was reached that
helium abundance is independent of existence of heavy neutrinos,
while deuterium is
quite sensitive to them. That would allow for interesting bounds on their masses,
life-times, and decay modes. This conclusion was corrected in ref.~\cite{kosc},
where more accurate calculations of the number density of massive neutrinos were performed based on numerical solution of eq.~(\ref{dotn}).
It was obtained
in particular that the maximum impact on $^4 He$ would have a neutrino with
$m=5$ MeV, which is equivalent to more than 4 light neutrino species. The
approach of the paper~\cite{kosc} was extended and somewhat improved
in refs.~\cite{ktcs,dr}. The calculations of the second work predicted
a somewhat larger value of the frozen energy density of $\nut$. But in the
translation of this result to the effective number of neutrino species, found
from the distortion of $^4He$ abundance, a numerical error was made that
resulted in an overestimation of the number of additional effective neutrino
species. Still, even with the error corrected, the results of
ref.~\cite{dr} are stronger than those of the pioneering
papers~\cite{kosc,ktcs}.
The calculations of both papers (see also a more recent paper \cite{rrw},
where a similar treatment was applied to the calculations of all light
element abundances and not only of $^4 He$) were done under the
following basic assumptions. It was assumed that the massive $\nut$ and the
two massless neutrinos, $\nu_e$ and $\nu_\mu$, are in complete
kinetic equilibrium so their energy distributions are
given by the canonical expression (\ref{ffb}).
Two more simplifying assumptions were made, namely that the chemical
potentials of the massless neutrinos are zero, and that the distribution
functions can be approximated by their Boltzmann limits:
\be{
f(E)= \exp[(\mu(t) - E) /T(t)] \ll 1,
\label{feb}
}\ee
which are accurate when the temperature is small in comparison with the
particle mass, $m>T$. In these approximations the problem was enormously
simplified technically. Instead of solving the system of
integro-differential kinetic equations (\ref{kin1}) for functions of
two variables, $f_j(t,p)$, one only had to solve an ordinary differential
equation (\ref{dotn}) for the total number density $n_{\nut} (t)$.

However in the case of $\mnt$ in the MeV range, nonequilibrium
corrections to the spectra of $\nut$ and even of massless neutrinos
happen to be quite
significant~\cite{dpv,ad9} and a more refined treatment of the problem had to
be developed. In ref.~\cite{fko} the simplifying assumption of
Maxwell-Boltzmann statistics was dropped in favor of the exact Fermi-Dirac one,
but it was assumed that kinetic equilibrium is maintained for all the species.
Nonequilibrium corrections have been treated
by one of the authors of the above quoted paper~\cite{fko} in the
update~\cite{kk},
who found that these corrections do not strongly change the original results of
ref.~\cite{fko}.

Exact numerical solutions of the full
system of kinetic equations for all neutrino species without any
simplifications have been done in refs.~\cite{hm1,dhsm}. In the latter work
a somewhat better numerical precision was achieved and in particular
an almost twice-higher cut-off in particle momenta was taken. Also,
expressions for matrix elements of some reactions with massive
Majorana neutrinos were corrected.
The amplitude squared of the relevant reactions are presented in
Table~\ref{table:amplitudes-nu-e}
for the case when the first particle is $\nu_e$ (or $\num$ with the indicated
there change of the coupling constants)
and in Table~\ref{table:amplitudes-nu-tau} when the first particle is
$\nu_\tau$. The entries in this Table are presented for the case of
massive Majorana $\nut$.

\begin{table}
\begin{center}
\begin{tabular}{|cccc|c|c|}
 \hline\hline
&&&&&\\
& {\bf Process }&&&{\bf S }& {\bf $2^{-5} G^{-2}_{F} S \left| A \right| ^{2}$}\\
&&&&&\\
 \hline\hline
&&&&&\\
$\nu _{e} + \nu_{e} $ & $\rightarrow$ & $  \nu _{e} + \nu_{e} $&&1/4&
$2 \left[ (p_{1} \cdot p_{4}) (p_{2} \cdot p_{3})
 + (p_{1} \cdot p_{3}) (p_{2} \cdot p_{4}) \right.$\\
&&&&& $\left. ~~~~~~~~~~~~~~~~~~~~~~
 +  (p_{1} \cdot p_{2}) (p_{3} \cdot p_{4}) \right]$\\
&&&&&\\
\hline
&&&&&\\
$\nu _{e}+ \nu_{e} $ & $\rightarrow$ &
$ \nu _{\mu} + \nu_{\mu} $&&1/4&
$ \frac{1}{2} \left[ (p_{1} \cdot p_{4}) (p_{2} \cdot p_{3}) +
(p_{1} \cdot p_{3}) (p_{2} \cdot p_{4}) \right]$\\
&&&&&\\
\hline
&&&&&\\
$\nu _{e}+ \nu_{e} $ & $\rightarrow$ &
$ \nu _{\tau} + \nu_{\tau} $&&1/4&
$ \frac{1}{2} \left[ (p_{1} \cdot p_{4}) (p_{2} \cdot p_{3}) +
(p_{1} \cdot p_{3}) (p_{2} \cdot p_{4}) \right.$\\
&&&&& - $\left. \mnt^2 (p_{1} \cdot p_{2}) \right]$\\
\hline
&&&&&\\
$\nu_{e} + \nu_\mu $ & $\rightarrow$&
$\nu _{e} + \nu_\mu  $&&1/2&
$ (p_{1} \cdot p_{2}) (p_{3} \cdot p_{4})
 + (p_{1} \cdot p_{4}) (p_{2} \cdot p_{3}) $\\
&&&&&\\
\hline
&&&&&\\
$\nu_{e} + \nu_\tau $ & $\rightarrow$&
$\nu _{e} + \nu_\tau  $&&1/2&
$  (p_{1} \cdot p_{2}) (p_{3} \cdot p_{4})
 + (p_{1} \cdot p_4) (p_{2} \cdot p_{3})  $\\
&&&&&
 $  +  \mnt^2 (p_{1} \cdot p_{3}) $\\
&&&&&\\
\hline
&&&&&\\
$\nu _{e} + \nu_{e} $ & $\rightarrow$ & $  e^{+} + e^{-}$ &&1/2&
$2 ( g_{L}^2 + g_R^2) \left\{ (p_{1} \cdot p_{4}) (p_{2} \cdot p_{3})
\right.$
\\&&&&&$ \left.+  (p_{1} \cdot p_{3}) (p_{2} \cdot p_{4}) \right\}
+ 4 g_{L} g_{R} m _{e}^{2} (p_{1} \cdot p_{2})$\\
&&&&&\\
\hline
&&&&&\\
$\nu _{e} +  e^{\pm} $ & $\rightarrow$ & $ \nu _{e} +  e^{\pm} $&&1/2&
$ 2 (g_L^2 +g_R^2) \left\{ (p_{1} \cdot p_{2}) (p_{3} \cdot p_{4}) \right.$ \\
&&&&&$ \left. + (p_{1} \cdot p_{4}) (p_{2} \cdot p_{3}) \right\}
 - 4 g_{L} g_{R} m _{e}^{2} (p_{1} \cdot p_{3}) $\\
&&&&&\\
\hline
\hline
\end{tabular}
\end{center}
\caption{Matrix elements squared for reactions with electron neutrino;
 $g_{L} = \frac{1}{2} + \sin^{2}\theta_{W}$ and
$g_{R} = \sin^{2} \theta_{W}$. Matrix elements for muon neutrino processes
are obtained by the substitutions $\nu_e \rightarrow \nu_\mu$ and
$g_{L} \rightarrow \tilde{g}_{L} =g_{L} - 1$.}
\label{table:amplitudes-nu-e}
\end{table}

\begin{table}
\begin{center}
\begin{tabular}{|cccc|c|c|}
 \hline\hline
&&&&&\\
& {\bf Process }&&&{\bf S }& {\bf $2^{-5} G^{-2}_{F} S \left| A
\right| ^{2}$}\\
&&&&&\\
 \hline\hline
&&&&&\\
$\nu _\tau + \nu_\tau $ & $\rightarrow$ & $  \nu _\tau + \nu_\tau $&&1/4&
$2 \left[ (p_{1} \cdot p_{4}) (p_{2} \cdot p_{3})
 + (p_{1} \cdot p_{3}) (p_{2} \cdot p_{4}) \right.$\\
&&&&& $
 +  (p_{1} \cdot p_{2}) (p_{3} \cdot p_{4})  + 3\mnt^4 $\\
&&&&& $
 \left. +  2 \mnt^2 \left\{ (p_{1} \cdot p_{3}) + (p_{1} \cdot p_{4}) -
(p_{1} \cdot p_{2}) \right\}   \right] $\\
\hline
&&&&&\\
$\nu _\tau+ \nu_\tau $ & $\rightarrow$ &
$ \nu _{e } + \nu_{e } $&&1/4&
$ \frac{1}{2} \left[ (p_{1} \cdot p_{4}) (p_{2} \cdot p_{3}) +
(p_{1} \cdot p_{3}) (p_{2} \cdot p_{4}) \right.$\\
&&&&& - $ \left. \mnt^2 (p_{3} \cdot p_{4}) \right]$\\
\hline
&&&&&\\
$\nu _\tau+ \nu_\tau $ & $\rightarrow$ &
$ \nu _\mu + \nu_\mu $&&1/4&
$ \frac{1}{2} \left[ (p_{1} \cdot p_{4}) (p_{2} \cdot p_{3}) +
(p_{1} \cdot p_{3}) (p_{2} \cdot p_{4}) \right.$\\
&&&&& - $ \left. \mnt^2 (p_{3} \cdot p_{4}) \right]$\\
\hline
&&&&&\\
$\nu_\tau + \nu_e  $ & $\rightarrow$&
$\nu _\tau + \nu_e   $&&1/2&
$ (p_{1} \cdot p_{2}) (p_{3} \cdot p_{4})
 + (p_{1} \cdot p_{4}) (p_{2} \cdot p_{3})  $\\
&&&&&
 $  +  \mnt^2 (p_{2} \cdot p_{4}) $\\
\hline
&&&&&\\
$\nu_\tau + \nu_\mu $ & $\rightarrow$&
$\nu _\tau + \nu_\mu  $&&1/2&
$ (p_{1} \cdot p_{2}) (p_{3} \cdot p_{4})
 + (p_{1} \cdot p_{4}) (p_{2} \cdot p_{3})  $\\
&&&&&
 $  +  \mnt^2 (p_{2} \cdot p_{4}) $\\
\hline
&&&&&\\
$\nu _\tau + \nu_\tau $ & $\rightarrow$ & $  e^{+} + e^{-}$ &&1/2&
$2 ( \tilde{g}_{L}^2 + g_R^2) \left\{ (p_{1} \cdot p_{4}) (p_{2} \cdot p_{3})
\right.$
\\&&&&&$ \left.+  (p_{1} \cdot p_{3}) (p_{2} \cdot p_{4})
- \mnt^2 (p_{3} \cdot p_{4}) \right\}$
\\&&&&& $ + 4 \tilde{g}_{L} g_{R} m _{e}^{2} \left\{ (p_{1} \cdot p_{2})
-2 \mnt^2 \right\}$\\
\hline
&&&&&\\
$\nu _\tau +  e^{\pm} $ & $\rightarrow$ & $ \nu _\tau +  e^{\pm} $&&1/2&
$2  (\tilde{g}_L^2 +g_R^2) \left\{ (p_{1} \cdot p_{2}) (p_{3} \cdot p_{4})
\right.$ \\
&&&&&$ \left. + (p_{1} \cdot p_{4}) (p_{2} \cdot p_{3})
 + \mnt^2 (p_{2} \cdot p_{4}) \right\}$
\\&&&&& $ - 4 \tilde{g}_{L} g_{R} m _{e}^{2} \left\{ (p_{1} \cdot p_{3})
+ 2 \mnt^2 \right\}$\\
\hline
\hline
\end{tabular}
\end{center}
\caption{Matrix elements squared for reactions with tau-neutrino;
 $\tilde{g}_{L} =g_{L} - 1 =
- \frac{1}{2} + \sin^{2}\theta_{W}$ and
$g_{R} = \sin^{2} \theta_{W}$.}
\label{table:amplitudes-nu-tau}
\end{table}

Numerical solutions of exact kinetic equations prove 
that nonequilibrium effects are quite significant, almost up to 50\%.
The assumption of kinetic equilibrium (\ref{ffb}) with an effective
chemical potential, equal for particles and antiparticles,
is fulfilled if
the rate of elastic scattering at the moment of annihilation freezing,
$\Gamma_{ann} \sim H$, is much higher than both the expansion rate, $H$, 
and the rate of annihilation, $\Gamma_{ann}$. This is correct 
in many cosmologically interesting cases. Indeed,
the cross-sections of annihilation and elastic scattering are usually of
similar magnitudes. But the rate of annihilation,
$\Gamma_{ann} \sim \sigma_{ann} n_m$ is suppressed relative to the rate
of elastic scattering, $\Gamma_{el} \sim \sigma_{el} n_0$, due to Boltzmann
suppression of the number density of massive particles, $n_m$, with respect
to that of massless ones, $n_0$. However in the case of MeV-neutrinos both
rates $\Gamma_{ann}$ and $\Gamma_{el}$
at the moment of freezing of annihilation are of the same order of
magnitude. Correspondingly, the assumption of kinetic
equilibrium at annihilation
freezing is strongly violated. Semi-analytic calculations of
deviations from kinetic equilibrium were performed in ref. \cite{ad9}, where a perturbative approach was developed. In the case of a momentum-independent
amplitude of elastic scattering, the integro-differential kinetic equation
in the Boltzmann limit can be reduced to the following differential equation:
\be
 JC'' + 2 J'C' =
- { 64\pi^3 H x^2 \over |A_0|^2 m} e^{y/2} \partial_y \left\{ e^{-y}
\partial_y\left[ e^{(u+y)/2} uy\partial_x \left(Ce^{-u}\right)\right]\right\}
\label{jc''}
\ee
where  $x=m/T$, $y=p/T$, prime means differentiation with respect to $y$,
$C(x,y)= \exp(\sqrt{x^2+y^2})f_m(x,y)$, $f_m$ is the unknown
distribution function of massive particles and
\be
J (x,y) =  {1\over 2} e^{y/2} \int^\infty_{u+y} dz e^{-z/2}
\left( 1- {x^2\over z^2}\right)-
{1\over 2} e^{-y/2} \int^\infty_{u-y} dz e^{-z/2}
\left( 1- {x^2\over z^2}\right)
\label{Jxy }
\ee
with $u=\sqrt{x^2 + y^2}$.

For the case of momentum-dependent weak interaction amplitude, an exact
reduction of integro-differential kinetic equation to a differential one
is unknown or impossible. But in this case one can make a polynomial
expansion in terms of momentum $y$, and reduce the problem to a sequence
of equations for partial amplitude~\cite{ad9,esposito00}.
This method greatly simplifies numerical calculations.

A direct application of perturbation
theory (with respect to a small deviation from equilibrium) to the
integro-differential kinetic equation (\ref{kin1}) is impossible or very
difficult because the momentum dependence of the anzats for the first
order approximation to $f(p,t)$ is not known. On the other hand,
eq. (\ref{jc''}) permits making a regular perturbative expansion around
the equilibrium distribution. The numerical solution of
exact kinetic equations \cite{dhsm} shows a good agreement with
the semi-analytic approach based on eq. (\ref{jc''}).

It can be easily shown that the spectrum of massive $\nut$ is softer (colder)
than the equilibrium one. Indeed, if elastic scattering of $\nut$, which would
maintain kinetic equilibrium is switched-off, the nonrelativistic $\nut$
cool down as $1/a^2$, while relativistic particles cool as $1/a$, where
$a(t)$ is the cosmological scale factor. Since the cross-section of
annihilation by the weak interactions is proportional
to the energy squared of the
annihilating particles, the annihilation of nonequilibrium $\nut$ is less
efficient and their number density becomes larger than in the equilibrium case.
Another nonequilibrium effect is the additional cooling of massless $\nue$ and
$\num$ due to their elastic scattering on colder $\nut$,
$\nu_{e,\mu}+\nut \rightarrow  \nu_{e,\mu}+\nut$. Because of that, the inverse
annihilation $\nu_{e,\mu}+\bar \nu_{e,\mu}\rightarrow\nut+\nut$ is weaker
and the frozen number density of $\nut$ is smaller. But this is a second
order effect and is relatively unimportant.

Considerably more important is an overall
heating and modification of the spectrum of $\nue$
(and of course of $\bar \nue$) by the late annihilation
$\nut+\nut\rightarrow \nue +\bar \nue$ (the same is true for $\num$ but
electronic neutrinos are more important for nucleosynthesis because they
directly participate in the reactions
(\ref{nnue},\ref{ne}) governing the
frozen $n/p$-ratio. It is analogous to the similar effect originating from
$e^-e^+$-annihilation, considered in section \ref{masslessdistr},
but significantly
more profound. The overfall heating and the spectral distortion work in the
opposite directions for $\mnt> 1$ MeV. An overall increase of the number and
energy densities of $\nue$ and $\bar \nue$ results in a smaller temperature of
neutron freezing and in a decrease of the $n/p$-ratio. On the other hand,
a hotter spectrum of $\nue$ shifts this ratio to a large value, as discussed
in the previous section. The latter effect was estimated semi-analytically in
ref. \cite{dpv}, where it was found that e.g. for $\mnt=20$ MeV the spectral
distortion is equivalent to 0.8 extra neutrino flavors for Dirac $\nut$ and to
0.1  extra neutrino flavors for Majorana $\nut$. The effect of overall heating
was found to be somewhat more significant \cite{fko,dhsm}.

The distortion of the spectrum of electronic neutrinos, found by
numerical solution of the exact integro-differential kinetic equations
in ref. \cite{dhsm}, is presented in fig. \ref{distr}.
\begin{figure}[htb]
 \begin{center}
    \psfig{figure=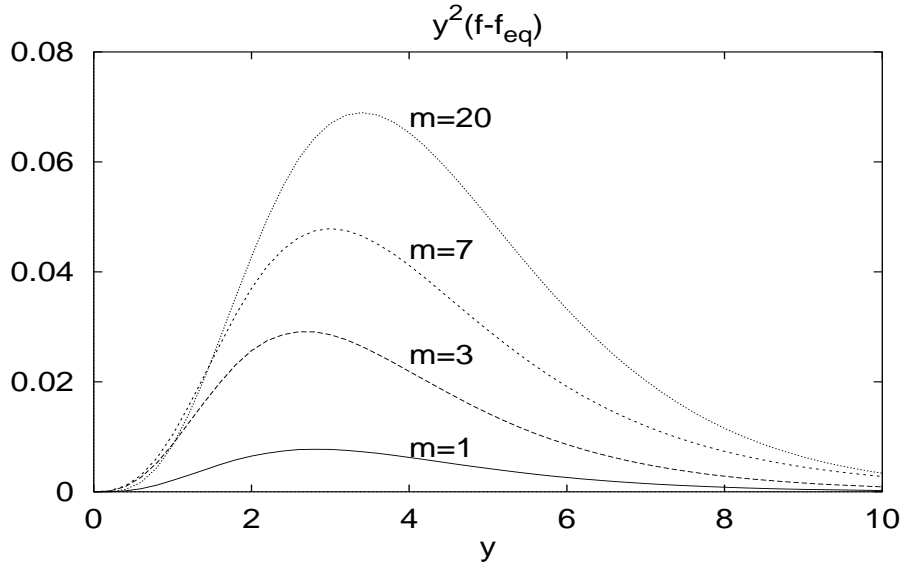,width=5in,height=3in}
  \end{center}
  \caption{Distortion of spectral distribution of electronic neutrinos
multiplied by $y^2$ as a function of dimensionless momentum $y$ for
several values of $\nut$ mass. }
 \label{distr}
\end{figure}
Though the frozen number density of $\nut$ obtained in ref. \cite{dhsm}
is larger than or equal to those obtained in any of
refs.~\cite{ktcs,dr,hm1}, (see fig. \ref{rm})
\begin{figure}[htb]
 \begin{center}
    \psfig{figure=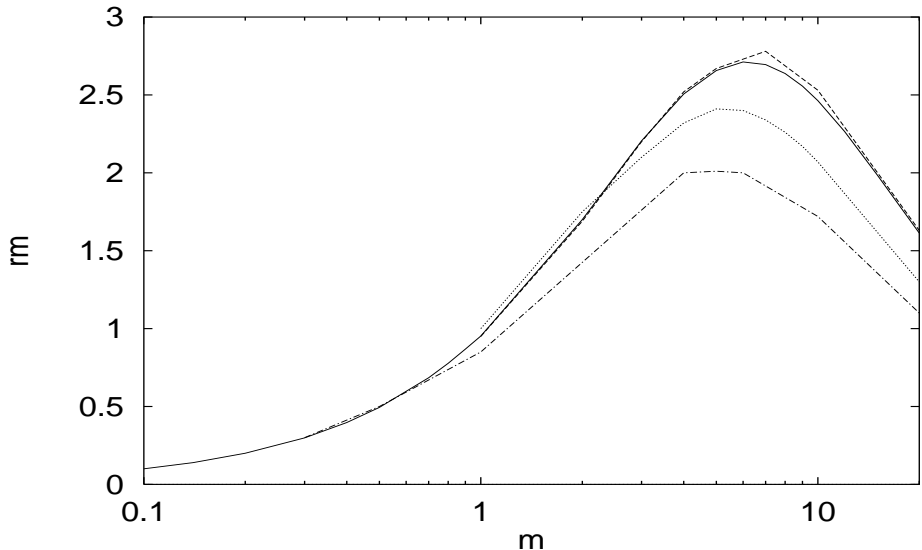,width=5in,height=3in}
  \end{center}
\caption{Relative energy density of massive  tau-neutrinos,
$rm = m_{\nut} n_{\nut} /n_{\nu_0}$, for asymptotically large
time as a function of
$\mnt$. The solid, dashed, dashed-dotted, and dotted lines are respectively
the results of refs.~\protect\cite{dhsm,hm1,ktcs,dr}.}
\label{rm}
\end{figure}
the influence of nonequilibrium corrections on nucleosynthesis found
in \cite{dhsm} is somewhat weaker than that found in \cite{fko,hm1}
in the mass range above 15 MeV.
It is possibly related to a larger momentum cut-off in numerical
calculations of ref.~\cite{dhsm}, which gives rise to a smaller
neutron freezing temperature.

The influence of a massive $\nut$ on the formation of 
light elements 
can be described by the effective number of extra massless neutrino species,
which gives the same abundance of the corresponding element as massive
$\nut$ does. This number is different for different elements and usually
$^4He$ is taken to this end.
In fig. \ref{heliumknu} the numbers of effective neutrino species,
obtained by different
groups from the mass fraction of primordial $^4 He$, are compared.
\begin{figure}[htb]
 \begin{center}
    \psfig{figure=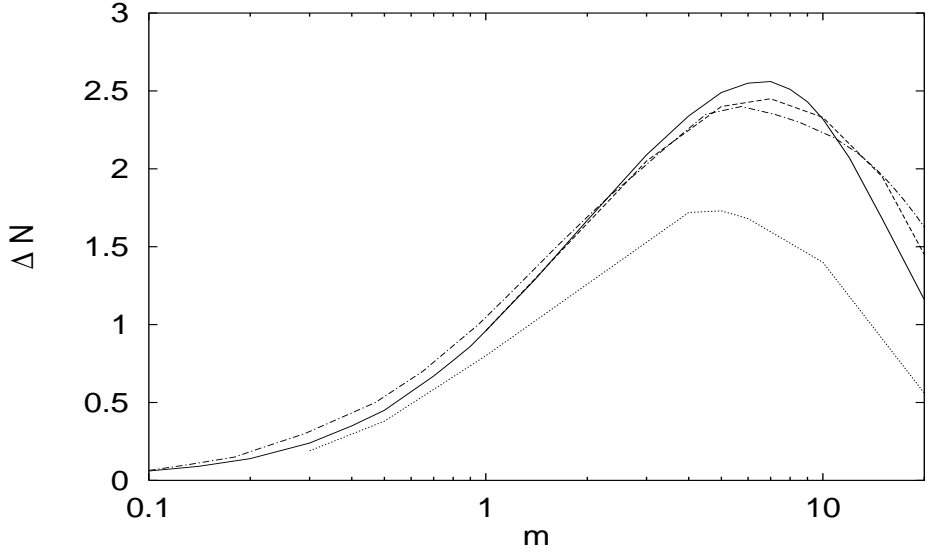,width=5in,height=3in}
  \end{center}
\protect\caption{The effective number of equivalent massless neutrino species
$\Delta N = N_{eff} -3$ found from $^4 He$ by different
groups~\protect\cite{dhsm,hm1,fko,ktcs}, correspondingly solid, dashed,
dashed-dotted, and dotted lines.}
 \label{heliumknu}
\end{figure}
All nonequilibrium calculations predict systematically, and
considerably, larger effects 
than earlier equilibrium calculations~\cite{kosc}-\cite{dr}.
These newer and more accurate works permit to close the window
in the mass range 10-20 MeV, which was not excluded by nucleosynthesis if
the permitted number of extra neutrinos flavors was 1. Now even if 1 extra
neutrino is permitted, the upper bound on $\mnt$ is about 1 MeV. If 0.3
extra neutrino flavors are allowed, the $\nut$ mass is bounded from above by
0.3 MeV. Though the accuracy in determination of $^4 He$ is the largest,
one can also include other light elements for obtaining the bound on
$\mnt$ mass. The effective number of extra neutrinos found in this way
in ref. \cite{dhsm} is presented in fig. \ref{elements}.
\begin{figure}[htb]
 \begin{center}
    \psfig{figure=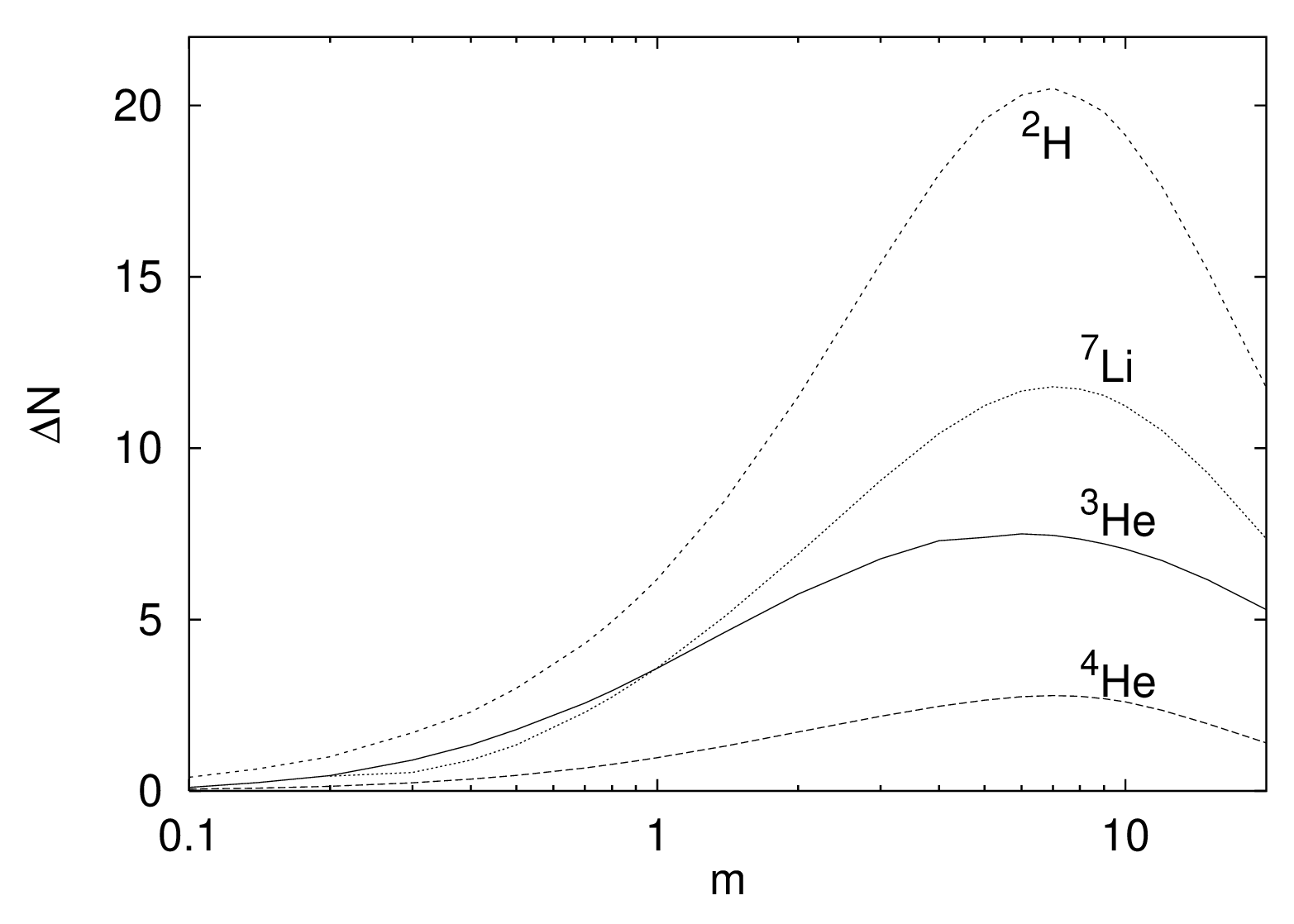,width=5in,height=3in}
  \end{center}
\caption{The effective number of equivalent massless neutrino species
$\Delta N = N_{eff} -3$ calculated from the abundances of deuterium
(dotted), $^7 Li$ (dense dotted),  $^3 He$ (solid), and $^4 He$ (dashed).
 }
\label{elements}
\end{figure}
The still existing confusion regarding the data on abundance of primordial
deuterium~\cite{tfb}-\cite{wcl}
makes it difficult to deduce a
reliable value for the ratio of the baryon-to-photon number densities,
$\eta_{10} = 10^{10} n_B/n_\gamma$, and to obtain a stringent bound on
$\Delta N $  (see discussion at the end of secs.~\ref{prns},\ref{knu}).
An independent determination of $\eta_{10}$ from the position of the
second acoustic peak in the angular spectrum of CMBR~\cite{2ndpeak}
may very much help in the near future.
It seems rather safe to conclude that $\Delta N <1$,
though quite probably a better limit $\Delta N<0.2$ is valid.
In this case the consideration of primordial nucleosynthesis
safely excludes the mass of $\nut$ in the interval $1- 22$ MeV.
Recall that it is valid for the
sufficiently long-lived $\nut$, i.e. for $\tau_{\nut} > 200$ sec.
Together with the direct experimental bound presented by eq.~(\ref{mnt}),
it gives $\mnt < 1$ MeV. This result is obtained for $\eta _{10}=3.0$. At
lower $\eta _{10} = 1-2$ the lower bound is slightly strengthen.
Hopefully a resolution of the
observational controversies in the light element abundances will permit to
shift this limit to even smaller values of $\mnt$. In particular, if
the limit on $\Delta N_\nu$ would return to the "good old" value,
$\Delta N_\nu < 0.3$, one could conclude from Fig.~\ref{heliumknu}
that $\mnt < 0.35$ MeV. In the case of the optimistic limit,
$\Delta N_\nu < 0.2$~\cite{bntt,tytler00} we find $\mnt < 0.2$ MeV.

The results obtained in the papers~\cite{hm1,dhsm}, where exact calculations
were performed, are valid for the Majorana $\nut$. The Dirac case
demands much more computer time, because an additional unknown
distribution function for right-handed massive $\nut$ should be taken 
into consideration. Simplified
calculations of refs.~\cite{ktcs,dr} have been also done for the Dirac case
under the assumption that for $\mnt > 1 $ MeV both helicity states are equally
populated. On the other hand,
the BBN bound on the Dirac mass of $\nut$ is considerably weaker
than the bound obtained from
SN1987~\cite{grifols90}-\cite{mayle93}, $\mnt < 10$ keV.

\subsection{Massive unstable neutrinos. \label{massunstbl}}

The bounds on $\mnt$ would be quite different if $\nut$ decayed during BBN
on the time scale  $0.1-10^3$ sec.
The corresponding effects were actively studied during last 20 years.
In the earlier papers several interesting physical effects were
observed but the accuracy of calculations was rather poor and the concrete
numerical results and bounds should be taken with care. The papers on this
subject written at the end of the 70th were briefly reviewed in ref.~\cite{dz}
but we will also discuss them here for the sake of completeness. The first
paper where the influence of both stable and unstable neutrinos on primordial
nucleosynthesis was considered was that by Sato and Kobayashi~\cite{sako}. It
was the only paper of the 1970s that correctly concluded that a
massive neutrino
with the mass in the range 1-20 MeV would noticeably change $Y_p$ -
the primordial abundance of $^4 He$. It was argued in other 
papers~\cite{dkt,dktw}
that $^4 He$ is not sensitive to a possible massive and decaying  neutrino.
It was also noticed in ref.~\cite{sako} that a radiative decay of neutrino,
$\nu_h \rar \nu_l +\gamma$ would change the baryon-to-photon ratio
$\eta$ at nucleosynthesis with respect to the present day value, and
correspondingly change the predicted fraction of primordial deuterium, 
which is very
sensitive to $\eta$. A similar conclusion was also
reached in ref.~\cite{dktw}, where the bound for life-time
$\tau_{\nu_h} < $ hours was derived. On the basis of considerations of
the total mass density in the universe and the combined
analysis of production of $^4 He$, as well as $^2 H$, and $^7 Li$
it was concluded~\cite{misa} that no neutrinos can exist in the mass range 
70 eV - 10 MeV if their life-time is bounded from below by
$\tau_{\nu_h} \geq 6\cdot 10^7 (1\,{\rm MeV}/m_{\nu_h})^5$ sec, i.e.
by the electroweak theory value.

It was noticed in ref.~\cite{lindl1} that neutrinos decaying into photons
or $e^\pm$ at a rather late time,
when nucleosynthesis was effectively completed, would change primordial
abundances
of previously produced light elements by their destruction through
photo-fusion.
However important secondary processes, as e.g. pair creation
$\gamma \gamma \rar e^+e^-$, or Compton scattering,
which led to thermalization of the decay products were neglected. The
process of pair creation was taken into account by the author 6 years
later~\cite{lindl2} and the following result was obtained: for
neutrino mass
in the interval 1-100 MeV its life-time should be below $3\cdot 10^3$ sec.
Considerations of this paper were extended in ref.~\cite{als}, where
photo-destruction of $^4 He$ was considered. Even if a minor fraction of
$^4 He$ were destroyed, the produced $^2 H$ and $^3 He$, which normally
constitute about $10^{-4}$ of $^4 He$, might strongly deviate from the
usual primordial abundances. This
effect was also mentioned in ref.~\cite{dz} but without any 
calculations. If that was the case, primordial nucleosynthesis 
would not constraint the
baryon density of the universe. However, all the results for
photo-disintegration of light elements (included those discussed below)
are valid only if the energy of the products of the decay are above
nuclear binding energy (2.2 MeV for deuterium and 28 MeV for helium-4).
We know now that the mass of $\nut$ is below 18 MeV and that there are
no other neutrinos with a larger mass.
Still we mention below other papers where photo-fission of light elements
was considered, partly because of historical reasons and partly because
physical effects could be of interest. Moreover these results with a slight
modification may be applied to other heavy long-lived particles, even
possibly to heavy neutrinos of the 4th generation.

The influence of radiative decays of a massive neutrino on light elements was
also discussed in the papers~\cite{kosc,sche}. In the first of them the main
emphasis was placed on the stable neutrino case, where quite accurate
results were
obtained (see section \ref{massstabnu}), but
the effects of neutrino decay in changing the value of $\eta$ and in
photo-dissociation of $^2 H$ were also considered (in the appendix).
However the consideration of photo-dissociation was based a on yet-uncorrected
paper~\cite{lindl1} and was numerically wrong. In a subsequent
paper~\cite{sche}
the photo-fission of deuterium by energetic photons coming from neutrino decay
was also considered and is also subject to the same criticism as above.
Nevertheless a new and
very interesting effect was noticed in this paper~\cite{sche}. Namely,
if a heavy neutrino decays into $\nue$, with the life-time
$\tau_{\nu_h} > 10^3-10^4$ sec,
then energetic electronic anti-neutrinos from the decay would produce
additional neutrons through the reaction $\bar \nu_e p \rar ne^+$.
These neutrons would catch
protons and form additional deuterium. This effect would permit to create
deuterium that would be consistent with observations and would allow to have a large cosmological baryonic number density, $\eta$.

The calculations of ref.~\cite{lindl2} were repeated with a better accuracy
in ref.~\cite{katesa}, where the spectra of photons and electrons coming from
radiative decays of massive neutrinos, with secondary processes included,
were calculated by numerical solution of
kinetic equation, and their role in photo-destruction of light elements was
re-estimated. The paper concludes that the life-time of a heavy
neutrino with mass in the interval 10 MeV and 1 GeV must be shorter than
$10^4 $ sec.

In ref.~\cite{tesa} the role played by non-radiatively decaying
neutrinos in
nucleosynthesis was considered. The authors took into account a change in the
expansion/cooling regime related to the contribution to the energy density
from a heavy $\nut$ and the effect found by Scherrer~\cite{sche} of creation
extra neutrons and ultimately deuterium by late produced $\bar \nue$.
For earlier decaying $\nut$ the spectral distortion of the $\nue$ caused
by the $\nue$ originating from the decay was also taken into account.
This effect would change the frozen neutron-to-proton ratio and subsequently
abundances of all light elements. It was noted~\cite{tesa} that for $\mnt<10$
MeV this effect could result only in a reduction of $n/p$-ratio. This
conclusion was not shared by ref.~\cite{dolkir}, where it was independently
found that a distortion of electronic neutrino spectrum by $\nue$ coming from
the decay of a heavy particle would have a strong influence on $n/p$ ratio but
this influence could go both ways. If the characteristic energy of the
produced $\nue$ is below the threshold of the reaction
$p \bar \nue\rar e^+ n $
($E_{th} \approx 1.8 $ MeV), then neutrons are not produced by excessive
$\bar\nue$, while they are efficiently destroyed in $\nue n\rar pe^-$. It
gives
rise to a smaller $n/p$-ratio. For $\nue$ above the threshold, neutron
creation is more efficient because protons are more abundant in the plasma and
$n/p$-ratio increases. As was shown in ref.~\cite{dolkir} for $\mnt=7-10$ MeV
and $\tau_{\nut} = 1 $ sec the $n/p$-ratio might be 25\% larger than the
canonical value, while for smaller masses, e.g. for $\mnt = 5$ MeV and
$\tau_{\nut} = 1$ sec the
$n/p$-ratio might smaller than the canonical one be by the same amount.

A generic and rather comprehensive study of nucleosynthesis
(though at that stage an approximate one) with
both invisible and electromagnetic decays of heavy particles was undertaken
in refs.~\cite{sctu1,sctu2}. An important effect discovered there is a
decrease of $^4 He$-production if the life-time of 
heavy particles is about 0.1 sec.
This phenomenon can be explained as follows. At $t>0.1$ sec the
temperature of the cosmic plasma is below 3 MeV so that $\num$ (and $\nut$)
are out of thermal contact with the plasma, while $\nue$ remain in 
strong contact. Hence the excess energy produced by the decay is almost
equally (in accordance with thermal equilibrium) shared between $\gamma$,
$e^\pm$ and $\nue$, while the other neutrinos would be under-abundant. It means
that the effective number of neutrino species becomes smaller than canonical
value 3 and $n/p$-ratio goes down. The results of ref.~\cite{sctu1} were
somewhat amended in a later paper~\cite{lesc} where neutrino heating
was not treated in instantaneous approximation. The improved calculations
diminishes a possible reduction of $^4 He$ down to $\Delta Y =-0.01$ but in
a larger range of life-times, $\tau = 0.1-0.7$ sec.

A detailed examination of the impact of radiative decays of neutrinos on BBN
was done in ref.~\cite{tekasa}, where all previously found effects were taken
into consideration with a somewhat better numerical precision:
1) an increase of the entropy due to decay and the
corresponding change of $\eta$; 2) the contribution into total cosmic energy
density from $\nu_h$ and its decay products; 
3) the destruction of light elements
by high energy photons created by the decay after light elements were
produced; 4) a shift of $n/p$-equilibrium by $\nue$ possibly produced in the
decay.
The conclusions of the paper are as follows. A heavy $\nut$ with
$\mnt > 30 $ MeV (now excluded by direct experiment), 
would induce too-strong photo-destruction of light elements 
if $\tau_{\nut} >10^4$ sec.
A lighter $\nut$ would contribute too much into the total energy density
and too much $^4 He$ would be produced. According to this paper, 
all pieces of data, including supernova bounds, permit the only
region for the radiative decays of $\nut$: $30\, {\rm MeV} < \mnt <
70\, {\rm MeV}$ and $10^2\, {\rm sec}  < \tau_{\nut} < 10^4\, {\rm sec}$.
The upper limit of 70 MeV comes from direct laboratory 
measurements of $\mnt$ available at that time. Since now we 
know that $\mnt < 18$ MeV, the results of
the paper~\cite{tekasa} imply that there is no space at all for radiatively
decaying massive $\nut$. Of course some small masses are not excluded, but
their limiting values are not presented in the paper. In particular, the
decay products of $\nut$ with $m_{\nut} <4.4$ MeV do not destroy deuterium,
to say nothing about $^4$He.

One more effect was pointed out in refs.~\cite{gss,lasc}. The authors
noticed that even if a massive neutrino did not directly produce photons
or electrons, energetic
neutrinos from the decay may interact with background neutrinos and
create $e^\pm$-pairs through the reaction $\bar \nu \nu \rar e^+e^-$.
In the second paper~\cite{lasc} thermalization of the decay products,
omitted in the first one~\cite{gss}, was taken into account, which
significantly changes the results for certain values of mass and life-time
of the decaying particle. This effect leads to some improvement of the
previous constraints on neutrino-producing heavy particles.
The secondary electrons and positrons could create energetic photons and
the latter in turn would destroy light elements or, if their energy is
higher than the binding energy of $^4 He$, would (over)produce $^2 H$ and
$^3 He$ as was indicated in refs.~\cite{dz,als}.
The results of the paper~\cite{lasc} are valid for a very heavy parent
particle, $ m=(1-1000)$ GeV, which could be a heavy lepton of the 4th generation. But, as we saw in sec. \ref{stabheavy}, the frozen number density of such
neutrinos could be too low (at least for some values of the mass) to produce
observable effects. The results found in paper~\cite{lasc} may be applicable
to supersymmetric partners and as such are not the subject of this review.
There is quite rich literature on cosmological constraints
for super-partners such as gravitino, neutralino, sneutrino. For the
discussion and a representative list of references one could address the
paper~\cite{hkkm}.

Decays of massive $\nut$ (with mass 17 keV) into invisible modes were
considered in ref.~\cite{mad}. It was argued there that due to the decay
$\nut \rar \nu_{e,\mu} + J$, where $J$ is a massless or light scalar boson,
light neutrinos, $\nu_{e,\mu}$ acquire chemical potentials and in the case
of decay into $\nue$ this changes the mass fraction of primordial $^4 He$
by $\Delta Y = 0.02-0.03$ for 
the life-time range $\taut = 3\cdot 10^{-4}-10^{-2}$ sec 
(this is the life-time of $\nut$ at rest; relativistic time
delay makes it much longer). The calculations of the paper have been
simplified by the assumption that all relevant particles are in kinetic 
equilibrium. Exact
calculations could noticeably change the results.

The next generation of papers treating BBN bounds on unstable massive
neutrinos appeared in the middle of the 1990s. The calculations, though still
approximate, were considerably more involved and included numerical
integration of kinetic equations, also approximate but more accurate than
previously.
The main contribution were done by Ohio~\cite{kks}-\cite{kkk} and
Chicago~\cite{dgt,gt} groups. In the first group of works the Boltzmann
kinetic equation was solved numerically under the following assumptions:
1) the products of the decay are in kinetic equilibrium; 2) their
distribution is described by pseudo-chemical
potential~\cite{adchem}-\cite{bern}:
\be
f = \left[ 1+ \exp \left( \xi + E/T \right) \right]^{-1}
\label{psdchem}
\ee
where $\xi$ and $T$ are functions of time only and do not depend on energy,
the pseudo-chemical potential $\xi$ has the same value for particles and
antiparticles if charge asymmetry is vanishing; 3) in some cases the validity
of Boltzmann statistics was assumed. In ref.~\cite{ksk} the inverse decay
was included into consideration for the first time. Under simplifying
assumptions described above, kinetic equations were reduced to ordinary
differential equations for functions of only one variable - time. These
equations were solved numerically. However, no accurate
calculations with the nucleosynthesis code were performed. The latter 
was included
in a subsequent paper~\cite{kkk}. The only decay mode that was considered
there was $\nut \rar \num + J$, where $J$ is a light or massless scalar. The
authors claim that they obtained, in particular, the strongest constraint for
the $\nut$ mass if $\nut$ is stable on BBN time scale.
This result disagrees, however, with the more precise calculations 
of refs.~\cite{hm1,dhsm} (see
discussion in sec. \ref{massstabnu}).

According to the paper~\cite{kkk}, in the case of decaying $\nut$,
if BBN permits 0.6 additional massless neutrino species,
the only range allowed for the mass and life-time is either
$\mnt \leq 0.1$ MeV for $\taut \geq 10^{-2}$ sec and
$\mnt \leq 0.1 (\taut /0.01 \,{\rm sec})$ MeV for
$\taut \leq 10^{-2}$ sec, or $(5-10)\,\, {\rm MeV} \leq \mnt \leq 31 $ MeV
provided that $\taut \leq 40$ sec;
$\nut$ with $\taut >40$ sec are excluded in the mass interval 0.1-50 MeV.
These results are compared to precise calculations of ref.~\cite{dhps}
below.

A much wider class of neutrino decays was considered in refs.~\cite{dgt,gt}.
The decays into electromagnetically interacting particles, $\nut\rar
\nu_{\mu,e} + \gamma$  or $\nut\rar \nu_{\mu,e} + e^+ + e^-$
as well as into sterile channels,
$\nut \rar \nu_{\mu,e} +J$, were discussed. The basic simplifying assumptions
were the following: 1) the number density of $\nut$ is assumed to be frozen;
2) inverse decay is not taken into account and low life-time limit is not
accurately treated; 3) Boltzmann approximation. An important improvement
with respect
to refs.~\cite{kks}-\cite{kkk} was an account of spectral distortion of light
neutrinos. The results of the papers~\cite{dgt,gt},
confirmed and quantitatively improved earlier statements, discussed above, that
in the case of decay $\nut\rar\nue$ the BBN constraint for the baryon
number density $\eta$ is about 10 times less restrictive than without decays,
so BBN would not prevent baryons to constitute all dark matter in the
universe. The results of refs.~\cite{dgt,gt} in the case of decay into
$\nu_{\mu,e}J$ are compared below with precise calculations of the
paper~\cite{dhps}.

A few papers related to the impact on BBN of electromagnetic decays of massive
particles, which are not necessary (but could be) heavy neutrinos, appeared
during the past few years, see e.g.~\cite{psb}-\cite{kamo},\cite{hkkm}
and references therein. A more precise treatment of  
electromagnetic cascades and correspondingly of the radiation spectrum
was developed. That permitted to improve the accuracy of the calculations of
photo-destruction of light elements.

In the case that such massive particles became non-relativistic and dominate
cosmic energy density before nucleosynthesis, rather strong constraints on
their properties could be derived from the condition that their decay products
must be thermalized and the universe must be reheated and come to thermal
equilibrium with $T_{reh}> 1 $ MeV, so that the
normal
BBN conditions would be created. However, as was noticed in
refs.~\cite{kakosu,kawasaki00}, thermalization
of neutrinos should be much slower than thermalization of other more strongly
interacting particles. The neutrinos are either non-thermally produced by the
decay, or created by reactions with secondary particles, as e.g.
$e^+e^- \rar \bar \nu \nu$. The effective number of neutrino species was
calculated in this paper by numerical solution of kinetic equation in Boltzmann
approximation and in the limit of $m_e =0$. It was found that if the
reheating temperature after decay is sufficiently high, 
$T_{reh}> 5-10$ MeV, then
$N_\nu \approx 3$ as in the standard model. However it does not mean that
smaller $T_{reh}$ are excluded. The authors demonstrated that for a smaller
$T_{reh}$, the number density of $\nue$ became smaller than in the standard
model and this resulted in a higher temperature of $n/p$-freezing and to
a lesser destruction of neutrons by $\nue$ after freezing. These two effects
could give rise to the normal primordial mass fraction of $^4 He$ even if
$N_\nu \ll 1$. The permitted value of $T_{reh}$ could be as small as 0.5 MeV.
In the subsequent paper by the same
authors~\cite{kawasaki00} the lower limit on the reheating temperature,
after late-time entropy production, was shifted to a slightly higher value,
$T_{reh} > 0.7$ MeV for leptonic and electromagnetic decay channels.
If the long-lived massive particles that create large additional 
entropy decay into hadrons with a branching ratio larger than 0.01, 
the reheat temperature should be larger than 2.5-4 MeV ~\cite{kawasaki00}.
These papers also mentioned that a constraint on the
effective number of neutrino species, or in other words, on the energy density
of relativistic matter can be found as well from the galaxy
formation~\cite{pedo} and from the future CMB measurements~\cite{ldht}
(see sec.~\ref{9cmb}).

A more accurate study of massive Majorana $\nut$ decaying into $\num+J$
was undertaken in ref.~\cite{kakosa}. The calculations were
done in non-relativistic approximation for $\nut$ and under assumption
of thermal equilibrium for $\nue$ and $\num$, so that the annihilation
could be treated in Boltzmann approximation. Scattering processes for
light neutrinos were neglected and only scattering of nonrelativistic
$\nut$ on equilibrium leptons were included. The effect of tau-neutrino
with mass and life-time in the intervals $10-24$ MeV and $10^{-4}-10^3$ sec
was studied. It was obtained, in particular, that for some values of $\mnt$
and $\taut$ the effect of decaying $\nut$ is to reduce the effective number
of neutrino species. For example, if $\mnt = 14 $ MeV and $\taut = 0.1$ sec,
the effective number of neutrinos is $N_\nu =2.5$; if $\mnt = 10 $ MeV
and $\taut = 1$ sec, $N_\nu =2.85$. These results are in a good
agreement with exact calculations of ref.~\cite{dhps}, see below
fig.~\ref{muknu}.

Numerical solutions of the complete system of kinetic equations without
any simplifying approximations were done in two
works~\cite{stha,dhps}. In ref.~\cite{stha} the decay
$\nut\rar \nue+J$ in the mass interval $0.1<\mnt<1$ MeV was studied,
while in ref.~\cite{dhps} both invisible decays
$\nut\rar \nue+J$ and $\nut\rar \num+J$
were discussed in the mass range 0.1-20 MeV. We will concentrate on the
last paper~\cite{dhps}, which is more complete and more accurate numerically.
It was assumed there that $\nut$ is a Majorana type fermion which is
coupled to a scalar boson $\phi$, possibly a majoron or
familon~\cite{maj,fam} (see also the papers~\cite{chikashige80,maj2}),
which is light or even
massless. The coupling of $\phi$ to neutrinos may have diagonal terms as e.g.
$g_1\bar \nut \nut \phi $ which are important for elastic scattering
$\nut + \phi \leftrightarrow  \nut + \phi$ and annihilation
$\bar \nut + \nut \leftrightarrow  2\phi$. The non-diagonal coupling
$g_a\bar \nut \nu_a \phi $ is responsible for the decay of $\nut$ into
lighter neutrinos, $\nue$ or $\num$ (correspondingly $a=e$ or $\mu$).
It is usually assumed
that one of these two couplings dominates, i.e. $\nut$ predominantly
decays either into $\nue \phi$ or $\num \phi$ and these two
possibilities are considered separately. It is also assumed
that both $\nue$ and
$\num$ are the usual active neutrinos. Since chirality is changed by the
coupling to a scalar field, the corresponding light neutrinos should also
be Majorana particles, otherwise new sterile states would be produced by
the decay. The scalar boson $\phi$ is supposed to be a weak singlet,
because the LEP measurements~\cite{pdg} of the total decay
width of $Z^0$ do not leave room for any other light weakly interacting
particles except those already known.

There are several possible ways of production of $\phi$ in the
primeval plasma.  The first and evident one is through the decay $\nut
\rightarrow \phi + \nu_a$.  Another
possibility is the annihilation $\nut+\nut \rightarrow \phi+\phi$ and
the third one is a possible non-thermal production in the course of a
phase transition similar to the production of axions at the QCD phase
transition. We neglect the last possibility, assuming that even if
(pseudo)goldstone bosons were created in the course of the phase
transition, the phase transition took place early enough so that the
created bosons were diluted by a subsequent entropy release in the
course of the universe cooling down. The rate of $\phi$-production in
$\nut$-annihilation can be estimated as:
\be
{\dot n_\phi \over n_{\nut} } = \sigma_{ann} v  n_{\nut} ,
\label{dotnphi}
\ee
where $v$ is the relative velocity and $ \sigma_{ann}$ is
the annihilation cross-section. In the limit of large energies,
$s=4E^2_{cm} \gg \mnt^2$ it is equal to:
$\sigma_{ann} \approx (g_1^4 /32\pi s)\ln (s/m^2_{\nut} )$
(see e.g.~\cite{dprv}). One can
check that this rate is small in comparison with the universe expansion
rate $H=\dot a /a$, if
$g_1 <10^{-5}$. In this case the production of Majorons through annihilation
can be neglected and they would dominantly be produced through the decay of $\nut$.
The opposite case of dominant production of $\phi$'s by $\nut$-annihilation
and their influence on nucleosynthesis was
approximately considered in ref.~\cite{dprv}.

The life-time of $\nut$ with respect to the decay into massless
particles $\phi$ and $\nu_a$ is equal to:
\be
\tau_{\nut} = {8\pi \over g^2_a \mnt } {\sqrt {\mnt^2 +9T^2} \over \mnt}
\label{tau}
\ee
where the last factor accounts for the relativistic time delay.
The decay would be faster than the universe expansion rate
at $T\sim \mnt$ if $T<0.3\cdot  10^{10} g_a\sqrt \mnt $,
where the temperature $T$ and $\mnt$ are
expressed in MeV. The interval of life-times of $\nut$, which we will consider
below - $\tau_{\nut} = (10^{-3}-10^3)$ sec -  corresponds to
$g_a \sqrt{\mnut}= \left( 4\cdot 10^{-9} - 6.3\cdot 10^{-12}\right)$.
Thus there is a large range of parameters (coupling constants and masses)
for which decay is essential while annihilation is not.
These parameter values are not in conflict with the astrophysical
limit $g_a \left( {\rm MeV} /\mnt \right)^{1/2}
< 3\cdot 10^{-7}$~\cite{raffelt96} (page 563).

In ref.~\cite{dhps} the BBN impact of unstable $\nut$
decaying into invisible channels $\nut \rar \nu_{e,\mu} + \phi$
was treated without any approximations through numerical solutions of
exact kinetic equations.
The basic equations governing the evolution of the distribution
functions $f_a$ ($a=\nue$, $\num$, $\nut$, and $\phi$) are discussed in
some detail in sec. \ref{masslessdistr}. Now there is a new unknown
function $f_\phi (p,t)$ and a new contribution to the collision integral
from the decay:
\be
(\partial_t - Hp_j\partial_{p_j}) f_j (p_j,\,t) = I^{scat}_{j}+I^{decay}_{j},
\label{dtf1}
\ee
where the collision integral for two-body reactions
$1+2 \rightarrow 3+4$ is given by the expression (\ref{icoll}) and the
"decay" parts of the collision integral for different initial particles
are:
\be
I^{decay}_{\nut} = -{ m \over E_{\nut} p_{\nut} \taut }
\int_{\left(E_{\nut}+p_{\nut}\right) /2}^{\left(E_{\nut}-p_{\nut}\right) /2}
dE_\phi F_{dec} (E_{\nut},\, E_\phi,\, E_{\nut}-E_\phi),
\label{decnt}
\\
I^{decay}_{\nu_a} = { m \over E_{\nu_a} p_{\nu_a} \taut}
\int^\infty_{\mid\left( m^2/4p_{\nu_a}\right) -p_{\nu_a} \mid}
{dp_{\nut}p_{\nut} \over E_{\nut} }
F_{dec} (E_{\nut},\,E_{\nut}-E_{\nu_a} ,\, E_{\nu_a} ),
\label{decnua}
\\
I^{decay}_{\phi} = {2  m \over E_{\phi} p_{\phi}\taut }
\int^\infty_{\mid\left( m^2/4p_{\phi}\right) -p_{\phi} \mid}
{dp_{\nut}p_{\nut} \over E_{\nut} }
F_{dec} (E_{\nut},\, E_\phi,\, E_{\nut}-E_{\phi} ),
\label{decphi}
\ee
where $m$ is the mass of $\nut$ (we omitted the index $\nut$) and:
\be
F_{dec} (E_{\nut},\, E_\phi,\, E_{\nu_a})=
f_{\nut}(E_{\nut})\left[1+f_{\phi}(E_\phi)\right]
\left[1-f_{\nu_a}(E_{\nu_a})\right]  \nonumber \\
-f_{\phi}(E_\phi)f_{\nu_a}(E_{\nu_a}) \left[ 1- f_{\nut}(E_{\nut}) \right].
\label{fdec}
\ee

The contribution of the decay term, $I^{decay}$, into the collision integral
of eq.~(\ref{dtf1}) is considerably simpler for numerical calculations than
the contribution of scattering, $I^{scat}$,
because the former is only one-dimensional, while the scattering terms
can be reduced to no less than two dimensions.
Technical details of the calculations and modification of the
nucleosynthesis code are described in the paper~\cite{dhps}.
Before discussing the results of the calculations, it is worth mentioning
that possible effects of neutrino oscillations on primordial
nucleosynthesis were not taken into account. According to a recent Super
Kamiokande result~\cite{sk98} $\num$ may be strongly mixed with $\nut$
with a very small mass difference $\delta m^2 = 10^{-2}- 10^{-3} {\rm eV}^2$.
If that is the case then $\mnt < 160 $ keV and the results obtained for
a larger mass of $\nut$ would be irrelevant. However if $\num$ is mixed
with a sterile neutrino (which is almost ruled out now) then the
mass difference between $\nut$ and $\num$ can be large, and
the oscillations may be unimportant. If this is the case then
the $\nut$ mass is only restricted by a loose
laboratory limit (\ref{mnt}) and BBN constraints are of interest.
On the other hand, if all three known
neutrinos are light, then the results presented here may be applicable to new neutrinos of a possible fourth generation.

The impact of decaying $\nut$ on BBN is significantly different for the decay
$\nut\rar \phi \num$ and $\nut\rar \phi \nue$. In the first case
the most important effect is an overall change in the total energy density
and a corresponding change of the universe cooling rate. Nonequilibrium
corrections to the spectra of $\nue$ are relatively weak for a small
life-time, so practically all $\nut$ have already
decayed at the moment of neutron-proton
freezing, $T\approx 0.6 $ MeV. For a larger life-time, some nonequilibrium
$\nue$ would come from annihilation $\nut +\nut \rightarrow \bar \nue + \nue$
and, as we have already discussed, would directly change the
frozen $n/p$-ratio. The
distortion of the $\nue$ spectrum is much stronger in the case of the decay
$\nut \rightarrow \nue +\phi$. Moreover, the electron neutrinos originating
from the decay at later times would over-produce deuterium, as found in
ref.~\cite{sche}.

\begin{figure}[htb]
\begin{center}
  \leavevmode
  \hbox{
    \epsfysize=3.0in
    \epsffile{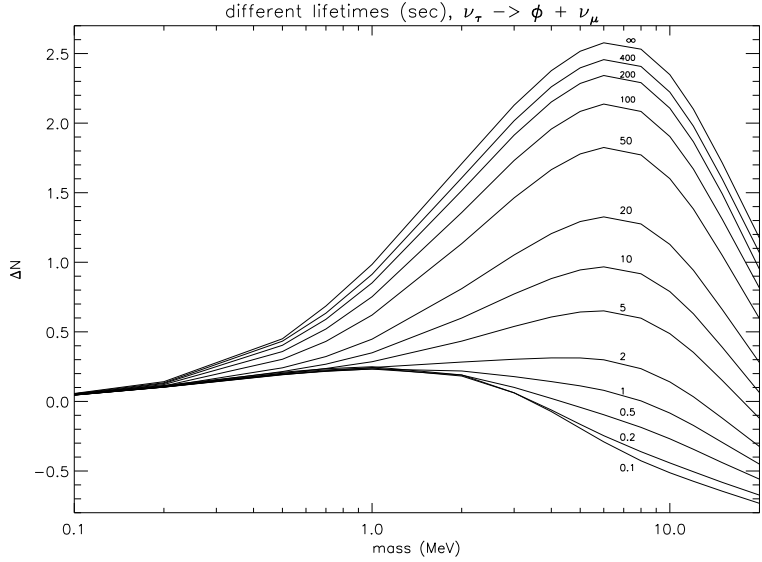}}
\end{center}
\caption{The number of equivalent massless neutrino species, $\Delta N =
N_{eqv} - 3$, as a function of $\nut$ mass and  lifetime $\tau$, found
from $^4He$ in the case of the decay $\nut\rar \num +\phi$.
\label{muknu}}
\end{figure}
First we present and discuss the results for a case of decay into
$\num\phi$-channel. In fig. \ref{muknu} the effective number of massless
neutrino species, which would give the same mass fraction
of $^4 He$ as a massive
$\nut$ decaying into $\num\phi$, is presented for different life-times as
a function of $\nut$ mass. For large masses and low life-times $\Delta N $
is negative. This is related to the decrease of the energy density if
all $\nut$ have completely decayed.
Because of that
$\Delta N=-1 $ which is partly
compensated by the production of scalars $\phi$ giving $\Delta N \approx 0.5 $
Thus if $\mnt = 10$ MeV and $\tau_{\nut} = 0.1$ sec,
the effective number of neutrino species at nucleosynthesis would be
only 2.5.

A comparison with the results of other groups shows a rather strong deviation.
We ascribe this to the simplifying approximations made in the earlier papers,
which have apparently given rise to a significant difference with the exact
calculations,
and to a better accuracy of our numerical calculations, which is typically
at the fraction of per cent level. For example in the case of
$\mnt = 14 $ MeV and $\taut = 0.1 $ sec we obtain for the energy density
$\rho/ \rho^{eq}_{\nu_0} = 2.9$, while the group~\cite{kakosa} obtained 2.5.
In the limit of small life-times and masses our result is 3.57
(this is the energy density of three light neutrinos and one scalar), while
the results of~\cite{kakosa} are close to 3.9. The effective number of
neutrino species found from $^4$He in our case is
$3+\Delta N=2.9$ for $\mnt = 10$ MeV and
$\taut = 1 $ sec, while that found in ref.~\cite{kkk} is 3.1. The difference is
also large for $\mnt = 10$ MeV and $\taut = 0.01 $ sec: we find
$3+\Delta N=2.66$ and the authors of~\cite{ksk} obtained 2.86. In view of
the approximations made in the latter paper, it can be considered good
agreement.

It was shown in ref.~\cite{sche} that late decaying $\nut$ with $\taut =
10^3-10^4$ sec and $\mnt > 3.6 $ MeV would strongly distort deuterium
abundance if the decay proceeded into electronic neutrinos.  These
$\nue$ would create excessive neutrons through the reaction $\nue + p
\rightarrow n + e^+$, which would form extra $^2$H.
This is seen clearly in fig.~\ref{eh2ofm},
\begin{figure}[htb]
\begin{center}
  \leavevmode
  \hbox{
    \epsfysize=3.0in
    \epsffile{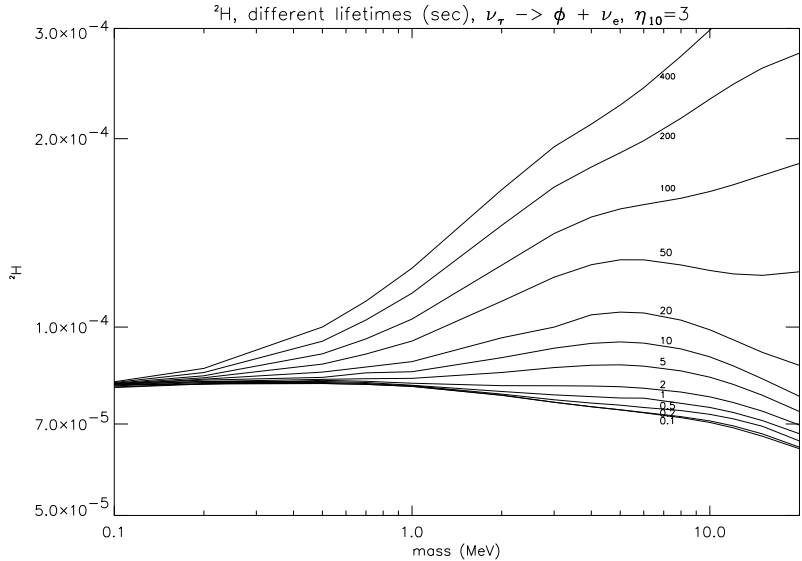}}
\end{center}
\caption{Primordial deuterium (by number) for different $\taut$ as a
function of tau-neutrino mass in the case of the decay $\nut\rar \nue +\phi$.
\label{eh2ofm}}
\end{figure}
where $^2H$ clearly increases
as a function of lifetime. The extra deuterium production goes up 
quadratically with the baryon density, and it was indeed 
observed in ref.~\cite{dhps}, that the effect 
is much less pronounced for low $\eta_{10}$.

The mass fraction, $Y_p$, of primordial helium-4 is presented in
fig.~\ref{ehe4ofm}.
\begin{figure}[htb]
\begin{center}
  \leavevmode
  \hbox{
    \epsfysize=3.0in
    \epsffile{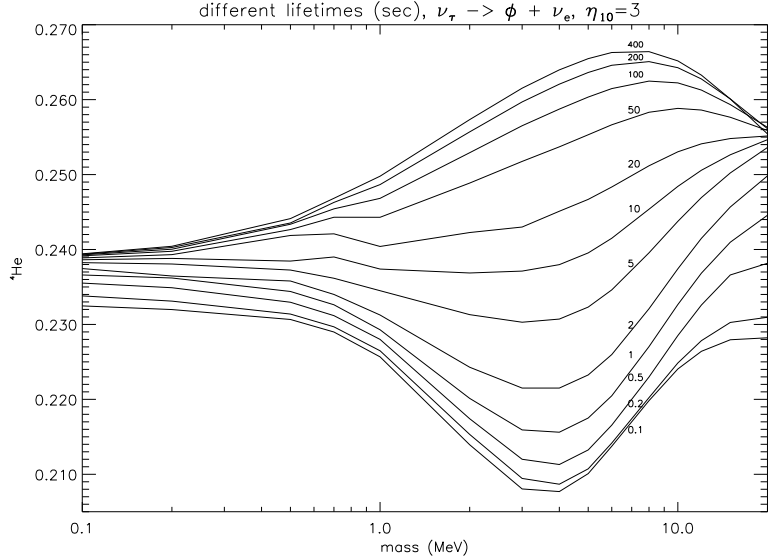}}
\end{center}
\caption{Primordial helium-4 (by mass) for different $\taut$ as a
function of tau-neutrino mass  for the case of decay
$\nut\rar \nue +\phi$.
\label{ehe4ofm}}
\end{figure}
For this channel as well, there is noticeable disagreement with
previous papers. E.g. for
$m=0.6$ and $\tau = 100$ sec we find $Y({^4 \mbox{He}}) \approx 0.244$,
whereas ref.~\cite{stha} obtains  $Y({^4 \mbox{He}}) \approx 0.20$.

More graphs showing various elements ($^2H$, $^4$He and $^7$Li) as
functions of mass and lifetime, for both channels
$\nut \rightarrow \nu_\mu + \phi$ and
$\nut \rightarrow \nu_e + \phi$ and for $\eta_{10} = 1,3,5,7,9$
can be found on the web-page:
{\tt http://tac.dk/\~{}sthansen/decay/}
together with plots of the n-p reaction rates.
The calculated abundances of light elements~\cite{dhps} for the case
of the decay into $\nu_e$ disfavor the low
and high values of life-time of the model of
ref.~\cite{dodelson94}, where tau-neutrino with the mass in the
interval 1-10 MeV and life-time 0.1-100 sec was invoked to remedy
the CDM model of large scale structure formation.

The previous results were obtained under assumption that the only source
of light scalars $\phi$ were the decays of $\nut$. We may consider
the opposite extreme, assuming that at the initial moment $x_{in}$
the majorons were in thermal equilibrium,
$ f_\phi (x_{in}) = 1/[\exp(y)-1]$.
This situation could be realized if majorons
were produced by some other mechanism prior to the $\nut$ decay
as discussed above. In the case of non-vanishing $f_\phi$
the inverse decay is evidently more efficient
than for $f_\phi (x_{in}) = 0$ and $\rho_{\nut}$
decreases slower.
The change in $\Delta N$, as compared to the case when
$f_\phi (x_{in}) = 0$, varies between 0.4 and 1.0,
$\Delta N _{(f_\phi = f_{eq})} = \Delta N _{(f_\phi = 0)} + (0.4 - 1.0)$,
depending on mass
and lifetime. In particular, for long lifetimes this difference goes to
0.57 for all masses, as can be expected.

\subsection{Right-handed Dirac neutrinos \label{ssec:nur}}

It is usually assumed that neutrinos are ``left-handed'', i.e. they have
only one helicity state, negative for neutrinos (spin is anti-parallel to
the momentum) and positive for antineutrinos. If they are strictly massless
and interact only with left-handed currents then another spin state
would never be excited. A non-zero mass permits to make a Lorenz boost
into neutrino rest frame and moreover to change the direction of its
momentum into the opposite one, thus kinematically changing a left-handed
$\nu$ into right-handed one. If neutrinos are massive then the population
of right-handed states in the primeval plasma should be non-vanishing
and they could influence primordial nucleosynthesis by enlarging the
effective number of neutrino species (see sec.~\ref{knu}). This is true
only for the Dirac mass, while Majorana neutrinos, massive or massless,
have the same number of degrees of freedom. The problem of mass-generated production of ``wrong-helicity'' Dirac neutrinos was first
discussed in refs.~\cite{dz,shws,ad1}. The probability of production
of right-handed neutrinos in weak interaction reactions with left
currents is suppressed at high energies as $(m_\nu /E)^2$. It was shown
that light right-handed neutrinos with masses below the Gerstein-Zeldovich
limit, $m_\nu < 30 $ eV (sec.~\ref{gerzel}), are never in thermal equilibrium
and their energy density at BBN is always negligible. A simple estimate
can be done as follows. The production rate of right-handed neutrinos generated by their mass is approximately given by
\be
\Gamma_R^m = (m_\nu/E)^2 \Gamma_L \sim H\,(m_\nu/T)^2 (T/T_W)^3
\label{gammar}
\ee
where $T_W$ is the decoupling temperature of ``normal'' left-handed
neutrinos and $H$ is the Hubble parameter. As follows from the estimates
given at the beginning of sec.~\ref{gerzel}, $T_W = 2-3$ MeV. Equilibrium
with respect to $\nu_R$ would be established if $\Gamma_R \approx H$.
Using formally eq.~(\ref{gammar}) we find that it could be achieved at
$T>10^7$ GeV. However, that is definitely incorrect because at
$T\geq M_{W,Z} $ the cross-sections of weak reactions with neutrino do not
rise with energy as $E^2$ but decrease as $1/E^2$ (see
sec.~\ref{sec:heavynu}).
The maximum contribution to the production rate of $\nu_R$ is given by
the decays of real $W$ and $Z$ bosons~\cite{ad1} (it is exactly the same
as the resonance contribution from the scattering:
${\rm all} \rar (W,\,Z) \rar \nu + ...$).

The $\nu_R$ production rate through decays of intermediate bosons can
be estimated as follows
\be
\Gamma_R^{(d)} = {\dot n_{\nu_R} \over n_{\nu_L}}
= \left({m_\nu \over T}\right)^2
{\Gamma_W^\nu n_W + \Gamma_Z^\nu n_Z \over 0.1 T^3}
\label{gammard}
\ee
where $n_{\nu_L} \approx 0.1\, T^3$ is the equilibrium number density of
left-handed neutrinos, $n_W \approx 3\,(m_W T/2\pi)^{3/2} \exp (-m_W/T)$
is the number density of $W$ (and the same for $Z$), and
$\Gamma_W^\nu = 0.21 $ GeV and $\Gamma_Z^\nu = 0.17$ GeV are the decay
widths of $W$ and $Z$ into the channels with a certain neutrino flavor.
We used non-relativistic expressions for the number densities of
$W$ and $Z$ (\ref{nb}) because, as we will see in what follows, the maximum
rate is achieved at $T<m_{W,Z}$. To create equilibrium
$\Gamma_R^{(d)}$ should
be larger than the Hubble parameter $H= 1.66 g_*^{1/2} (T) T^2 /m_{Pl}$.
Correspondingly the equilibrium could be established if
\be
m_\nu > T \left[{ 0.166 g_*^{1/2} T^5 \over m_{Pl}
\left(\Gamma_W^\nu n_W + \Gamma_Z^\nu n_Z \right)}\right]^{1/2}
\label{mnur}
\ee
The maximum production rate is reached at $m_{W,Z}/T \approx 3.5$. 
Estimating the r.h.s. at this temperature we find that the equilibrium
could be achieved if $m_\nu > 2$ keV. In other words, for $m_\nu < 30$ eV
equilibrium is never established even with quite efficient production
of $\nu_R$ by $W$ or $Z$ decays~\cite{ad1}. Similar results were
obtained in refs.~\cite{antonelli81,enqvist96}, 
though in the last paper the two-body
resonance-dominated reactions were considered and it was claimed, in
contrast to the above arguments, that the two-body $W$ and $Z$ decays
are negligible.
Right-handed neutrinos could be produced in equilibrium amount only at
Planck temperature (if such high temperature state ever existed in the
universe history) by gravitational interaction which is helicity blind.
However, the entropy dilution by massive particle annihilation diminishes
the relative energy density of $\nu_R$ at BBN by the factor
$(g_* (T=1\, {\rm MeV}) /g_* (T\approx 100\,{\rm GeV})^{4/3} \approx 0.05$
and even more for the case of production at the Planck epoch. The
entropy suppression might be not so strong in the case of multi-dimensional
theories with the Planck scale as low as TeV~\cite{arkani98} or
even lower~\cite{dvali01}.

Another possible way of creating (even massless) right-handed neutrinos
is through direct interactions of $\nu_R$ with right-handed currents. If one
assumes that the right-handed interaction has the same form as the
left-handed one but with heavier intermediate bosons, one can obtain from
BBN a lower limit on their mass. This was first done in
refs.~\cite{steigman79,olive81} where the limit obtained was
$m_{W'} > 53 m_W$, if the allowed number of extra neutrino species at BBN
was one. The
calculations go as follows. The relative rate of $\nu_R$ production through
new integrations with right-handed $W'$-bosons scales as:
\be
\Gamma'_R/H = \left( T /T_W \right)^3\,\left ( m_W /m_{W'}\right)^4
\label{gamma'R}
\ee
Thus at some high temperature $\nu_R$ would be abundantly produced. However
their energy density would be diluted at BBN by the factor
$[g_* (T_{prod})/g_* (1\,{\rm MeV})]^{4/3}$ with $g_* (1\,{\rm MeV}) =10.75$.
For $T\sim 100$ MeV but below the QCD phase transition $g_* = 17.25$ and the
suppression factor is 0.53. Taking into account that there are 3 neutrino
flavors, there would be 1.6 extra neutrino flavors at BBN - and that is excluded by the data. So the decoupling of $\nu_R$-production should be
above the QCD phase transition when $g_* \geq 58.25$. In this case the
energy density of $\nu_R$ is suppressed at BBN by the factor
$3\times 0.105 = 0.315$. Thus the limit can be found
\be
m_{W'}/m_W > \left( T_{QCD} /T_W \right)^{3/4}
\label{mw'mw}
\ee
In ref.~\cite{steigman79} the temperature of the QCD phase transition
was assumed to be 200 MeV and the decoupling temperature of left-handed weak
interaction was taken as $T_W=1$ MeV. That's how the limit
$m_{W'}/m_W >53$ was obtained.
However the production of $\nu_R$ goes through the annihilation
of right-handed charged leptons and not through the much stronger elastic
scattering, which conserves the number density of participating particles.
Correspondingly, for the decoupling temperature of weak interactions one
should take a larger value - $T_W=3-5$ MeV. In this case the limit
would be considerably weaker, $m_{W'} > 1 $ TeV~\cite{ad1}. On the other
hand, the scaling assumption~(\ref{gamma'R}) is not precise and at
$T>T_{QCD}$, the new reaction channels including quark annihilation would
be open. This would result in a stronger lower limit. If the allowed
extra number of neutrino species $\Delta N< 0.3$, then the decoupling
of $\nu_R$ should take place at $T> 1$ GeV and $m_{W'} > 10$ TeV. If
the limits on $\Delta N$ are further improved to $\sim0.15$, so that
decoupling
moves to temperatures higher than that of the electroweak phase transition, then the limit becomes
$m_{W'} > 300 $ TeV. These results differ somewhat from those presented
in ref.~\cite{olive99st}. For example, for the same bound $\Delta N_\nu <0.3$
the authors of that paper requested the decoupling temperature of $\nu_R$
to be higher than 300 MeV and correspondingly $m_{Z'} >2.8$ TeV. The
difference is related to a different choice of decoupling temperature,
1 GeV in the first case and 300 MeV in the second case. The number of degrees
of freedom in the last case is $g_* = 58.25$, while in the first 
it is 68.75, due to addition of the charmed quark. The entropy suppression
factors are respectively 0.105 and 0.084. So three right-handed neutrinos
correspond effectively to 0.315 or 0.25 normal neutrino species. The example
shows the strong sensitivity of the result to the bound on $\Delta N_\nu$.

If neutrinos are unstable on the cosmological time scale, then their mass
is not restricted by 30 eV and the right-handed partners
could be noticeably produced in the early universe. The condition
that they would not strongly disturb BBN permits to put an upper
limit on their mass. This question was first raised
in ref.~\cite{fuller91}, where an approximate limit
$m_\nu < 300\, {\rm keV}$ was obtained assuming that the QCD transition
temperature was 100 MeV and $\Delta N_\nu < 0.4$. In ref.~\cite{lam91} a
different bound was obtained, $m_\nu < 430\, {\rm keV}$, under the assumption
that the dominant mechanism of production of $\nu_R$ is the decay
$\pi^0 \rar \nu_R \bar\nu_R$ and that $\Delta N_\nu < 0.3$. In contrast to
the previous bound this one does not depend upon the value of the QCD
transition temperature. Significantly weaker limits were obtained in
ref.~\cite{enqvist93}: $m_{\nut} < 740\, {\rm keV}$ and
$m_{\num} < 480\, {\rm keV}$ for $T_{QCD} = 200$ MeV, but several important
processes of creation of $\nu_R$ were overlooked there. The limits were
strongly improved in ref.~\cite{dolgov95a} where all essential processes
of production of right-handed neutrinos were taken into account and,
in particular, the decay $\pi^{\pm} \rar \mu^{\pm} \nu_R$ not included in
the earlier research. The limits depend upon the $T_{QCD}$ and read:
$m_{\num} < 170\, {\rm keV}$ and $m_{\nut} < 210\, {\rm keV}$
for $T_{QCD} = 100$ MeV and
$m_{\num} < 150\, {\rm keV}$ and $m_{\nut} < 190\, {\rm keV}$
for $T_{QCD} = 200$ MeV, all for $\Delta N_\nu < 0.3$. If a very optimistic
limit, $\Delta N_\nu < 0.1$, is taken, then the right-handed neutrinos should
decouple at or before the electroweak phase transition and the masses
should be bounded by 10 keV. In this case the limit is similar to that
found from the consideration of the cooling of supernova
SN87~\cite{grifols90,burrows92,mayle93}.
The results of ref.~\cite{dolgov95a} were further improved in the
paper~\cite{fko} where more accurate calculations of $\nu_R$ production
and of their impact on BBN were performed. The bounds are roughly 30\%
stronger and are as follows. For $\Delta N_\nu < 0.3$:
\be
m_{\nu_\mu} \: \leq \:
\left\{
\begin{array}{lll}
&130~{\rm keV} &
\;  \; T_{QCD}=100~{\rm MeV} \\
&120~{\rm keV} &
\;  \; T_{QCD}= 200~{\rm MeV} \nonumber
\end{array}
\right. \, \nonumber \\
m_{\nu_\tau} \: \leq \:
\left\{
\begin{array}{lll}
&150~{\rm keV} &
\;  \; T_{QCD}=100~{\rm MeV} \\
&140~{\rm keV} &
\;  \; T_{QCD}= 200~{\rm MeV.}
\end{array}
\right.
\label{mnur03}
\ee
while for $\Delta N_\nu < 1.0$, they are
\be
m_{\nu_\mu} \: \leq \:
\left\{
\begin{array}{lll}
&310~{\rm keV} &
\;  \; T_{QCD}=100~{\rm MeV} \\
&290~{\rm keV} &
\;  \; T_{QCD}= 200~{\rm MeV} \nonumber
\end{array}
\right. \, \nonumber \\
m_{\nu_\tau} \: \leq \:
\left\{
\begin{array}{lll}
&370~{\rm keV} &
\;  \; T_{QCD}=100~{\rm MeV} \\
&340~{\rm keV} &
\;  \; T_{QCD}= 200~{\rm MeV.}
\end{array}
\right.
\label{mnur10}
\ee

These limits are much stronger than laboratory limits for $\nut$ mass
and comparable to the limit on ${\num}$ mass. They are applicable if the
neutrino life-time is longer than the characteristic time of nucleosynthesis
but shorter than the universe age.

An unusual case of right-handed neutrinos that are heavier than the
right-handed intermediate bosons, was considered in ref.~\cite{adhya99}.
The authors calculated the frozen number density of such heavy neutrinos
as discussed in sec~\ref{sec:heavynu}. From the condition
$\rho_{\nu_R} < \rho_c$ they
found an upper limit on $\nu_R$ mass of approximately
700 GeV. This result is incompatible with the initial assumption of
$m_{\nu_R} > m_{W'}$, because as we see in this section $W'$ should be
heavier than $\sim$TeV. Hence, cosmology forbids right-handed neutrinos
with $m_{\nu_R} > m_{W'}$. However this conclusion would be invalidated
if there exists an additional anomalous interaction of $\nu_R$
or in the obvious case of an unstable heavy neutrino~\cite{adhya00}.

Another type of right-handed neutrinos coupled to the usual intermediate
bosons but with a weaker strength was considered in
ref.~\cite{fargion84-r}. The authors derived mass/life-time limits
from the total cosmological energy density, CMBR, and BBN. According
to the author's conclusion, such neutral fermions could exist in
the mass and life-time ranges of $0.1 - 1$ GeV and $10^{-4}-10^6$ sec
respectively. A similar question was raised in ref.~\cite{masso94} a
decade later. It was assumed that there existed a new
neutrino-neutrino interaction where both left- and right-handed neutrinos
participated. Assuming that the interaction is described by the 4-fermion
coupling, $F(\bar\nu_R \nu_R)(\bar\nu_L \nu_L)$, the authors concluded
from BBN that the coupling strength is bounded by
$ F < 3\cdot 10^{-3} G_F$. In the case that this interaction is mediated
by a massless boson exchange the Yukawa coupling constant of this boson
to neutrinos should be smaller than $2\cdot 10^{-5}$ (see also
sec.~\ref{sec:majbbn}).

\subsection{Magnetic moment of neutrinos. \label{ssec:magnmom}}

If neutrinos are massive with Dirac mass,
they should have a non-zero magnetic moment. On
the other hand, the existence of a magnetic moment does not strictly
imply a non-zero neutrino mass, though its absence in this case would
be highly unnatural because chiral invariance, which prevents from
generation of mass, is broken by magnetic moment and interaction
with an electromagnetic field through magnetic moment
always changes helicity and excites right-handed neutrino states.
In the standard electroweak model modified only by a non-zero mass of
neutrino, with $\nu_R$ being $SU(2)$-singlet, the
magnitude of neutrino magnetic moment is given by eq.~(\ref{munu})
and is extremely small~\cite{lee77,marciano77,fujikawa80,lynn81}.
If $\mu_\nu$ is that small, the magnetic interaction of neutrinos 
would be unnoticeable in cosmological phenomena. However, in
some extensions of the standard model the magnitude of $\mu_\nu$ might be
much larger, up to $(10^{-10}-10^{-11}) \mu_B$ (see
e.g.~\cite{voloshin86}-\cite{mclaughlin99} and references
therein). In this case $\mu_\nu$ could be cosmologically interesting.
Direct experimental limits on diagonal magnetic moments of different types
of neutrinos are given by expressions~(\ref{muexsp}).
A consideration of stellar evolution permits imposing more stringent limits
at the level $(10^{-10}-10^{-12})\,\mu_B$, see the book~\cite{raffelt96}.
Cosmology and, in particular, big bang nucleosynthesis give similar bounds.
As we have seen in the previous section, the excitation of additional
right-handed neutrino states would change primordial abundances of light
elements. If $\mu_\nu$ is non-vanishing, then neutrino interactions with 
electromagnetic field would excite $\nu_R$ because the coupling
$\bar \nu \sigma_{\alpha\beta} q^\alpha \nu$ mixes $\nu_L$ and $\nu_R$.
There are two possible types of processes in the early universe in which
neutrino spin-flip might take place: first, the production of $\nu_R$ in
helicity changing processes, either in particle collisions,
$e^\pm + \nu_L \rar e^\pm + \nu_R$ and
$e^-+e^+ \rar \nu_{L,R} +\bar \nu_{R,L}$ or in the
plasmon decay, $\gamma_{pl} \rar \bar \nu_{L,R} + \nu_{R,L}$; second,
the classical spin rotation of neutrinos in large scale primordial magnetic
fields that might exist in the early universe. The former mechanism was
first considered in ref.~\cite{morgan81}, while the second one in
refs.~\cite{shapiro81,lynn81}.

In ref.~\cite{morgan81} the production of $\nu_R$ through the
process $e^\pm + \nu_L \rar e^\pm + \nu_R$ was estimated. It was found
there that the predictions of BBN would not be strongly disrupted by
the excitation of the additional ``wrong'' helicity states if
$\mu_\nu < (1-2)\cdot 10^{-11} \mu_B$. The calculations of this work were
further elaborated in ref.~\cite{fukugita87} and
an about thrice weaker limit was obtained
$\mu_\nu < 5.2\cdot 10^{-11} \mu_B (T_d/ 200\,{\rm MeV})^{1/2}$, where
$T_d$ is the decoupling temperature of the magnetic interactions of neutrinos;
$T_d$ should be taken smaller than the QCD phase transition temperature,
otherwise the energy density of $\nu_R$ would be strongly diluted and would
not effect BBN even if $\nu_R$ were abundantly produced at higher
temperatures.

The cross-section of $\nu_R$ production by $e^+e^-$-annihilation is equal to:
\be
\sigma \left(e^-+e^+ \rar \nu_L +\bar \nu_R \right) =
{\pi \alpha^2 \kappa^2 \over 12 m_e^2 }
\label{sigmaeenunu}
\ee
where $\kappa = \mu_\nu /\mu_B$ and $\alpha = 1/137$. This process is
sub-dominant with respect to the quasi-elastic scattering with the
cross-section:
\be
\sigma \left(e^\pm + \nu_L \rar e^\pm + \nu_R\right) =
{\pi \alpha^2 \kappa^2 \over  m_e^2 }\,
\ln\left( q^2_{max} \over q^2_{min} \right)
\label{sigmaenu}
\ee
where $q_{max}$ is the maximum value of the momentum transfer which is
determined by the particle spectral density. In thermal equilibrium
it is close to the average momentum $\langle q \rangle \approx 3 T$. The
logarithmic infrared cut-off $q_{min}$ is related to the long-range
nature of (electro-)magnetic interactions between $\nu$ and $e^\pm$.
In ref.~\cite{fukugita87} $q_{min}$ is
taken as the inverse Debye screening length, $q_{min}= 2\pi /l_D$ with
$l_D =  \left( T /4\pi n\alpha \right)^{1/2}$ and $n\approx 0.1 T^3$ (the
latter is the equilibrium number density of massless fermions).

A more accurate treatment of plasma effects was performed in
refs.~\cite{elmfors97,ayala99}. According to the first paper, the
production rate of $\nu_R$ is equal to $\Gamma_R = 0.0132 \mu_\nu^2 T^3$
and from the usual condition $\Gamma_R <H$ one obtains
$\mu_\nu< 6.2 \cdot 10^{-11}\mu_B$ for $T_d = 100$ MeV, which is rather
close to the estimate of ref.~\cite{fukugita87}. In the second
paper~\cite{ayala99} a much weaker production rate of $\nu_R$ was found,
$\Gamma_R = 5.8\cdot 10^{-4} \mu_\nu^2 T^3$. As stated by the authors, the
difference is due to a more precise treatment of the thermal photon
polarization function. Correspondingly, the bound on magnetic moment of
neutrinos is 5 times weaker:
\be
\mu_\nu< 2.9 \cdot 10^{-10}\mu_B
\label{munufin}
\ee
(also for $T= 100$ MeV).

The limits discussed above are applicable to light neutrinos, with
$m_\nu \ll 1$ MeV. If the mass is larger than MeV (in principle, it might
be true for $\nut$), such neutrinos would be non-relativistic at BBN and
their energy density would be significantly different from the energy
density of light neutrinos. If the magnetic moment is large,
$\mu_{\nut} \sim 10^{-6} \mu_B$, then the electromagnetic annihilation 
of $\bar\nut \, \nut$ would be strong enough so that $\nut$ would
decouple when they are nonrelativistic~\cite{giudice90}. For neutrinos with
such a large mass both helicity states would be equally populated, and to
avoid contradiction with BBN their number density at the decoupling should
be sufficiently Boltzmann suppressed. On the other hand, the energy density
of decoupled nonrelativistic $\nut$ rises as $m_{\nut}/T$ with respect
to the energy density of relativistic species. These effects have been
analyzed in refs.~\cite{kawano92,grasso96}.
For the case of cosmologically stable
$\nut$ the universe age constraint demands 
$\mu_{\nut} > 5\cdot 10^{-7} \mu_B $.
If $\nut$ is unstable but decays after the nucleosynthesis epoch, its
magnetic moment should be larger than roughly $(6-7)\cdot 10^{-9} \mu_B$
(more precisely, the limit depends upon the $\nut$ mass and presented
in refs.~\cite{kawano92,grasso96}). Otherwise the $\nut$-annihilation
would not be efficient enough to reduce their number density at BBN
(compare with sec.~\ref{massstabnu}). The limits are valid up to
$m_{\nut} \approx 30 $ MeV. For larger masses the
annihilation of $\nut$ in the standard electroweak model
is sufficiently strong to suppress their
abundance at BBN (see sec.~\ref{massstabnu}).
On the other hand, large values of $\mu_{\nut}$,
about $10^{-6}\mu_B$, could also be excluded because in this case
massive $\nut$'s
would be effectively absent at BBN and the total number of neutrino
species would be 2 instead of normal 3. The case of $\nut$ being unstable on
BBN scale is discussed in sec.~\ref{massunstbl}. Especially dangerous is
the electromagnetic decay $\nut \rar \nue e^+e^-$ open for
$m_{\nut}> 1$ MeV because electrons and positrons produced from this
decay would induce disintegration of deuterium (see e.g. the
paper~\cite{tekasa}). An additional argument against stable $\nut$
with MeV-mass was presented in ref.~\cite{fargion96kk} where it was
argued that the annihilation of $\nut$ in Galactic halo would produce
too high flux of cosmic ray electrons and positrons.

Another group of papers used reasonable assumptions about the magnitude
of magnetic field on the early universe to estimate the neutrino spin-flip
due to a possible magnetic moment. In the pioneering
works~\cite{shapiro81,lynn81} the spin-flip rate was estimated in the
following way. The energy difference between two neutrino states moving
parallel (anti-parallel) to the direction of magnetic field ${\vec B}$ is
$\Delta E_{magn} = 2\mu_\nu\, B$, if one neglects
the difference between effective potentials of $\nu_L$ and $\nu_R$ in
the plasma (see below). Correspondingly the spin precession frequency in
magnetic field (cyclotron frequency) is equal to:
\be
\omega = 2\mu_\nu B = 1.76\cdot 10^7\,(B/{\rm G})
(\mu_\nu /\mu_B)\,{\rm rad/sec},
\label{omegamu}
\ee
where G is ``Gauss''.
In particular, for neutrino with magnetic moment given by eq.~(\ref{munu})
the characteristic time for spin-flip in magnetic field $B$ is:
\be
\tau_{flip} =
0.55\cdot 10^{19}\,{\rm sec}\,(10^{-7} {\rm G} /B )({\rm eV}/m_\nu)
\label{taumunu}
\ee
We assume, following refs.~\cite{shapiro81,lynn81}, the flux-freezing
model of cosmological evolution of magnetic field, such that the magnitude 
of magnetic field at red-shift $z$ scales as $B_z=B_0 \,(z+1)^2$, where
$B_0$ is its present-day value. The latter could be as large as
$(10^{-10}-10^{-7})$ G (for reviews on cosmic magnetic fields see
e.g.~\cite{asseo87,grasso00}. The red-shift is
$(z+1) = T/2.73 {\rm K} = 4.25\cdot 10^9 (T/{\rm MeV})$. Using the relation
$tT^2 = 0.74 {\rm MeV}^2\,{\rm sec}$~(\ref{tT2}) we find for the angle of
the spin rotation:
\be
{\delta \theta } = 7\cdot 10^7\, \mu_\nu B_0
\ln \left(t_{max} /t_{min} \right)
\label{thetabbn}
\ee
where $t_{max}\sim (1-2)\,{\rm sec}$ is close to the time of neutron-proton
freezing. The value of $t_{min}$ will be discussed below, but it is
clear that at it should be at least larger than $\sim 10^{-4}$ sec
corresponding to $T\sim 100$ MeV because all wrong-helicity neutrinos
produced at that time or earlier would be diluted by the entropy release
at QCD phase transition. Demanding $\delta \theta <\pi$ one can obtain a
bound for the product of $\mu_\nu B_0$.

Another model, discussed in ref.~\cite{lynn81}, is based on the assumption that
the energy stored in magnetic field is proportional to the kinetic energy
of electrons (equipartition model). At BBN this model envisages magnetic field three orders of magnitude larger than the previous one. However, the estimates of the magnitude of
primordial magnetic fields suffer from serious uncertainties (in
particular, the field could be dynamo-amplified at later stages, its
size of homogeneity could be small, the mechanism of generation of the seed
field is unknown, etc) and the limits obtained this way should be taken
with caution. Moreover, a very important neutrino refractive effects were
neglected in the papers~\cite{shapiro81,lynn81}. Left-handed neutrinos
have the usual weak interaction with plasma, while the right-handed ones
are (practically) sterile. The difference between effective potentials
of $\nu_L$ and $\nu_R$ in primordial plasma could strongly suppress the
magnetic spin-flip. This effect and corresponding modification of BBN bounds
was first considered in ref.~\cite{fukugita88}.
Neutrino refraction index without external magnetic field in connection
with cosmological neutrino oscillations is discussed in sec.~\ref{refr}.
The effective potential is given by eq.~(\ref{nref}). Magnetic field $B$ can change refraction properties of the plasma, and
an extra term proportional to $B$ may arise in effective potential. 
This phenomenon was studied in the
papers~\cite{semikoz87}-\cite{elizalde00}. A large contribution to $V_{eff}$
found in ref.~\cite{semikoz94} originated from the an error later
corrected in ref.~\cite{elmfors96} (see also erratum to the
paper~\cite{semikoz94}). It is agreed now that cosmic magnetic field causes such small correction to the refraction index of neutrinos in the primeval plasma that the index is approximately given by
the expression (\ref{nref}) found in the limit of $B=0$.

Below we will derive the probability of neutrino spin conversion in
external magnetic field in cosmological plasma.
The Lagrangian of neutrino interaction with electromagnetic field
has the form
\be
{\cal L}_{magn} =-{\cal H}_{magn}= - {1\over 2} \munu F_{\mu\nu}
\bar \psi \sigma_{\mu\nu} \psi =
 - {1\over 2} \munu \epsilon_{ijk} B^k
\bar \psi \gamma_i\gamma_j \psi
\label{lmagn}
\ee
The last equality is true if the the external Maxwell field $F_{\mu\nu}$ is
reduced to magnetic one $\vec B$. The neutrino wave operator $\psi$
is a solution of the Dirac equation, so it has the form
$\psi = [1, \vec \sigma \vec p /(E+m)]^T\phi$, where upper $T$ means
``transpose'', $\vec \sigma$ are Pauli matrices, $E$ and $\vec p$ are
the energy and momentum of neutrino, and $m$ is the neutrino mass
(sub-$\nu$ is omitted for simplicity of notations). After straightforward
manipulations with Dirac gamma-matrices this expression can be rewritten as
\be
{\cal H}_{magn} =  \munu B_{tr} \left( \phi^*_- \sigma_{tr} \phi_+ +
 \phi^*_+ \sigma_{tr} \phi_-\right)
\label{hmagn}
\ee
where $\phi_{\pm}$ are the eigenfunctions of the helicity operator
$(1\pm \vec \sigma \vec n)/2$, $\vec n =\vec p /p$, and $tr$ means transverse
to the direction of the neutrino momentum. One can see that indeed magnetic
field induces helicity flip.

The free part of the Hamiltonian has the usual form:
\be
{\cal H}_{free} = \psi^* \gamma_0 \left( \vec \gamma \vec p +m \right)
\psi
\label{hfree}
\ee
and it is diagonal and proportional to the unit matrix in the
$\phi_{\pm}$ basis.

The last essential contribution to the Hamiltonian describes interaction
with medium and has a simple form in the chiral basis. It has the only
non-zero entry in the upper left corner ${\cal H}_{LL}$, where $L$ means
left-handed chirality state. In the massless limit helicity and chirality
bases coincide, but for $m\neq 0$ they are a little rotated against
each other. The eigenstates of chirality are obtained by the projector
$(1\pm \gamma_5)/2$ and proportional to
$[1 \pm \vec\sigma \vec p /(E+m)] \phi$. One can decompose e.g.
\be
1 - {\vec \sigma \vec p \over E+m} =
a_- \left( 1 - \vec \sigma \vec n\right ) +
a_+\left( 1 + \vec \sigma \vec n\right)
\label{chirhel}
\ee
and find $a_+ = (E+m-p)/(E+m) \approx m/2E$ and $a_- = 1-a_+ \approx 1$.
The second equations are valid for $m/E \ll 1$. For ultrarelativistic
neutrinos the rotation angle is very small and we neglect it in what
follows, assuming that helicity and chirality bases coincide. The
effects of non-zero $m$ and the relative rotation of bases
could be essential
in the case when the production of $\nu_R$ is solely due to the mass term,
as considered in the previous subsection.

Taking together all the contributions to the Hamiltonian  and including
the coherence breaking terms in the same way as is done for the case
of the usual neutrino oscillations (see sec.~\ref{nuosceu}) we obtain
the following evolution equations for the density matrix elements:
\be
\dot \rho_{LL} &=& -2\tilde I - \Gamma_0 \left(\rho_{LL} - f_{eq} \right)
\label{dotrhoLL}\\
\dot \rho_{RR} &=& -2\tilde I
\label{dotrhoRR}\\
\dot{\tilde R} &=& -(1/2)\Gamma_0 \tilde R + V_{eff} \tilde I
\label{dottilR}\\
\dot {\tilde I} &=& -(1/2)\Gamma_0 \tilde I - V_{eff} \tilde R +
B^2_{tr}\left( \rho_{LL} - \rho_{RR}\right)
\label{dottilI}
\ee
where $\tilde I =B_x I - B_y R$, $\tilde R =B_x R +B_y I$, and $R$ and
$I$ are the real and imaginary parts of the non-diagonal elements of
the neutrino density matrix, $\rho_{LR} = \rho_{RL}^* = R+iI$;
$V_{eff}$ is given by eq.~(\ref{nref}).

Neglecting the charge asymmetric term in $V_{eff}$, we can estimate the
latter as $V_{eff} \approx 0.8\cdot 10^{-20} T^5$ for $\nu_{\mu,\tau}$ and
$V_{eff} \approx 3.1\cdot 10^{-20} T^5$ for $\nu_{e}$ (everything here and
below, $V_{eff}$, $T$, $H$, $\Gamma$, are
in MeV). The Hubble parameter is $H\approx 4.5\cdot 10^{-22} T^2$,
and the reaction rate $\Gamma_0 \approx  4.5\cdot 10^{-22} T^5 T_W^{-3}$
(see eq.~(\ref{gammaj1})),
where $T_W$ is the temperature when neutrinos are effectively decoupled
from the plasma. By different estimates $T_W = 1-3$ MeV
(see discussion in sec.~\ref{sec:nuoscbbn}). For $T<T_W$
one can neglect $\Gamma_0$ and the total number of $\nu_L +\nu_R$ would
be conserved as follows from eqs.~(\ref{dotrhoLL},\ref{dotrhoRR}). Thus
at these low temperature the new species are not created and in the case
of $\num$ and $\nut$ there is no influence on BBN from this period.
It is not so for the spin-flip of $\nue$ because the decrease of the
number (energy) density of $\nu_e^L$, though accompanied by the same increase
in $\nu_e^R$, would lead to lower efficiency of $n-p$ transformation
(\ref{nnue},\ref{ne}) and as a result to a higher temperature of
$n/p$-freezing and to a higher mass fraction of $^4 He$.

In the limit $V_{eff} \gg \Gamma_0 \geq H$ the equations
(\ref{dotrhoLL})-(\ref{dottilI}) can be solved in the same way as
oscillation equations in non-resonance case, sec.~\ref{sec:nonres}.
In this limit the stationary point approximation works pretty well
and one can find $\tilde R$ and $\tilde I$ from
eqs.~(\ref{dottilR},\ref{dottilI}) algebraically, assuming that their
right hand sides vanish. Substituting $\tilde I$ into eq.~(\ref{dotrhoRR})
one finds
\be
\dot \rho_{RR} = {\Gamma_0 \munu^2 B_{tr}^2 \over V^2_{eff}}
\left( \rho_{LL} -\rho_{RR}\right)
\label{rhoRRsol}
\ee
As follows from this equation the rate of $\nu_R$-production can be
estimated as
\be
{\Gamma_{R}\over H}={\Gamma_0 \over H}{\munu^2 B^2_{tr}\over V^2_{eff}}
\label{gammarr}
\ee
Demanding that $\nu_R$ are not produced in equilibrium amount, we come
to the condition:
\be
\left( B_{tr}/{\rm gauss}\right) \left(\mu_{\nu_a}/\mu_B\right) <
10^{-6}\, C_a\, T^{7/2} T_W^{3/2}
\label{bmunu}
\ee
where $C_{\mu,\tau} = 1.8$ and $C_e =7$ and
temperatures are measured in MeV. This result is close to those
obtained in refs.~\cite{fukugita88,elmfors96} if one takes
$T \approx T_W \approx 1$ MeV. The smaller $T$ is, the stronger is the
limit. A weaker limit used in refs.~\cite{enqvist92,enqvist93c}
is a result of an incorrect conclusion that the rate of spin conversion
grows with rising temperatures. The suppression of spin oscillations
at high temperatures kills the rise of the transition probability
demonstrated by eq.~(\ref{thetabbn}), which is a result of neglected
refraction. Thus the strongest limit on the product
$\munu B$ could be obtained at the lowest essential $T$ even if one
takes into account the possibility that the magnetic field decays in
the expanding universe as inverse scale factor, $B\sim 1/a^2\sim T^2$.
As we have already noted, one cannot go below $T_W$ for $\num$ and $\nut$
because after they are decoupled from the plasma the spin-flip in magnetic
field does not change the total number (energy) density of $\nu_L$
plus $\nu_R$. For $\nue$ one could go below $T_W$ and obtain a stronger
bound because a decrease of the energy density of $\nu_e^L$, due to
their transformation into sterile right-handed partners, would result
in an earlier freezing of $(n-p)$-transformation and to higher mass
fraction of $^4 He$.

Several comments are worth making after this result.
First, it was assumed above that there is no
resonance transition or, in other words, the potential $V_{eff}$ never
vanishes. However, this is not so and for $E= 0.4\,\, (1.4) /T$ (all in MeV)
there is a resonance in $\nue$ ($\nu_{\mu,\tau}$) channel. Its impact on
spin-flip might be significant. Another chance for resonance conversion
is in possible non-diagonal magnetic magnetic moments that would
induce transitions between different neutrino flavors with non-vanishing
mass difference. A large lepton asymmetry in the sector of left-handed
neutrinos could be generated by the resonance (see sec.~\ref{sec:leptas}).
Second, the magnitude of magnetic field
and its coherence length in the early universe is poorly known. One
could make a more or less reasonable guess about that by extrapolating into
the past the present-day observed galactic or intergalactic magnetic fields (see
e.g. refs.~\cite{enqvist92,enqvist93c} for discussion and literature). This
extrapolation is subject to uncertainty of the magnetic field evolution,
in particular, due to unknown dynamo 
amplification. Short scale random magnetic
fields of large amplitudes could be generated in the early universe by
e.g. electroweak phase transition~\cite{vachaspati91}.
Neutrino spin-flip in such fields was considered in several
papers~\cite{enqvist93b}-\cite{pastor96} and interesting limits on the
magnitude of the neutrino magnetic moment were obtained under reasonable
assumptions about the magnitude of magnetic field near BBN epoch.

Let us note at the end that one may obtain from BBN as well a
bound on the electric charge radius of neutrinos,
$\langle r_\nu^2\rangle $ if the latter are massive and
correspondingly right-handed states exist~\cite{grifols87}. The electric
interaction proceeds through the chirality conserving coupling
\be
F(q^2) A_\alpha \bar\nu \gamma_\alpha \nu
\label{chradnu}
\ee
where $F(q^2)$ is the electric form-factor of neutrino. Since neutrino
electric charge is supposed to be zero, $F(0)=0$, but the first term in the
expansion, $F(q^2) \approx (1/6) \langle r_\nu^2\rangle  q^2$
is generally non-vanishing.
The right-handed neutrino states can be produced in the early universe
through the reaction $e^+e^- \rar \bar\nu_R \nu_R$. According to
ref.~\cite{grifols87} its cross-section is
$\sigma = \pi\alpha^2 \langle r^2\rangle^2 q^2/54$.
Imposing the condition that this reaction was never in equilibrium one can obtain the
bound:
\be
r^2 < 7\cdot 10^{-33}\,\,{\rm cm}^2
\label{r2nu}
\ee

\subsection{Neutrinos, light scalars, and BBN  \label{sec:majbbn}}

If neutrinos are coupled to a new light boson, as suggested by some models
of particle physics~\cite{maj,fam}, the existence of such bosons could be
observable through BBN. One should distinguish the cases when the new
interaction excites right-handed neutrino states or involves only the
usual left-handed neutrinos. The first case is discussed in
sec.~\ref{ssec:nur} and here we will consider the second one. A general
discussion of a possible presence of new light particles at BBN,
is done in ref.~\cite{kolb86}. If the latter are produced in equilibrium
amount then a single bosonic state is equivalent to $4/7 = 0.57$ extra
neutrino species, while one fermionic state is equivalent to 0.5 extra
neutrino species. If both spin state of the new fermions are excited, this
number turns into 1 and if the fermions go together with their antifermions,
the number of equivalent neutrino species becomes 2. These results are true
if the mass of new particles is below 1 MeV. In the opposite case the result
depends upon their annihilation rate. If the latter is strong enough so that
the equilibrium is maintained, then the contribution of massive particles into
the cosmological energy density would be smaller than that of massless
particles. If annihilation is frozen, then the number density of
non-relativistic particles could be larger
than the equilibrium one and the effective contribution into $N_\nu$ would
be larger by the factor $(m/T)(n_m/n_0)$ where $m$ is the mass of particles
in question.

In ref.~\cite{bertolini92} the new interaction of neutrinos with majorons
was discussed in connection with the 17 keV neutrino hypothesis that was
supposed to exist at that time. The authors obtained an upper bound on the
coupling constants of new neutrino interactions from the conditions that
$\Delta N_\nu <0.3$ at BBN. Depending upon the model of interaction this
limit is satisfied either if new neutrinos and majorons are decoupled at
$T>(\sim 100)$ MeV and their number density is entropy diluted at BBN, or if
they come to equilibrium already after BBN (the latter could take place
if the cross-section rises with decreasing $T$ as e.g. $1/T^2$).
In addition, the problem of depleting the cosmological number density of
17 keV neutrinos was studied and it was found that the annihilation of such
neutrinos into pair of majorons is not sufficiently strong to make their
number/energy density cosmologically acceptable, but the decay
$\nu_h\rar \phi \nu_l$ could be efficient enough. Similar arguments
but used for the usual neutrinos with additional interaction with
majorons where applied in ref.~\cite{chang94} to obtain the upper limit
on the majoron-neutrino Yukawa coupling, $ g <10^{-5}$,
found from the condition that $\Delta N_\nu < 0.3$.

In the case that an active neutrino has a mass in MeV range (a few years ago
it was the usual assumption for $\nut$, now cast into doubt by the solar and
atmospheric neutrino data) the BBN bounds on majoron coupling to $\nut$ were
derived in ref.~\cite{dprv}. The difference with respect to the previous cases
is that a heavy neutrino could efficiently annihilate before or during BBN
and though this process creates additional particle species (majoron) it
simultaneously diminishes the number density of $\nut$, so the net outcome
could even be $N_\nu <3$ and the Yukawa coupling about $10^{-4}$ is
allowed.

\subsection{Heavy sterile neutrinos: cosmological bounds and direct
experiment. \label{ss:nush}}.

It is assumed usually that possible sterile neutrinos, related to  
ordinary ones through a small mixing angle, are relatively light, 
with masses in eV region (or even smaller) or with keV masses. In the
latter case these particles could form cosmologically interesting warm
dark matter (see sec.~\ref{sterilenu}). Another possibility of much
heavier sterile neutrino originated from the observation of the KARMEN
anomaly in the time distribution
of the charged and neutral current events induced by neutrinos from
$\pi^+$ and $\mu^+$ decays at rest~\cite{karmen95}. A suggested
explanation of this anomaly was the production of a new neutral particle
in pion decay
\be
\pi^+ \rar \mu^+ + x^0~,
\label{piondecay}
\ee
with the mass 33.9 MeV, barely permitted by the phase space, so this
particle moves with non-relativistic velocity. Its subsequent
neutrino-producing decays could be the source of the delayed neutrinos
observed in the experiment. Among possible candidates on the role of
$x^0$-particle was, in particular, a 33.9-MeV sterile
neutrino~\cite{barger95}.

In refs.~\cite{barger95,dolgov00krm} cosmological and astrophysical bounds
on such neutrino were considered. According to ref.~\cite{dolgov00krm},
cosmology and astrophysics practically exclude the
interpretation of the KARMEN anomaly by a 33.9~MeV neutrino mixed
with $\nut$. According to a statement of the KARMEN collaboration
made at Neutrino 2000~\cite{karmen2000} the anomaly
was not observed in upgraded detector KARMEN 2, but the question
still remains which area in the mass-mixing-plane for heavy sterile
neutrinos can be excluded. These issue was addressed recently by NOMAD
collaboration in direct experiment~\cite{nomad00} and in
ref.~\cite{dolgov00hv} by considerations of big bang nucleosynthesis and
the duration of the supernova (SN)~1987A neutrino burst.

We assume that the sterile neutrino mixes predominantly with only one
active flavor $\nu_a=\nue$, $\num$ or $\nut$.  Mixed flavor states
are expressed in terms of the mass eigenstates and the
mixing angle $\theta$ as
\be
\nu_a &=& \cos\theta ~\nu_1 + \sin \theta ~\nu_2\,,  \nonumber \\
\nu_s &=& -\sin \theta ~\nu_1 + \cos\theta ~\nu_2\,,
\label{nuash}
\ee
where $\nu_1$ and $\nu_2$ are assumed to be the light and heavy mass
eigenstates, respectively. Since the mixing angle is supposed to be small
$\nu_1$ is mostly an active flavor and $\nu_2$ is dominantly the sterile
one. This mixing couples the heavier neutrino to the $Z^0$, allowing for
the decay:
\be
\nu_2 \rightarrow \nu_1 + \ell + \bar \ell \, ,
\label{dech}
\ee
where $\ell$ is any lepton with the mass smaller than the mass $m_2$
of the heavy neutrino. If  $m_2 < 2 m_\mu$ the decay into $\bar\mu \mu$
and $\bar\tau \tau$ is kinematically forbidden.
If $\nus$ is mixed either with  $\num$ or $\nut$, the life-time is
expressed through the mixing angle as:
\be
\tau_{\nu_s}  \equiv \Gamma_{\nu_2}^{-1} = {
1.0~{\rm sec} \over (M_s/\mbox{10 MeV})^5\, \mbox{sin}^2 2\theta}  \, .
\label{taumuh}
\ee
For the mixing with $\nue$ the numerator is 0.7~sec; the
difference is due to the charged-current interactions.

A sterile neutrino mixed with $\nu_a$ could be observed in direct
experiments, in particular in those where upper bounds on neutrino masses are
obtained (see the list of references in~\cite{pdg}). The most accurate
limit exists for $\nue$, roughly $m_{\nue} <3$~eV~(\ref{mne}).
However, these experiments are not helpful
in eliminating a heavy sterile neutrino because they are not sensitive
to the mass range $M_s > 10$~MeV which we consider. Such heavy
neutrinos are not produced in beta-decays because of a lack of phase
space and their impact is only indirect, e.g. they could renormalize
vector and axial coupling constants.

There are several effects operating in different directions, by which
a heavy unstable sterile neutrino could influence big-bang
nucleosynthesis. First, their contribution to the total energy density
would speed up the expansion and enlarge the frozen neutron-to-proton
ratio. Less direct but stronger influence could be exerted through the
decay products, $\nue,\, \num$, and $\nut$, and $e^\pm$ and through
the change of the temperature ratio, $T_\nu/T_\gamma$.
The impact of $\num$ and
$\nut$ on BBN is rather straightforward: their energy density increases
with respect to the standard case and this also results in an increase
of $r_n$. This effect can be described by the increased number of
effective neutrino species $N_\nu$ during BBN.
The increase of the energy density of $\nue$, due to decay of
$\nus$ into $\nue$, has an opposite effect on $r_n$. Though a larger energy
density results in faster cooling, the increased number of $\nue$
would preserve thermal equilibrium between neutrons and protons for a longer
time and correspondingly the frozen $n/p$-ratio would become smaller.
The second effect is stronger, so the net result is a smaller $n/p$-ratio.
There is, however, another effect of a distortion of the equilibrium
energy spectrum of $\nue$ due to $e^\pm$ produced
from the decays of $\nus$. If the spectrum is distorted
at the high-energy tail, as is the case, then creation of protons in
the reaction $n+\nue \rar p + e^-$ would be less efficient than neutron
creation in the reaction $ \bar \nue + p \rar n + e^+$. We found that
this effect is quite significant. Last but no the least, the decays of
$\nus$ into the $e^+e^-$-channel will inject more energy into the
electromagnetic part of the primeval plasma and this will diminish
the relative contribution of the energy density of light neutrinos and
diminish $r_n$.

In refs.~\cite{dolgov00krm,dolgov00hv} the distribution functions of
neutrinos were calculated from kinetic equations in Boltzmann
approximation and in a large part of parameter space they significantly
deviate from equilibrium. The distributions of electrons and positrons
were assumed to be very close to equilibrium because of their very
fast thermalization due to interaction with the photon bath. However,
the evolution of the photon temperature, due to the decay and annihilation of
the massive $\nus$ was different from the standard one,
$T_{\gamma} \sim 1/a$, by an extra factor $(1+\Delta) >1$ where $a$
is the cosmological scale factor and the correction $\Delta$
was numerically calculated from the energy balance 
condition\cite{dolgov00hv}.
At sufficiently high temperatures, $T>T_W \sim 2$~MeV,
light neutrinos and electrons/positrons were in strong contact, so 
the neutrino distributions were also very close to the equilibrium ones. If
$\nus$ disappeared sufficiently early, while
thermal equilibrium between $e^{\pm}$ and neutrinos remained, then
$\nus$ would not have any observable effect on primordial abundances,
because only the contribution of neutrino energy density relative to
the energy density of $e^{\pm}$ and $\gamma$ is
essential for nucleosynthesis. Hence a very short-lived $\nus$ has a negligible
impact on primordial abundances, while with increasing lifetime the
effect becomes stronger. Indeed at $T<T_W$ the exchange of energy between
neutrinos and electrons becomes very weak and the energy injected into the
neutrino component is not immediately redistributed between all particles.
The branching ratio of the decay of $\nus$ into $e^+e^-$ is approximately
1/9, so the neutrino component is heated much more than the
electromagnetic one. As we mentioned above, this leads to faster cooling
and to a larger $n/p$-ratio.

In the early universe sterile neutrinos were produced through their
mixing with the active ones. The production rate for relativistic
$\nus$ (i.e.~for $T_\gamma \geq m_2$) is given by eq.~(\ref{sprod})
below (note the factor 1/2 difference with respect to the standard
estimate). The mixing angle in matter is strongly suppressed at high
temperatures, $T_\gamma>1.5\,\mbox{GeV} (\dm/\mbox{MeV}^2)^{1/6}$
due to refraction effects~(\ref{nref},\ref{sinthetam}). Correspondingly
the  production rate reaches maximum at
$T_{\rm max} = 1.28 (\dm/\mbox{MeV}^2)^{1/6}$~GeV. For the
masses, $10<m_{\nus}<150$ MeV, that are considered below,
$T_{\rm max}$ is well above the neutrino mass.

If the equilibrium number density of sterile neutrinos is reached, it
would be maintained until $T_f \approx 4 (\sin 2\theta)^{-2/3}$~MeV.
This result does not depend on the mass of heavy neutrinos because they
annihilate with massless active ones, $\nu_2 + \nu_a
\rar all$. The heavy neutrinos would be relativistic at decoupling and
their number density would not be Boltzmann suppressed if, say,
$T_f>M_s/2$. This gives
\be
\sin^2 2\theta (\dm/\mbox{MeV}^2)^{3/2} < 500~.
\label{relath}
\ee
If this condition is not fulfilled the impact of $\nu_s$ on BBN
would be strongly diminished. On the other hand, for a sufficiently
large mass and non-negligible mixing, the $\nu_2$ lifetime given by
Eq.~(\ref{taumuh}) would be quite short, so they would all decay
prior to the BBN epoch. (To be more exact, their number density would
not be frozen, but would follow the equilibrium form $\propto e^{-M_s/T_\gamma}$.)

Another possible effect that could diminish the impact of heavy
neutrinos on BBN is entropy dilution. If $\nu_2$ were decoupled while
being relativistic, their number density would not be suppressed
relative to light active neutrinos. However, if the decoupling
temperature were higher than, say, 50~MeV pions and muons were still
abundant in the cosmic plasma and their subsequent annihilation would
diminish the relative number density of heavy neutrinos. If the
decoupling temperature is below the QCD phase transition the dilution
factor is at most $17.25/10.75 =1.6$.  Above the QCD phase transition
the number of degrees of freedom in the cosmic plasma is much larger
and the dilution factor is approximately 5.5. However, these effects
are only essential for very weak mixing, for example the decoupling
temperature would exceed 200~MeV if $\sin^2 2\theta < 8\times
10^{-6}$. For such a small mixing the life-time of the heavy
$\nu_2$ would exceed the nucleosynthesis time and they would be
dangerous for BBN even if their number density is 5 times diluted.

 Sterile neutrinos would never be abundant in the
universe if $\Gamma_s/H < 1$. In fact we can impose a stronger
condition demanding that the energy density of heavy neutrinos
should be smaller than the energy density of one light neutrino
species at BBN ($T\sim 1$ MeV). Taking into account a possible entropy
dilution by factor 5  we obtain the
bound:
\be
\left( \dm/\mbox{MeV}^2\right)\,\sin^2 2\theta
< 2.3\times 10^{-7} \, .
\label{dmsinh}
\ee
Parameters satisfying this conditions cannot be excluded by BBN.

If $\nus$ mass is larger than 135 MeV, the dominant decay mode becomes
$\nu_2 \rar \pi^0 + \nu_a$. The life-time with respect to this decay can
be found from the calculations~\cite{fishbach77,kalogeropoulous79}
of the decay rate $\pi^0 \rar \nu \bar \nu $ and is equal to:
\be
\tau = \left[{ G_F^2 M_s (M_s^2 -m_\pi^2) f_\pi^2 \sin^2\theta
\over 16\pi} \right]^{-1}
= 5.8 \cdot 10^{-9} \,{\rm sec} \left[ \sin^2 \theta { M_s (M_s^2-m_\pi^2)
\over m_\pi^3} \right]^{-1}
\label{taupi}
\ee
where $M_s$ is the mass of the sterile neutrino, $m_\pi = 135$ MeV is
the $\pi^0$-mass and $f_\pi= 131 $ MeV is
the coupling constant for the decay $\pi^+ \rar \mu+\nu_\mu$.
The approximate estimates of ref.~\cite{dolgov00hv} permit one to conclude
that for the life-time of $\nu_2$ smaller than 0.1 sec, and corresponding
cosmological temperature higher than 3 MeV, the decay products would
quickly thermalize and their impact on BBN would be small. For a life-time
longer than 0.1 sec, and $T<3$ MeV, one may assume that thermalization of
neutrinos is negligible and approximately evaluate their impact on BBN.
If $\nu_s$ is mixed with $\num$ or $\nut$ then electronic neutrinos are
not produced in the decay $\nus\rar \pi^0 \nu_a$ and only
the contribution of the
decay products into the total energy density is essential. As we have
already mentioned, non-equilibrium $\nue$ produced by the decay would
directly change the frozen $n/p$-ratio. This case is more complicated
and demands a more refined treatment.

The $\pi^0$ produced in the decay $\nus\rar \nu_a+\pi^0$ immediately decays
into two photons and they heat up the electromagnetic part of the plasma,
while neutrinos by assumption are decoupled from it. We estimate the
fraction of energy delivered into the electromagnetic and neutrino
components of the cosmic plasma in the instant decay approximation.
Let $r_s=n_s/n_0 $ be the ratio of the number densities of the heavy
neutrinos with respect to the equilibrium light ones,
$n_0 = 0.09 T_\gamma^3$. The total energy of
photons and $e^+e^-$-pairs including the photons produced by the
decay is
\be
\rho_{em} = {11\over 2}{\pi^2 \over 30} T^4 + r_s n_0 {M_s \over 2}
\left( 1+{m_\pi^2 \over M_s^2}\right)\, ,
\label{rhoemh}
\ee
while the energy density of neutrinos is
\be
\rho_{\nu} = {21\over 4}{\pi^2 \over 30} T^4 + r_s n_0 {M_s \over 2}
\left( 1- {m_\pi^2 \over M_s^2}\right) \,.
\label{rhonuh}
\ee
The effective number of neutrino species at BBN can be defined as
\be
N_\nu^{(eff)} = {22\over 7} {\rho_\nu \over \rho_{em}} \, .
\label{keffnuh}
\ee
Because of the stronger heating of the electromagnetic component of the
plasma by the decay products, the relative role of neutrinos diminishes
and $N_\nu^{(eff)}$ becomes considerably smaller than 3. If $\nu_s$ are
decoupled while relativistic their fractional number
at the moment of decoupling would be
$r_s=4$ (two spin states and antiparticles are included). Possible
entropy dilution could diminish it to slightly below 1. Assuming that the
decoupling temperature of weak interactions is $T_W = 3$ MeV we find that
$N_\nu^{(eff)} =0.6 $ for $M_s = 150$ MeV and
$N_\nu^{(eff)} =1.3 $ for $M_s = 200$ MeV, if the frozen number density
of $\nu_s$ is not diluted by the later entropy release and $r_s$ remains
equal to 4. If it was diluted down to 1, then the numbers would
respectively change to $N_\nu^{(eff)} =1.15 $ for $M_s = 150$ MeV and
$N_\nu^{(eff)} =1.7 $ for $M_s = 200$ MeV, instead of the standard
$N_\nu^{(eff)} =3 $.
Thus a very heavy $\nu_s$ would result in under-production of $^4 He$.
There could, however, be some other effects acting in the opposite direction.

Since $\nue$ decouples from electrons/positrons at smaller temperature
than $\num$ and $\nut$, the former may have enough time to thermalize.
In this case the temperatures of $\nue$ and photons would be the same
(before $e^+e^-$-annihilation) and the results obtained above would be
directly applicable. However,
if thermalization between $\nue$ and $e^\pm$ was not efficient, then
the temperature of electronic neutrinos at BBN would be smaller than in
the standard model. The deficit of $\nue$ would produce an opposite
effect, namely enlarging the production of primordial $^4 He$, because it
results in an increase of the $n/p$-freezing temperature. This effect
significantly dominates the decrease of $N_\nu^{(eff)}$ discussed above.
Moreover even in the case of the decay
$\nu_2 \rar \pi^0 + \nu_{\mu, \tau}$, when $\nue$ are not directly
created through the decay,
the spectrum of the latter may be distorted at the high energy tail
by the interactions with
non-equilibrium $\nut$ and $\num$ produced by the decay. This would
result in a further increase of $^4He$-production. In the case of
direct production of non-equilibrium $\nue$ through the decay
$\nu_2 \rar \pi^0 + \nu_e$ their impact on $n/p$ ratio would be even
much stronger.

To summarize, there are several different effects on BBN
from the decay of $\nu_s$
into $\pi^0$ and $\nu$. Depending upon the decay life-time
and the channel these effects may operate in opposite directions.
If the life-time of $\nu_2$ is  larger  than 0.1 sec but smaller than
0.2 sec, so $e^\pm$ and $\nue$ establish equilibrium, the production
of $^4He$ is considerably diminished and this life-time interval
would be mostly excluded. For life-times larger than 0.2 sec the
dominant effect is the decrease of the energy density of $\nue$ and this
results in a strong increase of the mass fraction of $^4 He$. Thus large
life-times should also be forbidden. Of course there is a
small part of the parameter space where both effects cancel each other
and this interval of mass/mixing is allowed. It is, however, difficult to
establish its precise position with the approximate arguments used in
ref.~\cite{dolgov00hv}.

Thus, in the case of $\nu_s \leftrightarrow \nu_{\mu, \tau}$ mixing and
$M_s>140$ MeV we can exclude the life-times of
$\nu_s$ roughly larger than 0.1~sec, except for a small region near
0.2~sec where two opposite effects cancel and
the BBN results remain undisturbed despite the presence of sterile neutrinos.
Translating these results into mixing angle according to
eq. (\ref{taupi}), we conclude that mixing angles
$\sin^2 \theta < 5.8 \cdot 10^{-8} m_\pi/M_s /[(M_s/m_\pi)^2-1]$
are excluded by BBN. Combining this result with eq. (\ref{dmsinh})
we obtain the exclusion region for  $M_s>140$ MeV:
\be
5.1 \cdot 10^{-8} ~\frac{\mbox{MeV}^2}{M_s^2}
<\sin^2 \theta < 5.8 \cdot 10^{-8}
~\frac{m_\pi}{M_s}\frac{1}{(M_s/m_\pi)^2-1}~.
\label{sin_pi_excl}
\ee

In the case of $\nu_s \leftrightarrow \nu_e$ mixing
and  $M_s>140 MeV$ the limits are possibly stronger,
but it is more difficult to obtain reliable estimates because
of a strong influence of non-equilibrium $\nu_e$, produced by the
decay, on neutron-proton reactions.

The constraints on the mass/mixing of $\nus$ from neutrino observation
of SN 1987A are analyzed in some detail
in ref.~\cite{dolgov00krm} and are based on the upper limit of the
energy loss into a new invisible channel because the latter would
shorten the neutrino burst from this supernova below the observed
duration.

The results are summarized in fig.~\ref{fig:all}.
The region between the two horizontal lines
running up to 100 MeV are excluded by the duration of the neutrino
burst from SN~1987A. A more accurate consideration would probably permit
to expand the excluded region both in the horizontal and vertical
directions.

\begin{figure}
\begin{center}
\psfig{file=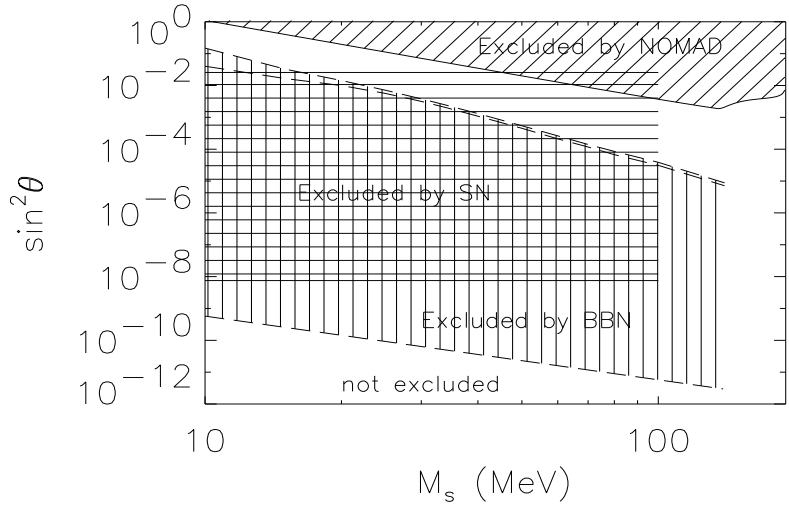,width=4.5in,height=3in}
\caption{Summary of the exclusion regions in the
$(\sin^2 \theta$-$M_s)$-plane.  SN~1987A excludes all
mixing angles between two solid horizontal lines. BBN excludes the
area below the two upper dashed lines if the heavy neutrinos were abundant
in the early universe. These two upper dashed lines both correspond to
the conservative limit of one extra light neutrino species permitted
by the primordial $^4$He-abundance. The higher of the two is for
mixing with $\nu_{\mu, \tau}$ and the slightly lower curve is for
mixing with $\nu_e$.
In the region below the lowest dashed curve the
heavy neutrinos are not efficiently produced in the early universe and
their impact on BBN is weak.
For comparison we have also presented here the region excluded by NOMAD
Collaboration~\cite{nomad00} for the case of
$\nu_s \leftrightarrow \nu_\tau$ mixing.}
\label{fig:all}
\end{center}
\end{figure}

\section{Variation of primordial abundances and lepton asymmetry of the
universe \label{varabund}}

Neutrinos may play an important role in a very striking phenomenon, namely
they may generate considerable chemical inhomogeneities on cosmologically
large scales while preserving a near-homogeneous mass/energy distribution.
It is usually tacitly assumed that the universe is chemically
homogeneous over all its visible part, though strictly speaking, the only
established fact is that the spatial variation of cosmic energy density
is very small. The observed smoothness of CMB and of the average matter
distribution at large scales strongly indicate that the universe is very
homogeneous energetically.
Surprisingly we know very little about the chemical content
of the universe at large distances, corresponding to red shifts
$z > 0.1$ to say nothing of $z\geq 1$. Of course it is quite natural
to believe that if the mass/energy density of matter is homogeneous
and isotropic over all
observed universe, the average chemical composition of the matter
should not vary over the same scales. Still, though the
energetic homogeneity of the universe is well verified up to
red shift $z=10^3$, which corresponds to the last scattering of CMB, its
chemical homogeneity remains an assumption, maybe quite natural, but
still an assumption. Recent observational
data~\cite{tfb}-\cite{tytler00} (see sec. \ref{prns})
of primordial deuterium at large
red-shifts $z> 0.5$ to some extend justify the  
hypothesis that primordial chemical composition of the universe may be
different in different space regions~\cite{dopa}.
From theoretical point of view it is an interesting challenge to find out whether
there exist (not too unnatural) cosmological scenarios consistent with
the observed smoothness of the universe but predicting large
abundance variations.
An example of such mechanism was proposed in ref.~\cite{dkad} (see
also~\cite {adbs}), where
a model of leptogenesis was considered which, first, gave a large lepton
asymmetry, which could be close to unity, and, second, this asymmetry
might strongly vary on astronomically large scales, $l_L$. The magnitude of
the latter depends on the unknown parameters of the model and can easily
be in the mega-giga parsec range. The model is based on the
Affleck-Dine~\cite{afdi} scenario of baryogenesis but in contrast to
the original one, it gives rise
to a large (and varying) lepton asymmetry and to a small baryonic one.
There are more models~\cite{ftv}-\cite{ccg} in recent literature,
where a large lepton asymmetry together with a small baryonic one is
advocated, though without any significant spatial variation of the
asymmetries. A varying lepton asymmetry of large magnitude due to
resonance amplification of neutrino oscillations was recently proposed
in the paper~\cite{dibari99b} (see sec.~\ref{sec:spatfluc}).

If we assume that the variation of deuterium abundance by approximately
an order of magnitude is indeed real, then, according to the data,
the characteristic scale $l_L$
should be smaller than a giga-parsec. The lower bound on this scale
may be much smaller. It might be determined by measurements
of the abundances of light elements at large distances in our neighborhood,
say, $ z\geq 0.05$.

A variation of deuterium abundance may be also explained by a variation of
the cosmic baryon-to-photon ratio. This possibility was explored in
refs.~ \cite{jf,cos}. The isocurvature
fluctuations on large scales, $l > 100$ Mpc, which are necessary to create
the observed variation of deuterium, are excluded \cite{cos} by the smallness
of angular fluctuations of the cosmic microwave background radiation (CMB).
Variations of baryonic number density on much smaller scales,
$M \sim 10^5 M_\odot$, are not in conflict with the observed smoothness of
CMB and in principle can explain the data subject to a potential conflict
with the primordial $^7$Li abundance \cite{jf}.

Exactly the same criticism of creating too large fluctuations in CMB
temperature is applicable to a simple version of the model with a varying
lepton asymmetry. One can ascertain that the necessary value of the chemical
potential of electron neutrinos $\xi_{\nu_e}$ should be close to $-1$
to explain the possibly observed variation of deuterium by roughly an
order of magnitude. Such a change in $\xi_{\nu_e}$
with respect to zero value, assumed in our part of the world,
would induce a variation in total energy density during the RD
stage at a per cent level, which is excluded by the smoothness of CMB. However,
this objection can be avoided if there is a conspiracy between different
leptonic chemical potentials such that in different spatial regions they have
the same values but with interchange of electronic, muonic and/or tauonic
chemical potentials. Since the abundances of
light elements are much more sensitive to the magnitude of the electron
neutrino chemical potential than to those of
muon and tauon neutrinos, the variation of $\xi_{\nu_e}$
(accompanied by corresponding variations of $\xi_{\nu_\mu}$ and
$\xi_{\nu_\tau}$) would lead to a strong variation in the abundance of
deuterium and other light elements.
The equality of, say,  $\xi_{\nu_e}$ at one space point to $\xi_{\nu_\mu}$ at
another point looks like very artificial fine-tuning, but this may
be rather naturally realized due the lepton flavor symmetry, $S_3$,
with respect to permutations
$e \leftrightarrow \mu \leftrightarrow  \tau$. It is interesting to note
that in the model of ref.~\cite{dibari99b} the fluctuations of the
cosmological energy density are very small despite large possible variations
of the lepton asymmetry, because it is compensated
by the equal but opposite sign variation in the energy
density of sterile neutrinos (see sec.~\ref{sec:spatfluc}).

If lepton asymmetry changes at large distances, then not only deuterium but
also $^4 He$ would not remain homogeneous in space. Playing with the
nucleosynthesis code~\cite{kaw} one can check that in the deuterium rich
regions the mass fraction of helium could be larger than 50\% (twice larger
than in our neighborhood). There may also exist the so called mirror regions
with a positive and large chemical potential of electronic neutrinos. In such
regions abundances of both deuterium and helium would be about twice smaller
than those observed nearby. For more detail see ref.~\cite{dopa}.
\begin{table}
\begin{center}
\begin{tabular}{|c|| r |r| r|| c| c| c|}
\hline
&&&&&& \\
$ \eta_{10}$ & $\xi_{\nue}$~~&~~$\xi_{\num}$~~&~~$\xi_{\nut}$
~&~$10^5\, {D\over H}$~&~$Y_p$~~&~~$10^{10}\, {^7 Li\over H}$~~\\
&&&&&&\\
 \hline
 4&0.1 & $-$1 & 1 & 5.35 & 0.229 & 1.61 \\
    &  $-$1 & 0.1 & 1 & 13.2 & 0.548 & 4.84 \\
 &  1 &$-$1  & 0.1 & 3.98  & 0.080 & 0.70 \\
\hline
 5&0.1 & $-$1 & 1 & 3.77& 0.231 & 2.54 \\
    &  $-$1 & 0.1 & 1 & 9.21 & 0.553 & 4.49 \\
 &  1 &$-$1  & 0.1 & 2.80  & 0.081 & 1.12 \\
 \hline

\end{tabular}
\bigskip
\caption{Abundances of light elements for
  $\eta_{10} = 10^{10} n_B / n_\gamma=4,\,\,5$ and different values of
 neutrino chemical potentials $\xi_{\nu_a}$.}

\end{center}
\end{table}

Surprisingly nothing is known about helium abundance at large distances.
All known accurate measurements of $^4$He based on emission lines
were done at $z \leq 0.045$
corresponding  to a distance of $140 h^{-1}$ Mpc \cite{hel}, whereas helium
line and continuum absorption measurements made at high red-shifts give the
abundance merely within ``a factor of a few" owing to uncertain
ionization corrections \cite{hog}. In regions with a large fraction
of $^4 He$ one would expect bluer stars with a shorter life-time, though
the structure formation there may be inhibited due to a less efficient
cooling. In the helium-poor regions the effects may be opposite. Properties
of supernovae could also be somewhat different in different regions, as was
noted by A. Kusenko (private conversation). This problem is very interesting
and deserves further and more detailed investigation.

Lepton conspiracy, mentioned above, would diminish energy density fluctuations
in first approximation. However there are some more subtle effects which
could be either dangerous for the model or observable in CMB. The first one
is related to the binding energy of $^4 He$ (7 MeV per nucleon).
Since the mass fraction
of $^4 $He may change by a factor of 2 in deuterium- (and helium-) rich regions
(from 25\% to more than 50\%),
this means that the variation in baryonic energy density may be as large as
$2\cdot 10^{-3}$. Rescaling the estimates of ref.~\cite{cos} one can
find~\cite{dopa} for the fluctuations of the CMB temperature:
${\delta T / T }\approx 10^{-5} \left({R_{hor} /  10\lambda } \right)$,
where $\lambda$ is the wavelength of the fluctuation and $R_{hor}$ is the
present day horizon size. The restriction on the amplitude of temperature
fluctuations would be satisfied if
$\lambda > (200 - 300)\, {\rm Mpc}/h_{100}$
($h_{100} = H/100$ km/sec/Mpc). Surprisingly, direct astrophysical effects of
such big fluctuations of the helium mass fraction at distances above 100 Mpc
cannot be observed presently, at least the evident simple ones.

Another effect which would induce energy inhomogeneities,
is the heating of neutrinos by $e^+e^-$-annihilation and the corresponding cooling of photons at
$T\leq 1$ MeV, when neutrinos practically decoupled from plasma.
(For the most recent and
precise calculations of this cooling see ref.~\cite{dhs0}).
The efficiency of the cooling depends upon the chemical potential
of neutrinos and would create fluctuations in CMB temperature at the
level of $2\cdot 10^{-5}$~\cite{dopa}.

Possible neutrino oscillations would have a very strong impact on the model.
First, they might drastically change predictions 
about light element abundances
because oscillations generically change the value of individual and
overall lepton asymmetries. In particular, if oscillations between all
neutrino flavors are fast enough to come to equilibrium by the time
of nucleosynthesis (NS), all different lepton asymmetries would be 
equal everywhere and no spatial
variation of primordial abundances would be created. 
The effect of asymmetry redistribution due to oscillations between
active neutrinos is investigated in ref.~\cite{dolgov02}, see 
sec.~\ref{ssec:active}. It is shown that if large mixing angle solution
to solar neutrino anomaly is realized then complete flavor equilibrium
would be established in the primeval plasma prior to BBN. If however
low mixing is true, then oscillations between active neutrinos might
not be efficient enough for flavor equilibration.

Second, neutrino mass differences, which must
be non-vanishing if oscillations exist, would
give rise to density inhomogeneities and to fluctuations in the CMB
temperature.
For nonrelativistic neutrinos the density contrast of
two mirror regions can be estimated as follows. For simplicity let us
consider only two participating neutrinos, $\nu_\mu$ and $\nu_\tau$. Their
energy densities in two regions are respectively:
\be
\rho_1 = n_0 m_{\nu_\mu} + n_1  m_{\nu_\tau}, \nonumber \\
\rho_2 = n_1 m_{\nu_\mu} + n_0  m_{\nu_\tau}.
\label{rho12}
\ee
We assume here that $n_0$ is the number density of neutrinos and anti-neutrinos
with zero chemical potential and $n_1$ is the sum of a pair with $\xi =1$
and $\xi =-1$. In this case the ratio $(n_0 -n_1) / n_0 = -0.4$.
The relative density contrast with respect to the {\it total}
cosmological energy density is equal to
\be
r_\nu = {\delta \rho_\nu \over \rho_{tot}} =
\Omega_\nu \,{m_{\nu_\mu}-  m_{\nu_\tau} \over m_{\nu_\mu}}
{n_0 -n_1 \over n_0}
\left( 1 + {n_1 \over n_0} \,\, {m_{\nu_\tau} \over m_{\nu_\mu}}\right)^{-1}
\label{drho12}
\ee
where $\Omega_\nu = \rho_1 /\rho_{tot}
\approx 10^{-2} (m_\nu /{\rm eV} )\, h_{100}^{-2}$, and $m_\nu$ is the
largest neutrino mass.  An increase in neutrino number density due to
a possible large degeneracy is neglected. This effect
would create approximately twice larger $\Omega_\nu$.

In the limit of large mass difference, i.e. $\delta m \approx m$, we
obtain
\be
r_\nu = 0.4 \Omega_\nu = 4\cdot 10^{-3}  h_{100}^{-2}
\left( m_\nu / {\rm eV}\right).
\label{rnu}
\ee
This large density contrast would evolve only
at the stage when neutrinos became nonrelativistic. For for $m_\nu < 0.1$ eV,
this happens after hydrogen recombination. So primary anisotropies of CMB
are reduced by the factor
$(m/p)^ 2 \approx (m/T_{{\rm rec}})^ 2\approx 10 (m_\nu /{\rm eV})^2$
and we obtain
\be
\left({\delta \rho_\nu \over \rho_{tot}}\right)_{rec}
\approx 4\cdot 10^{-2}  h_{100}^{-2} \left( m_\nu / {\rm eV}\right)^3.
\label{drrec}
\ee
With $m_\nu =0.1 $ eV, this may contradict the observed smoothness of CMB,
but there is no problem with a twice smaller mass difference.

The other limit of small
mass difference, $ \delta m \ll m$, is less dangerous. The relative density
contrast in this case is $r_\nu \approx 0.4 \Omega_\nu (\delta m^2 /2m^2)$
and with e.g.
$m_\nu = 5$ eV and $\delta m^2 = 0.01 {\rm eV}^2$ this ratio is
smaller than  $10^{-5}$.
Thus, the
existing data on neutrino oscillations do not restrict or eliminate
our model but may lead to serious bounds in future.

A variation of the 
mass fraction of primordial $^4 He$ could be observed in the
future high precision measurements of CMB anisotropies at small angular
scales. There are two possible effects~\cite{hssw}. First, a slight
difference in recombination temperature which logarithmically depends on
hydrogen-to-photon ratio, and second, a strong suppression of high
multipoles with an increase of $R_p$. The latter is related to the
earlier helium recombination with respect to hydrogen and correspondingly to
a smaller number of free electrons at the moment of hydrogen recombination.
This in turn results in an increase of the mean free path of photons in
the primeval plasma and in a stronger Silk damping~\cite{js}. The position
and the magnitude of the first acoustic peak remains practically
unchanged~\cite{hssw}.

This effect seems very promising for obtaining a bound on or an
observation of a possible variation of primordial helium mass fraction. If
this is the case then the amplitude of high multipoles at
different directions on the sky would be quite different.
The impact of the possible variation of primordial abundances on the
angular spectrum of CMB anisotropy at low $l$ is more model dependent. It
may have a peak corresponding to the characteristic scale
$R > 200-300$ Mpc or a plateau, which would mimic the effect of
hot dark matter.

\section{Decaying neutrinos \label{decnu}}
\subsection{Introduction \label{ss:intr-dec}}

If neutrinos are massive then they may decay into a lighter neutral
fermion and something else, which could be either a photon, or a light
scalar boson, or a pair of light fermions. In this case, as was mentioned
above, the cosmological limits on neutrino mass both from above and below
would be invalid. It was first mentioned in ref.~\cite{gost}
that Gerstein-Zeldovich limit (it was called Cowsik-McLelland in
that paper) would not be applicable if neutrinos were unstable and,
depending on the model of their decay, they might be as heavy as 25 keV.
Goldman and Stephenson~\cite{gost}, used the condition $\tau_\nu > t_U$ and a
concrete model of the decay, so that the relation between the mass
and life-time did not contain any arbitrary parameters.
Similar conclusions but for heavy neutrinos with mass bounded from
below~\cite{lw}-\cite{sako} (see sec. \ref{stabhvnu}),
was made both in the papers where the bound
was derived~\cite{vdz,sako} and in papers that appeared immediately
after derivation of this bound~\cite{dikote,dkt}. Concrete mechanisms of
neutrino decays and the violation of cosmological mass limits
were discussed in the papers~\cite{chikashige80,fam} in majoron
and familon models respectively. Calculations of neutrino radiative
decays $\nu' \rar \nu \,\gamma$ were pioneered in ref.~\cite{petkov77}
(the update of the works on radiative decays prior to 1987 can be
found in the review~\cite{bilenky87p}).

There are four different kinds of cosmological arguments that permit
constraining mass/life-time of unstable neutrinos. The first is
very close to the one used for stable particles: the energy of both 
decaying
particles plus the red-shifted energy of the relativistic decay products
should not over-close the universe. This argument could be made more
restrictive if one takes into account that the theory of large 
scale structure
formation demands that transition from radiation domination stage
to the matter domination stage should happen sufficiently early.

The second argument is
applicable if neutrino (or in fact any other particle) decays into photons
or $e^+e^-$-pairs. Photons produced as a result of these
decays could be directly observable and should not contradict
astronomical data on cosmic electromagnetic radiation with different
frequencies. If the decay took place before hydrogen recombination, i.e.
$\tau_\nu < 10^{13}$ sec, then decay products could be traced through
a possible distortion of the Planck spectrum of cosmic microwave radiation
(CMB). Decay at a later epoch would produce electromagnetic radiation in
all parts of the spectrum from microwave to visible light, ultraviolet and
even higher, depending on the mass of the decaying neutrino. The present
data permit putting rather strong restrictions on the mass of the decaying
particle and on its life-time.
A good description of these issues and the state of the art of the 1990s
can be found in the review paper~\cite{retu}.

The third set of arguments is based on the consideration of primordial
nucleosynthesis and is discussed in section~\ref{massunstbl}.
Rather interesting
limits on neutrino decays into sterile channels (i.e. into channels
which produce unobservable particles, $\nu_h\rar 2\nu_l\, \bar \nu_l$ or
$\nu_h \rar \nu_l \phi$) can be obtained from a study of the spectrum
of angular fluctuations of CMB. This is the fourth and last subject
discussed in this section. One can find e.g.
in ref.~\cite{hime91} how such arguments can be used to restrict the
properties of 17 keV neutrino, though the existence of latter is
now ruled out.

\subsection{Cosmic density constraints \label{cosden}}

If there are relic stable massive particle species with the
number density $n_h = r_h n_\gamma$ then, as we have seen above, the
particle mass is bounded by the condition that the present-day
mass/energy density of these particles is below the critical energy
density, $\rho_c\approx 10 h^2 {\rm keV/cm^3}$ (see eq.~(\ref{rhoc})).
From this it immediately follows that
$m_h < 25h^2 {\rm eV} /r_h$. If the particles
in question are unstable and if they decay into light or massless species,
their energy is red-shifted in the course of the universe expansion and
the limit becomes weaker. A rather crude condition that must be fulfilled
is that the energy density of decay products today is also smaller than
$\rho_c$~\cite{sako,dikote,vdz,dz}. An approximate
bound can be easily obtained in the instant
decay approximation and under assumption that the universe is flat and is
dominated by usual matter, i.e.  $\rho_{vac}=0$ and $\Omega_{matter} =1$.
Now we believe that the universe is flat, i.e. $\Omega =1$ but that it is
dominated by vacuum or vacuum-like energy, while normal matter contributes
about 30\% into $\Omega$. Correspondingly the limit on the mass density of
neutrinos and their decay products would be 3 times more restrictive.

If unstable particles decayed at an early stage when universe was
dominated by relativistic matter, i.e. $\tau_h < t_{eq}$, then at
the moment of decay, at $t=\tau_h$, the temperature of the universe was
roughly $T/{\rm MeV}=1/ \sqrt {\tau_h/{\rm sec} }$. Keeping in mind that
today temperature is approximately $T_{CMB} \approx 2\cdot
10^{-4} {\rm eV}$, we conclude that the cosmological energy density of
decay products of an unstable particle would be smaller than
that of a stable one by the factor
$5\cdot 10^9 /\sqrt{\tau_h/{\rm sec}}$ (recalling that this is true for
$\tau_h < t_{eq}$).
If $\tau_h > t_{eq}$, then the scale factor during matter dominated stage
behaves as $a(t) \sim t^{2/3}$ and the red-shift at decay is
$z_{dec}+1 \approx (t_U /\tau_h)^{2/3}$. The cosmological bound on the
mass would be weakened by this factor. For example for $\nut$ with the mass
10 MeV, the frozen energy density in accordance with calculations of
ref.~\cite{dhsm} is $\mnt n_{\nut} = 2.5\,n_{\nu_0}$ MeV,
where $n_{\nu_0}=110 /{\rm cm}^3$
is the standard number density of massless neutrinos
(see fig.~\ref{rm}). In order not to
overcome $\rho_c$, decay products must be red-shifted roughly by
the factor $10^5$. It means that $\taut$ should decay at RD stage with
$\taut < 2.5 \cdot 10^9$ sec. A similar bound,
$\taut< 2 \cdot 10^{10} (100\,{\rm keV}/\mnt)^2$ sec, was obtained in
ref.~\cite{pal} from the condition that the universe, which was
radiation dominated by the relativistic products of neutrino
decay, must be older than $10^{10}$ years. This means in particular that
if neutrinos decayed after hydrogen recombination but before the present epoch,
$10^{13}\,{\rm sec} <\tau_\nu < 3\cdot 10^{17}\,{\rm sec}$, neutrino
mass must be smaller than 5 keV and the decay into $e^+e^-$ is impossible.

On the other hand, the limit can be strengthened~\cite{gllss,sttu}
by approximately $10^4$ if one takes into account the fact that that
the structures in the universe do not form at RD stage~\cite{mesz}. The
limits would be somewhat different in the exotic case that the universe
now is radiation dominated as a result of heavy particle decay, while
recently it was matter dominated and structures were formed at MD
stage~\cite{sttu}; for details one can address the above quoted paper.
Knowing that the fluctuations of CMB temperature are very small,
$\delta T/T < {\rm a\,\, few} \times 10^{-5}$, we must conclude that
structures started to form at least at $z=10^4$. It means that now
$\rho_{rel}/\rho_m< 10^{-4}$.  Of course for more precise
limits one should accurately take into account the universe expansion law
with arbitrary $\Omega_{matter}$ and $\Lambda$ and to relax instant decay
approximation. All this is straightforward but not very simple
and we will not go into more
detail. There could be some complications if the decay products
are relativistic but massive. In this case one should reconsider the above
estimates of the cosmic energy density, including massive decay products
and take into account the fact that
they became non-relativistic. One more complication would arise if 
heavy particle decays at an earlier
MD-stage but the universe became dominated
by relativistic matter as a result of decay into relativistic particles.
This feature may produce interesting signatures in the large scale
structure  of the universe (see sec.~\ref{normalnu}).

There is an essential point that should not be overlooked. The calculated
abundance of heavy particles is normalized to the number density of
photons in CMBR that existed prior to decay (to be more precise, the
normalization is made with respect to cosmic entropy density which
is conserved in comoving volume in the case of thermal equilibrium,
but this difference is not important
for our purpose). The products of the decay could be quickly thermalized
with background radiation. It could happen e.g. if decays proceeded into
photons and/or electrons at sufficiently early stage (see the next
subsection). In this case the ratio $n_h/n_\gamma$ at the present day
would be different from that at the time of decay and the mass constraint
for unstable particle would be changed. The condition
$\rho_{decay}<\rho_c$ would be useless for obtaining a constraint on
mass/life-time of a decaying particle if all decay products were thermalized
and disappeared in the thermal bath of CMBR. However in the case of neutrino
decay at least one particle among decay products must be a light
neutrino and the arguments presented above are applicable to it with
an evident weakening of the bound because this neutrino carries only a
fraction of the total energy of the initial particle.

\subsection{Constraints on radiative decays
from the spectrum of cosmic microwave background radiation
\label{cmbr}}

If neutrinos decay into particles which possess electromagnetic interactions
their life-time can be further bounded. There are two possible
effects that may either restrict the properties of decaying particles or
explain some observed features on the sky: for an early decay, before
hydrogen recombination, the electromagnetic decay products would
be thermalized through strong interaction with cosmic electrons and photons,
but the thermalization might be incomplete and the decay would distort
the Planckian spectrum of CMBR.
In case of a late decay, $z<10^3$ or $\tau_\nu > 10^{12}-10^{13} $ sec,
after hydrogen recombination, when the universe became transparent to
photons, decay products remained undisturbed
by secondary interactions and may be directly
observed in cosmic electromagnetic radiation in all frequency ranges,
depending upon the mass and life-time of the decaying particle. In this
subsection we consider the early decays, $z>10^3$, while the late ones
are discussed in the following subsection.

The possibility of constraining late-time (but prior to the recombination)
electromagnetic energy release in the primeval plasma through the
limits on spectral distortion of CMB was probably first considered in
refs.~\cite{zesu,suze}. Early works on the subject are reviewed in
ref.~\cite{szcmb}, where an extensive list of references is provided
(see also the books~\cite{zn,peeb}). Implications of CMB spectrum constraints
for the electromagnetic decays of massive neutrinos were originally
considered in refs.~\cite{sako,dkt,gllss,dz}. If neutrino decays into
electromagnetic channels, $\nu\rar \gamma + ...$ or $\nu\rar e^+e^- + ...$,
then decay products would interact with cosmic electromagnetic background
and would distort its spectrum. If decay takes place sufficiently early,
the distortion would be washed out by subsequent interaction with
plasma and the Planck spectrum would be restored with a different
temperature. However, for a late decay thermalization might be incomplete
and traces of the decay would be imprinted on the spectrum.

The process of thermalization includes essentially two mechanisms.
One establishes kinetic equilibrium, which is realized by elastic
scattering, $\gamma e \lrar \gamma e$, without changing particle number,
and the other involves inelastic processes with different number 
of photons in initial
and final states, which reduce chemical potential down to zero. It is
known from the observations ( see e.g. ref~\cite{cmbchem}) that the latter
is bounded by $\mu_\gamma /T_\gamma < 10^{-4}$.

A crude estimate of the characteristic time of restoration of kinetic
equilibrium
of photons can be done as follows. The number of collisions per unit time
is given by $\dot n_\gamma /n_\gamma = \sigma_T n_e$, where $\sigma_T=
8\pi\alpha^2/3m^2_e $ is the Thomson cross-section and the number densities
of photons, $n_\gamma$, and electrons, $n_e$, are related as
\be
{n_e \over n_\gamma } \approx {n_B \over n_\gamma } =
{\Omega_B \rho_c \over m_p n_\gamma} \approx 2.5\cdot 10^{-8}
\Omega_B h^2
\label{nengamma}
\ee
where sub-$B$ means baryonic and the preferred value of $\Omega_B h^2$ is
around $10^{-2}$. In a single collision a photon may change the magnitude
of its momentum on average by $\omega v_e$
where $\omega$ is the photon energy and $v_e\sim \sqrt{T/m_e}$ is the
electron velocity. To ensure the relative momentum change
is of the order of unity, approximately
$1/v_e$ collisions are necessary. Thus the characteristic time is
\be
\tau_{elastic} = {m_e^{1/2} \over T^{1/2} \sigma_T n_e} \approx
4\cdot 10^{12} {\rm sec} (T/{\rm eV})^{-7/2} (\Omega_B h^2 /10^{-2})^{-1}
\label{tauel}
\ee
Comparing this expression with cosmological time (at RD stage),
$t_c/{\rm sec} = (T/{\rm MeV)}^{-2}$, we find that kinetic
equilibrium would be
restored at $t < 10^{11}\, {\rm sec} (\Omega_B h^2 /0.01 )^{4/3}$.
This result is close to the one obtained in ref.~\cite{sako}, where
a conclusion was made that the life-time of a heavy neutrino with
respect to electromagnetic decays should be smaller than $10^{10}$ sec.
(Presumably the authors of this paper used a different value of
$\Omega_B $.) However this
limit is too weak because only the restoration of kinetic equilibrium
was considered, while spectral distortion due to a non-zero chemical
potential was not taken into account. Indeed if only elastic scattering is
operative, the photons would acquire thermal Bose distribution with a
non-zero chemical potential. The latter could be washed out by
reactions where the number of photons in initial and final states are different.
These are the following two processes: double Compton scattering,
$\gamma e \lrar 2\gamma e$, and Bremsstrahlung, $eX\lrar eX\gamma$,
where $X$ is a proton or an ion.
For a large number density of baryons, roughly for $\Omega_b >0.1$
Bremsstrahlung dominates, while for a smaller and more realistic
$\Omega_b <0.1$ the double Compton plays a more important role in
establishing complete thermal equilibrium.

In most of the earlier works (in the 1970s), only Bremsstrahlung was taken
into account because it was believed that the cosmic baryon number density
is large, $\Omega_b h^2 \sim 1$. According to calculations of
refs.~\cite{zesu,suze}, Bremsstrahlung would wash out all distortion
in CMB spectrum if a significant energy release (around 10\%) took
place before the red-shift $z=  10^8(\Omega_B h^2)^{-4}$ ((to avoid
possible confusion one should keep in mind that in different papers
the parameter $h$ is normalized in different ways; here we use
$h =H /100 {\rm km/sec/Mpc}$, while in many earlier papers it is
normalized to 50 km/sec/Mpc). Based on that result
the limit on the life-time of neutrino in the case of radiative
decays was obtained~\cite{gllss,dz},
$\tau_\nu^{rad} < (2-3)\cdot 10^3$ sec.
The characteristic time of Bremsstrahlung can be estimated as follows.
The cross-section of scattering of non-relativistic electron on a heavy
target with emission of photon with energy $\omega\ll m_e$ is given
by~\cite{akhber}
\be
d\sigma_{BS} = {32\alpha^3 \over 3p m_e} {d\omega \over \omega}
\ln \left(\sqrt{{p^2 \over 2m_e \omega}}+
\sqrt{{p^2 \over 2m_e \omega}-1} \right)
\label{dsigmabs}
\ee
where $p$ is the electron momentum; in thermal equilibrium $p^2\sim 2m_e T$.
We assume that the $\log^2$-factor coming from integration over $\omega$
is close to unity (it is difficult to make a closer estimate by this naive
approach). The number of photon producing collisions per unit time is
given by
\be
\left({ \dot n_\gamma \over n_\gamma}\right)_{BS} =
 {\sigma_{BS}v n_e^2 \over n_\gamma}
\label{dotngambs}
\ee
where $v=p/m_e$ is the velocity of electrons. It is assumed that the
plasma is electrically neutral, so the number density of protons
is the same as that of electrons. A possible presence of neutral
helium-4 atoms is neglected because the relevant time is
far smaller than the recombination time. For electron momentum we take the
``thermal'' value $p^2 = 2m_e T$ and assume that the temperatures of
photons and electrons are the same. Substituting the numbers we find
\be
\tau_{BS} = {5.5\cdot 10^{18} {\rm sec} \over (\Omega_B h^2 /0.01)^2
(T/{\rm keV})^{5/2} I}\, ,
\label{taubs}
\ee
where $I$ is an unknown value of the integral over $\omega$. If $I=1$,
this result is approximately twice higher than that given by more accurate
considerations~\cite{light,husi}. In refs.~\cite{husi} the factor
$4\pi$, omitted in the paper~\cite{light}, was corrected. Taking $I=2$
and $\Omega_B h^2 =1$
we find that Bremsstrahlung is faster than expansion, $\tau_{BS} < t_c$,
for $T> 0.07$ MeV. In other words, radiative decays of neutrino with a 
life time below 200 sec would not disturb the CMB spectrum. For a realistic
number density of baryons, $\Omega_B h^2 = 10^{-2}$, 
the Bremsstrahlung seems ineffective, but
this is not exactly so. In fact, for
temperatures higher than $\sim m_e/20$ the number density of electrons and
positrons is given by thermal equilibrium and is much larger than their
asymptotic value, $n_e/n_\gamma \sim 3\cdot 10^{-10}$. The high number
density of $e^\pm$ made Bremsstrahlung very efficient when the universe
was younger than a few thousand seconds. Thus we come to the estimates
quoted above.

The effect of Bremsstrahlung for restoring the equilibrium spectrum
of CMB for the radiatively decaying right-handed neutrinos which decoupled
at high temperatures, when number of species was about 50, was considered
in ref.~\cite{sist}. However, the conclusion that the photons from the decays
would be unobservable if the life-time is below $50 {\rm y} \approx
1.5\cdot 10^9$ sec, seems to be too strong, possibly because the omitted
factor $4\pi$ mentioned above (the shift of life-time is proportional
to $(4\pi)^4$). Subsequent more accurate calculations~\cite{husi}
showed that Bremsstrahlung could not be that efficient even for
$\Omega_b h \approx 1$.

The importance of double Compton (DC) reaction for thermalization of CMB
was mentioned in several early papers~\cite{wey}-\cite{dazo1}. It was
shown in ref.~\cite{dkt} that the distortion of CMB spectrum would be
smoothed down if the energy was released before $z= 4\cdot 10^6 (\Omega_B
h^2)^{1/3}$. It is close to the result presented in ref.~\cite{szcmb})
that DC is efficient before $z=10^7$. It permits the restriction of 
neutrino life-time in case of predominantly radiative decays,
by $\tau > 10^5 -10^6 $ sec~\cite{dkt,dz,saco}.
The characteristic time of thermalization by the double Compton process
can be found as follows. The cross-section of this reaction, in the limit
of low photon energies and if the energy of one of the photons is much
smaller than the other, can be approximately taken as~\cite{jaro}:
\be
\sigma_{DC} = {32\alpha^3 \over 9}{\omega^2 \over m_e^4}
\label{sigmadc}
\ee
The characteristic time is
\be
\tau_{DC} \equiv\left({\dot n_\gamma \over n_\gamma}\right)^{-1}
=\left( \sigma_{DC} n_e \right)^{-1} = {6\cdot 10^{22} \over
(\Omega_B h^2 /0.01) (T/{\rm eV})^5 }\, {\rm sec}
\label{taudc}
\ee
This time is smaller than the cosmological time,
 $t_c = 10^{12} /(T/{\rm eV}))^2$ sec, if
$T> 4\cdot 10^3 {\rm eV} /(\Omega_B h^2 /0.01)^{1/3}$.
This permits to obtain the limit
\be
\tau_\nu^{rad} < 6\cdot 10^4 (\Omega_B h^2 /0.01)^{2/3}\, {\rm sec}
\label{taunudc}
\ee
We substituted here thermal average $\langle \omega^2 \rangle \sim 10$.
The bound (\ref{taunudc}) is essentially the same as that
found in ref.~\cite{dkt}.

The simple estimates presented above are not very precise and can be trusted
within a factor of few. More accurate results strongly depend upon the photon
frequency and can be found only through a solution of kinetic equations.
Fortunately for non-relativistic electrons and in the limit of low photon
energies, $\omega \ll m_e$, the
system of integro-differential kinetic equations (\ref{kin1},\ref{si}),
which are very difficult to treat numerically, can be approximately reduced
to ordinary differential equations. The essential condition that allows this simplification is that the relative frequency change in a
single collision be small,
$\delta \omega /\omega \sim (T/m_e)^{1/2} \ll 1$.
In zeroth approximation one may neglect the frequency shift and
take the latter into account perturbatively expanding $\delta$-function
which gives energy-momentum conservation, $\delta (\sum p_i - \sum p_f)$.
For the case of elastic $\gamma\,e$-scattering such equation was derived
by Kompaneets~\cite{kom}. It reads
\be
t_{el}{\partial \fg(t,y) \over \partial t}\mid_K =\left( {T\over m_e y^2}
\right) {\partial \over \partial y} \left[ y^4 \left({\partial \fg \over
\partial y} + \fg +\fg^2 \right) \right]
\label{komp}
\ee
where $y= \omega /T$ and $t_{el} = (\sigma_T n_e)^{-1}$. In the
equation above it is assumed that the temperatures of photons and electrons
are the same, $T_e=T_\gamma$, but in the original version of the Kompaneets
equation there was no such assumption and the factor $T_e/T_\gamma$ in
front of the first term in the r.h.s. was present.
This equation describes diffusion of photons in frequency space with a
conserved number of photons. The impact of cosmological expansion on
this equation
was considered in refs.~\cite{wey2,bedo,husi} and was shown to be weak.

The equilibrium solution of this equation is the
Bose-Einstein distribution, $\fg = [\exp (\xi + y) -1]^{-1}$ with the
dimensionless chemical potential $\xi =\mu/T$, which is determined by the
initial conditions. If only elastic scattering is operative, chemical
potential does not relax down to zero. Such relaxation may be achieved
only by inelastic processes with a different number of photons in initial
and final states, which go in higher order in the fine structure coupling
constant $\alpha$. As we have already noted, there are two such processes
in the leading (third) order in $\alpha$: Bremsstrahlung and double Compton.
The analog of the Kompaneets equation for these processes was derived
in ref.~\cite{light}. Bremsstrahlung thermalization was considered earlier
in the papers~\cite{fere,ilsu} and the equation for the double Compton
was derived independently in ref.~\cite{thorn} and in a simplified version
in ref.~\cite{gould}. These two processes create new photons, mostly at low
energy $\omega$. Then, these photons diffuse upward in energy in accordance
with eq. (\ref{komp}) in a much shorter time. Altogether, black body spectrum
is created if the characteristic time scale of the reactions is smaller
than the cosmological time. If relativistic electrons or photons are
injected into plasma, the relaxation time would be longer because the
corresponding cross-sections are suppressed roughly by $m^2_e/s$, where
$s$ is the total center-of-mass energy squared. However relativistic
electrons or positrons are very quickly thermalized by scattering on a
large number of photons. Thermalization of non-equilibrium photons is
much slower because it is achieved by scattering on electrons with
a very small number density $n_e/n_\gamma =10^{-9}-10^{-10}\ll 1$.

Let us now consider inelastic processes that can reduce chemical potential
down to zero.
The contribution of Bremsstrahlung into thermalization is described by the
equation:
\be
t_{el}{\partial \fg(t,y) \over \partial t}\mid_{BS}=
Q{g(y) \over y^3 \exp (y)} \left[ 1-\fg \left(e^y -1\right) \right]
\label{brems}
\ee
where
\be
Q={4\pi \alpha\over (2\pi)^{7/2}} \left({m_e\over T}\right)^{1/2}
{\sum_i n_i Z_i^2 \over T^3},
\label{Q}
\ee
and $n_i$ and $Z_i e$ are respectively the number density and charge of ions.
The function $g(x)$ is given by $g(y)= \ln (2.2/y)$ for $y\leq 1$ and
$g(y)= \ln 2.2/\sqrt y$ for $y\geq 1$. In accordance with ref.~\cite{husi}
the factor $Q$ presented here is larger by $4\pi$ than that in
paper~\cite{light}.

The contribution of double Compton into evolution of $\fg$ is given by
\be
t_{el}{\partial \fg(t,y) \over \partial t}\mid_{DC}=
\left({4\alpha \over 3\pi}\right)\left({T\over m_2}\right)^2
{I(t)\over y^3}\left[1-\fg(y,t) \left(e^y -1\right)\right]
\label{doubcom}
\ee
where
\be
I(t)=\int dy y^4 \fg (y,t)\left[1+\fg(y,t)\right]
\label{ioft}
\ee
The total evolution of $\fg$ is determined by the sum of all three
contributions
(\ref{komp},\ref{brems},\ref{doubcom}). Based on these equations
the efficiency of thermalization
of CMB due to double Compton was studied in ref.~\cite{dazo2}. It was
shown there that double Compton efficiently smoothed down any spectrum
distortion for the red-shift $z> 6\cdot 10^6/(\Omega_B h^2 /0.01)^{1/3}$.
This is quite close to the naive estimates presented
above (\ref{taudc},\ref{taunudc}).

Detailed calculations of the impact of decaying neutrinos
on CMB spectrum were made in refs.~\cite{kasa,husi}, where all three
contributions into evolution of $\fg$ were taken into account. However the
contribution of Bremsstrahlung was underestimated in ref.~\cite{kasa} by
the factor $4\pi$, as was noticed in ref.~\cite{husi}.  In ref.~\cite{husi}
both numerical and approximate analytical solution to the evolution
equation have been found under the simplifying assumption that the injection
of energy from the decay was instantaneous. It was found that any significant
energy injection is ruled out for a red-shift smaller than
\be
z ={ 5\cdot 10^5 \over \left( \Omega_B h^2 \right)^{2/5}} =
{3\cdot 10^6 \over\left( \Omega_B h^2 /0.01 \right)^{2/5}}
\label{zhusi}
\ee
practically for all (except a very small class) injection scenarios. It
corresponds to the bound on neutrino life-time
\be
\tau_\nu^{rad} < 2.5\cdot 10^6 \left( {\Omega_B h^2 \over 0.01}
\right)^{4/5} {\rm sec}
\label{tauradfin}
\ee
For further details and discussion of energy dependence and effective
chemical potential one might address the paper~\cite{husi}.

Though the above calculations are quite accurate, the
underlying assumptions which permit to reduce the complete system of
kinetic equations to simpler differential equations, may be invalid
or not very precise. In particular, they are not true for relativistic
electrons or energetic photons. The commonly used assumption that photons
and electrons are described by thermal distribution with the same or
different temperatures may also be inaccurate. In view of that, it is
very desirable to do precise calculations without any simplifying
assumptions by numerical solution of exact integro-differential kinetic
equation in the similar way as it was done for the impact of neutrinos on
nucleosynthesis~\cite{hanmad,hm1,dhs0,dhsm,dhps}. Of course the solution
of the integro-differential equation is much more difficult technically,
and what's more, the matrix elements of the reactions are not polynomial as in the
case of local weak interactions. It can be demonstrated that the collision
integral for elastic Compton scattering can be reduced down to two
dimensions~\cite{dhps2d}, even for the exact non-polynomial matrix element.
So at least for Compton scattering the problem seems to be tractable.
The numerical solution of the complete set of kinetic equations
found in ref.~\cite{dhps2d} agrees well with the solution of Kompaneets
equations for nonrelativistic electrons.
However for inelastic processes the collision integral hardly can be
reduced down to two or even three dimensions without any approximations,
and the direct numerical solution of kinetic equations looks extremely
difficult.

\subsection{Cosmic electromagnetic radiation, other than CMBR
\label{cer}}

If massive neutrinos live longer than hydrogen recombination time, then,
depending on their mass, either they would distort the
CMBR spectrum (if they
are very light) - or, in case of a larger
mass, the flux of the decay photons would be observable in
more energetic cosmic photon backgrounds at different frequencies: gamma-
and X-rays, ultraviolet optical, infrared and, for very small masses,
radio. Cowsik~\cite{cow} was the first
to point out that the life-time of a massive neutrino $\nu_h$ decaying
into a lighter one, $\nu_l$, and a photon, $\nu_h\rar \nu_l +\gamma$, can
be restricted based on these considerations. He discussed a relatively light
neutrino and found that for $m_\nu \approx 10^{-3}$ eV its life-time
must be larger than $10^{19}$ sec, otherwise they would contribute too
much into CMBR. Neutrino with $m_\nu = 1$ eV would produce optical
photons and from the limit on the background star light flux,
$f= 3\cdot 10^8 /{\rm cm^2/sec}$,
one can conclude that neutrino should live longer than $10^{23}$ sec.
These limits are somewhat overestimated because the number density of
cosmic neutrinos was taken approximately as 6 times bigger
than the actual value (see discussion in sec. \ref{gerzel}).

Cosmic electromagnetic radiation created by possible electromagnetic
decays of massive neutrinos was estimated for neutrinos with any
(small or large) mass in a slightly later paper~\cite{gllss}.
It was roughly concluded that if neutrinos live more
than $10^5$ years (this is the approximate time of recombination), then
they must live longer than $10^{18}$ years.
The range of neutrino masses from 10 to 100 eV, i.e. approximately
satisfying the Gerstein-Zeldovich limit, was considered in
refs.~\cite{rugl,steck,kiboja}, where the contribution from
neutrino radiative decays into cosmic UV (ultraviolet) background
was calculated. The conclusion of ref.~\cite{rugl}, that no bound on
life-time can be derived from the known astronomical data on UV,
contradicts the other two papers~\cite{steck,kiboja} and is possibly
related to a numerical error, as stated in ref.~\cite{kiboja}.
In ref.~\cite{steck} the hypothesis was investigated that an observed
feature in the spectrum of UV background might be explained by the
decay $\nu_h\rar \nu_l +\gamma$. If this were the case, the mass of
$\nu_h$ would have been around 14 eV. Presently this spectral signature
disappeared and does not give an indication of the existence of 14-eV neutrino.
A systematic study of the constraints
on the life-time of radiative decays of light neutrinos with mass 10-100 eV
was performed in ref.~\cite{kiboja}, where it was shown that the observations
of electromagnetic radiation from infrared to extreme ultraviolet excludes
electromagnetic decay in the life-time interval $10^{13}-10^{23}$ sec.
The results of refs.~\cite{steck} and ~\cite{kiboja} agree in the overlapping
region of mass values. The spectral density of electromagnetic radiation
originating from the decay $\nu_h\rar \nu_l +\gamma$ can be calculated as
follows. In the absence of absorption, when the photon energy is smaller
than 13.6 eV, the photon distribution function $\fg (t,\omega)$ obeys
the equation:
\be
\left( \partial_t - H\omega\partial_\omega \right) \fg =
{1\over 2\omega} \int {d^3 p \over 2E_p\,(2\pi)^3}
{d^3 q \over 2E_q\,(2\pi)^3} f_{\nu_h} |A|^2 (2\pi)^4 \delta^4 (p-q-k)
\label{dtfgam}
\ee
where $f_{\nu_h}$ is the distribution function of the heavy neutrino and
$p$ is its momentum; $q$ and $k$ are the momenta of the light
neutrino and photon respectively. In this simplified kinetic equation the inverse
decay as well as Fermi suppression for neutrinos and Bose amplification
for photons are neglected. The amplitude $A$ is related to the decay
width as
\be
\Gamma = {1\over 2m} \int {d^3 q \over 2E_q\,(2\pi)^3}
{d^3 k  \over 2 \omega\,(2\pi)^3} |A|^2 (2\pi)^4 \delta^4 (p-q-k)
= {|A|^2 \over 16\pi m}
\label{gammanuh}
\ee
It is convenient to introduce the variables $x=ma(t)$ and $y=\omega a(t)$
where $a(t)$ is the cosmological scale factor, normalized so that $a=1$
at the present day (This normalization is different from the one used
in previous subsections). In terms of these variables
the l.h.s. of equation~(\ref{dtfgam}) takes the form $Hx\partial_x \fg$.
The calculations are very much simplified if the heavier neutrino is
non-relativistic, so that $E_p \approx m$, while the light one is massless
or very light. After some simple algebra one obtains
\be
Hx\partial_x \fg (x,y) = {16\pi^2 \Gamma n_{\nu_h} \over m^3} \,
\delta \left( 1 - {2y\over x}\right)
\label{hxdxfg}
\ee
where
$ n_{\nu_h} (x) = \int {d^3 p f_{\nu_h} / (2\pi)^3}$ is the number
density of the heavy neutrinos. It decreases due to decay and the
universe expansion, so that
\be
n_{\nu_h} (x) = n_{\nu_h}^{(0)}\exp [-\Gamma(t-t_0) ]/a^3
\label{nnuh}
\ee
where $t_{0}$ is the universe age
and $ n_{\nu_h}^{(0)}$ is the number density of heavy neutrinos
at the present time. To proceed further we need to know the time
dependence of the scale factor which is determined by the Einstein
equation (\ref{dota2}). In the simple case of matter-dominated flat
universe (i.e. $\Omega =1$) the expansion law is $a(t) = (t/t_0)^{2/3}$
and $H= 2/3t$. We will make one more simplifying assumption that the
life-time of $\nu_h$ is large in comparison with the universe age $t_0$.
After that, equation~(\ref{hxdxfg})
is easily integrated and we obtain for the intensity of the radiation
in the interval of wave length $d\lambda$:
\be
dI = {\Gamma n_{\nu_h}^{(0)}\over H_0} {\lambda_{min}^{3/2}
 \over \lambda^{5/2}}
d\lambda
={\Gamma n_{\nu_h}^(0)\over H_0} {\omega^{1/2} d\omega \over
\omega_{max}^{3/2}}
\label{ioflambda}
\ee
where $\lambda_{min} = 4\pi /m$ is the minimal-wave length of emitted
photons and $\omega_{max}=m_\nu /2$ is the maximum energy of the photons.
This is essentially the result obtained in refs.~\cite{steck,kiboja}, where
a more general expression, valid for a non-flat universe, was derived.
Calculations for a more general case of neutrinos decaying with an arbitrary
energy, not necessarily at rest, were done relatively
recently in ref.~\cite{mato}.

If $\nu_h$ is more massive so that the photon energy is larger than 13.6 eV,
they can ionize hydrogen, and the universe becomes opaque to such photons.
However the red-shifted low energy tail of the spectrum still remains
dangerous. The analysis made in ref.~\cite{kiboja} permits
to exclude life-times smaller than $10^{22}$ to $10^{23}$ sec in the mass
interval 10-100 eV.

In ref.~\cite{berezhiani87} cosmological restrictions on the decay
$\nu_h \rar \nu_l + \gamma$ induced by a large non-diagonal
magnetic moment, $\mu_{lh} = (10^{-8}-10^{-10}) \mu_B$, were considered.
Such a large $\mu_{lh}$ would be allowed by the data on cosmic
electromagnetic radiation if the decay has a very small branching ratio
$(<10^{-6})$ and the dominant mode is invisible or the mass of $\nuh$ is
rather high ($>100$ keV) so the photons from $\nu_h$ decay were thermalized 
with CMBR at high red-shift.

A more complicated chain of decays was discussed in 
ref.~\cite{berezhiani90}: $ \nu_h \rar \nu_l + \gamma$, $a\rar 2\gamma$,
where $a$ is an axion. Such decays are consistent with cosmology for
the taken in the paper axion mass about 3 eV. If the decays took place near
red-shift $z\sim 10^3$ it would lead to a significant reionization of matter,
which in turn would smooth down angular fluctuations of CMBR.

For a smaller neutrino mass, $m=0.01 - 1$ eV, the bound on their possible
radiative decays can be found from the extra-galactic infra-red (IR)
background. The bounds~\cite{retu} found from direct IR observations are
roughly $\tau_\nu > ({\rm a \,\, few}) \cdot 10^{21}$ sec for
$m_\nu \approx 1$ eV,  $\tau_\nu >3 \cdot 10^{18}$ sec for
$m_\nu \approx 0.1$ eV, and $\tau_\nu >3 \cdot 10^{19}$ sec for
$m_\nu \approx 0.03$ eV. Recently considerably stronger
limits~\cite{bibubu} on the density of IR background
were found from the observations of high energy (TeV) cosmic photons.
Since those energetic photons should produce $e^+e^-$-pairs through scattering
on IR background, the interstellar medium should become opaque to them and distant
sources would be unobservable.
The idea was first formulated in the paper~\cite{gosh} and later
considered in detail in ref.~\cite{stja}. The effect can be estimated in the
following way. The cross-section of the pair production
$\gamma +\gamma \rar e^+ + e^-$ is
\be
\sigma(\gamma\gamma \rar e^+ e^-) = {\pi \alpha^2 \over 2 m^2_e }
(1-\beta^2) \left[ (3-\beta^4)\ln {1+\beta \over 1-\beta} +2\beta (\beta^2
-2) \right]
\label{sigmagg}
\ee
where $\beta = \sqrt { 1-4m_e^2/s}$ is the electron velocity in center-of-mass
frame and $s=(k_1+k_2)^2$, with $k_j$ being the momenta of colliding photons.
In the limit of small red-shift, $z\ll 1$,
the kinetic equation for the distribution function $f_1$ of the high energy
photons with the energy $\omega_1$ can be written as
\be
{\dot f_1 \over f_1} = - {1\over 2\omega_1} \int { d^3k_2 \over (2\pi)^3\,
2\omega_2}\, {s f_2  \over 2}\, \sigma(\gamma\gamma \rar e^+ e^-)
\label{dotf1}
\ee
The function $f_2$ is expressed through $dI/d\omega$ (\ref{ioflambda}) in
an evident way: $\omega^2 f_2 /(2\pi)^2 = dI/d\omega$. Near threshold,
$s\approx 4m_e^2$, the product $\sigma s/2$ can be approximately taken as
$ 2\pi \alpha^2 \beta$. After that the integration in eq. (\ref{dotf1})
is straightforward and we obtain
\be
{\dot f_1 \over f_1} = - {\pi \alpha^2 n^{(0)}_\nu \Gamma \over 2H_0 m_e^2}
F \left(m_\nu \omega_1 \over 2m_e^2 \right)
\label{dotf12}
\ee
where
\be
F(x) =  x^{-3/2} \int^x_1 dy \sqrt{y-1 \over y}
\label{fforf1}
\ee

For $m_\nu =1$ eV and correspondingly $\omega_2 = 0.5$ eV the threshold for
pair production is reached if high energy photons have the energy above
0.5 TeV. For 10 TeV photons the pairs are produced on IR background with 
energy larger than 0.025 eV. The corresponding optical depth is given by
$d = (\dot f_1 /f_1)^{-1} \approx 10^{23} H_0 \tau_\nu F^{-1}$ cm.
The TeV photons are observed from active galaxies
Mrk 421 and Mrk 501 both at red-sifts slightly above 0.03 or at the distance
$\sim 100/h$ Mpc. No spectral features that may correspond to attenuation of
TeV photons at this distance were observed. This permits us to obtain an upper
limit on the intensity of IR background and the lower limit on
possible radiative decays of neutrino. It was found in ref.~\cite{bibubu}
that the radiative life-time of neutrino should be larger than $10^{14}$ years
for $m_\nu = 1$ eV and $\tau_\nu >2\cdot 10^{13}$ years for $m_\nu = 0.1$ eV.
A substantial improvement in the strength of these limits is expected for the
next generation of instruments.
At the present time, however, astrophysics permits putting stronger limits on
$\tau_\nu$ with respect to radiative decays~\cite{raf98}.

The bounds discussed above tested the hypothesis that cosmic neutrinos are
uniformly distributed in space and, because of that, their decays create a diffuse
electromagnetic background. More stringent limits
can be obtained from the observations of discrete sources
rather than from background measurements if neutrinos are accumulated in
galaxies or their clusters. However, such limits are intrinsically uncertain
because they depend upon unknown fraction of clustered neutrino
dark matter. Under the assumption that the entire
(virial) masses of Coma and Virgo clusters are composed of neutrinos the
conclusion~\cite{shco} was made that
$\tau_\nu > ({\rm a\,\, few})\cdot (10^{23} - 10^{24})$ sec
from the observation of the ultraviolet (UV) spectrum by the ``Voyager 2''
in the 912-1200~{\AA} range and that $\tau_\nu > 10^{25}$ sec for roughly
twice longer wave length.
A slightly stronger limit in the different
wave length range 1240-1550~{\AA} from Apollo 17 UV spectrometer was
obtained in ref.~\cite{hefe},
$\tau_\nu > 2\cdot (10^{24} - 10^{25})$ sec. Later a considerable
improvement of the results of the paper~\cite{shco} was
achieved in ref.~\cite{hoba} in practically the same wave length
interval 912-1150~{\AA},
based on new series of measurements of UV radiation from Coma cluster by
Voyager 2.
For $\lambda = 912 {\AA}$ the upper bound on diffuse line emission is
$J^{(line)}< 6.3 \cdot 10^3\, {\rm photons/cm^2/sec/sr}$ and for continuum
emission $dJ^{(ce)}/d\lambda < 75\, {\rm photons/cm^2/sec/sr/{\AA}}$. This
permits us to obtain the limit $\tau_\nu > 2.4\cdot 10^{25}$ sec.
For  $\lambda = 1150 {\AA}$ the bounds are $J^{(line)}< 2.7 \cdot 10^4$
and $dJ^{(ce)}/d\lambda < 300$ (in the same units as above). Correspondingly
$\tau_\nu < 7.1\cdot 10^{24}$ sec. For the intermediate values of wave
length the limit on $\tau_\nu$ smoothly changes between these two results.

Except for possible direct observations of photons from the decay
$\nu_h \rar \nu_l + \gamma$, they may be observed through ionization
of interstellar hydrogen if neutrino mass is larger than 27.2 eV
and the photon energy is higher than hydrogen ionization threshold.
From the requirement that the ionization level of high velocity clouds
of neutral hydrogen in the Galaxy does not exceed observational limits,
it was found~\cite{mesc} that neutrino life-time should be larger than
$10^{24}$ sec. This limit is independent of the discussed above bounds
based on UV and other backgrounds.
A similar limit was found from the
observation of neutral hydrogen in the nearby galaxy M 31~\cite{resz}
for neutrinos with masses in the range 30-150 eV.
Radiative decays of neutrinos with a
shorter life-time would practically destroy such neutral clouds.
However the accuracy of both results are roughly an order of magnitude,
so that $\tau_\nu$ in the range $10^{23}-10^{25}$ sec can still be
considered as a possibility~\cite{resz}.

On the other hand, the photons from the decay could serve as a missing
ionization factor explaining a high level of ionization of matter in
the universe~\cite{resz,sc82,sc90} (for a detailed discussion and the list of
earlier references see the book~\cite{scibook}). In the standard cosmological
model the density of diffuse neutral hydrogen in the intergalactic medium
should be much higher than the actual upper limits.
The latter are obtained by
the Gunn-Peterson test~\cite{gupe}, i.e. by absorption of the quasar
radiation at the Lyman alpha resonance, where no significant continuum
absorption was registered. Thus there is a very strong indication that
intergalactic medium is highly ionized up to red shifts $z=5$~\cite{gia}.
A recent analysis of the ionization level up to redshift 6 and references
to new observations can be found in the paper~\cite{lidz01}.

The flux of ionizing UV photons from the conventional stellar sources
(mostly from quasars themselves) seems to be insufficient for the observed
high level of ionization (see e.g. ref.~\cite{madau}). However neutrinos
with masses about $m_\nu \approx 27.5$ eV and life-time
$\tau_\nu = (1-2)\cdot 10^{23}$ sec~\cite{sc97} could produce the necessary
photons to maintain the required near-complete ionization.
It was suggested by Melott~\cite{mel84} that radiative decays of neutrinos
producing photons with the energy $\omega \geq 13$ eV  with life-time
around $10^{24}$ sec could be responsible for the sharp hydrogen ionization
edges observed in many galaxies.
Moreover, the same decays could simultaneously
account for the ionization level of hydrogen found in HI
regions, local interstellar medium, and in pregalactic medium (for details
and references see the book~\cite{scibook}).
The role of ionization induced by electromagnetic neutrino decays
in establishing equilibrium between cold and hot phases in the interstellar
medium was recently
studied in ref.~\cite{sanchez00}. It was shown, in particular, that
an increase of neutrino flux (e.g. due to supernova explosion) might
induce condensation of cold clouds stimulating star formation
processes.

The hypothesis of radiative decays of neutrinos was
actively studied in the recent years and seems to be on verge of
exclusion. The search~\cite{dakrfe} of 14-15 eV line from the
galaxy cluster Abel 665 produced a negative result corresponding to the
lower life-time limit $\tau_\nu > 3\cdot 10^{24}$ sec. However, strictly
speaking, one cannot exclude that the line of sight to Abel 665 as well as
to Coma and Virgo clusters (discussed above) is blocked by an
unknown amount of absorbing matter~\cite{ovwebo} and one has to turn to
diffuse extra-galactic UV background. The study of the latter in
ref.~\cite{ovwebo} and in the corrected version~\cite{ovwe} still leaves
some, though rather narrow, room for this hypothesis. Moreover,
the arguments presented in ref.~\cite{bow95} showed that the observational
constraints depended on the distribution of neutrinos in clusters of galaxies
and for some distributions the Sciama scenario was not ruled out.
However as is argued in
the recent paper~\cite{bow99}, this window is closed by their measurements
of diffuse extreme UV emission in the wave range $\lambda = 890-915$ {\AA}.
The measurements made in this work are approximately an order of magnitude
below the level predicted by the Sciama model. But the model of
Melott~\cite{mel84} with a longer neutrino life-time which may explain only 
part of the cosmic ionization pattern, namely
the sharp ionization edges, is not excluded.

Some other bounds for neutrino radiative decays are the following.
A restrictive {\it upper} limit from the Gunn-Peterson test was derived in
ref.~\cite{doju}: $\tau_\nu <10^{23}$ sec for $m_\nu \approx 27$ eV
and $\tau_\nu <5\cdot 10^{23}$ sec for $m_\nu > 28.5$ eV. Taken together
with optimistically strong {\it lower} limits from UV-data, these results
would exclude the Sciama's preferred values $m_\nu =27.4\pm 0.2$ eV and
$\tau_\nu = (1-2)\cdot 10^{23}$ sec~\cite{scibook,scipref}.
If, however, decaying neutrinos are not the only source of HI ionization,
then their life-time could be larger than $10^{24}$ sec. Moreover,
as mentioned above,
the conclusions derived from the UV-data may be weakened due to an
uncertainty in opacity by interstellar dust and an unknown fraction
of neutrino dark matter in galactic clusters (see the paper~\cite{doju} for
details and references).

Growing pressure from accumulated observational data demanded a
modification~\cite{sci95} of Sciama's scenario with a smaller ionizing
contribution from the neutrino decay and with a larger contribution from
conventional sources: stars and quasars. Some more restrictions on the model
would be also eased if one neglects a possible role of neutrinos in large
scale structure formation and concentrates only on the properties of
re-ionization. An analysis performed in
ref.~\cite{masa} under the assumptions that QSOs ionize HI, HeI, and HeII,
stars ionize just HI and HeI and decaying neutrinos ionize only HI,
shows that it is possible to avoid contradictions with Gunn-Peterson test
if additional sources (e.g. stars at large red-shifts $z=2-4$) are
sufficiently strong.

Possible cosmological sources of ionizing photons and the present constraints
on their intensity inside the Local Group were reviewed recently in
ref.~\cite{mabl}. As described there, the cosmological sources of ionizing
photons fall into two categories: standard (active galactic nuclei and
stellar ionizing photons from galaxies) and exotic (decaying particles).
However the recent $H\alpha$-observations in spiral galaxy NGC3198
(see~\cite{mabl} for the references) indicate that the observed emission
is an order of magnitude weaker than requested by decaying neutrino theory.

One more test of the model of radiatively decaying neutrinos can be done
by a study of the angular fluctuations of CMBR~\cite{adams98} and is discussed
below in sec.~\ref{9cmb}. The decays $\nu_h \rar \nu_l + \gamma$ would
significantly suppress the level of angular fluctuations. This seems to
contradict already existing data.

Thus the radiative decays of 27.5 eV neutrinos probably cannot explain the
observed level of ionization in interstellar and intergalactic media.
Since it is also difficult to find an explanation through conventional sources,
the mystery of ionization remains unsolved and at the present state it is
unclear if one has to invoke new physics (e.m. decays of long-lived
particles, mirror photon oscillations, or something even more unusual)
for the resolution of the problem. One more argument against 27.5 eV
neutrinos is that, according to contemporary data, the mass fraction
of matter in the universe is relatively small, $\Omega_m \leq 0.3-0.4$.
Neutrinos should contribute into that no more than 0.1. Correspondingly
their mass should be smaller than 10 eV, according to eq.~(\ref{mnu1}).
Consideration of large scale structure formation imposes even stronger
upper bound on $m_\nu$ (see sec.~\ref{normalnu}).

An extension of the model to include possible mixing of active and 
sterile neutrinos could help. It was considered in refs.~\cite{sci98,mosc}.
The number density of sterile neutrinos with mass 27.4 eV could be
much smaller than the density of normal neutrinos. If e.g. $\nu_s$ were
produced at an early stage above QCD phase transition their number density
would be suppressed by the entropy release. In addition, the mixing angle
between $\nu_s$ and active neutrinos should be very small, otherwise
they would be produced by oscillations at low temperatures.
An extra free parameter, the number density of sterile neutrinos,
permits in some cases to weaken discussed above contradictions between the
decaying neutrino model and observations.

In ref.~\cite{sethi95} the arguments were inverted. The author derived a
limit on possible radiative neutrino decay from the observation of
singly ionized helium in diffuse intergalactic medium. It was assumed
that neutrinos predominantly decay into invisible channels with a small
branching into radiative mode. The observed amount of singly ionized
helium~\cite{jacobsen94} is the lower bound on its abundance and it gives
an upper bound on the amount of doubly ionized helium. If there exists
a radiatively decaying neutrino with the mass twice larger than the
ionization potential of singly ionized helium, $m_\nu > 108.8$ eV,
one can put an upper limit on the radiative decay probability.
According to the paper~\cite{sethi95} for neutrino life-time bounded
from below by $\tau_\nu > 10^{18}\,(1\, {\rm eV} /m_\nu)^2$ sec,
the magnetic transition moment of a heavier neutrino with respect to
the decay $\nu_h \rar \nu_l + \gamma$ is quite strongly bounded by
$\mu_{hl} < (4-8) \cdot 10^{-17} \mu_B$ for
$110\,{\rm eV} < m_\nu < 10\, {\rm keV}$.


\section{Angular anisotropy of CMBR and neutrinos. \label{9cmb}}

The spectrum of angular fluctuations of cosmic microwave background
radiation (CMB) is very sensitive to the fraction of relativistic matter
in the universe, to a possible neutrino mass in eV range, and to decays of
neutrino with life-time around $10^{12}-10^{13}$ sec.
If neutrino decays create photons or $e^+e^-$-pairs, the decay products
could distort the perfect Planckian spectrum of CMB (as discussed in
sec.~\ref{cmbr}), but the spectrum of angular fluctuations would be
distorted even by
decays into invisible modes, such as $\nu_h \rar \nu_l +\phi$
or $\nu_h \rar 3\nu_l$, where $\phi$ is a light or massless scalar,
$\nu_h$ is a heavy neutrino, and $\nu_l$ are some light ones.
We will briefly describe basic physical effects leading to this
distortion and then present existing and potential bounds on $m_\nu$
and $\tau_\nu$ that can be deduced from the existing and mostly coming
measurements of angular fluctuations of CMBR.
The following presentation is by necessity oversimplified. It can be
considered a set of
intuitively simple rules which give basic physical features of the phenomena
A more detailed discussion and a list of references
can be found in review papers~\cite{hu95}-\cite{gawiser00}.

The spectrum of angular fluctuations of CMB depends, first, upon the initial
spectrum of metric and density perturbations and, second, upon physical
processes governing the evolution of these perturbations in cosmological
Friedman background. The evolution on the second stage depends upon the
geometry (curvature) of the universe, its matter content (relativistic
versus non-relativistic, vacuum energy) and expansion regime, amount of
baryons, etc. This dependence permits to determine in principle the
corresponding cosmological parameters.

It is assumed that some primordial density fluctuations existed
in the universe. These density
fluctuations are necessary seeds for formation of large scale structure of
the universe. In a perfectly smooth universe no structures can be formed.
There could be different mechanisms of creating density perturbations
on astronomically large scales, e.g. inflation or topological defects, but
we will not discuss the concrete mechanisms. It is assumed that the
spectrum of primordial perturbations has a very simple one-parameter form:
\be
\langle \left(\delta \rho_k / \rho\right)^2 \rangle \sim k^n
\label{spectr}
\ee
where $k$ is the wave number of the fluctuations (inverse wave length) and
the parameter $n$ is the power index. In what follows we assume a special
case of $n=1$ corresponding to
the so called scale-free (or flat) Harrison-Zeldovich spectrum. Similar
spectra appear in simplest inflationary models.

If the wave length of perturbation is longer than the horizon size,
$L_h \sim t$, then the amplitude of the so called adiabatic or curvature
perturbations (to be more precise of the rising
mode) evolves kinematically, as dictated by General Relativity:
\be
{\delta \rho \over \rho} \sim \left\{ \begin{array}{ll}
a(t)^2\sim t ,\,\, \,{\rm at\,\,\,  RD- stage}, \\
a(t)\sim t^{2/3} ,\,\, {\rm at\,\,\, MD-stage}
 \end{array}\right.
\label{drhokin}
\ee
where $a(t)$ is the cosmological scale factor; the presented time dependence
of $a(t)$ is true for the case of $\Omega_m =1$. To illustrate the derivation
of this result we can proceed as  follows. The energy density $\rho$,
the Hubble parameter $H$, and the curvature $c$ are related by one of the
Einstein equations (\ref{dota2}):
\be
\rho = {3H^2 m^2_{Pl}\over 8\pi } + {c\over a^2}
\label{rhoofc}
\ee
Let us choose a coordinate frame in which the Hubble parameter is
independent of space points. Then the density fluctuations are 
proportional to
curvature fluctuations, $\delta \rho = \delta c /a^2$. Keeping in mind that
$\rho \sim a^{-4}$ at RD-stage and $\rho\sim a^{-3}$ at MD-stage, we will
obtain the expressions (\ref{drhokin}).

Since the wave length of perturbation rises as $\lambda \sim a(t) $, at some
moment it becomes shorter than the horizon, $L_h \sim t$,
 and dynamics comes into play. The
evolution of perturbations at this stage is determined by competition between
attractive forces of gravity and the pressure resistance. If the magnitude of
a perturbation is sufficiently large, the pressure could not resist gravity and
the excessive density regions would collapse indefinitely or until equation of state is changed to a more rigid one. Until that happens, such density perturbations
would keep on rising. For smaller perturbations the pressure of
compressed fluid (plasma) could stop gravitational contraction, the rise
of perturbation would be terminated, and acoustic
oscillations would be produced. The boundary between these two regimes is
given by the so-called Jeans  wave length
$\lambda_J = c_s\sqrt{\pi m_{Pl}^2/\rho} $ where $c_s$ is the speed of
sound. Waves shorter than $\lambda_J$ oscillate, while those with
$\lambda > \lambda_J$ are unstable against gravitational collapse. For
relativistic gas $c_s = 1/\sqrt 3$ and thus the waves shorter than horizon
are stable. After hydrogen
recombination, which took place at $T\approx 3000$ K, the photons of CMBR
propagated freely and temperature fluctuations which
existed at this moment
(the moment of last scattering) were imprinted in the angular distribution
of CMB. So the waves whose phase corresponded to maximum of compression or
rarefaction at the moment of last scattering would create peaks in the angular
spectrum of CMB.

Thus one can understand the basic features of the angular spectrum of
CMB presented in fig.~\ref{figcmb}(a). This figure is taken from
ref.~\cite{hssw} where a very good explanation of different physical effects
leading to the structure in the CMB spectrum is presented. The spectrum
is given in terms of $C_l$, the squares of the amplitudes in the decomposition
of the temperature fluctuations in spherical harmonics:
\be
{\Delta T \over T} = \sum_{l,m} a_{lm} Y_{lm} (\theta, \phi)
\label{spherdt}
\ee
and
\be
C_l = {1\over 2l+1} \sum_{m=-l}^l |a_{lm}|^2
\label{C-l}
\ee
Very long waves which were outside horizon during
recombination retain a constant amplitude (for flat spectrum of
perturbations) because for them acoustic oscillations were not important
and the relative density contrast rises as $\delta \rho /\rho \sim a (t)$
at MD-stage, both for waves inside and outside horizon. This result is
true for the universe dominated by nonrelativistic matter. In the case
of dominance of vacuum energy (lambda-term) perturbations do not
rise and there should be some decrease of the amplitudes from
quadrupole to higher multipoles. For shorter waves
which were inside horizon at recombination and whose phase reached $\pi$
at the moment of last scattering (in other words, the mode had time
to oscillate for exactly one half of the period),
the temperature fluctuations should reach maximum creating the first peak
in the figure. (The concrete value of the phase depends on the form of
fluctuations and may differ from $\pi$.) The second peak is created by
the mode that had time to
oscillate a full period, etc. Since the speed of sound in photon dominated
cosmic plasma is $c_s \approx 1/\sqrt 3$ the wave length corresponding to
the first maximum is
\be
\lambda_1 \approx l_h^{rec} /\sqrt 3
\label{lambda1}
\ee
where $l_h^{rec}$ is the cosmological horizon at recombination. The latter
is determined by the expansion regime and in particular by the competition
between contributions of relativistic and non-relativistic matter. This
is why the position of the peak depends upon the fraction of relativistic
matter. (This peak is often called "Doppler" peak,
but this name is quite misleading; the Doppler effect has nothing to do with
this peak.) The decrease of $C_l$ at large $l$'s is related to the
Silk damping~\cite{js}, the diffusion of photons
from the hotter regions, which is more efficient at small scales.

\begin{figure}[htb]
\begin{center}
  \leavevmode
  \hbox{
    \epsfysize=3.0in
    \epsffile{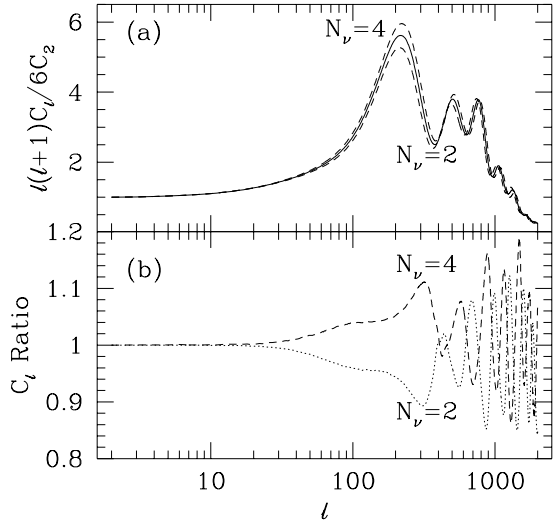}}
\end{center}
\caption{(a) An example of an angular spectrum of CMB anisotropies
with varying
number of neutrino species, $k_\nu =2,3,4$. (b) The ratio of $C_l$ for
$k_\nu =2,\,4$ relative to $k_\nu =3$ (from ref.~\protect\cite{hssw})}
\label{figcmb}
\end{figure}

The amplitude of acoustic oscillations depends on the temporal evolution of
the gravitational potential. In a static potential the amplitude remains
constant because the blue-shift due to infall into potential well is
compensated by a red-shift when the wave emerges from the well. On the
other hand, in a time varying potential a resonant amplification of the
amplitude may take place. The potential varies at RD-stage while it
remains constant at MD-stage. Indeed from the Newtonian equation
\be
a^{-2} \partial^2 \psi = G_N \delta \rho
\label{potmd}
\ee
we find $\psi \sim (\delta \rho /\rho) a^2 \rho$. At MD-stage
$\rho \sim a^{-3}$ and $\delta \rho /\rho \sim a$, hence $\psi$ is time
independent. At RD-stage or because of non-negligible contribution of
relativistic matter at an early MD-stage, the potential changes and
oscillations could be enhanced.
Thus the position and the height of the peaks, roughly speaking, depend upon
the moment of equality, $t_{eq}$, between matter and radiation,
$\rho_{nonrel} = \rho_{rel} $.
One should keep in mind, however, that the position of the peaks is much
more sensitive to the geometry of the universe. The same physical size on
the sky would correspond to different angular scales depending on the space
curvature. This is why the position of the first peak measures the total
energy density of the universe, i.e. $\Omega_{tot} = \rho_{tot}/\rho_c$.

As we have already mentioned, after recombination of hydrogen, which took
place at red-shift $z\approx 1300$ (see e.g.~\cite{zn,sw,peeb}), the
interactions of CMB photons became very weak and they propagate freely over
cosmologically large distances. So the temperature fluctuations observed
today present the picture that existed at the last scattering
surface up to some secondary anisotropies (see below). The photons
"last scattered" slightly after recombination at $z\approx 1065$. In fact
the switch-off of the interactions is not an instantaneous process so that
the last scattering surface has a finite thickness,
$\Delta z \approx 80$~\cite{jones85}. The anisotropies created by the acoustic
oscillations with wave length shorter than this thickness are strongly damped
because the observed signal is averaged over several peaks and througs.
Evidently an increase of $\Delta z$ would result in a suppression of the
angular fluctuations of CMB. Correspondingly, reionization of the intergalactic
medium would lead to a suppression of the of the angular fluctuations of CMBR
at the scales smaller that horizon at reionization epoch~\cite{bond84e}.

This explains the statement made in
sec.~\ref{cer} that possible radiative decays of neutrino would
significantly suppress the level of angular fluctuations. Indeed
neutrinos with masses about 27.5 eV and life-times $\sim 10^{23}$ sec would
ionize universe not only at the present day but also at earlier periods,
in particular during recombination epoch. The UV photons produced by the
decay would reionize hydrogen making the last scattering surface
significantly thicker. This in turn would result in a strong suppression
of acoustic peaks in the angular fluctuations of CMB~\cite{adams98}.
The resulting
anisotropy of CMB is presented in fig.~\ref{figadams}, taken from
ref.~\cite{adams98}. The suppression of the level of angular fluctuations
is quite strong and seems to be disfavored by the data.

\begin{figure}[htb]
\begin{center}
  \leavevmode
  \hbox{
    \epsfysize=3.0in
    \epsffile{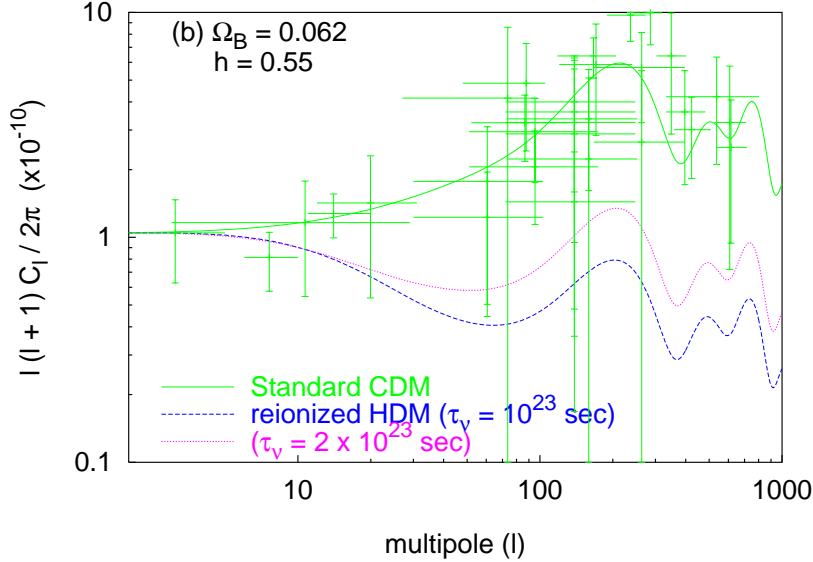}}
\end{center}
\caption{Angular power spectrum of CMB anisotropy in the decaying
         neutrino cosmology for
 $\Omega_B=0.062$, with $\tau_\nu=10^{23},\,2\times10^{23}$~sec.}
\label{figadams}
\end{figure}

One of the mechanisms that could create secondary anisotropies, essential
for the subject of this section, is a possible variation of gravitational
potential during the propagation of light
ray from the surface of the last scattering to the observer. In a static potential
the blue shift of the radiation, when it enters the potential well, is
canceled out by the red shift when it escapes the potential. However, if
the potential changes during the time of propagation, some frequency shift
must arise. This effect is called integrated Sachs-Wolfe (ISW) effect.

Now, bearing this simple picture in mind, we can discuss how neutrino
properties would influence the angular spectrum of CMBR. A review on the
interplay between CMBR and particle, and in particular neutrino physics,
can be found in ref.~\cite{kamionkowsky99}. It is evident that
the shape of the angular spectrum of CMB depends on the number of massless
neutrino species, $k_\nu$. In the standard model $k_\nu =3$ and a deviation
from this number would change $t_{eq}$ and, in turn, the angular
spectrum~\cite{hu95,jungman95} (see fig.~\ref{figcmb}).  As is argued
in refs.~\cite{gg,ldht}, even much smaller non-equilibrium corrections
to the energy
density of massless neutrinos discussed in sec.~\ref{masslessdistr}, could be
in principle observable in the future MAP and especially PLANCK missions if
the latter reaches the planned accuracy of 1\% or better.
A very serious problem of accuracy and
degeneracy (when the same effect in the spectrum is created by different
physical phenomena, for example the position of the peaks depends not only
on $k_\nu$ or $t_{eq}$ but, as mentioned above and to a much larger 
extent, on
the curvature of space) were analyzed e.g. in ref.~\cite{bond97,bond98}.
The problem of degeneracy in connection with detection of
cosmological neutrino background was discussed in ref.~\cite{hu98etw}.
The authors argued that detection of neutrino background radiation
requires detecting the anisotropies of the latter due to degeneracy in
CMBR acoustic peaks. Anisotropies of neutrino background radiation are
potentially observable through their effect on CMBR anisotropies.

The CMBR is most sensitive to the matter radiation
ratio rather than to the individual energy densities.
However combining the measured CMBR anisotropies and the data on the
galaxy power spectrum with the additional information on the baryonic
fraction, $\Omega_b/\Omega_m$, in the galaxy clusters permits to determine
$\Omega_r$ individually~\cite{hu98etw}, but still the accuracy at the
per cent level necessary to observe the details of neutrino
decoupling~(\ref{omegarel}) seems questionable. On the other hand, the
observation of neutrino anisotropies looks feasible with the future
Planck mission.

Assuming the standard cosmological model, and using the available CMBR
data, Hannestad~\cite{hannestad01} found a rather loose limit on the
effective number of neutrino species, $N_\nu <17$ (95\% confidence level)
for the Hubble parameter $h = 0.72\pm 0.08 $ and
$\Omega_b h^2 = 0.020\pm 0.002$. Larger values of these parameters
allow for a larger relativistic energy density or larger $N_\nu$.
This bound is not competitive
with BBN at the present time. However it could be such in future.
Moreover, the BBN and CMBR limits are sensitive to different forms of
relativistic energy. In particular, the BBN limit could be modified by
a non-zero chemical potential of electronic neutrinos, while the CMBR
is not sensitive to that. Furthermore, additional relativistic species
produced by decays of some new particles or heavy neutrinos would
contribute to relativistic energy density at recombination but not
during BBN. An additional consideration~\cite{hannestad01} of the data
on the large scale structure (see discussion in the previous
paragraph and in sec.~\ref{11dm})
permitted to arrive to an interesting lower limit,
$N_\nu > 1.5$. Thus an indication of non-vanishing cosmological background
of massless or very light neutrinos is obtained. A combined analysis of
CMBR and BBN data
ref.~\cite{hansen01} gives a somewhat better limit, $N_\nu < 7$ at the
same $2\sigma$ level. According to the results of this group the neutrino
chemical potentials, $\xi = \mu /T$, are bounded as $-.01< \xi_e < 0.2$ and
$|\xi_{\mu,\tau} | \leq 2.6$.
The idea of using CMBR data for extracting a fraction of relativistic
energy at recombination was discussed in several papers but the
earlier ones~\cite{hannestad00}-\cite{mangano00},
where the bound on $N_\nu$ was derived,
used less precise earlier data and their results were subject to uncertainty
related to the values of other cosmological parameters~\cite{kneller01}.
An impact of additional relativistic background on determination of
cosmological parameters from CMBR anisotropies is studied in the
paper~\cite{bowen01}. It is shown that $\Omega_{rel}$ is nearly degenerate
with the fraction of energy in non-relativistic matter, $\Omega_m$,
at small $l$ but
this degeneracy can be broken on smaller scales available to Planck 
mission.

An imprint of active-sterile neutrino oscillations (in non-resonant
case) on CMBR angular spectrum was studied in ref.~\cite{hannestad98r}.
Light sterile neutrinos, produced by the mixing with active ones, could
contribute into relativistic matter at the epoch of matter-radiation
equality as well as into the cosmological hot dark matter.
The signature of sterile neutrinos cannot
be unambiguously observed in the CMBR spectrum, and they could add an extra
problem with extracting the value of the cosmological parameters from the
data.

The constraint on the number
of neutrino species recalls a similar one obtained from BBN (see
sec.~\ref{knu}). However, the BBN bounds are sensitive to neutrinos with mass
in or below MeV range, while CMB considerations are valid
for very light neutrinos with mass around recombination temperature, i.e.
$m_\nu \leq(\sim 1)$ eV. Putting it another way, BBN considerations permit
setting a limit on neutrino mass in MeV scale while CMB would permit
reaching higher accuracy in eV scale. If neutrinos are massive and
contribute into hot component of dark matter, their presence can be traced
through CMB~\cite{dodelson95}. Both effects mentioned above, a shift of the
peak positions and a change of their heights, manifest themselves depending
on the fraction of hot dark matter $\Omega_{HDM}$. Moreover the angular
spectrum of CMB is sensitive also to the value of neutrino mass
because the latter shifts $t_{eq}$,
the moment of the transition from radiation dominance to matter dominance.
According to the paper~\cite{dodelson95} the amplitude of
angular fluctuations of CMB is 5-10\% larger for $400<l<1000$
in the mixed hot-cold dark matter (HCDM) model with
$\Omega_\nu =0.2-0.3$ in comparison with the pure CDM model.
A detailed analysis of the latest data~\cite{2ndpeak} on CMBR angular
spectrum was performed in ref.~\cite{wang01} and the best-fit range of
neutrino mass was found, $m_\nu = 0.04 - 2.2$ eV.

The influence of unstable neutrinos on the CMB anisotropies in connection
with large scale structure formation was considered in
refs.~\cite{dodelson94,white94,bopi,pibo}. It was shown that the first peak
in models with decaying particles is noticeably higher than that in the standard
CDM model, and the secondary
peaks are strongly shifted to the right (toward higher $l$'s). As we have
already noted it is related to the change of the sound horizon at the moment
of last scattering and to the integrated Sachs-Wolfe effect.

Similar arguments can be used to put rather tight constraints on neutrino
mass/life-time~\cite{hann98}-\cite{lopez99a} in the case of
decays into invisible channels. For
sufficiently small life-times, $\tau_\nu < {\rm a\,\, few}\times 100$ sec,
and large masses, $m_\nu \sim O({\rm MeV})$,
the consideration of big bang nucleosynthesis rather strongly restricts
parameter space (see sec.~\ref{massunstbl}). However for a much longer
$\tau_\nu$ the nucleosynthesis does not help. On the contrary CMB angular
spectrum is sensitive to $m_\nu$ as small as a few eV and a life-time close
to the time of recombination, $t_{rec} \sim 10^{13}$ sec. The idea to rely
on the CMB spectrum (and in particular on the change of the height of
acoustic peaks due to ISW effect) for derivation of bounds on $m_\nu/\tau_\nu$
was first formulated in ref.~\cite{hann98}. Approximate calculations has been
done in refs.~\cite{hann98,lopez98,hann99} and an improved treatment,
correcting previously found results, has been presented in
ref.~\cite{kaplinghat99,lopez99a}. For low neutrino masses
and large life-times the
distortion of the angular spectrum of CMB was found to be much weaker than
in the earlier papers but still the obtained bounds remain quite restrictive.
The already existing CMB data permit to exclude the range $\mnu >100$ eV
and $\taunu >10^{12}$ sec. The bound on 
the life-time becomes less stringent with
decreasing $\mnu$. Future more precise measurements could significantly
enlarge the excluded area in $\mnu/\taunu$-plane permitting to reach the
accuracy in eV scale and, with measuring polarization, the accuracy
reached by Planck could be about 0.3 eV. The results however depend upon
the concrete model of neutrino decay.

An effect of unstable neutrinos on the position and height of the second
acoustic peak was discussed in ref.~\cite{hansen00}. The authors proposed
the decay of a heavier neutrino into a lighter one and a scalar boson,
$\nu_h \rar \nu_l + \phi$, to explain the inconsistency between BBN and
earlier CMBR data~\cite{boom,max1} on the value of $\Omega_b h^2$.
The new results~\cite{2ndpeak}, however, show much better agreement with
BBN.


\section{Cosmological lepton asymmetry. \label{sec:asym}}
\subsection{Introduction. \label{ss:intr-lasym}}

It is normally assumed that cosmological lepton charge asymmetry, i.e the
difference between the number densities of neutrinos and antineutrinos, is
vanishingly small. Of course relic neutrinos are not observed directly
but the asymmetries that can be observed are very small; baryon asymmetry
is $\beta_B = (n_B -n_{\bar B}) /n_\gamma = ({\rm a\,\, few})\times 10^{-10}$
and electric asymmetry is probably exactly zero. So by analogy, the
asymmetry between leptons and antileptons
$\beta_L = (n_L -n_{\bar L})/n_\gamma$ is assumed to be also small.
Moreover, there are some theoretical grounds for a small lepton asymmetry
( for a review see e.g.~\cite{adbs}). In $SU(5)$ grand unification models
the difference of leptonic and baryonic charges, $(B-L)$, is conserved,
so lepton and baryon asymmetry must be the same. Even in $SO(10)$, where this
conservation law is not valid, the asymmetries have similar magnitude in
simple versions of the theory. Despite that, it was suggested in
ref.~\cite{dz} that a large lepton asymmetry together with a small baryonic
one might be generated in grand unified theories. A model which permitted
to realize generation of a small $\beta_B$ and a much larger $\beta_L$
in the frameworks of $SO(10)$-symmetry was proposed in ref.~\cite{harvey81}.
On the other hand $(B-L)$ is conserved in
electroweak theory, and thus if electroweak
baryogenesis is operative, then after electroweak phase transition any
preexisting baryon or lepton asymmetry would be redistributed in more or
less equal shares between baryons and leptons.

Nevertheless, a few theoretical models predicting a large difference
between $\beta_B$ and $\beta_L$ have been proposed during the past decade.
To avoid electroweak "equalization" one has to assume that either 
generation of lepton asymmetry took place after electroweak phase transition
or that the electroweak washing-out of preexisting asymmetries is not
effective. A possible mechanism to suppress electroweak non-conservation of
baryons and leptons is triggered by lepton asymmetry itself. As
was pointed out in ref.~\cite{linde74} a large charge asymmetry suppresses
symmetry restoration at high temperatures. The suppression of symmetry
restoration or even symmetry breaking at high $T$, induced by large chemical
potentials, was found in several papers in different
theories~\cite{haber82}-\cite{riotto97}. It means in particular that
due to this effect  electroweak
non-conservation of baryonic and leptonic charges in strongly
asymmetric background would always be
exponentially small~\cite{liu94}. As was shown in ref.~\cite{bajc98}
electroweak symmetry in the minimal standard model is not restored
at high temperatures if $\xi_\nu = 2.5 -5.3$ and the masses of the Higgs
bosons lie in the range 100-800 GeV.

Another logically possible, though rather unnatural, way to avoid
contradiction with electroweak baryogenesis is to assume
that the total lepton asymmetry is small,
\be
\beta_L = \beta_e +\beta_\mu + \beta_\tau \sim \beta_B \approx
({\rm a\,\,\, few}\times 10^{-10} )
\label{betal}
\ee
while individual $\beta_j$ could be much larger, even of the order of unity.
A rather interesting argument in favor of this was found recently in
ref.~\cite{march99}: if electron number and lepton number are equal and
opposite, then baryon asymmetry produced by electro-weak processes
in the standard model is equal to the observed one within a factor of 2 and
has the correct sign.

A model predicting a large (even of order unity) lepton asymmetry
together with a small baryonic one was proposed in refs.~\cite{dkad,adbs}
in the frameworks of Affleck and Dine baryogenesis scenario~\cite{afdi}.
Other models in the same frameworks were suggested recently
in~\cite{ccg,mcdonald99}. A possible way to create an overpopulated,
though not necessarily asymmetric, cosmological neutrino density through
decays of a heavier particle was considered in ref.~\cite{mad}.
A possibility of generation of a large asymmetry by active/sterile neutrino
oscillations was advocated in ref.~\cite{ftv} and in many subsequent papers
(see sec.~\ref{nuosceu} for discussion and references).
Thus, there are plenty of mechanisms of efficient leptogenesis
and it is not excluded that cosmological
lepton asymmetry is large, and it is worthwhile to discuss its
observational manifestations. The earlier papers on
the subject are reviewed e.g. in~\cite{dz}.

\subsection{Cosmological evolution of strongly degenerate neutrinos.
\label{degennu}}

The usual thermal history of neutrinos
(see secs.~(\ref{thermdyn},\ref{gerzel})) is written under the assumption that
their chemical potentials, $\mu$, are not essential. It would be quite
different if the degeneracy is strong, i.e. $\xi = \mu /T \gg 1$.
The energy density of massless degenerate neutrinos in thermal equilibrium
is 
\be
\rho_\nu +\rho_{\bar\nu} &=&
{1\over 2\pi^2} \int_0^\infty dp p^3 \left[{1\over e^{p/T -\xi}+1} +
{1\over e^{p/T +\xi}+1}\right]  \nonumber \\
&=&{7\over 8}\,{\pi^2 T^4 \over 15}\left[1  +
{30\over 7}\,\left({\xi \over \pi}\right)^2 +
{15\over 7}\,\left({\xi \over \pi}\right)^4 \right]
\label{rhoxi}
\ee
and for a large $\xi$ may be considerably larger than the energy density
of non-degenerate ones. The magnitude of charge asymmetry is given by
\be
\eta_L = {n_\nu -n_{\bar \nu} \over n_\gamma } =
\left( {T_\nu \over T_\gamma}\right)^3 \, {\xi^3 +\pi^2 \xi \over
12 \zeta (3)}
\label{etal}
\ee

The cosmological evolution of strongly degenerate neutrinos was considered
in~\cite{freese83,kang92}. However these papers are in some disagreement
and
here we will reconsider and correct their results. It was noticed in
ref.~\cite{freese83} that in the case of strong degeneracy neutrino
decoupling would take place much earlier than in the usual case of
non-degenerate neutrinos. That statement is partly true. Indeed, the
reactions changing the number of neutrinos, e.g.
\be
\nu +\bar \nu \lrar e^+ + e^-
\label{nunuee}
\ee
would be frozen at much higher temperatures than the usual 2-3 MeV in the
standard case. However, as we see in what follows,
elastic neutrino scattering which would maintain equal temperatures
of neutrinos and the rest of the primeval plasma remains efficient down to
almost the same temperatures as in non-degenerate case. However the
efficiency of elastic scattering in the case of degenerate neutrinos is
strongly momentum-dependent and the spectrum would be distorted anyhow
(see below eq.~(\ref{hxdfnu})).

The estimates of the freezing temperature, $T_d$, for
annihilation (\ref{nunuee})
are different in papers~\cite{freese83,kang92} so "to find the truth"
we will perform the calculations of $T_d$ in some detail here.
We will use kinetic equation
in the form (\ref{hxdfdx}) with collision integral given by (\ref{icoll}) and
matrix element taken from table~\ref{table:amplitudes-nu-e0}.
We assume that the occupation numbers of
neutrinos and antineutrinos are given by $f_1$ and $f_2$ respectively, and
the latter have the equilibrium form (\ref{ffb}) with equal by magnitude
and opposite by sign dimensionless chemical potentials $\xi =\mu/T$. The
electron-positron occupation numbers, $f_{3,4}$,
 are given by the same expressions but
with vanishing $\mu$. The product of $f_j$ that enters kinetic equations
can be written as
\be
f_1 f_2 (1-f_3)(1-f_4) = f_1 f_2 f_3 f_4
\exp{\left[ \left(E_3+E_4\right)/T \right] }
\label{ffff}
\ee
If we assume Maxwell-Boltzmann statistics for $e^{\pm}$, then
$f_3 f_4 \exp{\left(E_3+E_4\right)/T} =1$. The corrections to this
approximation can be found with the help of expansion
\be
f \approx e^{-E/T} - e^{-2E/T} + ...
\label{maxbolexp}
\ee
and are evidently small. Integration over $d^3p_3 d^3p_4$ is trivial in this
approximation (it is just the usual phase space integral) and gives:
\be
Hx \partial_x f_1(x,y) = -{ 2^3G_F^2 (g_L^2 +g_R^2) m_0^5\over 9\pi^3 x^5}
f_1(x,y) y \int_0^\infty dy_2 y_2^3 f_2(x,y_2) + ...
\label{hxdfnu}
\ee
where multi-dots indicate contribution of inverse reaction.
The integral can be easily estimated, again using expansion similar to
(\ref{maxbolexp}), $f_2 = \exp(-y-\xi) +...$,
and we obtain:
\be
f_1 \sim \exp\left(- {0.01 y\over x^3} e^{-\xi}\sqrt{{10.75 \over g_*}}\,
{g_L^2 + g_R^2 \over 0.5858}\right)
\label{f1exp}
\ee
If two other neutrino species are not degenerate then
the contribution of the
annihilation of $\nue\,\bar\nue $ into $\num\,\bar\num$ and $\nut\,\bar\nut$
should be also taken into account and this changes the factor
$g_L^2+g_R^2 $ to $g_L^2+g_R^2 +1/2$
(see table~\ref{table:amplitudes-nu-e0}). The exponential suppression of
the annihilation rate, $\Gamma \sim \exp (-\xi)$, 
is related to a small number
density of antineutrinos, so it is difficult to find a partner for
a neutrino to annihilate. On the other hand, the annihilation rate for
antineutrinos is not suppressed. Thus the variation of the
number density of antineutrinos keeps pace with the universe expansion
to rather low temperatures, while the variation of neutrino
number density stopped at a rather high $T$, see below. (For a negative
chemical potential the situation is opposite.)

Freezing temperature, $T_f = 1/x_f$, is determined by the condition
that the power of the exponent in this expression is unity:
\be
T_f = 4.64 \,{\rm MeV}\, e^{\xi/3} y^{-1/3}
\left({g_* \over 10.75}\right)^{1/6}
\left( 0.586 \over g_L^2 + g_R^2\right)^{1/3}
\label{tfnu}
\ee
where the effective number of species, $g_*$, depends upon $\xi$ as
\be
g_* = 10.75\left[1 +0.3488 \sum_j \left(2\left({\xi_j \over \pi}\right)^2
+\left({\xi_j \over \pi}\right)^4 \right) \right]
\label{gstar}
\ee

The freezing temperature of course depends on the momentum of neutrino
$y= p/T$. Usually thermal averaging is performed so that
$\langle y \rangle \approx 3$. In this way we recover the known results
for the freezing of the annihilation of
non-degenerate neutrinos into $e^+e^-$: $T_f^{\nue} = 3.2$ MeV and
$T_f^{\num} = 5.3$ MeV.
The dependence on $\xi$ in this result is the same as in ref.~\cite{kang92},
and does not contain the preexponential factor $\xi^{-2/3}$ found in
ref.~\cite{freese83}, while the numerical coefficient is approximately 20
times bigger than that in ref.~\cite{kang92}. The numerical value of
$T_f$ obtained here is approximately twice larger than $T_f$ found in
ref.~\cite{orito00} from somewhat different considerations.

For $\xi \geq 7$ the freezing temperate would be higher than 50 MeV. At
such temperatures the primeval plasma contained in addition to $e^\pm$,
photons, and three types of neutrinos at least $\pi^\pm$, $\pi^0$, and
$\mu^\pm$, so $g_* \geq 17.25$, even without contribution from
degenerate neutrinos. In the course of expansion and cooling
down, massive particles would annihilate, and as a result the temperature
would drop slower than $1/a$. Usually the ratio $T/a^{-1}$ is calculated
with the help of entropy conservation (\ref{dtent}), which is true in
the case of vanishing chemical potentials. In particular, this is
how the well known ratio $T_\nu/T_\gamma = \left(4/11\right)^{1/3}$ after
$e^+e^-$-annihilation is obtained (see sec~\ref{gerzel}).

The calculation of the freezing temperature of elastic scattering is not
so simple. The rate of elastic scattering $\nu_1 + l_2 \lrar \nu_3 + l_4$,
where $l$ is a  lepton, can be found from the equation
\be
Hx\partial_x f_{\nu 1} (x,y_1) = -{f_{\nu 1}\over 2E_1}
\int \tilde {dl_2}\tilde {d\nu_3}\tilde {dl_4}
(2\pi)^4 \delta^4(p_1+p_2-p_3-p_4) |A|^2 f_{l1}(1-f_{\nu 4})
\label{rateel}
\ee
where $ \tilde {dl_2} = d^3p_2 /2E\,(2\pi)^3 $ and we assumed that the
leptons $l$ obey Boltzmann statistics. We also assume for simplicity sake that the amplitude $|A|^2$ can be substituted by its average value
$|A|^2 = 2^6 G_F^2 E_1^2 T^2$. Integration over the phase space is
first done over $d^3 p_4$ with the help of spatial $\delta$-function.
Then the integration over $d^3 p_2$ can be performed, in particular,
the integration over $\cos \theta$ is achieved with the energy
$\delta$-function. And ultimately we are left with the integral over
the energy of the degenerate neutrino in the final state:
\be
Hx\partial_x f_{\nu 1} (x,y_1) = -{|A|^2 f_{\nu 1} T^3 \over
64 \pi^3 E_1^2} \int_0^\infty {dE_3 \over T} (1-f_{\nu 3})
e^{{E_1-E_3\over 2T}}
\left[ e^{-{|E_1-E_3|\over 2T}} - e^{-{E_1+E_3)\over 2T}} \right]
\label{rate2}
\ee
where $f_{\nu 3} = [1+ \exp (y-\xi)]^{-1}$.
The remaining integral can be taken analytically:
\be
I = (\xi -y)e^{y-\xi} -\xi\,e^{-\xi} -
\left(1+ e^{-\xi}\right)\, \ln \left(1+ e^{-\xi}\right)
+ \left(1+ e^{y-\xi}\right)\, \ln \left(1+ e^{y-\xi}\right)
\label{fxiy}
\ee
Note that this function is not exponentially suppressed in $\xi$ near
$y=\xi$ where the bulk of degenerate neutrinos ``lives'':
$I(\xi=y) \sim 1$.

Integrating over $x$ and we obtain for the reaction rate:
\be
\Gamma_{el} = 4.6\,{\rm MeV}\, I^{-1/3}
\left({ g_* \over 10.75}\right)^{1/6}
\label{gammael}
\ee
Hence the freezing of elastic scattering takes place at a much lower
temperature than annihilation.
Numerical calculations of the freezing of degenerate neutrinos have been
done in ref.~\cite{esposito00} for relatively small values of the asymmetry,
$0\leq \xi_{\nue}\leq 0.5$ and $0\leq \xi_{\num,\nut} < 1$, where the
results are presented in the form of interpolating polynomials.

After annihilation of muons the distribution functions of neutrinos,
$f_\nu$, evolve in the usual way, i.e. they preserve the form (\ref{ffb})
with a constant ratio $\xi =\mu/T$ and $T$ decreasing as $1/a$. To the moment
of $e^\pm$-annihilation neutrinos were already completely decoupled from the
plasma so their evolution continued in the same way. If
$\bar\nu\,\nu \lrar e^+e^-$ was frozen before $\mu^+\mu^-$-annihilation,
then the dimensionless chemical potential of neutrinos $\xi$ did not stay
constant until all muons annihilated. The evolution of neutrino chemical
potentials (in the case of $\xi_{\nue} >0$) can be found from the
conservation of number density of neutrinos in the comoving volume, which
became true after freezing of neutrino annihilation~(\ref{nunuee}):
\be
a^3 T^3 \int {dy y^2 \over exp \left( y-\xi\right) +1 } = const
\label{conservedn}
\ee
If $Ta = const$ then the solution to this equation is $\xi = const$. For a
non-constant $Ta$ chemical potential $\xi$ cannot remain constant in the
course of expansion in contrast to the common assumption.
The solution $\xi(R)$ can be easily found in the limit of
large chemical potentials. For a large and positive $\xi$ the solution is
\be
\xi_1 (a) = {\xi_0\over R} \left[ 1 - {\pi^2\over 3 \xi_0^2}\left( R^2 -1
\right) \right]
\label{xi1}
\ee
and for a negative $\xi$:
\be
\xi_2 (a) = -\xi_0 - 3 \ln R
\label{xi2}
\ee
where $\xi_0$ is an initial value of $\xi$ and $R =Ta/ T_0 a_0 \geq 1$.

The evolution of antineutrinos is different from the evolution of
neutrinos. The number density of the former is small and they can easily
find a partner for annihilation so their distribution keeps the
equilibrium form until low temperatures, even slightly smaller that
the temperature of decoupling of non-degenerate neutrinos. Their number
density is not conserved in the comoving volume (if $Ta \neq const$)
and, even if initially $\xi +\bar \xi =0$, this relation would
not hold in the course of evolution. Thus chemical potentials of neutrinos
and antineutrinos during nucleosynthesis may have different absolute
values. Numerical calculations of the evolution of effective
chemical potentials of degenerate neutrinos were done in
ref.~\cite{esposito00}. Their results for $\nue$ and $\bar\nue$
are presented in fig.~\ref{f:xinue}.
\begin{figure}
\psfig{file=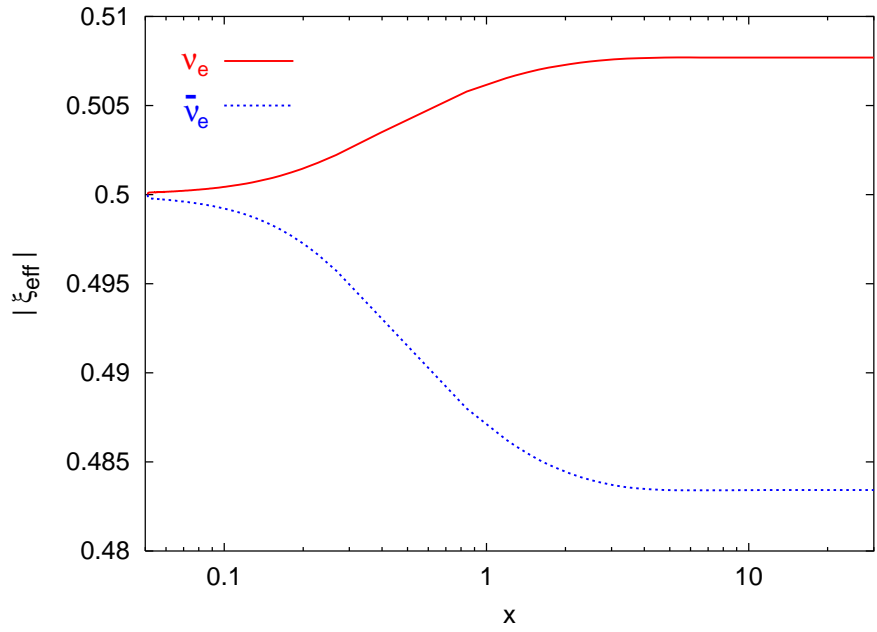,angle=-90,width=.8\textwidth}
\caption{The evolution of the absolute value of
the effective degeneracy parameter $\xi$ (if the distribution function is
written in the equilibrium form) for electron neutrinos and antineutrinos
with momentum $y=5$ ($\xi_e=0.5 , \xi_x=1$).}
\label{f:xinue}
\end{figure}

The variations of temperature of the cosmic plasma in the case of strong
degeneracy cannot be calculated on the basis of entropy conservation
because entropy is not conserved if chemical potentials are
non-vanishing. To this end one should use
the covariant energy conservation law~(\ref{dotrho}). The energy density of
neutrinos with negative chemical potential is exponentially suppressed,
$\rho \approx (3T^4/4\pi^2) \exp (-|\xi|)$, and can be neglected.
The total energy density of a certain neutrino flavor is given by
\be
\rho_{tot} = {T^4 \over 8\pi^2}
\left(\xi^4+2\pi^2 \xi^2+{7\pi^4 \over 15}\right)
\approx {T^4 \xi_0^4 \over 8\pi^2 R^4}
\left[ 1-{4\pi^2\over 3\xi_0^2\left( R^2-1\right)} \right]
\left( 1+{2\pi^2R^2 \over \xi_0^2} \right)
\label{rhonu1}
\ee

If we take into account only the leading, for large $\xi$, term in
this expression then
$\rho_1 \sim 1/a^4$ and automatically satisfies the conservation
law~(\ref{dotrho}). In this
case the remaining matter (photons, muons, electrons and positrons, etc)
also satisfy this law and since their chemical potentials by assumption are
vanishing, their entropy is conserved if they are decoupled from
neutrinos, and the ratio $R=Ta/T_0a_0$ varies because of the annihilation
of massive species.
At the nucleosynthesis epoch it would be $R=(g_*^{0} /10.75)^{1/3}$. If
$\nu \bar\nu$-annihilation is frozen at 100 MeV then $g_*^{(0)}=17.25$ and
$R= 1.17$; in the case of decoupling above QCD phase transition
$g_*^{(0)}=47.75$ and $R= 1.64$.

However the non-leading terms in $\rho_1$ that are of order of
$1/\xi^2$ are quite essential and their
presence destroys entropy conservation, so to find $R$ one has to solve
numerically differential equation~(\ref{dotrho}). The effect of these terms
is to diminish $R$, so the results presented above can be considered the
upper limits for $R$.


\subsection{Degenerate neutrinos and primordial nucleosynthesis.
\label{degnubbn}}

A possible role of neutrino degeneracy in big bang nucleosynthesis was
noted already in the pioneering paper by Wagoner, Fowler, and
Hoyle~\cite{wagoner67}. Even earlier the effects of neutrino degeneracy
on $\beta$-reactions~(\ref{nnue},\ref{ne}) were studied by 
Zeldovich~\cite{zeldovich63dg} in
old style cold universe model. 
After these works the effects of lepton degeneracy in BBN
were analyzed in a number of 
papers~\cite{doroshkevich71}-\cite{kim97a}, as well
as in the quoted above ref.~\cite{kang92} (it is
probably an incomplete list of references). The papers on this
list differ chronologically by an increasing precision of essential physical
parameters (in particular neutron life-time) and by increasingly accurate
astronomical data. So in what follows we will quote only the results
of the most recent works. The earlier papers are reviewed e.g. in
ref.~\cite{dz,malaney93,sarkar}

There are two physical effects from degenerate
neutrinos that could influence primordial abundances. First, an increase of
the energy density in comparison with non-degenerate case results in a
faster expansion which in turn leads to a larger $n/p$-ratio at the onset
of nucleosynthesis. Second, an asymmetry in the sector of electronic
neutrinos/antineutrinos would strongly shift the equilibrium value of
$n/p$-ratio, $n/p \sim \exp (-\xi_{\nue})$.
That's why the bounds on $\xi_{\nue}$
from BBN are much stronger than those for $\xi_{\num,\nut}$. The ratio of
the extra energy densities of degenerate 
$\nu$ and $\bar\nu$ to the energy density
of non-degenerate $\nu+\bar\nu$, under the assumption that $\xi =-\bar \xi$,
is (see eq.~(\ref{rhoxi})):
\be
\Delta N_\nu = {15\over 7}\left[ \left({\xi \over \pi}\right)^4 +
2 \left({\xi\over \pi}\right)^2 \right]
\label{deltanuxi}
\ee
If $\xi >2$ then one degenerate neutrino
(plus antineutrino) species are equivalent to more than three "normal"
neutrinos and should be rejected. However, a positive (and rather small) chemical
potential of $\nue$ would work in the opposite direction in BBN,
so it is difficult
to exclude a "conspired" degeneracy of $\nu_{\mu,\tau}$ and $\nue$.

First we neglect a possible conspiracy and consider the effect of
asymmetries of different neutrino families separately. It follows from the arguments
presented above that chemical potentials of $\nu_{\mu,\tau}$ are
bounded by the condition
\be
\xi_{\mu,\tau} < 1.5
\label{ximutau}
\ee
if one extra neutrino species is permitted by the data on light element
abundances (see sec.~\ref{knu})

A possible role of electron asymmetry was studied in relatively recent
works~\cite{kohri96,kim97}. The allowed range according to ref.~\cite{kim97}
is $|\xi_{\nue}|<0.1$ with 
$4\leq \eta_{10}\leq 6$. A more sophisticated statistical
analysis of ref.~\cite{kohri96} gives somewhat different numbers
$\xi_{\nue} = 0.043\pm 0.040$ and
$ \eta_{10}= 4.0^{+1.5}_{-0.9}$. 
However there are still some discrepancies in the
data on light element abundances (in particular, high versus low deuterium
controversy) and their interpretation, so
possibly these numbers will be changed in future. One can expect
a better accuracy in determination or restriction of the magnitude of
neutrino degeneracy if the baryon number density is fixed by the CMBR data,
independently from BBN.

A more interesting case is when all three chemical potentials are permitted
to influence BBN. In that case the theory has two additional parameters
in comparison with the standard model
(the roles of $\xi_{\num}$ and $\xi_{\nut}$ are the same because the
essential quantity is the total energy density of $\num$ and $\nut$)
and the bounds on their values become much less restrictive. In particular,
prior to measurements of the baryon density of the universe by CMBR,
the freedom in the values of chemical potentials permitted
the baryon-to-photon ratio to be much larger than in the standard BBN
theory, allowing baryon dominated universe. Now it looks unlikely with
any values of chemical potentials. Some decrease of $\eta$ with
respect to the standard value is also possible,  diminishing the conflict
between visible and invisible baryons. The analysis made in
refs.~\cite{olive91,kang92} permits to conclude that $\eta$ can be as large
as a hundred, so that even $\Omega_B =1$ is permitted. The appropriate
values of chemical potentials are $\xi_{\nue} \sim 1$ and
$\xi_{\num,\nut}\geq 10$. For that large values of $\xi_{\num,\nut}$
the freezing of reaction~(\ref{nunuee}) takes place above 100 MeV and,
according to eqs.~(\ref{xi1},\ref{xi2}), primordial values of
$\xi_{\num,\nut}$ differ from those at BBN, while neutrino temperature at
BBN remains equal to $T_\gamma$. If indeed $\xi >1$, then the energy
density of degenerate neutrinos would be very large and the 
neutrinos could have an
important impact on large scale structure formation. This problem and
corresponding bounds on $|\xi|$ are discussed in the next subsection.

A combined analysis of the effect of simultaneous variation of all three
chemical potentials on BBN was performed in the
papers~\cite{bianconi91,orito00,esposito00b}. 
As is stressed in ref.~\cite{orito00},
the authors carefully reexamined the decoupling temperature of
neutrinos (see discussion in the previous subsection). They have found
that the allowed range of variation of chemical potentials is
\be
0\leq \xi_{\nue} \leq 1.4 \nonumber \\
0\leq |\xi_{\num,\nut}| \leq 40
\label{xuiemutau}
\ee
for the baryonic density confined in the interval
$ 0.1 \leq \Omega_b h_{50}^2 \leq 1$. These results are somewhat less
restrictive than those found in ref.~\cite{esposito00b}. For
low deuterium abundance the
electronic chemical potential can roughly change in the interval
$-0.25\leq \xi_{\nue} \leq 0.25$ for the total number of neutrino species
changing from 1 to 16. The last number can be related to the muonic
or tauonic asymmetry through the equation~(\ref{gstar}). The results
are slightly different for high deuterium. The best fit values are
$\xi_{\nue} = 0.06$, $N_\nu = 3.43$, and $\eta_{10} = 5$ for low D
and $\xi_{\nue} =0.35$, $N_\nu = 13$, and $\eta_{10} =4.2$ for high D.
Some inconsistency between the papers~\cite{orito00}
and~\cite{esposito00b} is possibly related to a different evolution of
the temperature of degenerate neutrinos. As is stated in
ref.~\cite{esposito00b} the effect of non-standard value of neutrino
temperature~\cite{esposito00} is negligible. The latter is closer
to the estimate of the evolution of neutrino temperature presented in
the previous subsection. But the estimate is rather approximate and
moreover, the spectrum of neutrinos may be noticeably distorted
in the course of expansion because the elastic reaction rate is strongly
energy dependent~(\ref{gammael}). A more accurate treatment of this problem
is desirable.

The recent analysis~\cite{hansen01}, based on new measurements of the
angular spectrum of CMBR, gives the limits:
\be
-0.01 <\xi_{\nue} < 0.2,\,\,\,\, |\xi_{\num,\nut}| < 2.6
\label{limitsxi}
\ee
under assumptions that the primordial fraction of deuterium is
$D/H = (3.0 \pm 0.4)\cdot 10^{-5}$~\cite{kirkman01}.

The results presented above are valid for homogeneous distribution of
neutrinos. Possible inhomogeneities in lepton asymmetry at
cosmologically large scale and strongly chemically inhomogeneous universe
is discussed in sec.~\ref{varabund}. Models of variation of lepton
asymmetry in the sector of active neutrinos induced by the oscillations
between active and sterile $\nu$'s are described in
sec.~\ref{sec:spatfluc}.
Small scale inhomogeneities in neutrino degeneracy and their impact on
big bang nucleosynthesis are considered in ref.~\cite{whitmire00}. The
scale of variation of leptonic chemical potentials are assumed to be
sufficiently large, so the fluctuations in $\xi$ were not erased
before BBN began (this corresponds to approximately 100 pc today). On
the other hand, the scale is assumed to be smaller than the matter
mixing scale so the resulting element distribution is homogeneous
today. A surprising result is that in the case of the inhomogeneous
scenario the total energy density of neutrinos is not bounded by BBN.
Indeed one may have a regions with a very large and positive value of
$\xi$ which gives a dominant contribution into the energy density but
does not participate in the element formation, because for a very large
$\xi$ the production of light elements is negligible. The model permits
to enlarge considerably the upper limit on the baryon number density
allowed by BBN, while the lower limit remains practically untouched:
$3.0 \cdot 10^{-10} < \eta < 1.1 \cdot 10^{-8}$ for
$\xi_{\nue} \gg \xi_{\num,\nut}$ and
$3.1\cdot 10^{-10}< \eta < 1.0 \cdot 10^{-9}$ for
$\xi_{\nue} = \xi_{\num}= \xi_{\nut}$. These upper limits correspond to
$\Omega_b h^2 = 0.4$ and $\Omega_b h^2 = 0.036$ respectively.

In some papers a "double" deviation from the standard scenario is
considered - in addition to neutrino degeneracy another non-standard
assumption is made. In ref.~\cite{kajino98} inhomogeneous nucleosynthesis
(induced by inhomogeneities in baryon distribution) with
degenerate neutrinos is applied to the solution of a possible
discrepancy between the observed low deuterium abundance in Lyman-$\alpha$
clouds and a possible overproduction of $^4He$. In ref.~\cite{kim97a}
primordial nucleosynthesis with varying gravitational constant and
degenerate neutrinos is discussed.

Some more bounds on the neutrino degeneracy (energy density) follow from
structure formation and cosmic microwave background, which are considered
in the following subsections.

\subsection{Degenerate neutrinos and large scale structure.
\label{degnustr}}

If degeneracy is large, the energy density of neutrinos would be much larger
than that of non-degenerate ones and it would have a very strong impact
on cosmological evolution. A trivial upper limit on the magnitude of
degeneracy follows form the condition that neutrinos should not over-close
the universe, $\Omega_\nu <1$. It gives
\be
|\xi| = 53 h^{1/2} \Omega_\nu^{1/4} \left( 2.73 {\rm K} / T_\gamma
\right)
\label{xiomega}
\ee
To obtain this limit we used eqs.~(\ref{rhoc0},\ref{rhogamma}) and
took neutrino temperature after
$e^+e^-$-annihilation equal to $T_\nu = 0.71 T_\gamma$. It
would be true if before the
annihilation the temperatures were equal as is argued in
subsection~\ref{degennu}. This limit is stronger than those
obtained in refs.~\cite{freese83,kang92} where a smaller
$T_\nu$ was used (see discussion in sec.~\ref{degennu}), but still
very weak. Even this rather
weak limit excludes very high values of $\xi$ discussed in the previous
section in connection with BBN.

A much stronger upper bound on $|\xi|$ is obtained from the condition that
the universe must become matter dominated sufficiently early so that there
would be enough time for large scale structure
formation~\cite{freese83,kang92}.
(At RD-stage perturbations grow at most logarithmically and structure
formation is ineffective~\cite{mesz}.)
Since at MD-stage perturbations rise as
the scale factor $a(t)$ and the primordial density fluctuations are below
$10^{-4}$, as is seen from temperature fluctuations of CMB, we assume that
the equilibrium between matter and radiation should be earlier than
red-shift $z=10^4$. It gives
\be
|\xi| < 5.3 h^{1/2} \Omega_m^{1/4} \left( 2.73 {\rm K} / T_\gamma \right)
\label{xirdmd}
\ee
This limit is valid for massless neutrinos. Massive neutrinos
have practically the same distribution as massless ones, i.e. the
equilibrium one before decoupling and the rescaled distribution
after decoupling:
\be
f_m={1 \over \exp \left(\sqrt{ p^2(z_d+1)^2 +m^2 }\,/T_d -\xi\right) +1}
\approx {1\over \exp (p/T_\nu -\xi) +1}
\label{fmas}
\ee
if their mass is much smaller than decoupling temperature,
$T_d \sim {\rm MeV}$. Here $z_d+1 = a(t)/a_d$ is the red-shift
at decoupling and $T_\nu = T_d/(z_d+1)$.
For $m_\nu\sim T_d$ nonequilibrium corrections to the spectrum
are essential and the distribution may very much differ
from the usually assumed rescaled one, see sec.~\ref{massstabnu} and
ref.~\cite{ad9}. In this section we are interested in neutrinos with a
small mass (in eV range or below) so for them we may use the distribution
function~(\ref{fmas}). Such neutrinos become
effectively non-relativistic when $m_\nu/T_\nu > 0.2$; at that moment
their pressure is about 0.1 of their energy density, while for relativistic
gas $p/\rho =1/3$. Degenerate neutrinos have a larger average momentum and
pressure, so they are more relativistic at the same $T/m_\nu$. Degenerate
neutrinos become nonrelativistic at $m_\nu/T \sim \xi$ (if $\xi>10$) and
the upper limit on their chemical potential, which follows from the
condition that RD-stage was earlier than $z=10^4$,
is $\xi < m_\nu/ {\rm eV}$ (for a large $\xi$). On the other hand,
neutrinos with masses larger than 10 eV and $\xi >2$ would over-close the
universe because their number density is 5.3 times larger than the number
density of non-degenerate neutrinos. So the bound (\ref{xirdmd}) can be
taken as a safe upper bound for both massive and massless neutrinos.
Correlated bounds on neutrino mass and degeneracy based on their
contribution into cosmological energy density were analyzed in
refs.~\cite{pal98,lesgour99}. It is indicated there that neutrinos
may be cosmologically interesting even if they have a very small mass,
$m_\nu < 0.1$ eV, as follows from the data on neutrino oscillations.
If the bound~(\ref{xirdmd}) is
satisfied the contribution of such neutrinos into cosmological
energy density could be as large as:
\be
\Omega_\nu < 0.037 h^{-1/2} \Omega_m^{3/4}\,(m_\nu/0.1 {\rm eV})
\label{omdegnu}
\ee

There are a few points however, where the results presented in
ref.~\cite{pal98} disagree with our analysis. In particular, it is
stated there that the decoupling temperature of degenerate neutrinos may be
lower than that of non-degenerate ones, it may be even smaller that the
electron mass. If this were the case, then the
temperatures of relic neutrinos and photons at the present day would be
equal. To come to this conclusion the authors of
ref.~\cite{pal98} estimated the decoupling temperature from the usual
condition of equality of expansion rate, $H$, and reaction rate,
$\sigma n$,
and substituted for $n$ the largest number density of participating
particles, i.e. the number density of degenerate neutrinos. However, the
reaction rate is given by $\dot n/n$ so the rate of elastic scattering of
degenerate neutrinos on electrons, that maintain the equality of their
temperatures, is proportional to electron number density as in the standard
non-degenerate case (compare with the discussion in sec.~\ref{degennu}).
As a result, the
authors of ref.~\cite{pal98} obtained a high value of $T_\nu$, while
in other papers a much lower value found in ref.~\cite{kang92} was
used. The estimates presented in the previous subsection give an
intermediate result and some more work is necessary to confirm which
value of $T_\nu$ is correct. Accordingly the limits on the values of
$\xi$ presented here should be taken with caution.

The impact of massive degenerate neutrinos on structure formation was
considered in
refs.~\cite{freese83,larsen95,lesgour99,lesgour99b,orito00}.
An extra free parameter, $\xi$
permits breaking rigorous connection between the neutrino mass, their
energy density, neutrino free streaming and Jeans mass. A larger mass density
of degenerate neutrinos permits having the same contribution of HDM into
$\Omega$ with a smaller neutrino mass or permits a larger Hubble parameter
for a fixed $m_\nu$. Degeneracy gives rise to somewhat larger free-streaming
for a fixed $m_\nu$ and $h$ (because the average momentum of degenerate
neutrinos is larger than that of non-degenerate ones).
As shown in ref.~\cite{larsen95} degenerate neutrinos may resolve
inconsistency between mixed HCDM (hot+cold dark matter, $\Lambda=0$)
model with observations, that appears if Hubble parameter is large, $h>0.5$.

An analysis of the power spectrum of density perturbations in a model with
$\Omega_\Lambda = 0.7$ was performed in ref.~\cite{lesgour99} both for
massless and massive ($\mnu = 0.07$ eV) degenerate neutrinos. With an
increasing $\xi$ the power at small scales is suppressed because a large
degeneracy postpones the matter-radiation equality and correspondingly the
fluctuations that enter horizon at RD-stage began to rise later. Another
effect of neutrino degeneracy is a larger free streaming mentioned above.
It leads to a further suppression of small scale matter fluctuations.
In figure~\ref{fig.WIN} taken from ref.~\cite{lesgour99} the
region in the plane of neutrino chemical potential $\xi$ and spectral
index of density perturbations $n$ is presented so that the model
agrees with the observed large scale structure and CMBR, the latter is
discussed in the next subsection.
\begin{figure*}[t]
\vspace{-0.5cm}
\begin{eqnarray}
\psfig{file=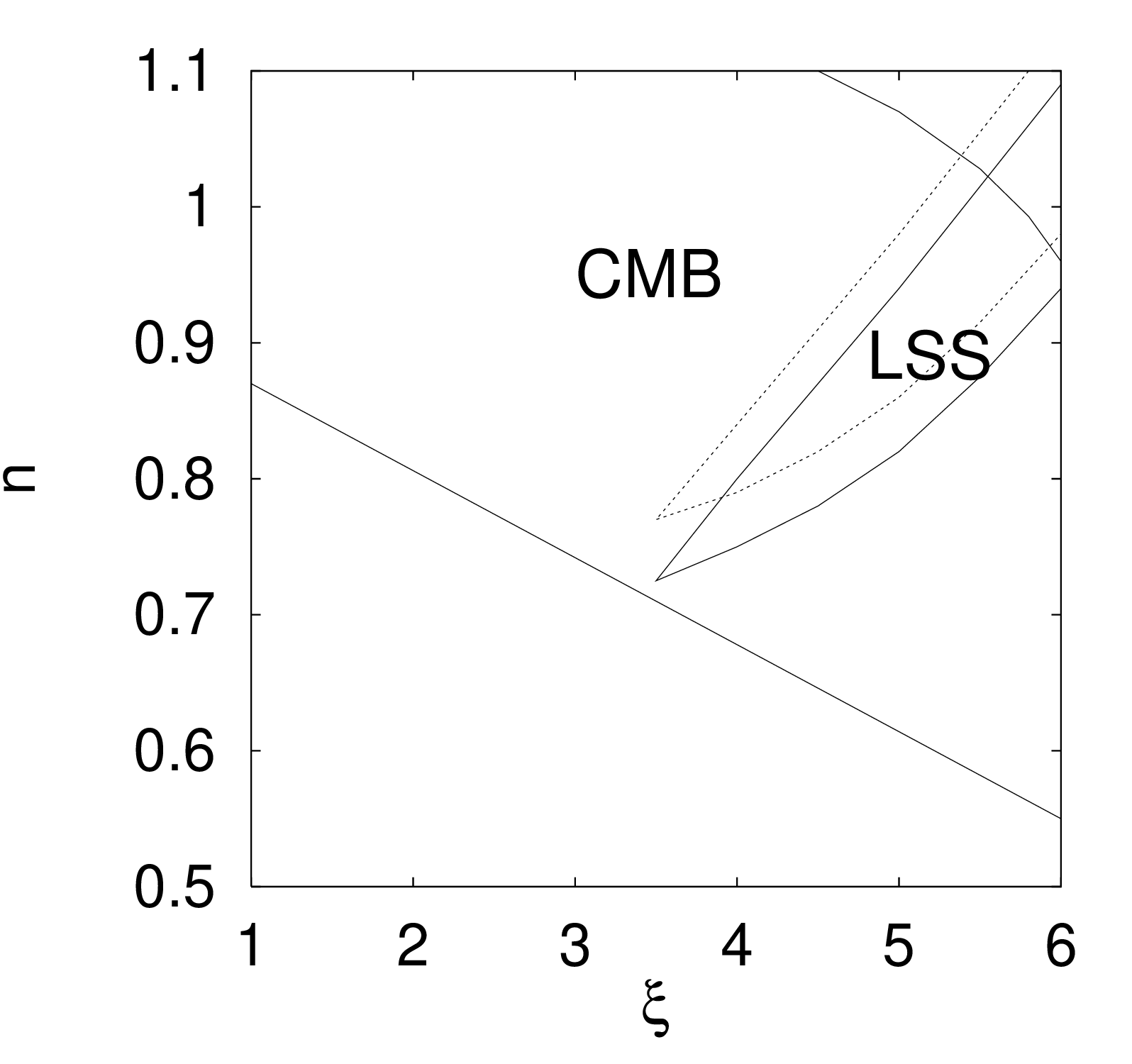,width=0.47\textwidth}~~~~
\psfig{file=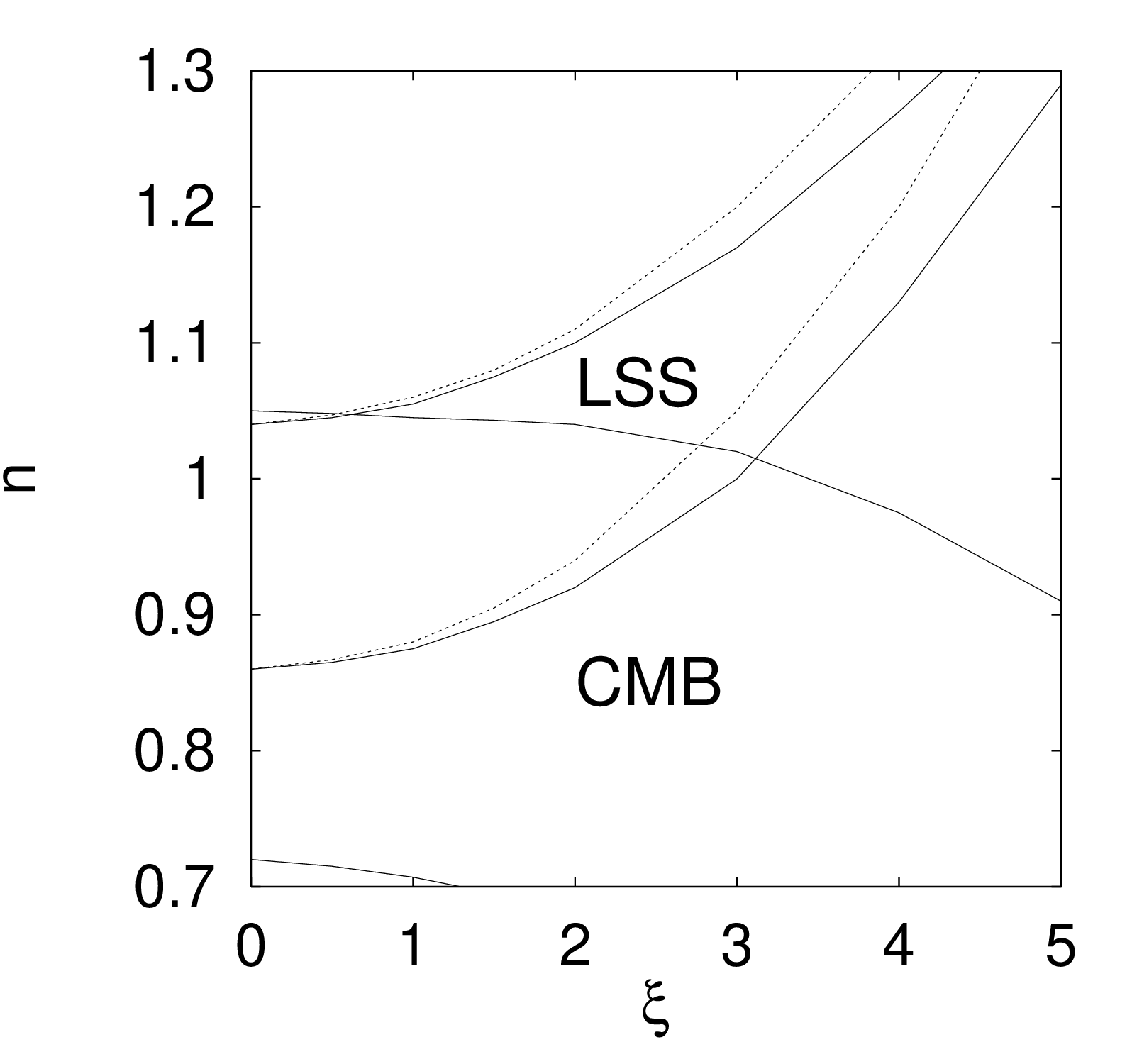,width=0.47\textwidth}
\nonumber
\end{eqnarray}
\vspace{-1.5cm}
\caption{LSS and CMB constraints in ($\xi$, $n$) space
for $\Omega_{\Lambda}=0$ (left) and $\Omega_{\Lambda}=0.6$ (right).
The underlying cosmological model is flat, with $h=0.65$,
$\Omega_b=0.05$, $Q_{rms-ps}=18~\mu$K, no reionization, no tensor
contribution. The allowed regions are those where the labels are. For
LSS constraints, we can distinguish between degenerate neutrinos with
$m_{\nu} =0$ (solid lines) and $m_{\nu} =0.07$ eV (dotted lines).}
\label{fig.WIN}
\end{figure*}

The idea to ``save'' the critical density universe with vanishing
vacuum energy and $\Omega_m =1$, using freedom in neutrino degeneracy,
was explored in ref.~\cite{lesgour01}. The authors concluded that
the model with massless neutrinos failed to fit the observational
data on large scale structure and CMBR anisotropies. If neutrinos have
the mass of order 1 eV, a much better agreement with observations can
be reached. However, with the latest results on the microwave
anisotropy~\cite{2ndpeak}, the model encounters a serious problem
with the observed baryon mass fraction in galactic clusters.

\subsection{ Neutrino degeneracy and CMBR. \label{degnucmb}}

The effects of neutrino degeneracy on the spectrum of angular fluctuations
of CMB is discussed in recent
papers~\cite{adams98-cmb,kinney99,lesgour99,lesgour99b}. In the
first one the analysis was done for massless neutrinos in $\Lambda=0$
cosmology, while in the other three the case of $\Omega_\Lambda =0.7$ was
considered for massless~\cite{kinney99} and for both massless and
$\mnu =0.07$ eV~\cite{lesgour99,lesgour99b} neutrinos. Another burst of
activity~\cite{lesgour00c,hannestad00,orito00,esposito00b,esposito00c}
in this area was stimulated by the
BOOMERanG~\cite{boom} and MAXIMA-1~\cite{max1} measurements of CMBR
anisotropies on sub-degree scales where a surprisingly small height of
the second acoustic peak was observed. The new data~\cite{2ndpeak},
however, do not support this result. Still these papers are of interest
because their arguments could be used to obtain the bounds on the neutrino
degeneracy from CMBR.

The main effect of neutrino degeneracy,
as  we have already mentioned, is to delay matter-radiation equality, which 
results in a larger amplitude of the first acoustic peak and in a shift in
the positions of the peaks toward higher $l$'s (see sec.~\ref{9cmb}).
However the dependence on $\xi$ is not monotonic and for a large $\xi$ the
hydrogen recombination may take place at RD-stage. This would give rise to
a suppression of fluctuations and to a decrease of the peak height. According
to ref.~\cite{lesgour99} this happens for $\xi> 7$. The location of the first
peak in this case would be at $l>450$, which disagrees with the data. Moreover
such big $\xi$ contradicts the bound~(\ref{xirdmd}).
Secondary peaks are influenced also by the damping at large $l$ and
their amplitude could decrease with rising $\xi$.
Another effect mentioned in these papers, a change of decoupling temperature
for degenerate neutrinos~\cite{freese83,kang92} and the corresponding
decrease of $T_\nu$, possibly is not effective, as discussed in
sec.~\ref{degennu}. The results of the calculations
of the angular spectrum of CMBR for different values of the degeneracy
parameter $\xi$ are presented in fig.~\ref{figcmbofxi} taken from
ref.~\cite{lesgour99}.
\begin{figure}[htb]
\begin{center}
  \leavevmode
  \hbox{
    \epsfysize=3.0in
    \epsffile{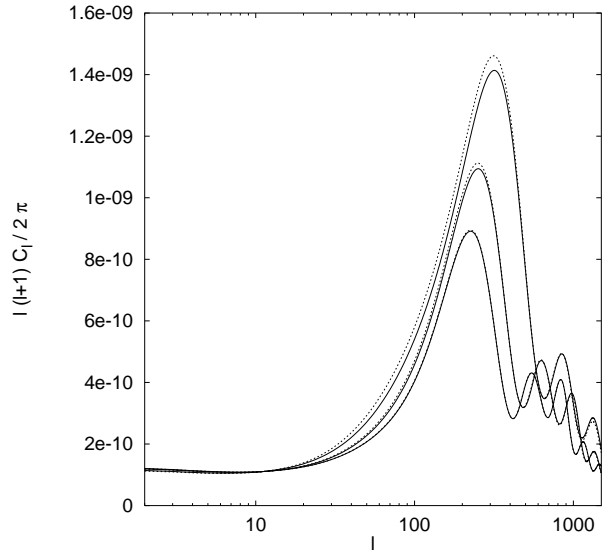}}
\end{center}
\caption{Spectrum of angular fluctuations of CMB for different values of
$\xi$ for one family of degenerate neutrinos; from bottom to top $\xi=0,3,5$.
Massless and massive ($\mnu=0.07$ eV) neutrinos are depicted by solid
and dashed lines respectively.
The appropriate cosmological parameters are chosen as
$h=0.65$, $\Omega_b=0.05$, $\Omega_{\Lambda}=0.70$,
$\Omega_{CDM}=1-\Omega_b-\Omega_{\nu}-\Omega_{\Lambda}$, $Q_{rms-ps}=18~\mu$K
$n=1$. Reionization and tensor contribution are neglected.
(from ref.~\protect\cite{lesgour99})}
\label{figcmbofxi}
\end{figure}

One can see from this figure that neutrino degeneracy has a very strong
impact on the height of the first peak. The
height is almost twice larger for $\xi =5$ in comparison with $\xi =0$. One
may conclude that $\xi >5$ does not fit the already existing
data~\cite{kinney99} if the data is interpreted in the frameworks of the
standard cosmological model. On the other hand, a large neutrino degeneracy
could help the survival of those cosmological models that predict a low first
peak~\cite{lesgour99}. The future missions, MAP and Planck, could significantly
improve the bounds on the neutrino degeneracy, roughly at the level
$\delta \xi < 0.1$, the accuracy strongly improving with rising
$\xi$~\cite{kinney99}. On the other hand, in the case of considerable
asymmetry, $\xi >3$, the future Planck mission will be able to measure
neutrino mass at the level of 0.1 eV~\cite{lesgour99b}.

A small amplitude of the second acoustic peak observed in the
earlier (preliminary?) data set~\cite{boom,max1}
in the frameworks of the standard model demanded a larger baryonic
number density than was allowed by big bang
nucleosynthesis~\cite{white00,tegmark00}. One possible way to diminish
the height of the peak is neutrino degeneracy; it is discussed in the
papers~\cite{lesgour00c,hannestad00,orito00,esposito00b,esposito00c}.
As we have seen in sec.~\ref{degnubbn}, neutrino degeneracy permits having
a much larger baryonic fraction than in the standard model. A larger value
of $\Omega_b$ moves acoustic peaks to the right, and simultaneously
enhances odd peaks against even ones.
The small height of the second peak indicates a large baryon fraction.
On the other hand, a larger fraction of relativistic matter, which can be
achieved by a large chemical potential of $\num$ or $\nut$, moves the
peaks in the opposite direction and also increases the amplitude of
the first peak, so the relative height of the second peak becomes
smaller. In this sense it could mimic an increase of the baryon number.
To ensure the necessary output of primordial abundances
one can vary a small chemical potential of $\nue$. An unnatural feature of
this scenario is an adjustment of two similar parameters
$\xi_{\num,\nut}$ and $\xi_{\nue}$ in such a way that the first
is large, while the second is small. Moreover, 
if there is mixing between
$\nue$, $\nu_{\mu}$, and/or $\nut$ then for the LMA solution for solar
neutrino anomaly all chemical potentials would acquire equal values
(see sec.~\ref{ssec:active}). On the other hand, for SMA solution
an initially large $\xi_{\num,\nut}$ might naturally give rise through
oscillations to a small $\xi_{\nue}$.

The values of chemical potentials and the baryon number density
that are necessary to reach an agreement
between the BBN and CMBR data are somewhat different in different papers
but an agreement about gross features is manifest.
According to the consistency plot between the number of effective neutrino
species $N_\nu$ and the baryonic number density $\Omega_b h^2$ presented in
the paper~\cite{esposito00c} agreement can be achieved approximately
for $N_\nu = 6-10$ and $\Omega_b h^2 = 0.025-0.030$, but with the
latest data~\cite{2ndpeak}
the baryonic density shifts to $\Omega_b h^2 = 0.022 \pm 0.004$.

\section{Neutrinos, dark matter, and large scale structure of the universe.
\label{11dm}}

\subsection{Normal neutrinos. \label{normalnu}}

Of all possible candidates for dark matter particles, neutrinos have a
definite advantage: they are certainly known to exist and may naturally
have a cosmologically interesting mass so that their contribution to
$\Omega$ would be close to 1 (see sec. \ref{gerzel}). All
other candidates for the role of dark matter particles remain
hypothetical. The suggestion that
$\sim 10$ eV neutrinos could solve the problem of the
missing mass of the universe was put forward in 1972 by Marx and
Szalay~\cite{masz} and by Cowsik and McClelland~\cite{cowsik72}, see
also~\cite{szma}. The idea was elaborated in many
subsequent publications~\cite{bisnovatyi80}-\cite{white83}
in the 1980s, especially by the Moscow group.
Detailed investigations showed that neutrino-dominated universe
would have large scale structure significantly different from the
observed one and, though neutrinos may still remain one of essential
components of dark matter, the dominant part most probably
consists of new unknown particles (or fields). The situation 
was summarized in the late 1980s in ref.~\cite{shandarin89}. For more
recent reviews on neutrino dark matter see e.g.
refs.~\cite{sarkar98,ma99,primack01}.

A strong argument against neutrino-dominated universe
is that light neutrinos ($m =O({\rm eV}$)) permit formation of
structures on a scale much larger then the galactic one, while
the formation of galaxies is strongly inhibited and could take place
rather late at red-shifts $z$ smaller or close to
1~\cite{peebles82,klypin83},\cite{doroshkevich83}-\cite{white83}.
The characteristic size of structures in neutrino-dominated universe
can be estimated as follows. Neutrinos were decoupled from the cosmic
plasma rather early, at $T\sim 2-3 $ MeV, when neutrinos were
relativistic. In the gas of non-interacting particles moving
practically with the speed of light all perturbations with 
wave lengths smaller than the distances traveled by these particles
until they become non-relativistic, are erased. Indeed, if in
some region the number (and energy) density of neutrinos is larger
than in nearby regions, the flux from neutrino-rich regions would be larger
than the inverse
flux from neutrino-poor regions and this evidently leads
to equalization. The distance that neutrino passed, before it became
non-relativistic, was roughly equal to $2t_\nu$, where $t_\nu$ was the
time when the neutrino temperature dropped down to $T_\nu = m_\nu/3$.
According to equations (\ref{tT2},\ref{g336}),
$t_\nu = 0.75 m_{Pl}/m_\nu^2 \approx
6\cdot 10^{12}\,{\rm sec} / (m_\nu/ {\rm eV})^2$.
Correspondingly, the mass contributed by neutrinos of one flavor
inside the free-streaming radius is
\be
M_\nu = {4 \pi (2t)^3\over 3} \rho_\nu = 0.135\, m_{Pl}^2 t_\nu
= 0.1\, {m_{Pl}^3 /m_\nu^2}
\label{Mnu}
\ee
where the numerical coefficient comes from the ratio of the
energy density of one neutrino species to the total relativistic
energy density: $\rho_\nu /\rho_{tot} = (1.36/3)/3.36 = 0.135$
with $\rho_{tot} = 3 m_{Pl}^2 /32 \pi t^2$.  
Thus we find that the smallest objects, which could be
initially formed, have the
mass~\cite{bisnovatyi80,doroshkevich80,bond80}:
\be
M_\nu \approx 1.5\cdot 10^{17} M_{\odot} \left( m_\nu / {\rm
eV}\right)^{-2}
\label{Mnu2}
\ee
where $M_{\odot} = 2\cdot 10^{33}$ g is the solar mass. In fact
the mass is even larger because neutrinos did not stop when
$T_\nu=m_\nu/3$.

The redshift when neutrinos became nonrelativistic is
$z_\nu =1.4 (m_\nu/3) / 2.73\, {\rm K} = 2\cdot 10^3 (m_\nu /{\rm eV})$.
Hence, the characteristic comoving size of the first formed structures
is given by
\be
l_\nu = 2\, t_\nu (1+z_{\nu})
\approx 250 \,{\rm Mpc} \left( {\rm eV} / m_\nu \right)
\label{lnuchar}
\ee
Neutrinos present an example of the so-called hot dark matter
(HDM), which gives
characteristic scale of the structures much larger than galactic
size. The opposite case of dark matter giving  $l \ll l_{gal}$
is called cold dark matter (CDM) and the intermediate one is naturally
called warm dark matter (WDM).

Initially small density inhomogeneities started to rise when
universe expansion became dominated by nonrelativistic matter.
In the case of neutrinos, the change of regime takes place at the
redshift given by the ratio of the present day energy densities
of nonrelativistic matter represented by one massive neutrino species
and relativistic matter containing cosmic microwave background radiation
and two massless neutrino species:
\be
z^{(eq)}_\nu =\left( {\rho_\nu \over \rho_{rel}} \right)_0 =
{ 112 m_\nu / {\rm cm}^3 \over 1.45 \cdot 0.261 \,{\rm eV /cm}^3}
= 3\cdot 10^2 \left( {m_\nu \over {\rm eV} }\right)
\label{znueq}
\ee
Thus the structures in the gas of cosmic neutrinos would be
formed before hydrogen recombination ($z_{rec} \approx 10^3$)
if $m_\nu > 5 $ eV.
The result (\ref{znueq}) would be trivially changed
if there are several (2 or 3) massive neutrino species.

It was initially believed that structures smaller than
$l_\nu$ (\ref{lnuchar}) which are observed on
the sky could be formed by fragmentation of the large sheets
(Zeldovich pancakes) and filaments into galaxies. In other
words, in universe dominated by neutrinos larger structures
formed first and smaller ones appeared later (top-bottom scenario).
However, as was argued in ref.~\cite{peebles84}, observations
point to the opposite picture: our Galaxy seems to be considerably
older than the Local Group.
Moreover, numerical simulations~\cite{white83}
showed that universe dominated by light
neutrinos is in disagreement with observations, or to quote
the authors of ref.~\cite{white83}, ``the conventional
neutrino dominated picture appears to be ruled out''.

The idea that neutrinos might have a much shorter free-streaming
length due to their self-interaction associated with the majoron
exchange (see sec. \ref{anomalous}), was proposed in
ref.~\cite{raffelt87} and further discussed in
ref.~\cite{atrio97}. In such model, neutrinos indeed could
behave similarly to cold dark matter at galactic scales.
However they could not provide dark matter in dwarf galaxies
(see the following paragraph).

Another strong blow to neutrino as the only form of dark matter was
dealt by the Tremaine-Gunn limit~\cite{tremaine79}. The latter is a
striking example of quantum phenomenon on cosmologically large scales.
Neutrinos are fermions, so only one particle could be
in a certain quantum state. Hence the total number of neutrinos in
a galaxy cannot be arbitrarily large, and in order to constitute the total
amount of galactic dark matter neutrinos must be sufficiently heavy.
We assume that neutrinos form strongly degenerate gas with Fermi
momentum $p_F = m_\nu V_F $, where $V_F$ is the neutrino velocity.
The number density of galactic neutrinos (plus
antineutrinos) is $n = 2p_F^3 /(6\pi^2)$. Correspondingly
the total mass of neutrinos in a galaxy with
radius $R$ is equal to $N=4\pi R^3 m_\nu n /3 = (2/9\pi) (Rp_F)^3 m_\nu$.
By virial theorem, velocity is related to gravitational potential
$V^2_F = G_N M_{gal} / R$. This set of relations permit us to express
$m_\nu$ through particle velocity around the galaxy and galactic
radius:
\be
m_\nu \approx 80\,{\rm eV} \left({300\,
{\rm km/sec}\over V_F }\right)^{1/4}
\left( { 1 {\rm kpc} \over R }\right) ^{1/2}
\label{mnutg}
\ee
Thus, to provide all dark matter in small galaxies with $R\sim 1$ kpc
and $V\leq 100$ km/sec neutrinos mush be heavier than $100$ eV, 
which goes against the Gerstein-Zeldovich limit (\ref{mnu1},\ref{mnu2}).

The validity of Tremaine-Gunn limit was questioned by Ruffini and
collaborators~\cite{ruffini88} who argued that numerical analysis does not
support this limit. However it is difficult to see a flaw in the
Tremaine-Gunn arguments presented above. Moreover, these arguments
were applied in ref.~\cite{salucci97s}
to a large number (1100) of galaxies  
with well measured rotational curves. It was
shown that in order to be clustered on galactic scale neutrino mass 
should violate the Gerstein-Zeldovich limit.

The large free-streaming length and the Tremaine-Gunn limit 
have made the idea of purely
neutrino dark matter rather unpopular and the attention
has shifted to cold dark matter models. Particle physics proposes
several possible candidates for CDM but none of them has been yet
experimentally observed (see e.g. review~\cite{dolgov99}). After
the magnitude of density fluctuations at large scales has been
normalized by the COBE measurements~\cite{cobe},
it has became clear that the simple CDM model with flat spectrum of primordial
density fluctuations predicts too much power at small scales and requests
some modifications. A possible resolution of the controversy was an
assumption of a mixed CDM+HDM
model~\cite{schaefer92}-\cite{klypin93} with about 30\% of
HDM and the rest in CDM.
\footnote {In fact, the pioneering suggestion
of mixed dark matter model with three flavors of neutrinos with mass
3-4 eV giving hot dark matter, and axions giving cold dark matter,
was made almost a decade earlier~\cite{shafi84}. Earlier references
also include~\cite{schaefer89,holzman89} }
The mixed model revived the role of neutrinos as building
blocks of the universe. The basic idea of why the model may work is easy to
understand. At very large scales there is no difference between HDM and
CDM. So the COBE data fixes their common contribution into density
fluctuations. At smaller scales neutrino perturbations disappear and
the remaining power becomes smaller than in the pure CDM model. Adjusting
the new free parameter, the  ratio $\Omega_{HDM}/\Omega_{CDM}$, one
can achieve agreement between the size of density fluctuations at very
large (horizon) scales and at galaxy cluster scales. Though both CDM and
HDM dark matter look quite natural from the point of view of elementary
particle physics, the similar magnitudes of their contributions
into total cosmological energy density remains a mystery. This
``cosmic conspiracy,'' which includes also a similar contribution of baryons
into $\Omega$ ($\Omega_B $ is at a per cent level) presents one of the
most interesting challenges in cosmo-particle physics. A possible way
to understand this conspiracy is discussed in a recent work~\cite{kane01}.
Now we also have
a contribution from vacuum or vacuum-like energy with
$\Omega_{vac} \approx 0.7$, which strongly increases the gravity of the
problem. 

Despite this unexplained conspiracy (which is a common shortcoming
of all models involving several types of dark matter), the mixed CDM+HDM
model was quite popular for several years.
With the fraction of mass density in HDM
$\Omega_{HDM} \approx 0.3$ and the rest in CDM,
$\Omega_{CDM}\approx 0.7$ (except for some small fraction in baryons),
the model successfully described the observed gross features of large
scale structures. In particular, the top-bottom scenario that was
a shortcoming of pure HDM models, is reversed into a bottom-top one
(i.e. smaller structures forming first) in the case of dominating
cold dark matter. However it was noticed almost immediately after
the mixed CDM+HDM model was proposed that the model had serious
problems with the description of structures at high
redshifts~\cite{mo94}-\cite{ma94}. The problem originated from
the fact that the hot component reduces the magnitude of perturbations
at small scales, and if one uses the COBE normalization at large scales
and assumes flat (Harrison-Zeldovich) spectrum of fluctuations,
there would be too little power at small scales and galaxy formation would be delayed. This phenomenon is
in disagreement with the observed abundances of damped
Ly$\alpha$-systems at high red-shifts, $z\sim 3$, and quasars
at $z\geq 4$. Numerical simulations of ref.~\cite{ma94}
(see also~\cite{ma93}) in the
frameworks of the reference model of that time with $h=0.5$, and
baryonic fraction $\Omega_B = 0.05$ lead to the conclusion that
$\Omega_\nu < 0.2$ and $m_\nu < 4.7$ eV. The CDM+HDM model
was defended in ref.~\cite{klypin94}, where it was argued that
$\Omega_\nu = 0.25$ is compatible with $z>3$  data.
However, subsequent
N-body and hydrodynamic simulations~\cite{ma97} indicate
that a model with $\Omega_\nu \geq 0.2$ predicts an amount of gas
in damped Ly$\alpha$-systems well below observations.
Reducing $\Omega_\nu$ down to 0.2 (from the originally proposed
0.3) gives rise to an overproduction of clusters~\cite{primack95},
because a smaller mass fraction of neutrinos results in a higher
power at cluster scales. In a better shape is a model with several
mass-degenerate neutrinos~\cite{primack95} with the same $\Omega$.
It was suggested that $\num$ and $\nut$ have almost equal masses
close to 2.4 eV, so that the total $\Omega_\nu$ remains the same
as in the model with a single massive neutrino with the mass 4.8 eV
but the neutrinos became nonrelativistic later and have a larger
free-streaming length. This leads to a lower abundance of clusters
and to better agreement with data. However, good agreement
was only reached for a rather low value of the Hubble parameter,
$h= 0.5$, while the observational data point toward a larger value, 
$h=0.65-0.7$.
A recent discussion of hot dark matter with 2 or 3 degenerate
neutrinos can be found in ref.~\cite{caldwell99}. A review of the
state of art of HDM at the end of 20th century can be found
in~\cite{primack00}.

As we have already mentioned, the description of the large scale
structure strongly depends upon the primordial spectrum of
density fluctuations. Usually it is assumed that the spectrum is
flat, i.e. perturbations of gravitational potential do not
contain any dimensional parameters. This is the simplest possible
spectrum and, moreover, it is predicted by the simplest
inflationary models. Usually deviations from the flat spectrum are
parameterized by a power law with the exponent $n$, see
eq.~(\ref{spectr}). The flat spectrum
corresponds to $ n=1$. The assumption of $n=1$ was relaxed in
refs.~\cite{bopi,pibo}, where the models of structure
formation with massive hot neutrinos and $n>1$ were considered.

The form of the evolved spectrum depends upon the relative
cosmological mass fraction of CDM and HDM. If HDM constituents are
neutrinos, their number density in the standard model is fixed,
$n_\nu = 112/{\rm cm}^3$ (see sec~\ref{gerzel}) and thus
$\Omega_{HDM}$ is determined by neutrino mass. The larger $m_\nu$
and $\Omega_{HDM}$ are,
the stronger the suppression of density fluctuations at small
scales is. This phenomenon is illustrated in fig~\ref{fig:f} taken from
reference~\cite{ma99}.
\begin{figure}
\epsfxsize=3.7truein
\epsfbox[0 244 482 653]{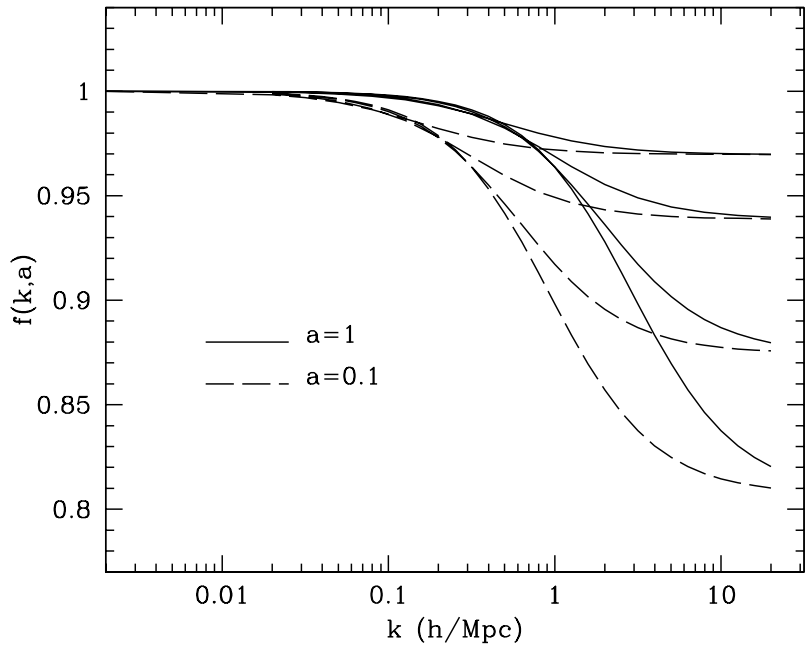}
\caption{Growth rate of the CDM density field, $f\equiv
d\log\delta/d\log a$, in four flat C+HDM models at cosmological
scale factor $a=1$ (solid) and
0.1 (dashed).  The four models assume different neutrino masses:
$m_\nu=1.2$, 2.3, 4.6, and 6.9 eV (from top down), corresponding to
$\Omega_\nu=0.05$, 0.1, 0.2, and 0.3.}
\label{fig:f}
\end{figure}
At large scales there is no difference between CDM and HDM, so that
all the curves coincide. They start to deviate at smaller scales
corresponding to neutrino free streaming length.
The deviation becomes weaker with time because neutrino velocities
drop in the process of expansion. One can see that a change in
neutrino mass by 1 eV has a visible effect on the spectrum of
density perturbations.

On the other hand,
astronomical data accumulated over the past several years
strongly indicates that cosmological constant is non-zero
with $\Omega_{vac} \approx 0.7$ and $\Omega_{CDM}\approx 0.3$
(for discussion and a list of relevant references see e.g. the
papers~\cite{perlmutter98}-\cite{bahcall99}).
If this is indeed the case then there remains much less room for hot
dark matter.
Even if we assume that HDM makes 100\% contribution into total
density of matter, i.e. $\Omega_{HDM} = 0.3$ and if we take the
currently accepted value $h = 0.7$, then from the
limit (\ref{mnu1}) follows $\sum m_\nu < 14$ eV. The data on
neutrino oscillations (see sec.~\ref{2prop}) show that the mass differences
between different types of neutrinos are much smaller than 1 eV.
If the oscillations take place between active neutrinos only, then
they are nearly mass degenerate and their masses should be below
roughly 5 eV. However, as we have already discussed, hot dark matter
could not be dominant because in that case, short wave length
perturbations would be efficiently erased and formation of small
scale structures would be suppressed. This effect is
stronger for a smaller $\Omega_{CDM}$. Hence in the models
with non-zero cosmological constant and/or with a low $\Omega_{matter}$
the upper limit on neutrino mass is more restrictive.

The cosmological limits on the neutrino mass from the Ly$\alpha$
forest have been reanalyzed in ref.~\cite{croft99} for a larger range
of values of the parameter $\Omega_m$ and with a recent
Ly$\alpha$ forest measurements. The conclusion was that
$m_\nu < 5.5 $ eV for all values of $\Omega_m$ and
$m_\nu < 2.4\, {\rm eV} \, \left( \Omega_m/0.17 -1 \right)$ for
small $\Omega$, $0.2 <\Omega < 0.5$.

It was argued in ref.~\cite{hu98} that galaxy red-shift surveys could
probe neutrino mass in eV range. The forthcoming data from the
high precision Sloan Digital Sky Survey (SDSS) will permit to measure
neutrino mass with an accuracy of 
\be
m_\nu \sim  0.65\, {\rm eV}\,\left(\Omega_m h^2/0.1N_\nu\right)^{0.8}
\label{exp-acc}
\ee
and even a mass as small as
0.01-0.1 eV is potentially observable in astronomy. Such strong
result was obtained under the assumption that all other relevant
cosmological parameters would be independently measured by CMB
experiments at 1\% level. To this end the standard theory of structure
formation with adiabatic density perturbations and with linear
scale independent bias have been used. The impact of 1 eV neutrinos
on the galactic power spectrum is illustrated in fig.~\ref{mnusdss}
taken from ref.~\cite{hu98} for high and low density of cosmic matter.
The effect is very strong for $k\sim 0.1/{\rm Mpc}$, while the impact
of such massive neutrinos on CMB spectrum is at the level of several
per cent. It is worthwhile to redo these calculations for cosmology with
non-zero Lambda. 

\begin{figure}[htb]
\begin{center}
  \leavevmode
  \hbox{
    \epsfysize=3.0in
    \epsffile{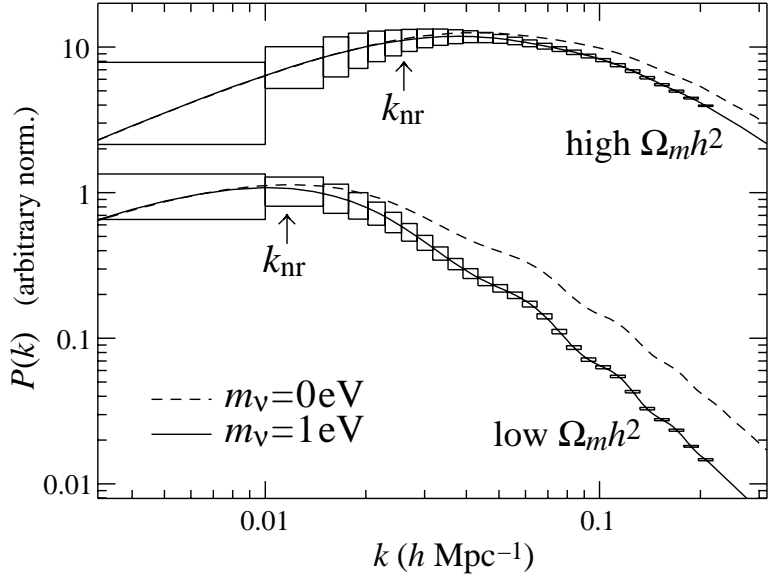}}
\end{center}
\caption{
Effect of a 1 eV neutrino on the bright red galaxy (BRG)
power spectrum compared with expected precision of the SDSS
BRG survey (1$\sigma$ error boxes). Upper curves:
an $\Omega_m=1.0$, $h=0.5$,
$\Omega_b h^2=0.0125$, $n=1$
model with and without a 1 eV neutrino mass.
Lower curves: the same but for an
$\Omega_m=0.2$, $h=0.65$ model.
\label{mnusdss}}
\end{figure}

As shown in ref.~\cite{fukugita00}, the cluster abundance
does not suffer from the biasing uncertainties and from the matching
condition of the observed fluctuation power at COBE scale (several
hundred Mpc) and at the cluster scale ($l_{cluster} = 8 h^{-1} $ Mpc).
It permits to obtain the
limit $m_\nu < 0.6$ eV for the flat universe with $\Omega_m = 0.3$
and $h< 0.8$. For this matching, the Harrison-Zeldovich spectrum of
fluctuations was assumed, $n=1$. If $n$ is significantly larger
than 1, the limit becomes weaker. For the same set of cosmological
parameters, and $n=1.2$ the limit is $m_\nu < 1.4 $ eV.

Another possible way of weighting neutrinos by astronomical means
is to use gravitational lensing effects on background galaxies
created by foreground large scale structures~\cite{cooray99}. The physics
behind this phenomenon is the same as discussed above,
namely suppression of power at small scale by massive neutrinos
and a reduction of the lensing signal. If all relevant cosmological
parameters are measured with an accuracy of 10 per cent expected
from the MAP mission, a weak lensing survey of 100 degrees squared
could be sensitive to neutrino mass about 3.5 eV. For the survey
of $\pi$ steradian down to 25th magnitude the limit could be
0.4 eV. With Planck satellite the limit could be further improved by
factor 3-4. As is argued in the paper ~\cite{cooray99},
the advantage of this method is that it is free of bias and 
evolutionary effects.

An analysis of all available data on CMBR and large scale structures was  performed in ref.~\cite{gawiser00a}. As was already mentioned,
the addition of hot dark matter reduces the power at small scales and
to compensate this effect for 
e.g. $\Omega_{HDM}= 0.05$ a blue tilt of primordial power
spectrum, $n\approx 1.3$, is necessary. On the other hand, large $n$
gives too-large angular fluctuations of CMBR at small scales.
They can be diminished by a significant tensor component or early
reionization. For the flat spectrum of perturbations, i.e. $n=1$,
and $h = 0.65$ in $\Lambda CDM$ scenario of the structure formation,
the upper bound on the amount of hot dark matter is
$\Omega_\nu < 0.05$ or $m_\nu <2$ eV.
With a possibility that $n$ can vary the limit is twice weaker.

A detailed study of all cosmological parameters, based on the large scale
structure observations at different scales and angular fluctuations
of CMBR together with the BBN data and measurements of the Hubble
constant, was performed in a series of
papers~\cite{novos99,durrer00,novos00}. The best fit model is
$\Lambda$-dominated, with $\Omega_\Lambda \approx 0.7$ and
with 8\% contribution of neutrinos into clustered dark matter, or
$\Omega_\nu = 0.03^{+0.07}_{-0.03}$. The upper limit on the neutrino
mass is $m_\nu < 4$ eV. A more recent analysis of ref.~\cite{arhipova01}
lead to the conclusion that $\Omega_\nu = (0-0.2)\,\Omega_{matter}$. 

To summarize, the study of details of large scale structure
permits obtaining an upper limit on neutrino mass that would be considerably
stronger than the Gerstein-Zeldovich limit, which is based solely on 
considerations of the total cosmological energy density.
However, the
latter is model-independent while the former demands
very accurate measurements of basic cosmological parameters.
Moreover, some essential assumptions about the structure formation
should be made. In particular, the simplest form of
the spectrum of primordial fluctuations, Gaussian statistic of the fluctuations, and their adiabatic character are assumed. In principle all these
assumptions can be tested by other astronomical observations,
especially by precise measurements of angular spectrum of CMB with
the Planck mission accuracy. Still the problem of degeneracy persists
and it is a difficult task to separate the effects of different
parameters. On the other hand, the best direct limit that can be expected from the total cosmological density could be hardly stronger
than 10 eV, while the bounds discussed above may be much better.

\subsection{Lepton asymmetry and large scale structure.
\label{leptaslss}}

The results of the previous subsection were obtained under the
standard assumption that the lepton asymmetry in neutrino sector
is negligibly small. Strictly speaking this is not known and the
data permit rather large chemical potentials of $\num$ and $\nut$,
see sec.~\ref{degnubbn}, $\xi = \mu/T \geq 1$. An extra parameter,
$\xi$ permits to break the rigid relation between neutrino mass and
their cosmological energy density:
\be
\rho_\nu (\xi) = \rho_\nu (0) \left[ 1 + { \xi^3 \over 9 \zeta (3)}
+ { \pi^2 \xi \over 9 \zeta (3)} -
{ 2 \left( 1-e^{-\xi}\right) \over 3\zeta (3)} \int_0^\infty {
dy\,y^2 \over \left( e^y +1\right)\left( e^{y+\xi} +1\right)}\right]
\label{rhoxi1}
\ee
where $\zeta (3) \approx 1.202$ and the dimensionless chemical
potential $\xi$ is positive. Note that this expression is
different from eq.~(\ref{rhoxi}), because the latter is valid for
relativistic neutrinos, while the former is true for non-relativistic
neutrinos - it is simply $\rho = m n$, where $n$ is the number
density. The energy density $\rho_\nu (\xi)$
is an increasing function of $\xi$, so for a degenerate neutrino
gas the same energy density could be achieved with smaller $\mnu$.
Another difference from the standard model is that the average
momentum of degenerate neutrinos is somewhat higher than that of
non-degenerate ones with the same mass, so degenerate neutrinos are more relativistic.
Correspondingly their free streaming path is larger.

A mixed dark matter model with non-vanishing chemical potentials of
neutrinos was considered in ref.~\cite{larsen95}. The introduction of
the additional parameter, $\xi$, helps alleviate some problems of
(H+C)DM models discussed in the previous subsection. A larger
$\Omega_\nu$ with the same $m_\nu$ permits higher values of the
Hubble parameter, $h> 0.5$, while a bigger free-streaming length
diminishes the destructive influence of neutrinos on the damped
Ly$\alpha$-system. For more detail see sec.~\ref{degnustr}.

Structure formation with degenerate neutrinos was considered in the
papers~\cite{adams98-cmb,lesgour99}. In the first one it was shown that
a model with massless neutrinos with chemical potential
$\mu_\nu = 3.4 T$ together with the usual cold dark matter gives
a good description of the large scale structure in $\Lambda =0$
universe and of the anisotropy of CMBR (see sec.~\ref{degnucmb}).
In the subsequent paper~\cite{lesgour99} the scenario was
generalized both to massless and massive neutrinos and cosmology
with non-zero Lambda-term.

\subsection{Sterile neutrinos. \label{sterilenu}}

A much richer zoo of possible forms of dark matter becomes
open if one permits the existence of sterile neutrinos, $\nu_s$, which
interact much weaker than the usual, active, ones, or 
if one allows usual neutrinos to have stronger, than 
normal weak, interaction (for the latter see the next
subsection). A recent review on physics of sterile neutrinos
can be found in~\cite{volkas01s}. 
The simplest way to produce sterile neutrinos is to
assume that neutrinos have both Majorana and Dirac masses.
The simultaneous existence of both types of mass excites right-handed
(sterile) neutrino states which are produced from the active ones
by oscillations~\cite{ad1}. If equilibration with respect to sterile
states was not achieved, then their number density could be smaller
than the number density of active neutrinos and correspondingly the
Majorana mass of $\nu_s$ could be around keV without violation of the
Gerstein-Zeldovich limit. However, sterile right-handed neutrino
states, $\nu_R$, are not necessarily
created through oscillations. They could be thermally produced at
an earlier hot stage in equilibrium abundance, and cosmology allows their
mass to be larger than the mass of their left-handed companions because
the number density of right-handed states is suppressed by the entropy
factor (see section~\ref{ssec:nur} where production of
right-handed neutrinos is discussed). This mechanism for making
warm dark matter from $\nu_R$ was considered in ref.~\cite{olive82}.

Another way to create warm dark matter from neutrinos was proposed in
ref.~\cite{giudice00b}, where the cosmological scenario with a very low
reheating temperature was considered. In this model the cosmological
number density of normal active neutrinos could be much smaller than
the usual 100 $/cm^3$ and their mass is not subject to
Gerstein-Zeldovich limit, sec.~\ref{gerzel}. A model with 4 neutrino
mixing, based on this scenario of low temperature reheating,
was considered in ref.~\cite{liu01}. It was assumed that $\num$ or
$\nut$ with keV mass form warm dark matter, while solar neutrino
anomaly is explained by the mixing between $\nue$ and a light sterile
neutrino. However, the new SNO data~\cite{CNO} disfavors the
$\nue-\nus$ solution of the solar neutrino problem, and though
some (even large) mixing of active and sterile neutrinos is not
excluded, the concrete values of the parameters used in the
paper~\cite{liu01} are not realistic.

The idea of ``using'' sterile neutrinos as warm dark particles
was further pursued  in ref.~\cite{dodelson94a} with production
of $\nu_S$ through oscillations.
WDM cosmology is considered in more detail in subsequent
research~\cite{colombi96}. Present day data indicate that warm dark matter
together with cold dark matter may resolve some problems with galaxy
properties that exist in CDM scenario (for discussion, literature,
and possible candidates for WDM see e.g. ref.~\cite{sommer99}). Thus,
WDM may become a respectable member of dark member community~\cite{bode00}.
Recently the properties of warm dark matter
particles were strongly constrained
by Lyman-alpha forest~\cite{narayanan00} and by 
cosmological reionization~\cite{barkana01}. The lower limits on their
mass are respectively 0.75 keV and 0.5 keV.

A dark matter model with sterile neutrinos but with an unusual,
non-thermal, spectrum was considered in paper~\cite{shi98}.
It was noticed some time before~\cite{kirilova97} 
that neutrino oscillations
can strongly distort the spectrum of active neutrinos and also
create sterile neutrinos with a non-thermal
spectrum. As was shown in ref.~\cite{shi98}, $\nu_s$ could
be produced by neutrino oscillations in the early universe,
but in contrast to the previous case, in the presence of a rather
large cosmological lepton asymmetry, about 0.001-0.1.
In this case the production proceeds mostly through resonance
conversion, and the resonance condition is fulfilled only for low
energy $\nu_s$ (see section~\ref{sec:leptas}). Thus, non-relativistic
$\nu_s$ are predominantly produced even though they are very light, with the mass in the interval of 0.1-10 keV. Because of the cold non-thermal spectrum
of $\nu_s$ they move more slowly than the usual warm dark matter
particles, so the authors of ref.~\cite{shi98} propose to call
such dark matter ``cool''. In this model the cut-off in the
spectrum of density perturbations could be around the dwarf galaxy
scale or even below.

A new heavy neutral fermion with several GeV mass was proposed
in ref.~\cite{gondolo98}, in a particular model with an extended
Higgs sector, as an explanation of the gamma-ray
emission from the galactic halo. This is completely analogous to
the annihilation of heavy leptons described in sec.~\ref{stabhvnu}.
The energy density
of such fermions could be cosmologically interesting
and they might contribute noticeably to cold dark matter.
However, these particles are not mixed with active neutrinos and calling them "neutrinos" is rather arbitrary.

Naturally light sterile neutrinos appear in a large class of
supersymmetric models with gauge mediated symmetry
breaking~\cite{dvali98}. These neutrinos have mixing with the
active ones at the level of $10^{-4}$ and mass in the interval of
10 eV - 1 keV. Their number density, created by oscillations in
the early universe, may be sufficiently high to make them
cosmologically interesting and to provide a warm component
to dark matter.

Sterile neutrinos may appear in our universe from mirror or shadow
worlds (see sec.~\ref{s:mirror}). Their implications for structure
formation are essentially the same as of ``normal'' sterile neutrinos.
Depending upon the mass difference between active and sterile $\nu$'s
the latter could constitute hot, warm, or even cold dark matter. On the
other hand, judging by the existing indications to neutrino oscillations,
the mass differences between different neutrino species
are small, though some heavy $\nu_s$ are not excluded.

The production of $\nus$ through oscillations could
result in a smaller number density of active neutrinos
because the total number of active plus sterile neutrinos was
approximately conserved at later stages, when the active neutrinos
decoupled from the plasma. The energy density of mirror/shadow particles
must be smaller than that of the usual ones, otherwise there
would be serious problems with nucleosynthesis, or a charge asymmetry
in $\nue-\bar\nue$ sector must exist to compensate the
effect of larger energy density of relativistic species at BBN.
In ref.~\cite{foot99c} an exact parity model (exact mirror symmetry)
was considered with a heavier $\nut$ and light mirror neutrinos.
The number density of $\nut$, which constitute hot dark matter particles
in this model, could be smaller than the standard one because of the
above mentioned effect of conversion of $\nut$ into mirror
neutrinos. With the parameters taken in that paper the authors obtained
$n_{\nut}=0.7 n_{\nu}^{standard}$. This means that effective number of 
neutrino generations participating in dark matter becomes non-integer
and allows for a larger
mass of HDM neutrinos. The particular example considered in~\cite{foot99c}
was $5\,\,{\rm eV}<\mnut < 10 \,\,{\rm eV}$.
Another possible effect related to the existence of sterile (mirror)
neutrinos that could be important for structure formation
is that the energy density of relativistic particles at the
moment when heavier neutrinos became non-relativistic
could be different from that of the standard
model~\cite{foot99c}. Thus, the model with sterile neutrinos allows
more freedom in comparison with the standard hot dark matter
model and may better agree with the data.

In the case of broken mirror parity~\cite{berezhiani95} it is natural
to expect sterile neutrinos with a mass in the keV range and with a
cosmological abundance of roughly two orders of magnitude below the
standard neutrino abundance (\ref {nnungamma}). Such neutrinos are
also good candidates for warm dark matter
particles~\cite{berezhiani96,berezhiani96a}.

Usually the production of sterile neutrinos in the cosmological plasma
proceeds through their mixing with active ones, so the equations derived
in secs.~\ref{sec:nonres}, \ref{sec:leptas},  \ref{sec:nuoscbbn}
for the mixing with light sterile $\nu$ are directly
applicable here.
The calculations of the production of heavy sterile (WDM) neutrinos through
mixing with active ones were performed in the paper~\cite{dodelson94a}.
The calculations were based on equation~(\ref{gammas}) but in contrast
to the previous works the production rate of active neutrinos was
not averaged over neutrino spectrum but taken with an explicit
energy dependence. It permits calculating the spectrum of the produced
$\nus$. The latter was found to be the same as the spectrum of active
neutrinos mixed with $\nus$~\cite{dodelson94a,dolgov00b}. In the second
paper the factor 2 was corrected, as shown in sec.~\ref{sec:nonres},
eq.~(\ref{sprod}). Hence the cosmological number density found in
ref.~\cite{dolgov00b} is twice smaller than that found in
ref.~\cite{dodelson94a}.

The relative number densities of sterile
neutrinos mixed with an active flavor $\nu_a$, $r_s^a = n_s/n_{eq}$,
according to the results of the paper~\cite{dolgov00b} are
\be
r_s^e &=& 1.8\cdot 10^5 \sin^2 2\theta \left( m/{\rm keV}\right)
( 10.75 /g_* (T_{prod} )^{3/2}
\, , \label{nse}\\
r_s^{\mu,\tau} &=& 2.5\cdot 10^5 \sin^2 2\theta \left( m/{\rm keV}\right)
( 10.75 /g_* (T_{prod} )^{3/2}
\label{nsm}
\ee
where the correction factor $( 10.75 /g_* (T_{prod} )^{3/2}$, due to
entropy generation, is included. Here $T_{prod}$ is the effective
temperature of production of sterile neutrinos given by
$T_s \approx 100 \,{\rm MeV} \, (m/{\rm keV})^{1/3}$
(see discussion in sec.~\ref{sec:nonres}, eq.~(\ref{tprodnus})).

If sterile neutrinos indeed constitute the dark matter, then their
number density can be found from
$\rho_{s} = 10\, \Omega_{DM} \,h^2 \,$keV/cm$^3$, which gives
\be
r_s \equiv {n_s \over n_a} = 1.2 \cdot 10^{-2} \,
\left( \frac{\mbox{keV}}{m} \right)
\left( \frac{\Omega_{DM}}{0.3} \right)
\left( \frac{h}{0.65} \right)^2 \, .
\label{prod}
\ee
Thus, comparing eqs.~(\ref{nse}, \ref{nsm}) with
eq.~(\ref{prod}) we find the necessary values of mass/mixing:
\be
\sin^2 2 \theta_{se}  \approx 6.7 \cdot 10^{-8} \left({
{\rm keV}\over m} \right)^{2}
\left( {g_*(T_{prod})\over 10.75 }\right)^{3/2}
\left( \frac{\Omega_{DM}}{0.3} \right)\left(\frac{h}{0.65} \right)^2,
\label{s2e}\\
\sin^2 2 \theta_{s\mu}  \approx 4.8 \cdot 10^{-8}  \left({
{\rm keV}\over m} \right)^{2}
\left( {g_*(T_{prod})\over 10.75 }\right)^{3/2}
\left( \frac{\Omega_{DM}}{0.3} \right)\left(\frac{h}{0.65} \right)^2,
\label{s2mu}
\ee

The mass eigenstate, the heavier neutrino, $\nu_2$, does not completely
coincide with $\nu_s$. It has an admixture of an active $\nu$,
proportional to $\sin \theta$. Because of this mixing $\nu_2$ is
coupled to the intermediate $Z^0$-boson and it allows for the decay:
\be
\nu_2 \rightarrow \nu_1 + \ell + \bar \ell \, ,
\label{dec}
\ee
where $\nu_1$ is mostly an active flavor and $\ell$ is any lepton with
a mass of less than half the mass of the heavy neutrino.
Following ref.~\cite{dolgov00b} we express the decay life-time as:
\be
\tau = \frac{10^{5} \, f(m) }{m(\mbox{MeV})^5 \,
\mbox{sin}^2 2 \theta}\,\, \mbox{sec} \, ,
\label{decaytime}
\ee
where $f(m)$ takes into account the open
decay channels (for $m<1$ MeV only the neutrino channels are open, and
$f(m)=0.86$, while for $m_s >2m_e$ the $e^+e^-$-channel is also open and
$f=1$). Now, for the sterile neutrino to be a dark matter particle
we must demand that it does not decay on cosmic time scales,
which means $\tau > 4 \times 10^{17}$ sec, and hence from
eq.~(\ref{decaytime}) we get
\be
\mbox{sin}^2 2 \theta < 2.5 \times 10^{-13} \,
\frac{f(m)}{m(\mbox{MeV})^5} \, .
\label{time}
\ee

We can obtain a stronger bound considering the radiative decay
\be
\nu_s \rightarrow \nu_a + \gamma \, ,
\label{nng}
\ee
where $\nu_a$ is any of the active neutrinos. This decay
will contribute with a distinct line into the diffuse photon
background near $m/2$. The branching ratio for
the reaction~(\ref{nng}) was found~\cite{barger95} to be: $BR \approx 1/128$.
The flux of electromagnetic radiation form the decay was calculated in
the papers~\cite{steck,kiboja} (see also
refs.~\cite{kt,drees00}). In the case of a life-time larger than the universe age, and of the matter dominated flat universe
the intensity of the radiation in the frequency interval $d\omega$
is equal to:
\be
dI = (BR)\,{n_s^{(0)} \over H \tau_s}\, {\omega^{1/2} d\omega \over
(m_s/2)^{3/2}}
\label{didom}
\ee
where $n_s^{(0)}$ is the present-day number density of $\nus$ and $H$ is
the Hubble constant (compare to eq.~(\ref{ioflambda}) of
sec.~\ref{cer}). We neglected here some corrections related to 
a possible dominance of the lambda-term in the latest history of the
universe.

In the relevant energy range a rather conservative upper limit on the
flux of electromagnetic radiation is (see e.g. ref.~\cite{retu}):
\be
\frac{d {\cal F}}{d \Omega} < 0.1 \, \left( \frac{1 {\rm MeV}}{E} \right)
{\rm cm }^{-2}{\rm sr }^{-1}{\rm sec }^{-1}
\label{dfdo}
\ee
Thus taking the accepted now values $\Omega_s = 0.3$  and $h=0.65$ we find:
$\tau > 4 \times 10^{22}$, which leads to the bound
\be
\mbox{sin}^2 2 \theta < 2.5 \times 10^{-18}
\, \frac{f(m)}{m(\mbox{MeV})^5} \, .
\label{time2}
\ee
The mass-mixing relation for warm dark matter consisting of sterile 
neutrinos is presented in fig.~\ref{fig-nuswdm} taken from 
ref.~\cite{dolgov00b}.

\begin{figure}[htb]
\begin{center}
\epsfig{file=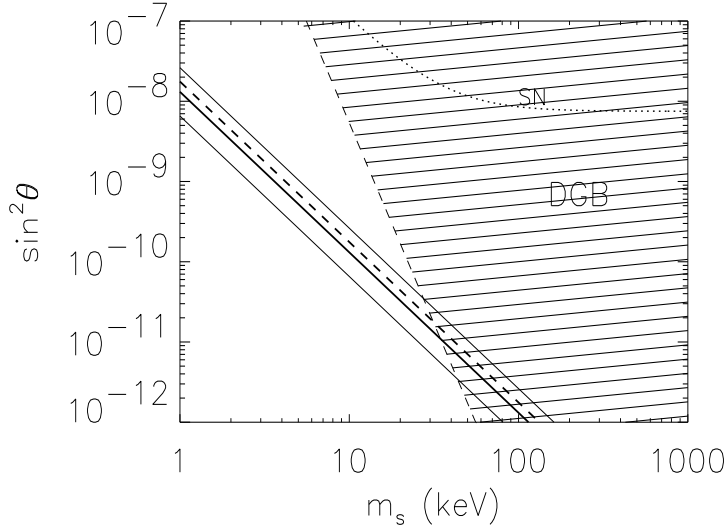,height=8cm,width=10cm}
\end{center}
\caption{
Bounds for $(\nu_\alpha - \nu_s)$-mixing. The middle full line describes
the mass-mixing relationship if sterile neutrinos are the dark matter for
$(\nu_\tau - \nu_s)$-mixing.
The two other full lines allow a factor 2 uncertainty in the amount
of dark matter, $\Omega_{DM} = 0.15 -0.6$.
The dashed line is for $(\nu_e - \nu_s)$-mixing.
The hatched region for big masses is excluded by the Diffuse Gamma Background.
The region above the dotted line is excluded by the duration of SN 1987A
for $(\nu_\tau - \nu_s)$-mixing. (for discussion and references see
the paper~\protect{\cite{dolgov00b}}) }
\label{fig-nuswdm}
\end{figure}

Recently there appeared a paper~\cite{abazajian01} where the
problems of production of sterile neutrinos and their role as possible
warm dark matter were
addressed both for resonance and non-resonance cases. The number
density of sterile neutrinos obtained in that paper for non-resonance
case is
\be
\Omega_\nu h^2 = 0.3 \, (\sin 2
\theta/10^{-5})^2 \, (m/100 {\mbox{keV}})^2.
\label{omegaaba}
\ee
It is about $(7-10)$ times smaller than the
results~(\ref{nse},\ref{nsm}) for $g_* (T_{prod} ) = 10.75$ and
differs by twice larger factor , i.e. by $(14-20)$ from
reference~\cite{dodelson94a}.
One could attribute the disagreement to different treatment of
of the cooling rate $T(t)$, entropy production, etc.
In view of the potential importance of sterile neutrinos as warm dark
matter particles, it is desirable to make an additional study to
resolve this discrepancy.

In a subsequent paper~\cite{abazajian01-X} direct detection of
sterile neutrino warm dark matter by the observation of the X-ray line
with the energy below 2.5 keV was suggested. The authors obtained the upper
limit $m_{\nus} < 5$ keV using the result (\ref{omegaaba}) of
ref.~\cite{abazajian01} for $\nus$ mass density
and the existing data on the X-ray background. However, the limit is
substantially weakened if $\nus$ are produced resonantly due to
a large primordial lepton asymmetry~\cite{shi98}.
The range of photon energies, where the line from the decay of $\nus$
can be observed, was narrowed in ref.~\cite{hansen01-X}. It was argued
there that the lower limits on $m_{\nus}$~\cite{narayanan00,barkana01}
should be strengthen by factor 3.4 for the particular case of $\nus$
produced with a smaller temperature, and correspondingly softer
spectrum than normal neutrinos. This
argument together with the result of ref.~\cite{abazajian01-X}
constraints the mass of $\nus$ in the range 2.6 - 5 keV and
the X-ray line, to be looked for, between 1.3  and 2.5 keV.

\subsection{Anomalous neutrino interactions and dark matter; unstable
neutrinos. \label{anomalous}}

As we have already mentioned, the cosmological upper limit on neutrino
mass (\ref{mnu1}) would not be applicable if the number density of
neutrinos becomes smaller than the standard value (\ref{nnungamma}).
There are several possible ways to achieve that. Neutrinos could
be very weakly coupled to ordinary matter, so they were
never abundantly produced. In the case of sterile neutrinos
considered above, the production proceeds through oscillations and they
never abundantly produced. Another
possibility is an early decoupling of sterile neutrinos, and though
they were in thermal equilibrium at high temperatures, their
present-day abundance is diluted by the entropy release i.e.
by the ratio of the number of particles species in the plasma
at the moment of decoupling of active $\nu$ to that at
decoupling of sterile $\nu$, $g_*^{(fin)}/g_*^{(in)}$. Since
$g_*^{(fin)} = 10.75$, one needs the decoupling of $\nu_s$ to take
place at electroweak scale or above to ensure suppression by an
order of magnitude.

Another possible way to diminish the number density of cosmic neutrinos
is to assume that they, on the contrary, have an additional,
stronger than weak, interaction. In that case neutrinos would
remain longer in thermal equilibrium and at the moment of
freeze-out their number density would be suppressed by the
Boltzmann factor, $\exp (-m_\nu /T)$. For this mechanism to be
operative, the interactions should be strong enough and
the neutrino mass, $\mnu$,
should be sufficiently high. Such a possibility is naturally
realized, for example, in the models of spontaneous breaking of
leptonic charge conservation~\cite{maj,chikashige80,maj2} or breaking
of family symmetry~\cite{fam}. As we have already mentioned, in this
case, due to the Goldstone theorem,
there appears a new light (or massless) (pseudo)goldstone
boson-majoron, $J$, (or familon) with the Yukawa coupling to
neutrinos:
\be
{\cal L}_{\nu,J} = g_{ik}\,J\, \bar \nu_i \gamma_5 \nu_k + ...
\label{gikj}
\ee
where $g_{ik}$ are the coupling constants of majoron to neutrino
species $i$ and $k$ and multi-dot stands for higher order terms
in the coupling constant.
This interaction not only could reduce the cosmological number
density of massive neutrinos but also induce the decay of
heavier neutrino $\nu_h$
to a lighter one and majoron, if $m_J < m_{\nu_h}$ (see
sec.~\ref{massunstbl},\ref{decnu}).

The idea that cosmological number density of neutrinos could
be strongly depleted through annihilation into light or
massless bosons was proposed and elaborated long
ago~\cite{georgi81}-\cite{carlson89}. Depending on
the parameters of the model and, in particular, on the value of
$m_\nu$, massive stable neutrinos could dominate cosmological
energy density and constitute either cold or warm dark matter.
If light scalar bosons do not exist, then one may diminish
cosmological number density of massive neutrinos assuming that
the latter have a large magnetic moment~\cite{giudice90},
$\mu_\nu \sim 10^{-6} $ Bohr magneton. However this value
contradicts the limits on the neutrino magnetic moment discussed in
sec.~\ref{ssec:magnmom} as well as the experimental
measurements~(\ref{muexsp}). Thus, as it was concluded in
ref.~\cite{cooper92}, tau-neutrino with MeV mass cannot constitute
cosmological dark matter.

The majoron model naturally opens two interesting possibilities
for neutrino dark matter: \\
1) a stable $\nu$'s providing warm or even cold dark
matter; neutrinos could be either rather heavy with a smaller
than normal cosmological number density or have a strong
self-interactions reducing their mean-free path, as mentioned
in subsection~\ref{normalnu}~\cite{raffelt87,atrio97};\\
2) an unstable dark matter either with a heavy $\nu$ decaying
into majoron and light $\nu$ or with majoron decaying into a pair of
neutrinos. \\
The role that unstable neutrinos might play in the large scale
structure formation was briefly mentioned in ref.~\cite{gllss},
where the bound on their life-time was derived from the condition
that the universe could not be radiation dominated during the
epoch of galaxy formation (see sec.~\ref{cosden}). A more
constructive idea to describe the observed structure of the universe
with pure neutrino dark matter,
using heavy unstable neutrinos and lighter stable ones
was proposed a little later
in the paper~\cite{davis81}. The authors argued that neutrinos
with masses about 100 keV would preserve and amplify initial
perturbations on galactic scales. Their life-time should be
small in comparison with the universe age to avoid the
Gerstein-Zeldovich bound. On the other hand the life-time
should be long enough so that the perturbations would not
be washed out after heavy neutrino decay. For further amplification
of perturbations the heavy neutrinos should decay
a little after light neutrinos (with $m_\nu \sim 100$ eV) became
nonrelativistic. An alternative possibility of preserving galactic scale perturbations until light neutrinos became
nonrelativistic is to introduce large amplitude fluctuations at small
wave length, such as primordial black holes with masses
$10^9 M_\odot$~\cite{davis81}.

Detailed works of the 1980s with the same idea
to save pure neutrino dark matter with heavier unstable $\nu$'s
can be found in refs.~\cite{doroshkevich84}-\cite{doroshkevich89}.
More general scenarios of decaying particle cosmology, when the
particles in question are not necessary neutrinos, are considered
in the papers~\cite{olive85,sttu}. In ref.~\cite{hut84} only the
decays of heavy neutrino into known particles were permitted:
$\nu_h \rar \nu_l \, \gamma$, $\nu\rar e^+\,e^-\,\nu_l$, or
$\nu_h \rar 3\nu_l$. Radiative decays are very strongly restricted
now, see secs.~\ref{cmbr},\ref{cer},\ref{sterilenu} and
fig.~\ref{fig-nuswdm} and permit the exclusion of a large range of
parameter values, while the decay into
invisible $3\nu$ channel might be a viable option. The life-time
with respect to this decay is given by eq.~(\ref{decaytime}).
With $m_\nu<2m_e$, so that the decay into $e^+e^-\nu$ is forbidden,
this life-time could be
cosmologically permitted and interesting for the structure formation.

There is considerably more freedom in the decaying particle model of
structure formation if new decay channels
are allowed, in particular, decays into lighter $\nu$ and a
(pseudo)goldstone boson majoron or familon. This idea was proposed
almost simultaneously in several
papers~\cite{doroshkevich84,turner84,gelmini84}.
A more complicated model with several decaying particles (heavy neutrinos),
and cold (axion) and hot (light neutrinos) dark matter is considered 
in ref.~\cite{berezhiani90dm}.

A new burst of interest to the structure formation with decaying
particles arose in the middle of 90th after COBE~\cite{cobe} fixed
the normalization of the power spectrum of density fluctuations at
large scales. If one assumes that the universe is dominated by cold
dark matter and that the spectrum of fluctuations is flat, then
one can use the COBE normalization to calculate the power at
galactic and cluster scales. The results were about twice larger
than the astronomical measurements. This meant that the simple
one-component dark matter scenario was ruled out and some
modifications of the latter need to be found.
{\footnote {In fact the conclusion
about the possible end of cold dark matter was made
earlier in ref.~\cite{davis92a} where a possible resolution of
existing discrepancies
with the help of decaying dark matter was mentioned.} }
Several papers appeared almost simultaneously that considered the decaying
neutrino model of structure formation. Essentially
two distinct possibilities were explored: tau-neutrino
with an arbitrary mass from a fraction of keV up to several
MeV~\cite{dgt,dodelson94},\cite{kikuchi94}-\cite{hannestad98prl}
and, somewhat earlier, a new 17 keV
neutrino~\cite{bond91,altherr91,mad,kim95a}, now dead.
The paper~\cite{bond91} is essentially based on ref.~\cite{bardeen87}
where the case of a neutrino with 
a mass in keV range and a life-time of
about $10^4$ years was considered. It is shown that such neutrinos
would provide the necessary extra power on galactic scale.
A detailed analysis of perturbation growth in the universe
with unstable neutrinos having the mass and life-time in the intervals
$ 30\,{\rm eV}<m_\nu< 10\,{\rm keV}$ and
$10^7\,{\rm sec} < \tau_\nu < 10^{16}\,{\rm sec}$
respectively, was performed in ref.~\cite{bharadwaj97}.

A new idea was proposed in
ref.~\cite{mad}, namely that the bosons, $B$,
from the decay of relativistic 17 keV neutrinos
through the channel
$\nu_{17} \rar \nu + B$ could form a Bose condensate producing
cosmologically interesting cold dark matter.
The model was further developed and corrected in
refs.~\cite{kaiser93}-\cite{hannestad97}
for an arbitrary type of decaying neutrino.
The energy spectrum of the bosons produced in the decay could
be quite different from the thermal one. A large
part of them is produced at small momenta, while some are
still relativistic.
The scenario permits to obtain simultaneously two forms of dark
matter: cold and hot ones with comparable contributions into
$\Omega$. However in the simplest version of the scenario only
35\% of matter is cold and about 60\% is hot. The fraction of
CDM could be somewhat enhanced if a sequence of the decays:
$\nut \rar \num+B$, $\nut \rar \nue + B$, and
$\num\rar \nue + B$ was effective.

The scenarios with decaying particles have some common basic features
that are illustrated below on the examples of models presented
in several different papers. There are two main effects important
for structure formation: an increase of a fraction
of relativistic
matter by the products of the decay and a possible earlier
MD stage prior to the
decay of a heavy original particle.

In the scenario of ref.~\cite{dodelson94} a tau-neutrino with a
mass in the interval of 1-10 MeV and a rather short life-time
0.1-100  sec was considered. The role of such neutrino in
primordial nucleosynthesis is discussed in sec.~\ref{massunstbl}.
Its impact on structure formation proceeds through an increase of
energy density in relativistic particles produced by the
decay of $\nut$. An excess of relativistic energy would shift the
moment of equality between matter and radiation to a
later time and
would
modify the spectrum of evolved perturbations. Indeed, perturbations
with the wave length $\lambda$
larger than horizon evolve in the same manner,
so that their spectrum is preserved. If a certain wave enters horizon
at MD stage, the perturbation continue rising as
they did before, so 
the
transfer function relating primordial perturbations to the evolved
ones may be taken as being
equal to unity, $f_{tr}(\lambda) = 1$
However, if perturbation enters
horizon earlier at RD stage, the amplitude of such 
perturbation 
essentially freezes so its relative magnitude becomes smaller. It
can be checked that
the transfer function can be taken as
\be
f_{tr} (\lambda) = (\lambda /\lambda_{eq})^2, \,\,\,
{\rm for}\,\,\, \lambda > \lambda_{eq}
\label{ftr}
\ee
where $\lambda_{eq}$ is the wave length that entered horizon at
a time when the
energy densities of nonrelativistic and relativistic matters were equal.
According to ref.~\cite{dodelson94}:
\be
\lambda_{eq} \approx 10\, {\rm Mpc} \, \left(\Omega h^2\right)^{-1}
\left( g_* /3.36\right)^{1/2}
\label{lambdaeq}
\ee
One can see from these arguments that an increase of $\lambda_{eq}$
would diminish the power at scales $\lambda < \lambda_{eq}$ and would
help resolve the discrepancy between COBE and cluster scales.

Another effect associated with unstable particles is that prior
to their decay they might dominate the cosmological energy density,
so the universe would be in an early temporary MD-stage.
Correspondingly, the scales that entered horizon during this
short MD period would have larger amplitudes and the formation of
structures at these scales would be enhanced. In the case that the
decay products are also massive, as e.g. decays of a heavy neutrino
into majoron (with keV mass) and light but still possibly massive
neutrino, they could contribute both to hot and/or
warm or cold dark matter (see
e.g.~\cite{gelmini84,berezinsky93,dolgov95b,kitazawa96,okada96}).
One more version of the idea of creation of hot dark matter from the
decay of heavier particles was considered in ref.~\cite{hannestad98prl}
in the version $\nu_h \rar \nu_l + \phi$. If the decay 
proceeded after
neutrino decoupling but before matter-radiation equality, the number
of light, but massive, neutrinos, $\nu_l$, would be twice larger than in
the standard model, as requested by two degenerate neutrino
scenario~\cite{primack95} discussed in sec.~\ref{normalnu}. However, the
models are not identical because the energy density of relativistic matter
in two models are different and the spectrum of $\nu_l$ produced in the
decay could be non-thermal.

Today, when cosmology is becoming more and more
precise, the new and forthcoming accurate data on large
scale structure and CMBR will
permit to check the models of structure
formation with great scrutiny and to make restrictive conclusions
about the properties of dark matter. To this end the calculations
of the impact of decaying dark matter on CMBR is of primary
importance~\cite{doroshkevich85a,doroshkevich88,white95,bopi,pibo}.
The considered models permit to vary quite many
parameters: the cosmological mass fractions of different forms of
dark matter, masses and life-times of possibly unstable particles
(neutrinos?), the value of the cosmological constant, and even
the spectral index $n$. However, 
the initial spectrum of perturbations is still assumed
to be of the simple power law form with
the power $n$. Hopefully the combined observational data will
be both accurate and detailed enough to resolve possible
ambiguities.

\section{Neutrino oscillations in the early universe. \label{nuosceu}}


\subsection{Neutrino oscillations in vacuum. Basic concepts.\label{vacosc}}

As we have already mentioned in the Introduction, the mass eigenstates
of neutrinos may be different from their interaction
eigenstates~\cite{pontos1}-\cite{bpvg},\cite{bahcall69} (for a review
and more references to the early papers see~\cite{bilenky78}).
In other words, the mass matrix of different neutrino species
is not diagonal in the basis of neutrino flavors: $ [\nue,\,\num,\,\nut]$.
The latter is determined by the interaction with charged leptons, so that
a beam of e.g. electrons would produce $\nue$ which is a mixture of
several different mass states. And since masses, as we believe, are
created by the Higgs mechanism, they know nothing about interactions
with $W$ and $Z$ bosons and it is only natural that mass matrix and
interaction matrix are not diagonal in the same basis.
An important condition is that the masses
are different, otherwise oscillations would be unobservable. Indeed,
if all the masses are equal, the mass matrix would be proportional
to the unit matrix which is diagonal in any basis.

Of course not only neutrinos are capable of oscillation. All particles
that are produced in the same reactions will do that, but usually the
oscillation frequency, $\omega_{osc} \sim \delta m^2 /2E$ is so huge
and correspondingly the oscillation length
\be
l_{osc}= 2 p /|\delta m^2|
\label{losc}
\ee
is so small, that
the effect is very difficult to observe. Here $E$ and $p$ are respectively
the energy and momentum of the particles under consideration
and $\delta m^2 = m_2^2 - m_1^2$. The expression
is valid in relativistic limit. Only for $K$-mesons and hopefully for
neutrinos the mass difference is sufficiently small so that $l_{osc}$ is,
or may be, macroscopically large.

The neutrino Lagrangian can be written as follows:
\be
{\cal L_\nu} = i \bar \nu \ds \nu + \bar \nu {\cal M} \nu
+ \bar \nu \Zs \nu + \bar \nu \Ws \,\,l
\label{lnu}
\ee
where the vector-column $\nu = [\nue,\,\num,\,\nut]^T$ is the operator of
neutrino field in interaction basis, $l = [ e,\, \mu,\,\tau]^T$ is the
vector of charged lepton operators; the last two terms describe
respectively  neutral and charge current interactions (with $Z$ and $W$
bosons). The upper index ``T'' means transposition.
The matrix ${\cal M}$ is the Dirac mass matrix and, by assumption, 
it is non-diagonal in the interaction basis. For the Majorana mass this term 
should be changed into $\bar \nu_c {\cal M} \nu$, where $\nu_c$ is the
charge conjugate spinor.

Transformation between mass and interaction eigenstates is realized by an
orthogonal, or to be more precise, unitary
matrix $U$ with the entries that are parameters which
should be determined from experiment. In the simplest case of only
two mixed particles the matrix $U$ has the form
\be
U= \left( \begin{array}{cc} \cost & \sint \\
-\sint & \cost \end{array} \right)
\label{u2}
\ee
If, for example, the only noticeable mixing is between electronic and muonic
neutrinos, then the flavor eigenstates are related to the mass eigenstates
$\nu_{1,2}$ as:
\be
\nue & =& \nu_1\, \cost +  \nu_2\, \sint\,, \nonumber \\
\num &=& - \nu_1\, \sint + \nu_2\, \cost
\label{nuemu}
\ee
Thus if electronic neutrinos are produced on a target by an electron beam, the wave function describing their propagation would have the form
\be
\psi_{\nue} (\vec r, t) = \cost \,|\nu_1\rangle\, e^{ik_1 x} +
\sint\, |\nu_2\rangle\, e^{ik_2 x}
\label{psinue}
\ee
where $kx = \om t - \vec k \vec r$ and sub-$\nue$ means that the initial
state was pure electronic neutrino. Below (in this section only)
we denote neutrino energy as $\om$
to distinguish it from the energies of heavy particles that are denoted as
$E$. We assume, as is normally done, plane wave representation of the wave
function.

If such a state hits a target, what is the probability of producing an
electron or a muon? This probability is determined by the fraction of
$\nue$ and $\num$ components in the wave function $\psi_{\nue}$ at
space-time point $x$. The latter can be found by re-decomposition of
$\nu_{1,2}$ in terms of $\nu_{e,\mu}$:
\be
\psi_{\nue} (\vec r, t) = \cost\,e^{ip_1x}\, \left( \cost\, |\nue \rangle -
\sint\, |\num \rangle \right)+
\sint\, e^{ip_2 x}\, \left( \sint\, |\nue \rangle +
\cost\, |\num \rangle \right)
\label{psinuemu}
\ee
One can easily find from that expression that the probabilities 
of registering
$\nue$ or $\num$ are respectively:
\be
P_{\nue}(\vec r, t) \sim
\cos^4 \theta +\sin^4\theta + 2\sin^2 \theta\, \cos^2\theta \cos\,
\delta \Phi\,,
\label{pnue}
\ee
\be
P_{\num}(\vec r, t) \sim 2 \sin^2\theta\, \cos^2\theta \left( 1 -
\cos\, \delta \Phi \right)
\label{pnum}
\ee
Here $\delta \om = \om_1 -\om_2$, $\delta \vec k = \vec k_1 - \vec k_2$,
and
\be
\delta \Phi\, (\vec r, t) =  \delta \om\, t - \vec \delta k\, \vec r
\label{deltaphi}
\ee
The energy difference between the mass eigenstates is
\be
\delta \om= {\partial\om \over \partial m^2}\, \delta m^2 +
{\partial \om \over \partial \vec k}\, \delta \vec k
\label{deltae}
\ee
Using this expression we find for the phase difference
\be
\delta \Phi\, (\vec r, t) = {\delta m^2 \over 2\om }\, t + \delta \vec k
\left( {\vec k \over \om}\, t - \vec r \right)
\label{deltaphi2}
\ee

The standard result of the neutrino oscillation theory is obtained if
one assumes that: 1) $\delta \vec k =0$, 2) $\vec k = \om \vec v$, and
3) $t=r/v$:
\be
\delta \Phi = {\delta m^2 r \over 2 k}
\label{deltaphist}
\ee
Each of these assumptions is difficult to understand, and moreover, some
of them, in particular, $\delta k=0$ may be explicitly incorrect
(see below). Both the second and the third conditions are fulfilled for a
classical motion of a point-like body, however their validity should be
questioned for a quantum mechanical particle (for a wave).
Despite all that, the final result (\ref{deltaphist})
is true and if there are some corrections, they can be 
easily
understood.

Basic features of neutrino oscillations were discussed in many papers. An
incomplete list of references
includes~\cite{nussinov76}-\cite{nauenberg99}.
One can find more citations and discussion in the above quoted papers and
in the books~\cite{bahcall89}-\cite{fukugita94}.
Still some confusion and 
suggestions of
possible controversies reappear
from time to time in literature,
so it seems worthwhile to present a consistent derivation
of eq.~(\ref{deltaphist}) from the first principles.
A large part of this section is based on discussions (and unpublished
work) with A.Yu. Morozov, L.B. Okun, and M.G. Schepkin and on the lecture
by the author~\cite{dolgov99-erc}.

Let us consider a localized source that produces oscillating neutrinos;
we keep in mind, for example, a pion decaying through the channel
$\pi \rar \mu + \nu_\mu$. The wave function of the source $\psi_s (\vec r,t)$
can be Fourier decomposed in terms of plane waves:
\be
\psi_s (\vec r,t) = \int d^3 p\,\, C(\vec p -\vec p_0)
e^{iEt -i \vec p \vec r}&\approx&
\nonumber \\
e^{iE_0 t - i\vec p_0 \vec r}
\int d^3 q\, C(\vec q) \exp \left[ -i\vec q \left(\vec r -
 \vec V_0 t \right)\right]
&=& e^{iE_0 t -i\vec p_0 \vec r}\, \tilde C \left( \vec r - \vec V_0 t \right)
\label{psis}
\ee
where $\vec V_0 = \vec p_0 /E_0$.
It is the standard wave packet representation. The function
$C(\vec p -\vec p_0)$ is assumed to be sharply peaked around the central
momentum $\vec p_0$ with dispersion $\Delta \vec p$. The particle is, by
construction, on-mass-shell, i.e. $E^2 = p^2 + m^2$. This is also
true for the central values $E_0$ and $p_0$.
As the last expression shows, the particle behaves as a plane wave,
with the frequency and the wave vector given respectively
by $E_0$ and $\vec p_0$ and with the shape function (envelope) given
by $\tilde C (\vec r- \vec V_0 t)$, which is
the Fourier transform of $C(\vec q)$. Evidently the envelope moves
with the classical velocity  $\vec V_0 $. 
The
characteristic size of the wave packet is $l_{pack} \sim 1/\Delta p$.

Let us consider the pion decay, $\pi \rar \mu + \nu$. One would naturally
expect $\delta \om \sim \delta k \sim \delta m^2 /E$. If this is true
the probability of oscillation would be
\be
P_{osc} \sim \cos \left[ { x + b\, \left(x -Vt\right) \over l_{osc}}\right],
\label{posc}
\ee
where $l_{osc}$ is given by expression (\ref{losc}) and
$b$ is a numerical coefficient relating $\delta p $ with $l_{osc}$.
For simplicity the one-dimensional expression is presented.

Thus to calculate
the probability of neutrino registration one should average
the factor $(x-Vt)$ over the size of the wave packet and for large
packets, if $b$ is non-negligible,
a considerable suppression of oscillations should be expected.
The size of the neutrino wave packet from the pion
decay at rest is macroscopically large,
$l_{pack} \approx c\, \tau_\pi \approx 7.8$ m,
where $c$ is the speed of light and $\tau_\pi = 2.6 \cdot 10^{-8}$ sec
is the pion life-time. The oscillation length is
$l_{osc} = 0.4\, {\rm m}\, ( E/ {\rm MeV} / (\delta m^2 /{\rm eV}^2) $,
so $l_{osc}$ could be smaller or comparable to $l_{pack}$ and the effect
of suppression of oscillations due to a finite size of the wave packet
might be significant. It is indeed true but only for the decay of a moving
pion, and this suppression is related to an uncertainty in the position
of the pion at the moment of decay. To 
check that
we have to abandon
the naive approach described above
and to work formally using the standard set of quantum
mechanical rules.

Let us assume that neutrinos are produced by a source with the
wave function $\psi_s(\vec x, t)$. This source produces neutrinos together
with some other particles.
We assume first the following experimental conditions:
neutrinos are detected at space-time point $\vec x_\nu, t_\nu$, while
the accompanying particles are not registered. The complete set of
stationary states of these particles is given by the wave functions
$\psi_n \sim \exp (iE_n t)$. The amplitude of registration of propagating
state of neutrino of type $j$ (mass eigenstate) accompanying by other
particles in the state $\psi_n$ is given by
\be
A_j^{(n)} = \int d\vec r_s\, dt_s \psi_s (\vec r_s, t_s)
\psi_n (\vec r_s, t_s) G_{\nu_j} \left(\vec r_\nu - \vec r_s, t_\nu - t_s
\right)
\label{ajn}
\ee
In principle one even does
not need to know the concrete form of $\psi_n$. The only necessary
property of these functions is the condition that they form a complete set:
\be
\sum_n \psi_n \left(\vec r, t\right) \psi^*_n \left(\vec r\,',t \right)
= \delta \left(\vec r - \vec r\,' \right)
\label{sumn}
\ee
However, in what follows, for simplicity sake, we will use the eigenfunctions of
momentum, $\psi_n \sim \exp (i \vec p\,\vec r -iEt)$.

For the subsequent calculations we need the following representation of the
Green function, which is obtained by the sequence of integration:
\be
G(\vec r, t) &=& \int {d^4 p_4 \over p^2_4 - m^2} e^{ip_4 x} =
\nonumber \\
&&2\pi \int_{-\infty}^{+\infty} d\omega e^{i\omega t} \int_0^{+\infty}
{dp p^2 \over \omega^2 - p^2 - m^2} \int_{-1}^1 d\zeta
e^{-i p r \zeta} =
\nonumber \\
&&{i\pi \over r}  \int_{-\infty}^{+\infty} d\omega e^{-i\omega t}
\int_{-\infty}^{+\infty}
{dp p^2 \over \omega^2 - p^2 -m^2 } \left( e^{ipr} - e^{-ipr} \right)
\label{green1}
\ee
Here $\omega$ and $p$ are respectively the fourth and space components of
the 4-vector $p_4$. We have omitted spin matrices because the final
result is essentially independent of them.
The integration over $dp$ was
extended over the whole axis (from $-\infty$ to $+\infty$) because the
integrand is an even function of $p$. This permits us to calculate this integral
by taking residues in the poles on mass shell:
$p = \pm \sqrt { \omega ^2 - m^2 +i\epsilon }$. Both poles give the same
contribution, so
skipping unnecessary numerical coefficients, we finally obtain:
\be
G( \vec r , t) = {1\over r} \int_{-\infty}^{+\infty}
d\om e^{-i\om t +i \sqrt{ \om^2 - m^2}\,r }
\label{green2}
\ee
As a source function $\psi_s$ we will take essentially
the
expression~(\ref{psis}) but assume that the source is a decaying particle
with the decay width $\gamma$,
that was born at the moment $t=0$. It corresponds to multiplication of
$\psi_s$ by $\theta (t) \exp (-\gamma t)$, where $\theta (t)$ is the
theta-function, i.e. $\theta (t<0) =0$ and $\theta (t>0) =1$.
Taking all together we obtain
the following expression for the amplitude:
\be
A_j^{(n)}(\vec r, t ) &=& \int_0^{\infty} dt_s
\int {d\vec r_s \over |\vec r - \vec r_s|}
\int d\vec p\, C\left( \vec p -
\vec p_0 \right) e^{i E t_s - i \vec p \vec r_s -\gamma t/2}\,
e^{i E_n t_s +i \vec p_n \vec r_s } \nonumber \\
&&\int_{-\infty}^{+\infty} d\om_j e^{i\om_j (t-t_s) -ik_j |\vec r -
\vec r_s | }
\label{ajn2}
\ee
Integrals over $d\vec r_s$ and $d\vec p$ are taken over all infinite space.
It is worthwhile to remind here that all the momenta are on mass shell,
$E^2 = p^2 +m^2_\pi$ (we assumed that the source is a decaying pion) and
$\omega_j^2 = k_j^2 + m_j^2$, where $m_j$ is the mass of $j$-th neutrino
eigenstate.

The integration over $dt_s$ is trivial and gives the factor
$\left( E-E_n -\om_j +i\gamma /2 \right)^{-1}$.
The integration over $d\vec r_s$ can be easily done if the registration
point is far from the source. In this case it is accurate enough to
take $1/|\vec r - \vec r_s| \approx 1/r$,
while the same quantity in the exponent should be expanded up to the first
order:
\be
|\vec r -\vec r_s| \approx r - \vec \xi\, \vec r_s
\label{r-rs}
\ee
where $\vec \xi = \vec r /r$ is a unit vector directed from the center of
the source taken at the initial moment $t=0$ to the
detector at the point $\vec r$.
In this limit the integral over $d\vec r_s$ gives $\delta \left(\vec p -
\vec p_n - \vec k_j \right)$, ensuring momentum conservation:
\be
\vec p = \vec p_n + \vec k_j \equiv \vec p_{\pi,j}
\label{pcons}
\ee
The vector of neutrino momentum is formally defined as
\be
\vec k_j = \vec \xi \, k_j = \vec \xi \sqrt{\om_j^2 - m_j^2}
\label{kj}
\ee
Ultimately we are left with the integral:
\be
A_j^{(n)} = {1\over r} \int_{-\infty}^{+\infty} d\om_j C\left( \vec p_n
+ \vec k_j - \vec p_0 \right) {e^{i\om_j t -ik_j r}
\over E_{\pi,j}-E_n -\om_j + i\gamma/2}
\label{ajn3}
\ee
where $E_{\pi,j} = \sqrt{(\vec p_n + \vec k_j )^2 + m_\pi^2 }$.
This integral can be taken in the 'pole approximation' and to do that
we need to expand the integrand around the energy conservation law
(see below eq.~(\ref{e0})) as follows.
The neutrino energy is presented as $\om_j = \om_j^{(0)} + \Delta \om_j$.
To avoid confusion
one should distinguish between the deviation of neutrino energy from the
central value given by the conservation law, $\Delta \om_j$, from the
difference of energies of different neutrino mass eigenstates,
$\delta \om = \om_1 - \om_2$.
The neutrino momentum is expanded up to the first order in $\Delta \om_j$:
\be
 k_j = \sqrt {\om_j^2 -m^2_j} \approx k_j^{(0)} +
\Delta \om_j /V_j^{(\nu)}
\label{kjexp}
\ee
where $V_j^{(\nu)} = k_j^{(0)} /\om_j^{(0)}$ is the velocity
of $j$-th neutrino.
The pion energy is determined by the momentum conservation~(\ref{pcons})
and is given by
\be
E_{\pi,j} =
\sqrt{ \left(\vec p_n + \vec k_j^{(0)} + \vec\Delta k_j \right)^2 +
m^2_\pi } \approx E^{(0)}_{\pi,j} + \vec V_{\pi,j} \vec \xi\, \Delta \om_j
\label{epi}
\ee
where the pion velocity is
$\vec V^{(\pi)}_j = (\vec p_n + \vec k_j^{(0)})/E^{(0)}_{\pi,j}$.
The neutrino energy, $\om_j^{(0)}$, satisfying the conservation law
is defined from the equation:
\be
E^{(0)}_{\pi,j} - E_n - \om_j^{(0)} =0
\label{e0}
\ee
Now the integral over $\om_j$ is reduced to
\be
A_j^{(n)} = {e^{i\om_j^{(0)} t - i k_j^{(0)} r }  \over r}
C\left( \vec p_n  + \vec k_j^{(0)} - \vec p_0 \right)
 \nonumber \\
\int_{-\infty}^{+\infty} d\Delta \om_j\,\,
{e^{i\Delta \om_j \left( t - r/V^\nu_j \right)} \over
\left( \vec V_j^{(\pi)} \vec \xi / V_j^{(\nu)}\right) \Delta \om_j -
\Delta \om_j +i\gamma/2 }
\label{ajn4}
\ee
The last integral vanishes if $t< V_j^{(\nu)} r$, while in the opposite
case it can be taken as the residue in the pole and we finally
obtain:
\be
A_j^{(n)} = {C\left( \vec p_{\pi,j} - \vec p_0 \right) \over r}\,
\theta \left (r - V_j^{(\nu)} t\right)
\exp  \left(i\om^{(0)} t -i k_j^{(0)}r
-{\gamma\over 2}{ V_j^{(\nu)} t - r \over V_j^{(\nu)} -
\vec V^{(\pi)}_j \vec \xi} \right)
\label{ajnfin}
\ee
We have obtained the neutrino wave packet moving with the
velocity $V^{(\nu)}_j$ with a well defined front (given by the
theta-function) and decaying with time in accordance with the decay law of
the source. 
A similar
wave packet, but moving with a slightly different
velocity describes another oscillating state $\nu_i$. It is evident from
these expressions that the phenomenon of coherent oscillations takes place only
if the packets overlap, as was noticed long
ago~\cite{nussinov76,ad1,kayser81}.

The probability of the registration of oscillating neutrinos at the space-time
point $(\vec r_\nu, t_\nu)$ is given by the density matrix
\be
\rho_{ij} = \int d\vec p_n\, A^{(n)}_i \left(\vec r_\nu, t_\nu \right)
A^{*(n)}_j \left(\vec r_\nu, t_\nu \right)
\label{rhoij}
\ee
The oscillating part of the probability is determined by the phase
difference~(\ref{deltaphi}) but now the quantities $\delta \om$ and
$\delta k$ are unambiguously defined. To this end we will use conservation
laws~(\ref{pcons},\ref{e0}). They give:
\be
\delta \om^{(0)} = \delta E_{\pi}\,\,\, {\rm and}\,\,\,
\delta \vec k^{(0)} = \delta \vec p_{\pi}
\label{deltaomk}
\ee
The variation of neutrino energy is given by
\be
\delta \om = V^{(\nu)} \delta k + \delta m^2 / 2 \om
\label{deltaom}
\ee
while the variation of the pion energy can be found from expression
(\ref{epi}):
\be
\delta E^{(\pi)} = \vec V^{(\pi)} \delta \vec k
\label{deltae1}
\ee
From these equations follows
\be
\delta \om = - {\delta m^2 \over 2\om} {\vec V^{(\pi)} \vec \xi
\over V^{(\nu)} - \vec V^{(\pi)} \vec \xi }\,\,\,
{\rm and}\,\,\,
\delta k = - {\delta m^2 \over 2\om} {1
\over V^{(\nu)} - \vec V^{(\pi)} \vec \xi }
\label{deltaomfin}
\ee
One sees that generally both $\delta \om$ and $\delta k$ are non-vanishing.
A similar statement was made in ref.~\cite{giunti00}.
Only in the case of pion decay at rest, $\delta \om = 0$ but $\delta k$ is
is invariably non-zero.
Inserting the obtained results into expression
(\ref{deltaphi}) for the phase difference we come to the standard
expression~(\ref{deltaphist}) if $V_\pi = 0$. This result
shows a remarkable stability with respect
to assumptions made in its derivation. However if the
pion is moving, then the oscillation phase contains an extra term
\be
\delta \Phi = {\delta m^2 \over 2\om}\,\, {\vec \xi \left(\vec r -
\vec V^{(\pi)} t\right) \over V^{(\nu)} - \vec \xi\, \vec V^{(\pi)} }=
{r\,\delta m^2  \over 2 k } + {(\vec \xi\, \vec V^{(\pi)})
(r-V^{(\nu)} t ) \over
V^{(\nu)} - \vec \xi \vec V^{(\pi)} }
\label{deltaphivpi}
\ee
This extra term would lead to
a suppression of oscillation after averaging over
time. This suppression is related to the motion of the source and reflects
the uncertainty in the position of 
the
pion at the moment of decay. So this
result can be understood in the framework of the standard naive
approach.

A similar expression can be derived for the case when both neutrino and
muon from the decay $\pi \rar \mu +\nu_j$ are registered in the space-time
points $\vec r_\nu, t_\nu$ and $\vec r_\mu, t_\mu$ respectively.
This case was considered in refs.~\cite{dolgov97,nauenberg99}.
Here we will use the same approach as described above when
the muon is not registered. The only
difference is that in eq.~(\ref{ajn}) for the oscillation amplitude
we have to substitute the Green's
function of muon $G_\mu (\vec r_\mu -r_s, t_\mu - t_s)$
instead of $\psi_n (\vec r_s,t_s)$.
The calculations are essentially the same and after some
algebra the following expression for the oscillation amplitude
is obtained~\cite{dolgov99-erc}:
\be
A_{\mu,\nu} &\sim& {V_\mu V_\nu \over r_\mu r _\nu}
\theta ( L_\mu + L_\nu ) \exp \left[ -{\gamma (L_\mu + L_\nu) \over
2 (V_\mu + V_\nu) } \right]
\tilde C (V_\mu L_\nu - V_\nu L_\mu ) \nonumber \\
&& \exp {\left[ i \left( k^{(0)}_\mu r_\mu + k^{(0)}_\nu r_\nu -
E^{(0)}_\mu t_\mu - E^{(0)}_\nu t_\nu \right)\right]}
\label{amunu}
\ee
where $L=Vt-r$. Each kinematic variable depends upon the neutrino state
$j$, so they should contain sub-index $j$. The upper indices ``0'' mean that
these momenta and energies are the central values of the corresponding wave
packets, so the classical relation $\vec k^{(0} = \vec V E^{(0)}$
holds for them. Here the direction of momenta are defined as above,
eq.~(\ref{kj}), along the vector indicated to the observation point.
However, the kinematics in this case
is different from the previous one and the change of the energy and
momentum of each particle for reactions with different sorts of
neutrinos are related through the equations:
\be
\delta E_\mu + \delta E_\nu =0\,\,\, {\rm and}\,\,\,
\delta \vec k_\mu + \delta \vec k_\nu =0.
\label{deltaek}
\ee
For the central values of momenta the following relations are evidently true,
$\delta k_\nu = \delta E_\nu - \delta m^2 / 2 k_\nu$, and the similar one for
the muon without the last term proportional to the mass difference.
Correspondingly the phase of the oscillations is given by the expression
\be
\delta \Phi = \delta k_\mu r_\mu - \delta E_\mu t_\mu + \delta k_\nu r_\nu
- \delta E_\nu t_\nu = \nonumber \\
\delta E_\nu \left( {r_\mu -V_\mu t_\mu \over V_\mu} -
 {r_\nu -V_\nu t_\nu \over V_\nu} \right) -{\delta m^2 \over 2 k_\nu} \, r_\nu
\label{deltaphimunu}
\ee
The first two terms in the phase are proportional to the argument
of $\tilde C$ in eq.~(\ref{amunu}) and thus 
their contribution is equal 
to the size of the the source, i.e. to the wave packet of the initial pion.
If the latter is small (as usually is the case) we again obtain the standard
expression for the oscillation phase.

\subsection{Matter effects. General description. \label{matter}}

Despite extremely weak interactions of neutrinos, matter may have a
significant influence on the oscillations if/because the mass difference
between the propagating eigenstates is very small.
A description of neutrino oscillations in matter was first done in
ref.~\cite{wolfenstein}. Somewhat later a very important
effect of resonance
neutrino conversion was discovered~\cite{mikheev}, when even with a very
small vacuum mixing angle, mixing in medium could reach the maximal value.

Hamiltonian of free neutrinos in the mass eigenstate basis has the form:
\be
{\cal H}_m^{(1,2)}= \left( \begin{array}{cc} E_1 & 0 \\
0 & E_2 \end{array} \right)
\label{hm}
\ee
where $E_j = \sqrt{ p^2 +m_j^2}$. In the interaction basis ${\cal H}_m$ is
rotated by the matrix~(\ref{u2}):
\be
{\cal H}_m^{(a,b)} = U {\cal H}_m^{(1,2)} U^{-1} =
 \left( \begin{array}{cc} \cos^2\theta\,E_1+ \sin^2\theta\, E_2
&g \sint\cost \\
g\sint\cost  & \sin^2\theta\, E_1 +\cos^2\theta\, E_2 \end{array} \right)
\label{hmab}
\ee
Here $g=\delta m^2 /2E$ and we returned to the more usual notation $E$,
for neutrino energy instead of $\omega$ used in the previous subsection.

The interaction Hamiltonian is diagonal in the interaction basis and if
only first order effects in the Fermi coupling constant, $G_F$, are taken
into account, then the Hamiltonian can be expressed through refraction
index, $n_a$, of flavor $a$-neutrino in the medium (recall that the
deviation of refraction index from unity
is proportional to the forward scattering amplitude and
thus contains $G_F$ to the first power):
\be
H_{int}^{(a,b)} =
 \left( \begin{array}{cc} E\,(n_a-1)  & 0 \\
0 & E\,(n_b - 1) \end{array} \right)
\label{hint}
\ee
where a small difference between $E_1$ and $E_2$ in front of small
factors $(n-1)$ was neglected.
This result is true for the forward scattering of neutrinos on electrons
or other active neutrinos that are not mixed with neutrinos in question.
For example, if $\nue$ and $\num$ are mixed between themselves but not
with $\nut$, refraction index in the equation above contains contributions
from $(\nue-\nut)$- and $(\num-\nut)$- scattering as well as from
$(\nue-e^\pm)$ and $(\num-e^\pm)$ ones. The contribution from
self-scattering, i.e.
$(\nue-\nue)$, $(\num-\num)$, and $(\nue-\num)$ is given by non-diagonal
matrix with off-diagonal entries. This point was noticed in
ref.~\cite{pantaleone92}. We will discuss these
non-diagonal terms in sec.~\ref{ssec:active}, where oscillations
between active neutrinos are considered. In the case of
oscillations between active and sterile neutrinos, which is especially
interesting for cosmology, the matrix~(\ref{hint}) has the diagonal
form presented above with $n_b = 1$.

Thus, up to a unit matrix, the total Hamiltonian in the interaction basis
can be written as
\be
H_{tot}^{(a,b)} =
 \left( \begin{array}{cc} f & g\stw /2 \\
g\stw/2 & 0 \end{array} \right)
\label{htot}
\ee
where $f=g\ctw + E \delta n$ and $\delta n = n_a -n_b$.
This matrix is easy to diagonalize. Its eigenvalues are
\be
\lambda_{1,2} = { f \pm \sqrt{ f^2 + g^2 \sin^2 2\theta} \over 2}
\label{lambda12}
\ee
and the eigenstates in matter (up to normalization factor) are
\be
|\nu_{1,2} \rangle = |\nu_a\rangle + {g \stw \over f\pm \sqrt{f^2+
g^2 \sin^2 2\theta}}\, |\nu_b\rangle
\label{nu12}
\ee

Refraction index may change with time, as happens in cosmology, or with
space point, if neutrinos propagate in inhomogeneous medium, for example
from the center of the Sun to its surface. If somewhere (or sometime) $f$
vanishes then the resonance transition of one neutrino species to another
is possible~\cite{mikheev}. Indeed let us assume that $\nue$ and $\num$
are mixed with a small vacuum mixing angle $\theta$
and that initially an electronic neutrino was
produced in vacuum. So the initial propagating state
would be mostly $\nue$:
\be
|\nu_1\rangle_{in} = |\nu_e \rangle + (1/2) \tan 2\theta
\,\, |\nu_\mu\rangle
\label{nu1in}
\ee
After propagation in the media where the function $f$ changes sign
passing through zero, the propagating state would become mostly $\num$:
\be
|\nu_1\rangle_{fin} = |\nu_e \rangle - (2 /\tan 2\theta)
\,\, |\nu_\mu\rangle
\label{nu1fin}
\ee
This effect of resonance conversion may play an important role in the
resolution of the solar neutrino problem and in cosmology.

\subsection{Neutrino oscillations in cosmological plasma \label{osccosm}}

\subsubsection{A brief (and non-complete) review \label{oscrev}}

Neutrino oscillations in the primeval plasma are significantly
different from e.g. solar neutrino oscillations in the
following two important aspects.
First, cosmological plasma in the standard model is almost charge
symmetric. The relative excess of any particles over antiparticles is
believed to be at the level $10^{-9}-10^{-10}$, while in stars
the asymmetry is of order unity. (Possible large cosmological
asymmetry is discussed below in
subsections~ \ref{sec:leptas} and \ref{ssec:active}.)
Neutrino oscillations in stars do not have a direct
impact on the stellar material, while on the other hand,
neutrino oscillations in the early universe 
may have changed
the
magnitude of cosmological charge asymmetry in the sector of active
neutrinos. This asymmetry has a strong influence on the oscillations
through the refraction index of the primeval plasma (see below). It leads
to an essential non-linearity of the problem and makes
calculations quite complicated. We will return to this effect a little
later.

Second important point is that the neutrino mean free path in the early
universe is quite small at high temperatures and hence breaking of
coherence becomes essential. Because of that one cannot use wave
functions to describe oscillations and should turn to the density
matrix formalism. It also leads to a greater complexity of equations.
Kinetic equations for density matrix with the account of neutrino
scattering and annihilation were discussed in the
papers~\cite{ad1},\cite{harris82}-\cite{lee00}.

In ref.~\cite{ad1}, where the impact of neutrino oscillations on
big bang nucleosynthesis (BBN) was first considered, only the second order
effects, proportional to $G^2_F$, were taken into account,
while the deviation of refraction index from unity was neglected.
This approximation is valid for a sufficiently high $\delta m^2$.
In an independent study~\cite{khlopov}
implications for BBN of possible CP-violating effects in oscillations
were discussed but matter effects were not taken into consideration.
Earlier works on neutrino oscillations in the early universe also include
refs.~\cite{fargion84}-\cite{langacker87}. In
ref.~\cite{mikheev86} the conversion of $\nue$ into sterile $\nus$
was considered and it was argued that for a large mixing angle
$\sin^2 2\theta > 0.05$ and $(-\dm) = 10^{-6}-10^{-9}$ eV$^2$ the
effect of oscillations on BBN can be significant due to a possible
asymmetry between $\nue$ and $\bar\nue$ created by the oscillations.
However subsequent works do not support the conclusion
about generation of asymmetry in the case of large mixing angles.

The case of $\nue-\num$ or $\nue-\nut$ oscillations was considered
in ref.~\cite{langacker87}. The authors concluded that the oscillations
could create asymmetry between $\nue$ and $\bar\nue$ prior to
primordial nucleosynthesis but the effect on primordial abundances would
be very weak. 
The asymmetry may only be generated due to deviations from thermal 
equilibrium. In the standard model it takes place due to different
heating of $\nue$ and $\nu_{\mu,\tau}$ by $e^+e^-$-annihilation (see
sec. \ref{masslessdistr}). According to ref.~\cite{langacker87} the
upper bound on the variation of primordial mass fraction of $^4 He$
by this effect is $\Delta Y_p < 1.3\cdot 10^{-3}$. 

The first calculations
of refraction index in cosmological plasma were performed in
ref.~\cite{nora}. The results of this work permitted to make a more
accurate description of neutrino oscillations early 
universe~\cite{barbieri90}-\cite{kostelecky96b}.

It was noticed in ref.~\cite{barbieri91} that the oscillations between
an active and sterile neutrinos for 
a
sufficiently small mixing angle
could generate an exponential rise of
lepton asymmetry in the sector of active neutrinos. The origin of this
instability is the following. Since lepton
asymmetry comes with the opposite sign to refraction indices of neutrinos
and antineutrinos (see below section~\ref{refr}),
it may happen that the transformation of antineutrinos
into their sterile partners would proceed faster than the similar transformation
of neutrinos, especially if resonance conditions are fulfilled.
It would lead to an increase of the asymmetry and through the refraction
index to more favorable conditions for its rise. However it was
concluded~\cite{barbieri91} that the rise was not significant and the
effect of the generated asymmetry on BBN was small. This conclusion
was reconsidered in ref.~\cite{ftv} where the arguments were
presented that the asymmetry
generated by this mechanism could reach very large values, close to unity,
and this effect, in accordance with an earlier paper by the same
group~\cite{foot95}, would have a significant influence on
primordial abundances. This result attracted great attention and was
confirmed in several subsequent
publications~\cite{foot97}-\cite{dibari01} where very serious arguments,
both analytical and numerical were presented. The rise of asymmetry
by 9 orders of magnitude was questioned in ref.~\cite{dolgov00} in the
frameworks of a certain analytical approximation. This work was criticized
in ref.~\cite{dibari00b} (see also refs.~\cite{buras00a,buras00b}) where
several drawbacks of the approximation were indicated. In
ref.~\cite{buras00b} the maximum possible value of the asymmetry was
calculated under the assumption that the resonance transition has 100\%
efficiency, so that all (anti)neutrinos with the resonance momentum
transform into sterile partners. It was shown that indeed the maximum
value of the asymmetry could be close to unity.
The method of analytical calculations of ref.~\cite{dolgov00} was
refined in ref.~\cite{dolgov01a}. The analytical results obtained there
show a large rise of asymmetry and are very close to the numerical results
obtained by other groups cited above.

Moreover, some works showed not only a rising and large asymmetry but also
a chaotic behavior of its sign~\cite{shi96}-\cite{buras00}. However,
this conclusion was based on the solution of
a simplified system of kinetic equations either averaged over momentum
or with a fixed momentum of neutrinos.
On the other hand,
numerical solution of momentum dependent kinetic equations performed
in ref.~\cite{dibari99c} showed no chaoticity, except for the region
where the authors could not exclude numerical instability. 
The analytical
solution of the complete kinetic equation found in ref.~\cite{dolgov01a}
is also not chaotic in the range of parameters where the rise is
observed. These two works contradict a recent numerical solution of the 
complete kinetic equations that indicates 
chaoticity~\cite{kainulainen01s}.
The problem remains open and deserves more consideration.

{\it If} the chaoticity indeed exists, then leptonic domains
in the early universe might be formed and the lepton number gradients
at the domain boundaries could enhance production of sterile neutrinos
by MSW resonance~\cite{shi99d1}. This phenomenon would have a noticeable
impact on big bang nucleosynthesis.

An interesting effect related to neutrino oscillations was found
in the paper~\cite{dibari99b} (see also ref.~\cite{enqvist00}).
It was shown that active-sterile neutrino oscillations in the
presence of small inhomogeneities in the baryon number could give
rise to large fluctuations of lepton number and 
the
formation of leptonic
domains in the universe. This phenomenon is induced by the neutrino
diffusion in initially inhomogeneous medium and has nothing to do
with the chaoticity mentioned above.

Though the set of kinetic equations for density matrix looks rather simple
(see below sec.~\ref{ssec:kineq}), its numerical solution is a difficult task
because the elements of density matrix are oscillating functions of
momentum and time. Due to complexity of equations some simplifying
approximations were made in their solutions, in particular, an
averaging over momentum and an approximate description of the loss
of coherence. In the exact system of equations the latter
is described by the
non-linear combinations of the elements of density matrix integrated
over phase space of interacting particles. In the simplified
approach these terms are mimicked by $\gamma (\rho_{eq} - \rho)$.
Both these approximations permit to reduce integro-differential system
of kinetic equations to the system of ordinary differential equations.
First accurate numerical solution of (almost) exact equations were done in
refs.~\cite{kirilova97}$-$\cite{kirilova00} for a rather small mass
difference, $\delta m^2 < 10^{-7}\, {\rm eV}^2$. Neither
chaoticity, nor a considerable rise of the asymmetry were found. For
a larger mass difference a strong numerical instability was observed.
However this result does not contradict other papers because
the latter found the above mentioned effects for much larger 
$\delta m^2$.

Neutrino oscillations in the presence of cosmic magnetic field were
considered in the papers~\cite{semikoz94,pastor95,semikoz96}. There
is an evident contribution to the process if
neutrinos have a noticeable diagonal or transition magnetic moment.
Moreover, there is a possibility of a medium-induced effective
interaction of left-handed neutrinos with magnetic field. There
is no consensus in the literature about a possible magnitude of
this interaction and its effect on the oscillations. For details
and references see above quoted papers and sec.~\ref{ssec:magnmom}.
Some complications of the process of oscillations may take place
at the Planck scale, in particular, quantum gravity effects might lead
to oscillation freezing~\cite{ahluwalia00}.

Further down in this section we will discuss 
these subjects in more detail. The
field is so vast that it possibly deserves a separate review, and an
interesting one has appeared recently~\cite{prakash01}.

\subsubsection{Refraction index \label{refr}}

In this section we derive the Schroedinger equation for neutrino wave
function in the primeval plasma. We will start with the neutrino
quantum operator $\nu_a (x)$ of flavor $a$ that satisfies the usual
Heisenberg equation of motion:
\be
\left( i \ds - {\cal M} \right) \nu_a (x)
+ {g \over 2\sqrt 2}\,\delta_{ae}\, W_\alpha (x) O_\alpha^{(+)} e (x)
+{ g \over 4\cos \theta_W} Z_\alpha (x) O_\alpha^{(+)} \nu_a (x) =0
\label{dirac}
\ee
where ${\cal M}$ is the neutrino mass matrix, $W(x)$, $Z(x)$ and $e(x)$
are respectively the quantum operators of intermediate bosons and electrons,
and  $O_\alpha^\pm = \gamma_\alpha \left( 1\pm \gamma_5\right)$.
We assume that the temperature of the plasma is in 
the
MeV range and thus
only electrons, photons, and neutrinos are present there.

Equations of motion for the field operators of $W$ and $Z$ bosons have the
form
\be
G_{W,\alpha\beta}^{-1} W_\beta (x) =
{ g\over 2\sqrt 2} \bar \nu_a (x) O^{(+)}_\alpha \nu_a (x)\,,
\label{gw}
\\
G_{Z,\alpha\beta}^{-1} Z_\beta (x) =
{g\over 4\cos \theta_W}\left[ \bar \nu_a (x) O_\alpha^{(+)} \nu_a (x) +
\right .
\nonumber \\
\left .
\left( 2 \sin^2 \theta_W -1\right) \bar e(x) O_\alpha^{(+)} e (x)
+2\sin^2 \theta_W \, \bar e(x) O_\alpha^{(-)} e (x) \right]
\label{gz}
\ee
where
the differential operators $G^{-1}_{W,Z}$ are inverse Green's functions
of $W$ and $Z$. In momentum representation they can be written as
\be
G_{\alpha\beta} = {g_{\alpha\beta} - q_\alpha q_\beta / m^2 \over
m^2 - q^2 }
\label{g}
\ee
It can be be shown that the term $q_\alpha q_\beta/m^2$ gives
contribution proportional to lepton masses and can be neglected.

In the limit of small momenta, $q\ll m_{W,Z}$, the equation (\ref{gw}) can be
solved as
\be
W_\alpha (x) = - {g \over 2\sqrt 2 \, m^2_W} \left( 1 -
{\partial^2 \over m^2_W} \right) \left( \bar e(x)
O_\alpha^{(+)} \nu_e (x)\right)
\label{wx}
\ee
A similar expression with an evident substitution for the r.h.s. can be
obtained for $Z_\alpha (x)$. These expressions should be inserted
into eq.~(\ref{dirac}) to obtain equation that contains only field operators
of leptons, $\nu_a (x)$ and $e(x)$:
\be
\left( i\ds - \cal M \right) \nu_a (x) =
{G_F \over \sqrt 2} \left\{ \delta_{ae} \left[\left( 1-{\partial^2
\over m_W^2}\right) \left(\bar e(x) O_\alpha^{(+)} \nu_e (x) \right) \right]
O_\alpha^{(+)} e(x)  + \right . \nonumber \\
\left .
{1\over 2} \left[\left( 1-{\partial^2  \over m_Z^2}\right)
\left( \bar \nu_b(x) O_\alpha^{(+)} \nu_b (x) +
\left( 2\sin^2\theta_W -1\right)
\bar e(x) O_\alpha^{(+)} e(x) +
 \right . \right. \right . \nonumber \\
\left . \left . \left .
2\sin^2\theta_W\,\, \bar e(x) O_\alpha^{(-)} e(x)
\right)\right] O^{(+)}_\alpha \nu_a (x) \right\}
\label{dsnua}
\ee

Neutrino wave function in the medium is defined as
\be
\Psi_a (x) = \langle A | \nu_a (x) |A + \nu^{(k)} \rangle
\label{psinu}
\ee
where $A$ describes the state of the medium and $\nu^{(k)}$ is a
certain one-neutrino state, specified by quantum numbers $k$,
e.g. neutrino with momentum $\vec k$.
The
equation of motion for this wave function can be found from
eq.~(\ref{dsnua}) after averaging over medium.
The theory is quantized perturbatively in the standard way. We define
the free neutrino operator $\nu_a^{(0)}$ that satisfies the
equation of motion:
\be
\left ( i \ds - \cal M \right ) \nu_a^{(0)} (x) = 0
\label{nufree}
\ee
This operator is expanded as usual
in terms of creation-annihilation operators:
\be
\nu^{(0)} (x) = \int {d^3 k \over (2\pi)^3 \sqrt{ 2 E_k} }
\sum_s \left( a_k^s u^s (k) e^{-ikx} + b_k^{s\,\dagger} v^s (k) e^{ikx}
\right)
\label{nu0x}
\ee
and one-particle state is defined
as $|\nu^{(k)}\rangle = a^\dagger_k | {\rm vac} \rangle$.

The equation of motion for the neutrino wave function $\Psi_a (x)$ can be
obtained from expression~(\ref{psinu}) perturbatively by applying
the operator $\left ( i \ds - \cal M \right )$ and using eq.~(\ref{dsnua})
with free neutrino operators $\nu_a^{(0)}$ in the r.h.s. After some
algebra, which mostly consists of using equations of motion for the free
fermion operators and (anti)commutation relations between the
creation/annihilation operators, one would obtain the equation of the form:
\be
i\partial_t \Psi (t) = \left( {\cal H}_m+ V_{eff} \right) \Psi
\label{dtpi}
\ee
where ${\cal H}_m$ is the free Hamiltonian; in the mass eigenstate
basis it has the form
$ {\cal H}_0 = {\rm diag} \left[ \sqrt{p^2 + m^2_j}\,\,\right]$.
The matrix-potential $V_{eff}$
describes interactions of neutrinos with media and
is diagonal in the flavor basis (see however the discussion after 
eq.~(\ref{hint})).
Up to the factor
$E$ (i.e. neutrino energy) it is essentially the refraction index of neutrino
in the medium. The potential contains two terms. The first comes
from the averaging of the external current $J\sim \bar l O_\alpha l$.
Due to homogeneity and isotropy of the plasma only its time component
is non-vanishing and proportional to the charge asymmetry
(i.e. to the excess of particles over antiparticles) in the plasma. This
term has different signs for neutrinos and antineutrinos. The second
contribution to effective potential comes from non-locality of
neutrino interactions. Indeed,
the interactions of neutrinos with the medium are not always of the
(current)$\times$(current) form due to non-locality related to the
exchange of $W$ or $Z$ bosons. If incoming and outgoing neutrinos interact
in different space-time points, the interaction with the medium cannot
be represented 
as an interaction with the external current. Such terms are inversely 
proportional to $m^2_{W,Z}$ but formally they are of
the first order in $G_F$. With these two types of contributions
the diagonal matrix elements of the
effective potential for the neutrino of flavor $a$ has the form:
\be
V_{eff}^a =
\pm C_1 \eta G_FT^3 + C_2^a \frac{G^2_F T^4 E}{\alpha} ~,
\label{nref}
\ee
where $E$ is the neutrino energy, $T$ is the temperature of the
plasma, $G_F=1.166\cdot 10^{-5}$ GeV$^{-2}$ is the Fermi coupling
constant, $\alpha=1/137$ is the fine structure constant, and the signs
``$\pm$'' refer to anti-neutrinos and neutrinos respectively (this
choice of sign corresponds to the helicity state, negative for $\nu$ and
positive for $\bar\nu$). According to ref.~\cite{nora} the
coefficients $C_j$ are: $C_1 \approx 0.95$, $C_2^e \approx 0.61$ and
$C_2^{\mu,\tau} \approx 0.17$ (for $T<m_\mu$).  
These values are true in the limit of
thermal equilibrium, but otherwise these coefficients are some
integrals from the distribution functions over momenta.
For oscillating neutrinos deviations from thermal equilibrium could be
significant and in this case
the expression (\ref{nref}) should be modified. However it is technically
rather difficult to take this effect into account
in numerical calculations and the simplified version (\ref{nref}) is used.
The contributions to the charge asymmetry $\eta^{(a)}$ from different
particle species are as follows:
\be
\eta^{(e)} =
2\eta_{\nue} +\eta_{\num} + \eta_{\nut} +\eta_{e}-\eta_{n}/2 \,\,\,
 ( {\rm for} \,\, \nue)~,
\label{etanue} \\
\eta^{(\mu)} =
2\eta_{\num} +\eta_{\nue} + \eta_{\nut} - \eta_{n}/2\,\,\,
({\rm for} \,\, \num)~,
\label{etanumu}
\ee
and $\eta^{(\tau)}$ for $\nut$ is obtained from eq.~(\ref{etanumu}) by
the interchange $\mu \lrar \tau$. The individual charge asymmetries,
$\eta_X$, are defined as the ratio of the difference between
particle-antiparticle number densities to the number density of photons:
\be
\eta_X = \left(N_X -N_{\bar X}\right) /N_\gamma
\label{etax}
\ee

\subsubsection{Loss of Coherence and Density Matrix \label{coher}}

Breaking of coherence appears in the second order in the Fermi coupling
constant $G_F$, so 
equations of motion for the operators of
{\it all} leptonic fields (including electrons) should be solved up to
the second order in $G_F$. Since the calculations are quite lengthy,
we only sketch the derivation here. In this approximation
the lepton operators $l(x)$, where $l$ stands for neutrino
or electron, in the r.h.s. of eq.~(\ref{dsnua}),
should be expanded up to the first
order in $G_F$. The corresponding expressions
can be obtained from the formal solution of eq.~(\ref{dsnua}) also 
up to first order in $G_F$. Their typical form is as follows:
\be
l = l_0  +  G_l * ({\rm r.h.s.}_0)
\label{l1}
\ee
where the matrix (in neutrino space) $G_l$ is the Green's function of the
corresponding lepton and ${\rm r.h.s.}_0$ is the r.h.s. of
eq.~(\ref{dsnua}) in the lowest order in $G_F$, i.e. with
lepton operators taken in the zeroth order, $l=l_0$.
The expression~(\ref{l1})
should be inserted back into eq.~(\ref{dsnua})
and this defines the r.h.s. up to the second order in $G_F$ 
expressed through the 
free lepton operators $l_0$.
Of course in the second approximation we neglect the non-local terms,
$\sim 1/m_{W,Z}^2$.

Now we can derive 
the kinetic equation for the density matrix of neutrinos,
$\hat {\rho}^i_j = \nu^i \nu_j^*$, where over-hut indicates that
$\hat \rho$ is a quantum operator. The $C$-valued density matrix is
obtained from it by averaging the matrix element over 
medium,
$\rho = \langle \hat \rho \rangle$. We should apply to it the differential
operator $(i\ds - {\cal M})$ and use eq.~(\ref{dsnua}).
The calculations of 
the
matrix elements of the free lepton operators
$l_0$ are straightforward and can be achieved by using the standard
commutation relations. There is an important difference between equations
for the density matrix and the wave function. The latter contains only terms
proportional to the wave function, $i\partial_t \Psi = {\cal H} \Psi$,
while the
equation for the density matrix contains source term that does not
vanish when $\rho = 0$. Neutrino production or destruction is described by
the imaginary part of the effective Hamiltonian. The latter is not hermitian
because the system is not closed. 
According to the
optical theorem the imaginary part
of the Hamiltonian is expressed through the cross-section of neutrino
creation or annihilation. Such terms in kinetic equation for the
density matrix are similar to the ``normal'' kinetic equation for the
distribution functions~(\ref{kin1}), where the matrix elements of the
density matrix enter the collision integral
but with a rather complicated algebraic and matrix structure. 
Let us
consider the case of mixing between active and sterile neutrinos.
The contribution to the coherence breaking terms by elastic
scattering of oscillating neutrinos on leptons in plasma
(i.e. on electrons, positrons and other active neutrinos which are
not mixed with the neutrinos in question) is given by~\cite{ad1}:
\be
\dot\rho=\left( {\partial \over \partial t} -
Hp {\partial \over \partial p} \right) \rho
&=& i\left[ {\cal H}_m + V_{eff}, \rho \right] +
\int d\tau (\bar \nu, l,\bar l) \left( f_l f_{\bar l} A A^+ -
{1\over 2} \left\{ \rho, A \bar\rho A^+ \right\} \right)+  \nonumber \\
&&\int d\tau (l,\nu',l') \left( f_{l'} B \rho' B^+ -
{1\over 2} f_l \left\{\rho, B B^+\right\} \right)
\label{decoh1}
\ee
where $\rho$ is the density matrix of the oscillating neutrinos,
$f_l$ is the distribution function of other leptons in the plasma,
$d\tau$ is the phase space element of all particles participating in
the reactions except for the neutrinos in question - it is given by
eqs.~(\ref{si},\ref{dnuy}). The first commutator term in the r.h.s.
is first order in the interaction. It is the usual contribution from
refraction index 
that
does not break coherence. The last two terms
are second order in the interaction and
are related respectively to annihilation, $\nu\bar\nu \lrar l\bar l$,
and elastic scattering, $\nu l \lrar \nu' l'$.
The quantum statistics factors, $(1-f)$, and $(I-\rho)$ are neglected
here. They can be easily reconstructed, see ref.~\cite{sigl93}.
In the interaction basis and in the case of active-sterile mixing
the matrices $A$ and $B$ have only one non-zero entry in the upper left
corner equal to the amplitude of annihilation or elastic scattering
respectively; the upper $'+'$ means Hermitian conjugate.

The contribution of 
the similar coherence breaking terms but related
to self-interaction of the oscillating neutrinos is presented in
ref.~\cite{sigl93}. It has
more complicated and lengthy matrix
structure.

The equation for the evolution of the diagonal components of 
the density
matrix, with the coherence breaking taken into account, has the same
form as the equation for the distribution function of non-oscillating
particles~(\ref{kin1}), while the role of coherence breaking terms in
the evolution of non-diagonal terms is essentially given by
$\dot \rho_{as}=-\Gamma\rho_{as}$~\cite{ad1,thomson90,enqvist92b,sigl93}
as follows from eq.~(\ref{decoh1}).

The complete form of the coherence breaking terms in kinetic equations
is quite complicated. It can be found e.g. in the paper~\cite{sigl93}.
However, in many cases a quite accurate description can be achieved if
the production and destruction are
mimicked by the anti-commutator:
\be
\dot \rho = ...- \{\Gamma, \left(\rho - \rho_{eq} \right)\}
\label{gammarho}
\ee
where the multi-dots denote contributions from
the neutrino refraction in
medium (see below, eqs.~(\ref{dotrhoaa}-\ref{dotrhosa})),
$\rho_{eq}$ is the equilibrium value of the density matrix, i.e.
the unit matrix multiplied by the equilibrium distribution function
\be
\rho_{eq} = I\,f_{eq} =
 I\,/\left[ \exp ( E/T  - \xi ) +1 \right],
\label{feq}
\ee
and the matrix $\Gamma$ that describes the interaction with the medium,
is diagonal in the flavor basis; it is expressed through the reaction
rates (see below). Such equation exactly describes evolution of non-diagonal
components and in many cases gives an accurate approximation to the
behavior of the diagonal ones.

\subsubsection{Kinetic equations for density matrix.\label{ssec:kineq}}

Density matrix is defined in the usual way as $\rho = \psi \psi^\dagger$.
It satisfies the standard equation
\be
\dot \rho =  {\cal H}\rho - \rho {\cal H}^\dagger
\label{denmat}
\ee
where ${\cal H}$ is the total Hamiltonian. Since we consider an open system
the Hamiltonian may be non-hermitian and the r.h.s. of this equation
would contain, together with the usual commutator, the anti-commutator of
the density matrix with imaginary part of the Hamiltonian. The latter, as
we have already mentioned, is often mimicked by the simplified
expression (\ref{gammarho}). The total time derivative $d\rho /dt$ in
the FRW metric is given by eq.~(\ref{dndt}) and we obtain:
\be
i(\partial_t -Hp\partial_p) \rho_{aa}
&=& F(\rho_{sa}-\rho_{as})/2 -i \Gamma_0 (\rho_{aa}-f_{eq})~,
\label{dotrhoaa} \\
i(\partial_t -Hp\partial_p)  \rho_{ss}
&=& -F(\rho_{sa}-\rho_{as})/2~,
\label{dotrhoss} \\
i(\partial_t -Hp\partial_p) \rho_{as} &=&
W \rho_{as} +F (\rho_{ss}-\rho_{aa})/2-
i\Gamma_1 \rho_{as} ~,
\label{dotrhoas}\\
i(\partial_t -Hp\partial_p) \rho_{sa} &=& -W \rho_{sa} -
F (\rho_{ss}-\rho_{aa})/2- i\Gamma_1 \rho_{sa}  ~,
\label{dotrhosa}
\ee
where $a$ and $s$ mean ``active'' and ``sterile'' respectively,
$H=\sqrt{8\pi \rho_{tot}/3M_p^2}$ is the Hubble parameter, $p$ is the
neutrino momentum and
\be
F &=&\dm\sin 2\theta / 2E \nonumber  \\
W &=& {\dm\cos 2\theta / 2E} + V_{eff}^a
\label{wf}
\ee
with $V_{eff}$ given by eq.~(\ref{nref}).

The antineutrino density matrix satisfies the 
a similar set of equations
with the opposite sign of the antisymmetric term in $V_{eff}^a$ and with
a slightly different damping factor $\bar\gamma$ (this difference is
proportional to the lepton asymmetry in the primeval plasma).

Equations (\ref{dotrhoaa}-\ref{dotrhosa}) account exactly for the
first order terms described by the refraction index, while the second
order terms describing the breaking of coherence are approximately modeled
by the damping coefficients $\Gamma_j$ in accordance with
eq.~(\ref{gammarho}). If we take for the latter the {\it total}
scattering rate, including both elastic scattering and annihilation,
we obtain in the Boltzmann approximation~\cite{enqvist92b}:
\be
\Gamma_0 = 2\Gamma_1  = g_a \frac{180 \zeta(3)}{7 \pi ^4}
\, G_F^2 T^4 p  ~.
\label{gammaj1}
\ee
In general, the coefficient $g_a(p)$ is a momentum-dependent
function, but in the approximation of neglecting $[1-f]$ factors in the
collision integral it becomes a constant~\cite{bell99} equal respectively
to $g_{\nu_e} \simeq 4$ and $g_{\nu_\mu,\mu_\tau} \simeq
2.9$ \cite{enqvist92b}. In ref.~\cite{dolgov00} more accurate values
are presented 
derived from the thermal averaging of the complete electro-weak rates 
(with factors $[1-f]$ included), which were calculated numerically by 
using the Standard Model code
of ref.~\cite{dhs0}. This gives $g_{\nu_e} \simeq 3.56$
and $g_{\nu_\mu,\mu_\tau} \simeq 2.5$.

There is disagreement in the literature concerning
what should be used for
$\Gamma$: the annihilation rate~\cite{barbieri90,barbieri91} or
the total reaction rate, which is approximately an order of magnitude
larger~\cite{enqvist90a,kainulainen90,cline92,shi93}. To resolve this
ambiguity one has to make an exact
description of coherence breaking and not an approximate
substitution~(\ref{gammarho}) used
in the equations~(\ref{dotrhoaa}-\ref{dotrhosa}) above. Kinetic
equations for the elements of density matrix
with the exact form of coherence
breaking terms could be
obtained from eq~(\ref{decoh1}). It can be checked that the equations
for non-diagonal matrix elements remain the same
as eqs.~(\ref{dotrhoas},\ref{dotrhosa}) with $\Gamma_1$ 
being half the total 
reaction width, including both annihilation and elastic
scattering, while equations for the diagonal components are modified.
In the case of mixing with sterile neutrinos only
equation~(\ref{dotrhoaa}) essentially changes, while others remain
practically the same:
\be
\dot \raa (p_1) &=& -F I -
\int  d\tau (l_2,\nu_3,l_4)
A^2_{el} \left[ \raa (p_1) f_l(p_2) - \raa (p_3) f_l(p_4)\right]-
\nonumber \\
&&\int d\tau (l_2,\nu_3,l_4)
A^2_{ann} \left[\raa(p_1) \bar\raa (p_2)-f_l(p_3)f_{\bar l}(p_4)
\right],
\label{draa} \\
\dot \rss (p_1) &=& F I,
\label{drss} \\
\dot R (p_1) &=& WI - (1/2)\, R(p_1) \left[
\int d\tau (l_2,\nu_3,l_4) A^2_{el} f_l(p_2)\, + \right.
\nonumber \\
&&\int d\tau(\bar\nu_2,l_3,\bar l_4) A^2_{ann} \bar\raa (p_2)
\left. \right],
\label{dR} \\
\dot I (p_1) &=& -WR - (F/2) \left(\rss -\raa \right)
- (1/2)\, I(p_1) \left[
\int d\tau (l_2,\nu_3,l_4) A^2_{el} f_l(p_2)\, + \right.
\nonumber \\
&&\int d\tau(\bar\nu_2,l_3,\bar l_4) A^2_{ann} \bar\raa (p_2)
\left. \right],
\label{dI}
\ee
where real and imaginary parts of non-diagonal components
of
neutrino density matrix are introduced:
\be
\ras = \rsa^* = R +i\,I
\label{rhoas}
\ee
and $a$ and $s$ mean respectively ``active'' and ``sterile''.
The integration is taken over the phase space according to 
eqs.~(\ref{si},{dnuy}).
The amplitude of elastic scattering and annihilation with proper
symmetrization factors can be taken from the 
table~\ref{table:amplitudes-nu-e0}.

It is convenient to introduce new variables $x$ and $y_i$
according to eq.~(\ref{xyi}). In terms of these variables the differential
operator $(\partial_t -Hp\partial_p)$ transforms to $Hx\partial_x$
(see eq.~(\ref{hxdfdx}). In many cases the approximation $\dot T =-HT$
is sufficiently accurate, so that we can take $R=1/T$. On the other hand,
it is straightforward to include
the proper dependence of temperature on
the scale factor but 
that would make
the calculations significantly more
difficult.

The system of kinetic equations for the elements of density matrix
presented in this section 
can be approximately solved analytically
in many interesting cases and 
these solutions
will be applied below to the analysis of
the role played by oscillating neutrinos in big bang nucleosynthesis,
to possible generation of lepton asymmetry by oscillations between
sterile and active neutrinos, and to creation of cosmological warm
dark matter (sec.~\ref{sterilenu}). First we will consider a more
simple non-resonant case.

\subsection{Non-resonant oscillations. \label{sec:nonres}}

A non-resonance case means that the function $W$~(\ref{wf}) never reaches
zero, but is always positive. This case is realized in cosmology
when the mass difference
between $\nus$ and $\nu_a$ (to be more rigorous, between $\nu_2$, which is
mostly $\nu_s$, and $\nu_1$, which is mostly an active $\nu_a$)
is positive,
and the charge asymmetry contributing into neutrino refraction index is
sufficiently small. 
In stellar interior, on the opposite,
a positive $\dm $ is necessary for the resonance transition.

Firstly, we will formally solve
equations~(\ref{dotrhoas},\ref{dotrhosa}) to express the real and
imaginary parts $R$ and $I$ of the non-diagonal components
through the diagonal ones. The relevant equations can be written as
\be
\dot R &=& WI - \Gamma_0 R/2 \label{aprdR} \,  , \\
\dot I &=& - WR + \frac{F}{2} \left( \raa - \rss \right) - \Gamma_0 I/2 \, .
\label{aprdI}
\ee
In the limit when the oscillation frequency
\be
\omega_{osc} = \left( F^2 +W^2 \right)^{1/2}
\label{omegaosc}
\ee
is much larger than the expansion rate, the solution is given by
stationary point approximation, i.e. by the condition of vanishing
the r.h.s.:
\be
R = { FW \over 2\left( W^2 + \Gamma^2_0/4\right)} \left(
\rho_{aa} -\rho_{ss} \right)
\label{rstp}\\
I = { F\Gamma_0 \over 4 \left( W^2 + \Gamma^2_0/4\right)} \left(
\rho_{aa} -\rho_{ss} \right)
\label{istp}
\ee
In the non-resonant case, when $W\neq 0$, usually the condition
$W^2 \gg \Gamma^2/4$ is fulfilled and
\be
R \approx (\sin 2\theta_m /2)\left(\rho_{aa} -\rho_{ss} \right)
\label{rappr}\\
I \approx ({\sin 2\theta_m \Gamma_0 / 4W})
\left( \rho_{aa} -\rho_{ss} \right)
\label{iappr}
\ee
where $\theta_m$ is the mixing angle in matter and in the limit of
small mixing
\be
\tan 2\theta_m \approx \sin 2\theta_m \approx {F\over W}
= {\sin 2\theta \over \cos 2\theta + \left( 2EV_{eff}/\dm \right)}
\label{sinthetam}
\ee
Now we can insert the expression for $I$~(\ref{istp}) into
eqs.~(\ref{dotrhoaa},\ref{dotrhoss}) 
and obtain a closed system of
equations for the two unknown diagonal elements of density matrix which
is easy to integrate numerically. If the number density
of sterile neutrinos is small and the active neutrinos are close to
equilibrium, as is often the case, we obtain the following equation
that describes the production of sterile neutrinos by the oscillations
\be
\dot \rho_{ss} \approx ({\sin^2 2\theta_m \Gamma_0 / 4}) f_{eq}
\label{sprod}
\ee

An important conclusion of this derivation is that this production rate
is by factor 2 smaller than the approximate estimates used in
practically all earlier papers (see discussion below in
sec.~\ref{sec:nuoscbbn}). An explanation of this extra factor $1/2$ is
that the time derivative of $\rho_{ss}$ is proportional to imaginary part
of the non-diagonal component of the density matrix and the latter is
proportional to $\Gamma_1 =\Gamma_0/2$.

Now we will obtain essentially the same results 
by using more rigorous
arguments. The method and the equations will be used also in some other
more complicated cases. Following ref.~\cite{dolgov00} we introduce one
more new variable, which is especially convenient for the description 
of neutrino oscillations in cosmological background:
\be
q= \kappa_a x^3
\label{q}
\ee
where the coefficients $\kappa_a$ are given by
\be
\kappa_e = 6.63\cdot 10^3 \left(|\dm|/{\rm eV}^2\right)^{1/2}, \,\,\,
\kappa_{\mu,\tau} = 1.26\cdot 10^4 \left(|\dm| {\rm eV}^2\right)^{1/2}
\label{xiemu}
\ee
Equations~(\ref{aprdR},\ref{aprdI}) can be analytically solved 
as~\cite{dolgov00}
\be
I &=&{K \sin 2\theta \over 2y}
\int_0^q dq_1 e^{-\Delta \Gamma_{int}} \mbox{cos} \Delta \Phi \,
 \left( \raa - \rss \right)  \label{Isol} \, ,\\
R &=& {K \sin 2\theta \over 2y}
\int _0^q dq_1 e^{-\Delta \Gamma_{int}} \mbox{sin} \Delta \Phi \,
 \left( \raa - \rss \right) \label{Rsol} \, ,
\ee
where $\Delta \Gamma_{int}=\Gamma_{int} (q,y) - \Gamma_{int} (q_1,y)$,
$\Delta \Phi = \Phi(q,y) - \Phi(q_1,y)$, and
$\Phi$ and $\Gamma_{int}$ obey
\be
\partial_q \Phi = \frac{K}{y}W \,  \,  \,  \,  \,
\mbox{and} \,  \,  \,  \,  \,
\partial_q \Gamma_{int} = \frac{K}{y} \gamma \, ,
\ee
with
\be
K_e = 5.63\cdot 10^4 \,(\dm/{\rm eV}^2)^{1/2},
\, \, \, \, \, {\rm and}\,\,\,\,
K_{\mu,\tau} = 2.97\cdot 10^4 \, (\dm/{\rm eV}^2)^{1/2}.
\label{Ka}
\ee
The dimensionless damping factor, $\gamma = 2E\Gamma/\dm$ (\ref{gammaj1}),
expressed in terms of new variables, reads:
\be
\gamma = \epsilon_a\, y^2 \,q^{-2}
\label{newgamma}
\ee
where  $\epsilon_a$ are small coefficients,
$\epsilon_e \approx 7.4\cdot 10^{-3}$
and $\epsilon_{\mu,\tau} \approx 5.2\cdot 10^{-3}$, if one takes for
the damping term the total scattering rate.

If we are interested in the period of sufficiently high temperatures,
when the production of sterile neutrinos is non-negligible, the
integrals in eqs.~(\ref{Isol},\ref{Rsol}) are dominated by the
upper limits and can be easily taken. In this way we obtain exactly
the same results as above~(\ref{rstp},\ref{istp}). In the low $T$ limit
when the exponential damping due to $\Gamma$ is not essential the
calculations are more tricky. But in this regime new states are not
produced because the total number of active plus sterile neutrinos
is conserved, though the spectrum of active neutrinos could be
distorted.

The number density of the produced sterile neutrinos is an essential
quantity for BBN and for the amount of dark matter, if $\nu_s$ has
a keV mass and forms
warm dark matter particles (see sec.~\ref{sterilenu}).
As we have seen above, eq.~(\ref{sprod}),
the production rate of $\nu_s$ is proportional to the
reaction rate $\Gamma$. For the calculations it 
is important to know what should be 
substituted for $\Gamma$:
annihilation rate or 
the total rate, which is a sum of annihilation and a much larger 
elastic rate.
An argument against substitution of elastic
scattering rate is that it conserves the particle number and could not
lead to an increase of the net number density of active plus sterile
neutrinos. This condition is not fulfilled by the approximate
equations~(\ref{dotrhoaa}-\ref{dotrhosa}) but is explicit in exact
equations~(\ref{draa}-\ref{dI}). On the other hand, if sterile neutrinos
are effectively produced at sufficiently high temperatures, when
annihilation is fast enough to maintain complete thermal equilibrium,
the limit $\Gamma_{ann}\rar 0$ is physically meaningless. On the
opposite, equilibrium corresponds to very large $\Gamma$. To find 
which case
is true
we will solve kinetic equations~(\ref{draa}-\ref{dI})
in a more detailed way.

In the limit of fast oscillations and small $\Gamma$, $\Gamma \ll W$,
one can see 
that $I = F \Gamma_{tot} (\rho_{aa}-\rho_{ss})/4W^2$
also for the exact equations~(\ref{dR}-\ref{dI}). In what follows we
neglected $\rho_{ss}$ in comparison with $\rho_{aa}$; it is often true
but it is easy to include this term. 
After this expression for $I$ is
inserted into 
eqs.~(\ref{draa}) a closed equation governing
evolution of $\rho_{aa}(x,y)$ is obtained. The next important assumption
is that active neutrinos are close to kinetic equilibrium, i.e.
\be
\rho_{aa} = C(x) \exp (-y)
\label{rhoaaeq}
\ee
Here $C(x)$ can be expressed through an effective chemical potential,
$C(x) = \exp [\xi(x)]$, the same for $\nu$ and $\bar\nu$. A justification
for this approximation is a much larger rate of elastic scattering,
which maintains the form~(\ref{rhoaaeq}) of $\rho_{aa}$, with respect
to annihilation rate that forces $\xi$ down to zero (or $\xi =-\bar\xi$
in the case of non-zero lepton asymmetry). We have also assumed validity
of Boltzmann statistics. This approach is similar to the calculations of
cosmological freezing of species discussed in sec.~\ref{stabhvnu}.
Now we can integrate both sides of eq.~(\ref{draa}) over $d^3y$
so that the contribution of elastic scattering disappears and the
following ordinary differential equation describing evolution of $C(x)$
is obtained:
\be
{dC\over dx} =-{k_l\over x^4} \left[ C^2 -1 +
C\,{10 (1+g_L^2+g_R^2)\over 24 (1+2g_L^2 +2g_R^2) }
\int dy\,y^3 e^{-y} \left( F\over W\right)^2
\right]
\label{dcdx}
\ee
The first term in the r.h.s. of this equation comes from 
annihilation
and the second one from oscillations; the contribution of elastic
scattering disappears
after integration over $d^3y_1$.

We have assumed above that $F/W \ll 1$ and thus the term
$\sim (F/W)^2 dC/dx$ has been
neglected. It is a good approximation even for not very weak mixing.
The constants $k_l$ are given by
\be
k_l={8G_F^2 \left( 1+2g_L^2 +2g_R^2\right) \over \pi^3 H\,x^2 }
\label{K}
\ee
so that $k_e = 0.17$ and $k_{\mu,\tau} = 0.098$.

The integral over $y$ in eq.~(\ref{dcdx}) can be written as
\be
I_n(x) \equiv \int_0^\infty dy\,y^3 e^{-y}\left({F\over W}\right)^n
=\left( \tan 2\theta \right)^n
\int_0^\infty {dy\,y^3 e^{-y}\over
\left( 1+ \beta_l\, y^2 x^{-6} \right)^n },
\label{In}
\ee
with
\be
\beta_e = {2.34\cdot 10^{-8} \over \dm \cos 2\theta}\,\,\,
{\rm and}\,\,\,
\beta_{\mu,\tau} = {0.65\cdot 10^{-8} \over \dm \cos 2\theta},
\label{beta}
\ee
We have neglected here the charge asymmetry term in the neutrino refraction
index. It is a good approximation in the non-resonant case if the asymmetry
has a normal value around $10^{-9}-10^{-10}$.

Eq. (\ref{dcdx}) can be solved analytically if
$|\delta|=|1-C| \ll 1$:
\be
\delta =\zeta_l k_l\,\int_0^x {dx_1 \over x_1^4}
\exp\left[-{2k_l\over 3}\left({1\over x_1^3}-{1\over x^3}\right)\right]
I_2 (x_1)
\label{delta}
\ee
where $\zeta_l = 10(1+g_L^2+g_R^2)/24(1+2g_L^2+2g_R^2)$, so that
$\zeta_e =0.304$ and $\zeta_\mu = 0.375$. Since
\be
\Gamma_{ann} = {8G_F^2 (1+2g_L^2+2g_R^2)\over 3\pi^3} {y\over x^5}\,\,\,
{\rm and}\,\,\,
\Gamma_{el} = {8G_F^2 (9+8g_L^2+8g_R^2)\over 3\pi^3} {y\over x^5},
\label{Gammaannel}
\ee
this quantity $\zeta_l$
is proportional to the ratio of the total rate to the annihilation rate.

The increase of the total number density ($\nu_a + \nu_s$) of oscillating
neutrinos
\be
\Delta n \equiv \int {d^3\, y \over (2\pi)^3}\Delta (\raa + \rss)
\label{dntot}
\ee
can be found from the sum
of equations (\ref{draa}) and (\ref{drss}) and is given by
\be
\left({\Delta n \over n_{eq}}\right)_l =
2k_l\,\int_0^\infty dx\, {\delta (x) \over x^4}
\label{deltan}
\ee
where $\delta (x)$ is given by eq.~(\ref{delta}).
Changing the order of integration over $dx$ and $dx_1$ we can 
integrate over
over $dx$ and $dy$ analytically. 
In the last integral over $dx_1$
we have to neglect $\exp(-2k_l/3x_1^3)$ in comparison with 1. After that
the remaining integration can be also done analytically and we
and obtain for the increase of the total number density of
sterile plus active neutrinos
\be
\left({\Delta n \over n_{eq}}\right)_l =
{\pi k_l\zeta_l \sin^22\theta \over 6\sqrt{\beta_l}}
\label{deltan-fin}
\ee
To find the effective number of additional neutrino species at BBN one
has to divide this result by the entropy dilution factor,
$10.75 / g_* (T^{\nus}_{prod})$, where $T^{\nus}_{prod}$ is the temperature
at which sterile neutrinos are effectively produced. The latter can be
estimated as the temperature at which the production rate given by
eq.~(\ref{sprod}) is maximal. This rate is proportional to
$\Gamma_s \sim (y/x^4)(1+\beta_l y^2/x^6)^{-2}$ and the maximum is reached
at:
\be
T^{\nus}_{prod} = (12,\,15)\, (3/y)^{1/3}\,
(\dm/{\rm eV}^2)^{1/6}\,\, {\rm MeV}
\label{tprodnus}
\ee
The first number above is for mixing of $\nus$ with $\nue$, while the second
one is for mixing with $\num$ or $\nut$. This result is very close to
the estimates of
refs.~\cite{barbieri90,kainulainen90,barbieri91,dodelson94a}

Thus we obtain the following result for the increase of the effective 
number of neutrino
species induced by mixing of active neutrinos with sterile ones:
\be
\Delta N_\nu = {1
\over 9\pi^2}\,
{\sin^2 2\theta_{vac} \over \sqrt {\beta_l}}{G_F^2 (1+g_L^2+g_R^2) \over
Hx^2} \,\, {g_* (T^{\nus}_{prod}) \over 10.75}
\label{dlnnunus}
\ee
Substituting numerical values of the parameters we obtain for the
mixing parameters between $\nus$ and $\nue$ or $\nu_{\mu,\tau}$
respectively:
\be
(\dm_{\nue\nus}/{\rm eV}^2) \sin^4 2\theta_{vac}^{\nue\nus} =
3.16\cdot 10^{-5} (g_*(T^{\nus}_{prod})/10.75)^3 (\Delta N_\nu)^2
\label{dmess2}\\
(\dm_{\num\nus}/{\rm eV}^2) \sin^4 2\theta_{vac}^{\num\nus} =
1.74\cdot 10^{-5} (g_*(T^{\nus}_{prod})/10.75)^3 (\Delta N_\nu)^2
\label{dmmuss2}
\ee
Here another factor $g_*$ comes
from the Hubble parameter.

These results can be compared to other calculations. They
are approximately 2 orders of magnitude stronger than
those presented in refs.~\cite{barbieri90,barbieri91}, where too high
freezing temperature for weak interaction rates was assumed and the limit
was obtained: $ \dm \sin^4 2\theta <6\cdot 10^{-3} \Delta N_\nu^2$.
In ref.~\cite{kainulainen90} the limit was
$ \dm \sin^4 2\theta <3.6\cdot 10^{-4} \Delta N_\nu^2$. (All these
are given for mixing with $\nue$.)
On the other hand, the limits obtained in ref.~\cite{enqvist92b}
are approximately 6 time stronger than those found
above~(\ref{dmess2},\ref{dmmuss2}). They are:
$ \dm \sin^4 2\theta <5\cdot 10^{-6} \Delta N_\nu^2$ for
mixing with $\nue$ and
$ \dm \sin^4 2\theta <3\cdot 10^{-6} \Delta N_\nu^2$ for mixing
with $\nu_{\mu,\tau}$. The difference by factor 6
between these results and
eq.~(\ref{dmess2},\ref{dmmuss2}) can be understood in part by the factor
2 difference in the interaction rate, according to eq. (\ref{sprod}),
which gives factor 4 difference in the limits. The remaining difference
by roughly factor 1.5 could possibly be ascribed to different ways of
solution
of kinetic equations or to the fact that the increase in the number
density of sterile neutrinos is accompanied by an equal decrease in
the number density of active neutrinos if the production of the latter by
inverse annihilation is not efficient. This phenomenon is missed
in kinetic equations (\ref{dotrhoaa}-\ref{dotrhosa}), which are mostly
used in the literature, while equations (\ref{draa}-\ref{dI})
automatically take that into account. However, for 
a sufficiently large
mass difference, $\dm$, the effective temperature of
production~(\ref{tprodnus}) is larger than the temperature of the
annihilation freezing, so the active neutrino states are quickly
re-populated and the said effect
could be significant only for
a small mass difference.
Much weaker bounds obtained in ref.~\cite{dolgov00c} resulted from an
error in the coherence breaking
terms in kinetic equations for the non-diagonal matrix elements of the
density matrix.

We have calculated above the increase of the total number density of
mixed sterile and active neutrinos. It is interesting sometimes to know the
separate number density of $\nus$ produced by oscillations. In particular,
it is important for calculation of the mass density of warm dark matter if
the latter consists of $\nus$ (see sec.~\ref{sterilenu}). The evolution
of the number density of sterile neutrinos is determined by
eqs.~(\ref{drss},\ref{iappr}). It can be easily solved if
$\rho_{ss} \ll \rho_{aa}$ and $\rho_{aa} \approx f_{eq}$ in accordance
with the calculations performed above. The obtained results for the
number density of sterile neutrinos are very close
to (\ref{nse},\ref{nsm}).

One last comment in this subsection is about the neglect of
$\exp (-2k_l/x_1^3)$ made before eq.~(\ref{deltan-fin}). It is easy
to see that this term is indeed small in comparison with unity if
the mass difference is small, $\dm <10^{-5}\, {\rm eV}^2$.
Otherwise the exponential term becomes non-negligible
and the bounds would be considerably weaker.
To find the size of the effect, more accurate calculations are necessary.

\subsection{Resonant oscillations and generation of lepton asymmetry
\label{sec:leptas}}

\subsubsection{Notations and equations. \label{sec:noteq}}

If the mass difference between $\nus$ and $\nu_a$ is negative then the
MSW-resonance transition might take place in cosmological background.
The resonance may also exist with an arbitrary sign of $\dm$ if the initial
value of the asymmetry is sufficiently large (see below,
sec.~\ref{ssec:active}). The analysis in the resonance case is much more
complicated and we will use 
a simplified version of kinetic
equations~(\ref{dotrhoaa}-\ref{dotrhosa}) with the approximate form of the
coherence breaking terms. Fortunately, the exact form of those terms is
not essential for the problem of lepton asymmetry generation, which
predominantly 
took place
at a relatively late stage when the breaking of
coherence was weak. We will start from a naturally small
value of the asymmetry, $\sim (10^{-9}-10^{-10})$ and will show
analytically that it may rise up to 0.375~\cite{dolgov01a}.
Our results are in good agreement with numerical calculations of the
earlier papers~\cite{foot97}-\cite{dibari01}, where a large rise of
asymmetry was discovered. Throughout this section we
assume that the mixing angle is very small, otherwise the resonance
transition would enforce a fast transition between $\nu_a \rar \nus$ and
$\bar\nu_a \rar \bar\nus$, so that thermal equilibrium would be quickly
established and no asymmetry would be generated.

We will normalize the elements of the density matrix
to the equilibrium distribution function $f_{eq}$:
\be
\rho_{aa} &=& f_{eq}(y) [1+a(x,y)],\,\, \rho_{ss} = f_{eq}(y) [1+s(x,y)]~, \\
\label{rhoaa}
\rho_{as} &=& \rho_{sa}^* = f_{eq}(y)[h(x,y)+i l(x,y)]~,
\label{hil}
\ee
and express the neutrino mass difference $\delta m^2$ in eV$^2$.

Written in terms of the variable $q$ (\ref{q}) the system of basic
kinetic equations takes a very simple form~\cite{dolgov00}:
\be
s' &=& -(K_a/y) \sin 2\theta\, l
\label{s'} \\
a' &=& (K_a/y) \left( \sv\, l - 2\gamma\, a \right)
\label{a'}\\
h' &=& (K_a/y) \left( W l - \gamma \, h\, \right)
\label{h'} \\
l' &=& (K_a/y) \left[\sv\,(s-a)/2 -Wh - \gamma\, l \right]
\label{l'}
\ee
where the prime means differentiation with respect to $q$ and the
constant coefficients $K_a$ are given by eqs.~(\ref{Ka}). It is
essential for what follows that these coefficients are large,
$K_a \gg 1$. This condition is valid for 
a sufficiently large mass
difference, $\dm > 10^{-8}$ and reflects high oscillation frequency
in comparison with the cosmological expansion rate. Big values of
$K_a$ permit solving kinetic equations analytically with the
accuracy of the order of $1/K_a$.

For the coefficient functions in the equations~(\ref{s'}-\ref{l'})
we use the same notation as in eqs.~(\ref{dotrhoaa}-\ref{dotrhosa})
but now they are all divided by the factor
$ 1.12\cdot 10^9\, |\delta m^2|\, {x^2/ y} $ and are equal to:
\be
W &=& U \pm y\,V\,Z, \,\,\, U = y^2\,q^{-2} -1, \nonumber \\
V &=& b_a\,  q^{-4/3}, \,\,\, \gamma = \epsilon_a\, y^2 \,q^{-2}
\label{coef}
\ee
where the signs $''-''$ or  $''+''$ in $W$ refer to neutrinos and
antineutrinos respectively;
$b_e = 3.3\cdot 10^{-3}|\dm|$,
$b_{\mu,\tau} = 7.8\cdot 10^{-3}|\dm|$,
and $\epsilon_a$ are small coefficients,
$\epsilon_e \approx 7.4\cdot 10^{-3}$
and $\epsilon_{\mu,\tau} \approx 5.2\cdot 10^{-3}$. Their exact
numerical values are not important.
The charge asymmetry term in $W$ is given by
\be
Z = 10^{10}\left[ {\eta_{o} \over 12}
+ \int_0^\infty \frac{dy}{8 \pi^2} ~y^2 f_{eq}(y) ~(a-\bar a)\right]~,
\label{Z}
\ee
where $\eta_o$ is the charge asymmetry of all particles except for
$\nua$ defined in accordance with eqs.~(\ref{etanue},\ref{etanumu}).
The normalization
of the charge asymmetry term~(\ref{Z}) is rather unusual and to
understand the numerical values of the coefficients one should keep
in mind the following. The coefficient $C_2$ in eq.~(\ref{nref}) is
obtained 
for the standard normalization of charge asymmetry with respect
to the present-day photon number density, which differs from $n_\gamma$ 
in the
early universe by the well-known factor 11/4, related to the increase of
the the number of photons
by $e^+e^-$-annihilation. On the other hand, lepton
asymmetry, $L_{\nua}$, induced by neutrino oscillations, which is
calculated in most papers,
is normalized to the number density of
photons that are in thermal equilibrium with neutrinos, so the factor
11/4 is absent. The photon number density is equal to
$N_\gamma = 2\zeta (3) T^3_\gamma /\pi^2$ with $\zeta(3)\approx 1.2$.
The charge asymmetry of neutrinos is
\be
\eta_\nu = {1\over 4\zeta (3)} \left({T_\nu \over T_\gamma}\right)^3
\int dy y^2 (\rho_{aa} - \bar\rho_{aa})
\label{etanu}
\ee
so that $\eta_{\nua} = 4 L_{\nua}/11$. The quantity $Z$ introduced in
eq.~(\ref{Z}) differs from $L_{\nua}$ by the factor
$2\cdot 10^{-10}\pi^2 /\zeta(3)$:
\be
L = 16.45\cdot 10^{-10}\, Z.
\label{lz}
\ee
The factor $10^{-10}$ is chosen so that initially $Z = O(1)$.
Noting that the charge asymmetry of neutrinos under study enters
expressions (\ref{etanue},\ref{etanumu}) with coefficient 2 one obtains
the coefficient $1/11.96\approx 1/12$ in eq.~(\ref{Z}).

It is worth noting 
that the charge asymmetric term enters eq.~(\ref{coef})
with a very large coefficient if expressed in terms of $L$:
$VZ \sim 10^7 q^{-4/3} L$, while in all other possible places,
$L$ (or chemical potential) enters with the coefficients of order 1, and
may be neglected.

Up to this point our equations 
have been essentially the same as those used by
other groups. The equations look rather innocent and at first sight one
does not anticipate any problem with their numerical solution.
However the contribution from $Z$ could be quite large with the
increasing magnitude of the asymmetry. The exact value of the latter is
determined by a delicate cancellation of the contributions from all
energy spectrum of neutrinos. The function $[a(x,y)-\bar a (x,y)]$
under momentum integral quickly oscillates
it takes very good precision
to calculate
the integral with desired accuracy.
Even a small numerical error
results in a large instability. To avoid this difficulty we analytically
separated fast and slow variables in the problem and reduced this
set of equations to a single differential equation for the asymmetry,
which can be easily numerically integrated. One can
find the details in ref.~\cite{dolgov01a}.

\subsubsection{Solution without back-reaction. \label{sec:nobr}}

One can solve analytically the last two kinetic equations
(\ref{h'},\ref{l'}) with respect to $h$ and $l$ in terms of $a$ and $s$
in the same way as in the previous section:
\be
l (q,y)=- ({K \sv / 2y}) \int_{q_{in}}^q dq_1
\left[ a(q_1) - s(q_1)\right] e^{-\Delta \Gamma} \cos \Delta \Phi,
\label{lqy} \\
h (q,y)=- ({K \sv / 2y}) \int_{q_{in}}^q dq_1 \left[ a(q_1) - s(q_1)\right]
e^{-\Delta \Gamma} \sin \Delta \Phi,
\label{hqy}
\ee
where $q_{in}$ is the initial ``moment'' $q$ from which the system started
to evolve, $\Delta \Gamma = \Gamma (q,y) - \Gamma (q_1, y)$,
$\Delta \Phi = \Phi (q,y) - \Phi (q_1, y)$, and
\be
\partial_q \Gamma= K\gamma /y,\,\,\,
\partial_q \Phi= KW /y.
\label{dgammaw}
\ee
We rewrite the first two equations (\ref{s'},\ref{a'}) in terms of
$\sigma = a + s $ and $\delta = a-s$:
\be
\sigma' &=& -(K\,\gamma /y) \, (\sigma + \delta)
\label{sigma'} \\
\delta' &=&  (2K\sv /y)\,l -(K\,\gamma /y) \, (\sigma + \delta)
\label{delta'}
\ee
The first of these equations can be solved for $\sigma$:
\be
\sigma (q,y) = \sigma_{in}(y)\, e^{-\Gamma(q,y) +\Gamma_{in}(y)} -
{K\over y} \int^q_{q_{in}} dq_1  e^{-\Delta \Gamma} \gamma (q_1,y)
\delta (q_1,y)
\label{sigma}
\ee
The first term in this expression proportional to the initial value
$\sigma_{in}$ is exponentially quickly ``forgotten'' and we obtain
the following equation that contains only $\delta$ (and another unknown
function, integrated charge asymmetry $Z(q)$ that is hidden in the
phase factor $\Delta \Phi$):
\be
\delta' (q,y) &=&  -{K\,\gamma(q,y) \over y} \, \delta +
{K^2\,\gamma (q,y) \over y^2} \int^q_{q_{in}} dq_1
e^{-\Delta \Gamma} \gamma (q_1,y) \delta (q_1,y)\\
&-&\left({K \sv \over y}\right)^2  \int_{q_{in}}^q dq_1
\delta (q_1,y) e^{-\Delta \Gamma} \cos \Delta \Phi
\label{deltaf'}
\ee
Up to this point 
we have been dealing with an exact equation
(with the omitted initial
value of $\sigma$, whose contribution is exponentially small). There is
also some uncertainty related to the choice of the form of $\rho_{eq}$ in
eq.~(\ref{gammarho}) either with zero or non-zero chemical potential,
see eq.~(\ref{feq}). We have chosen here $\mu =0$ and, as is argued
below, the choice $\mu \neq 0$ does not lead to 
noticeably different results. This ambiguity could be rigorously
resolved if one uses exact collision integrals instead of
eq.~(\ref{gammarho}). 

Let us now take a similar equation
for antineutrinos and consider the
sum and difference of these two equations for charge symmetric and
antisymmetric combinations of the elements of density matrix,
$\Sigma = \delta +\bar \delta$ and $\Delta =\delta -\bar \delta$.
The equations have the following form:
\be
\Delta' + {K \gamma \over y} \Delta &=&
{K^2 \gamma \over y^2} \int_{q_{in}}^q dq_1\, e^{-\Delta \Gamma}
\gamma_1 \, \Delta_1 \nonumber \\
&-&\left({K \sv \over y}\right)^2 \int_{q_{in}}^q dq_1\,e^{-\Delta \Gamma}
\left( \Sigma_1 {c-\bar c\over 2} +\Delta_1 {c+\bar c\over 2}
\right)
\label{Delta'}\\
\Sigma' + {K\gamma \over y} \Sigma &=&
{K^2 \gamma \over y^2} \int_{q_{in}}^q dq_1\, e^{-\Delta \Gamma}
\gamma_1 \, \Sigma_1 \nonumber \\
&-&\left({K \sv \over y}\right)^2 \int_{q_{in}}^q dq_1\,e^{-\Delta \Gamma}
\left( \Sigma_1 {c+\bar c\over 2} +\Delta_1 {c-\bar c\over 2}
\right)
\label{Sigma'}
\ee
where sub-1 means that the function is taken at $q_1$, e.g.
$\gamma_1 = \gamma (q_1,y)$, etc; $c = \cos \Delta \Phi$, and
$\bar c = \cos \Delta \bar\Phi$.
Using expressions~(\ref{coef}, \ref{dgammaw}) we find:
\be
{c-\bar c \over 2} = \sin \left[ K \left(q-q_1\right)
\left(-{1 \over y} + {y \over q \,q_1}\right)\right]\,
\sin \left[K\, \int_{q_1}^q dq_2 V(q_2)\,Z(q_2) \right]
\label{cminusc}\\
{c+\bar c \over 2} =\cos \left[ K \left(q-q_1\right)
\left(-{1 \over y} + {y \over q \,q_1}\right)\right]\,
\cos \left[K\, \int_{q_1}^q dq_2 V(q_2)\,Z(q_2) \right]
\label{cplusc}
\ee
Here $V(q)$ is given by 
the expression~(\ref{coef}) and does not depend on $y$.

At this stage we will make some approximations to solve the
system~(\ref{Delta'},\ref{Sigma'}). First, let us consider the terms
proportional to $\gamma$. They are definitely not important at large
$q$ (or low temperature). Let us estimate how essential 
they become at
low $q$ ($q \sim 1$). Integrating by parts the first term in the r.h.s.
of eq.~(\ref{Delta'}), using expression~(\ref{dgammaw}), and
neglecting the exponentially small
contribution of the initial value,
we find:
\be
{K\gamma \over y} \Delta -
{K^2 \gamma \over y^2} \int_{q_{in}}^q dq_1\, e^{-\Delta \Gamma}
\gamma_1 \, \Delta_1 ={K \gamma \over y}
\int^q_{q_{in}} dq_1 e^{-\Delta\Gamma}\, {d\Delta_1 \over dq_1}
\label{byparts}
\ee
The remaining integral can be easily evaluated in the limit of
large $K\gamma /y = K\epsilon y/q^2$. Indeed,
$\Delta \Gamma = K \epsilon y (q-q_1)/q\,q_1$ and for $q\leq 1$ the
coefficient in front of the exponential $(q-q_1)$ is larger than 400 for
$\nue$ and 300 for $\num$ and $\nut$. So the integral strongly sits on
the upper limit and, together with the coefficient in front, it gives
just $\Delta' (q)$. Thus, when the $\gamma$-terms are large, they simply
double the coefficient of $\Delta'$ in eq.~(\ref{Delta'}):
$\Delta' \rar 2\Delta'$. A possible loophole in this argument is a very
strong variation of the integrand, much stronger than that given by
$\exp (\Delta \Gamma)$. However one can see from the solution found
below that this is not the case.

Thus the role of $\gamma$ terms in eq.~(\ref{Delta'}) is minor,
they could only change the coefficient in front of $\Delta'$ from 1 to
2, and become negligible for large $q$ where the bulk of asymmetry is
generated. So let us neglect these terms in the equation. This
simplification does not have a strong impact on the solution.

Let us make one more approximate assumption, namely let us neglect the
second term, proportional to $\Delta_1$, in the last integral of the
r.h.s. of eq.~(\ref{Delta'}). Initially $\Sigma =2$ and
$ \Delta = 10^{-9}-10^{-10}$ and the neglect of $\Delta$ in comparison
with $\Sigma$ is a good approximation, at least at the initial stage.
We
will check the validity of this assumption after we find the solution.
And last, we assume that $\Sigma$ changes very slowly
$\Sigma \approx \Sigma_{in} = 2$.
The latter is justified by the smallness of the missing angle
 $\sv \sim 10^{-4}$. In the limit of
zero mixing, the solution of eq.~(\ref{Sigma'}) is $\Sigma = const$.
We will relax both these assumptions below.

As the last step we need to find a relation between
$\Delta= a-s-\bar a+\bar s$ and charge
asymmetry $Z$. To this end one may use the conservation of the total
leptonic charge:
\be
\int^\infty_0 dy\, y^2\, f_{eq}(y) \left( a+s-\bar a -\bar s\right)
= const
\label{cl}
\ee
Using this conservation law we find:
\be
10^{10}\,{d\over dq} \left[\int_0^\infty dy\,y^2 \,f_{eq}(y)
\Delta (q,y) \right] = 16\pi^2 {dZ\over dq}
\label{deltaz}
\ee

Keeping all these assumptions in mind we can integrate both sides of
eq.~(\ref{Delta'}) with $dy y^2f_{eq}(y)$ and obtain a closed
ordinary differential equation for the asymmetry $Z(q)$, valid in the
limit of large $K$. Integration over $y$ 
results in:
\be
Z'(q) &=& {10^{10} K^2 (\sv)^2 \over 8\pi^2}
\int_0^\infty dy f_{eq}(y) \nonumber \\
&&\int_{q_{in}}^q dq_1\exp \left[ - {\epsilon y \zeta \over qq_1} \right]
\sin\left[ \zeta \left({1\over y}-{y\over qq_1}\right)\right]
\sin\left[bK\int_{q_1}^q dq_2 {Z(q_2) \over q_2^{4/3}}\right]
\label{Z'}
\ee
where $b$ is defined in eq.~(\ref{coef}) and $\zeta = K(q-q_1)$.
Integration over $y$ here can be done explicitly and the result is
expressed through a real part of the sum
of certain Bessel functions of
complex arguments. To do that one has to expand
\be
f_{eq} = \sum_n (-1)^{n+1} \exp (-ny)
\label{feqexp}
\ee
and integrate each term analytically~\cite{gradshtein94}.
One can see from the result 
(it is more or
less evident anyhow) that the integral over $q_1$ is saturated in the
region $\zeta \sim 1$. So we can take $q q_1 \approx q^2$ and
\be
K\int_{q_1}^q dq_2 {Z(q_2) q_2^{-4/3}} \approx \zeta Z(q)/q^{4/3}
\label{kint}
\ee
The correction to this expression
is of the order of $Z' (q) /K$. 
One can ascertain 
using the solution
obtained below, that the correction terms are indeed small.

Keeping this in mind we can 
integrate
over $\zeta$ in
the r.h.s. of eq.~(\ref{Z'}). Since $\epsilon \sim 10^{-2}$ is a small
number it may be neglected in comparison with 1 in intermediate
expressions (for the details see ref.~\cite{dolgov01a}) and we obtain:
\be
Z'(q) = {10^{10}K(\sv)^2 \over 8\pi^2} q^2\chi (q) \sum_n (-1)^{n+1}
\int_0^\infty {dt\,t^2 \cos (nqt) \over (1+t^2)^2 + t^2\,\chi^2 (q)}
\label{Z'3}
\ee
where $\chi (q) = b Z(q) q^{-1/3}$. Both the integral over $t$
and the summation over $n$ can be done explicitly leading to the result:
\be
\kappa\, Z' =  {10^{10}K(\sv)^2 \over 16\pi}
{ q^2 \over \sqrt{\chi^2 +4}} \left[ t_2 f_{eq} (qt_2) -
t_1 f_{eq} (qt_1) \right]
\label{Z'fin}
\ee
where we introduced the coefficient $\kappa$, such that $\kappa =1$
for large $q$ and $\kappa = 2$ for $q\sim 1$. It reflects the role
of decoherence terms, proportional to $\gamma$, see discussion after
eq.~(\ref{byparts}). The quantities $t_{1,2}$ are the poles of the
denominator in eq.~(\ref{Z'3}) in the complex upper half-plane 
of $t$ (resonances):
\be
t_{1,2} = {\sqrt{\chi^2 +4} \pm \chi \over 2}
\label{t12}
\ee
It is an ordinary differential equation which can be easily integrated
numerically. This equation quite accurately describes the evolution of the
lepton asymmetry in the limit when back reaction may be neglected: we
assumed above that $\Sigma = 2$ and $\Delta \ll \Sigma$.

Before doing numerical integration let us consider two limiting cases
of $q$ close to the initial value
when asymmetry is very small and the
case of large $q$. When $q$ is not too large, $q\sim 1$, the r.h.s.
of eq.~(\ref{Z'fin})
can be expanded in powers of $\chi$ and we obtain a very simple
differential equation that can be integrated analytically:
\be
Z'= {10^{10}K(\sv)^2 \over 64 \pi} q^2 f_{eq}(q) \chi(q)
\left[ -1 + q \left( 1- f_{eq}(q) \right)\right]
\label{Z'small}
\ee
(we took here $\kappa =2$). One can see
that for
$q< q_{min}=1.278$
the asymmetry exponentially decreases and reaches the minimum value
\be
{Z_{min} \over Z_{in}} =
\exp \left[ -{10^{10} K(\sv)^2 b \over 64\pi}
\int_0^{q_{min}} dq \,q^{5/3} f_{eq}(q) \left(1 - {q\over 1+\exp(-q)}
\right)\right].
\label{zmin}
\ee
The integral in the expression above is equal to 0.0754 and e.g.
for $(\nue-\nus)$-oscillations, the initial asymmetry drops by 3 orders
of magnitude in the minimum. The drop would be significantly stronger
even with a mild increase of the mixing angle
or mass difference.
The temperature, when the minimum is
reached (corresponding to $q_{min} = 1.278$) is
\be
T_{min}^e = 17.3 \left( \dm \right)^{1/6}\,{\rm MeV},\,\,\,
T_{min}^{\mu,\tau} = 23.25 \left( \dm \right)^{1/6}\,{\rm MeV}
\label{tmin}
\ee
These results 
agree rather well
with ref.~\cite{dibari00} for
$\nue$, while agreement for $\num$ and $\nut$ case (see e.g.
the papers~\cite{foot97b,volkas00,lee00,dibari00b})
is not so good.

For $q>q_{min}$ the asymmetry starts
to rise exponentially and this
trend lasts
until $\chi$ becomes larger than unity and the asymmetry
reaches the magnitude $Z\sim 10^3$ or $L \sim 10^{-6}$. After
that the asymmetry starts
to rise as a power of $q$.
For large $q$ and $\chi$ the term containing $t_2$ dominates the r.h.s.
of eq~(\ref{Z'fin}) and now it takes the form:
\be
Z^2\,Z'= {10^{10} K(\sv)^2 \over 16\pi \, b^2} q^{8/3} f_{eq} (1/VZ)
\label{Zlarge}
\ee
where $V$ is given by eq.~(\ref{coef}). Assuming that $VZ$ is a slowly
varying function of $q$ we can integrate this equation and obtain:
\be
Z(q) \approx 1.6\cdot 10^3 q^{11/9}\,\,\,
{\rm or}\,\,\, L(q) \approx 2.5\cdot 10^{-6}q^{11/9}
\label{zlappr}
\ee
The concrete values of numerical coefficients above are taken for
$(\nue-\nus)$-oscillations with $\dm =-1 {\rm eV}^2$ and $\sv= 10^{-4}$.
This result 
agrees well with 
the numerical solution of
eq.~(\ref{Z'fin}) and the functional dependence,
$L\sim q^{11/9}\sim T^{-11/3}$, agrees with that found in
ref.~\cite{buras00a} and somewhat disagrees with the results of
refs.~\cite{ftv,foot95,foot97b,dibari00,dibari00b}
where the law $L\sim q^{4/3}\sim T^{-4}$ was advocated. If the
temperature changes by two orders of magnitude the difference between
these two results 
becomes significant. The results of ref.~\cite{shi99}
demonstrate a milder rise, $L\sim T^{-3.6}$, which are quite close
to $ T^{-11/3}$.

The numerical solution
of eq.~(\ref{Z'fin}) is straightforward. It
agrees well with the simple analysis presented above. In the power law
regime, where the bulk of asymmetry is generated, it is accurately
approximated by the found above law~(\ref{zlappr}):
\be
L_e = 2.5\cdot 10^{-6} C_e\, q^{11/9}
\label{Leofq}
\ee
For $\nue-\nus$ mixing with $(\sv)^2 = 10^{-8}$ and $\dm =-1$
the correction coefficient $C_e$ is 0.96, 0.98, 1, 1.01, and 0.997
for $q=6630,\,\, 1000,\,\, 100,\,\, 10,$ and $5$ respectively.
The results of numerical solution well agree with those of
ref.~\cite{dibari00} in the temperature range from 10 down to 1 MeV
for $\nue-\nus$ case.

For the $(\num-\nus)$-mixing with $\dm = -10$ and $(\sv)^2 = 10^{-9}$
the solution can be approximated as
\be
L_\mu \approx 1.2\cdot 10^{-6} C_\mu \, q^{11/9}
\label{Lmuofq}
\ee
with the correction coefficient $C_\mu = 0.84,\,0.9,\, 0.98,\,1.02,\,
1.05,\, 1$
for $q = 4\cdot 10^4,\, 10^4$, $10^3,\, 10^2,\,10,\,5 $ respectively.
These results reasonably well agree with the calculations of
ref.~\cite{foot97b} in the
temperature range from 25 down to 2 MeV. At 
lower temperatures
this power law generation of asymmetry must stop and it abruptly does,
according to the results of the quoted papers. But the solutions
of eq.~(\ref{Z'fin}) continue rising because back reaction effects
have been neglected
there. We will consider these effects in the next
subsection.

\subsubsection{Back reaction. \label{sec:br}}

The solution obtained above should be close to the exact one if
$\Delta (q,y) \ll \Sigma (q,y)$  and $\Sigma \approx 2 = const$,
see eq.~(\ref{Delta'}). Since now we know the function $Z(q)$ we
can find $\Delta (q,y)$ and 
find out
when these assumptions are
correct. We will consider the region of sufficiently large $q$
when the second pole (resonance) $t_1$ in eq.~(\ref{Z'fin}) is
not important. Its contribution is suppressed as $\exp (-q^2VZ)$
and it may be neglected already at $q>5$ (we assume for
definiteness 
that the initial value
of the asymmetry is positive,
otherwise the roles
of the two poles would interchange).
In terms of the oscillating coefficients, $\cos \Delta\Phi$ or
$\cos \bar\Delta\Phi$, entering eq.~(\ref{Delta'}), it means that
only one of them is essential. It has a saddle point where the
oscillations are not too fast, while the other quickly oscillates
in the significant region of momenta $y$. With our choice of the
sign of the initial asymmetry only $\cos \bar\Delta\Phi$ has an
essential saddle point and in this approximation the
equation~(\ref{Delta'}) can be written as:
\be
\Delta' (q,y) = - { K^2 (\sv)^2 \over y^2 } \int_0^q dq_1
\cos \bar\Delta\Phi
\label{Delta'2}
\ee
For large $q$ the phase difference is equal to
\be
\Delta \Phi = K\left(-{q-q_1 \over y} +
\int_{q_1}^q dq_2 V(q_2) Z(q_2) \right)
\label{DeltaPhi}
\ee
This integral can be taken in the saddle point approximation. To this
end let us expand:
\be
\Phi (q,y) = \Phi (q_R,y) + {(q-q_R)^2\over 2} \Phi'' (q_R,y)
\label{Phiqy}
\ee
where the saddle (resonance) point $q_R$ is determined by the
condition
\be
\Phi' (q_R,y) = K\left( VZ - {1\over y}\right) = 0
\label{Phyres}
\ee
For $q<q_R$ the integral in the r.h.s. of eq.~(\ref{Delta'2}) is
negligibly small, while for $q>q_R$ it is
\be
\int_0^q dq_1 \cos \bar\Delta\Phi \approx
\theta (q - q_R)\, {\cal R}e\,\left\{ \sqrt{{2\pi \over |\Phi''(q_R,y)|}}
\,\,e^{\left[ \Phi(q,y) -\Phi(q_R,y) -i\pi/4 \right] }\right\}
\label{intPhi}
\ee
where $\Phi'' = (VZ)'$.

Now repeating similar integration in eq.~(\ref{Delta'2}) we can easily
find $\Delta (q,y)$:
\be
\Delta (q,y)= \theta(q-q_R){\pi\,K(\sv)^2\over y^2|(VZ)'|_R }
\label{Deltaqy}
\ee
Note the factor 1/2 that comes from the theta-function in
expression~(\ref{intPhi}). It permits integration only over positive
values of $(q-q_R)$.

From the saddle point condition follows that
\be
(VZ)' = VZ \left( {V'\over V} + {Z'\over Z}\right)_R =
{1\over y}\left( {V'\over V} + {Z'\over Z}\right)_R =
-{1\over 9yq_R}
\label{VZ'}
\ee
In the last equality the solution $Z\approx 1.5\cdot 10^3\,q^{11/9}$
and $V=b q^{-4/3}$ were used. From the condition $VZ = 1/y$ we find
\be
q_R \approx (5y)^9
\label{qr}
\ee
where we took  $\dm =-1$ and $b_e = 3.31\cdot 10^{-3}$.

However, the magnitude of $\Delta (q,y)$ is too large. For example at
$q=6630$ (corresponding to $T = 1$ MeV for $(\nue-\nus)$-oscillations)
we find $\Delta = 9\pi K (\sv)^2 q_R /y \approx 200 \gg 1$. It
violates the Fermi exclusion principle, which is automatically
enforced by kinetic
equations if it is fulfilled initially. This
follows from the equation:
\be
\partial_q \left( a^2 + s^2 + 2h^2 +2l^2 \right) =
-4\gamma (K/y)\,\left( a^2 +h^2 +l^2\right)
\label{daslh}
\ee
so that the quantity in the r.h.s. may only decrease. The violation of
this condition originates from the assumption that $\Delta \ll \Sigma$
which is not true for the solution we have obtained.
So one would expect that
the back reaction should be important when $\Delta \sim 1$ and that
the asymmetry should be suppressed by two orders of magnitude, but as we
will see below this is not the case. The evolution of $Z(q)$ changes, due
to the back reaction, and the behavior of the resonance $y_R(q)$ also
becomes different from $y_R = q^{1/9}/5$ found above.

If only one of two resonances is essential then $(\Sigma +\Delta)$ is
conserved, as is seen from eqs.~(\ref{Delta'},\ref{Sigma'})
with $\cos \Delta \Phi \rar 0$.
It corresponds to the conservation
of the total leptonic charge
if oscillations are efficient only in neutrino (or antineutrino)
channel. In this case we arrive at the equation:
\be
\Delta' (q,y) = - { K^2 (\sv)^2 \over y^2 } \int_0^q dq_1
\cos \bar\Delta\Phi \left[ 1- \Delta (q_1,y)\right]
\label{Delta'br}
\ee
This can be integrated
in the same way as above in the saddle
point approximation, and we get:
\be
\Delta (q,y)=\theta(q-q_R) \lambda \left[ 1- \Delta (q_R,y)\right]
\label{Deltaqy1}
\ee
where the last term describes the back-reaction and
\be
\lambda = {\pi K (\sv)^2 \over y^2 |(VZ)\,'|_R}
\label{lambda}
\ee
The derivative of $VZ$ is taken over $q$ at $q = q_R (y)$ found from
the resonance condition $V(q_R)Z(q_R) = 1/y$.

Since $\theta (0) =1/2$, we find
\be
\Delta (q,y) = {2\lambda \over 2+\lambda}\,\theta (q-q_R)
\label{Deltalam}
\ee
With this expression for $\Delta (q,y)$ we can find the integrated
asymmetry
\be
Z(q) ={10^{10}\over 16\pi^2} \int_0^{y_R} dy y^2 f_{eq}(y) \,
{2\lambda \over 2+\lambda}
\label{Zbr}
\ee
where $y_R(q) = 1/[V(q)Z(q)]$. This equation can be reduced to an
ordinary differential equation in the following way. Let us introduce
the new variable
\be
\tau = {1 \over VZ}
\label{tau-VZ}
\ee
and consider the new unknown function $q= q(\tau)$. Correspondingly
$ Z = q^{4/3}(\tau) /(b\tau)$. The derivative
over $q$ should be rewritten as:
\be
{d(VZ) \over dq} = - {1\over \tau^2} \left({dq\over d\tau}\right)^{-1}
\label{dvzdq}
\ee
Under the sign of the integral over $y$ one should take $\tau = y$,
while the upper integration limit is $y_{max} = \tau$. Now we can
take derivatives over $\tau$ of both sides of the equation~(\ref{Zbr})
and obtain:
\be
{d\over d\tau}\,\left[ {q^{4/3} (\tau) \over b\tau }\right]=
{10^{10}\over 16\pi^2}\, {2\lambda \over 2+\lambda}\,
\tau^2 f_{eq}(\tau).
\label{dqdtau}
\ee
where now $\lambda = \pi K (\sv)^2\, dq/d\tau$.
This is the final equation for determination of the integrated
asymmetry with the account of the back reaction. As initial condition
we take the magnitude of asymmetry obtained from
the solution of
eq.~(\ref{Z'fin}) at $q=5$. At this $q$ the back reaction is still
small but already the regime of one-pole dominance
begins. Under the
latter assumption the above equation~(\ref{dqdtau}) is derived.

The numerical solution of eq.~(\ref{dqdtau}) is straightforward.
In particular, if one neglects $\lambda$
with respect to 2 in the denominator of the r.h.s. of this equation, then
its numerical solution gives exactly the same result for the asymmetry
as was found above from eq.~(\ref{Z'fin}). An account of back
reaction is not essential at high temperatures but it is quite important
in the temperature region of Big Bang Nucleosynthesis. In particular, at
$T=1$ MeV the asymmetry is $L= 0.0435$ and at $T=0.5$ MeV it is
$L = 0.25$. These values are approximately 3 times smaller than those
found without back reaction. Asymptotic constant value of $L$ is reached
at $T< 0.3$ MeV and is equal to 0.35. These numerical values are found
for electronic neutrinos with 
$\dm =-1$ eV$^2$ and $(\sv)^2 = 10^{-8}$. 
The asymptotic value well agrees with that presented
in the paper~\cite{dibari00}, the same is true for the magnitude
of the asymmetry in the nucleosynthesis region as well. Another important
effect for BBN is the shape of the spectrum of electronic neutrinos
that may noticeably deviate from the simple equilibrium one
$f_{eq} = 1/[\exp (y-\xi) +1]$ even with a non-zero chemical
potential $\xi$.

For $\num$ or $\nut$ with $\dm = -10$ eV$^2$
and $(\sv)^2 = 10^{-9}$ the
asymmetry $L_\mu$ asymptotically tends to 0.237 in good agreement
with ref.~\cite{foot97b}. For non-asymptotic values of the temperature,
the corrected by back reaction asymmetry is
$L_\mu=6.94\cdot 10^{-3}$ for $T=3$ MeV (1.24 smaller than
non-corrected one), $L_\mu= 0.025$ for $T=2$ MeV (1.48 smaller),
and $L_\mu= 0.164$ for $T=1$ MeV (2.6 times smaller).

There is an easy way to find the asymptotic constant value of the
asymmetry, $Z_0$ or $L_0$. To this end it is convenient to use
eq.~(\ref{Zbr}) with a constant $Z$:
\be
\lambda = {3\pi K (\sv)^2 b^{3/4} Z_0^{3/4} \over 4 y^{1/4}}=
71 (\dm/ {\rm eV}^2)^{1/4} (\sv/10^{-4})^2 L_0^{3/4}\,ay^{-1/4}
\label{lambdaz0}
\ee
This result is the same both for $\nue$ and $\nu_{\mu,\tau}$.

Since the upper limit of the integral over $y$ is
$y_{max} = q^{4/3} /(bZ_0)\gg 1 $  for large $q$, we obtain the
following equation for $L_0$:
\be
L_0 = 0.208 \int_0^\infty dy y^2 f_{eq}(y)\,
{\lambda \over 2+\lambda}
\label{z0}
\ee
Numerical solution of this equation gives $L = 0.35$ for $\dm =-1$ eV$^2$
and $(\sv)^2 = 10^{-8} $ and $L = 0.27$ for $\dm =-10$ eV$^2$ and
$(\sv)^2 = 10^{-9} $ in a good agreement with the solution of
differential equation~(\ref{dqdtau}). The absolute maximum value of the
asymmetry is 0.375 corresponding to 
$\lambda\gg 1$.
In the case of $\lambda \ll 1$ the asymptotic value of the asymmetry
is $L_0= 1.2\cdot 10^4 (\dm /{\rm eV}^2) (\sv /10^{-4})^8$.

The evolution of $\nue$-asymmetry according to calculations
of ref.~\cite{dibari00} for $\sin^2 2\theta = 10^{-8}$ and several
values of mass difference is presented in fig.~\ref{fig:dbfoot}. Notice that
the authors presented the quantity which enters the refraction
index of
$\nue$. i.e. essentially twice the asymmetry. Their numerical results
are in good agreement with analytical calculations of
ref.~\cite{dolgov01a} described above.
\begin{figure}[ht]
\begin{center}
\hspace{10cm}
\vspace{-1.0cm}
\epsfig{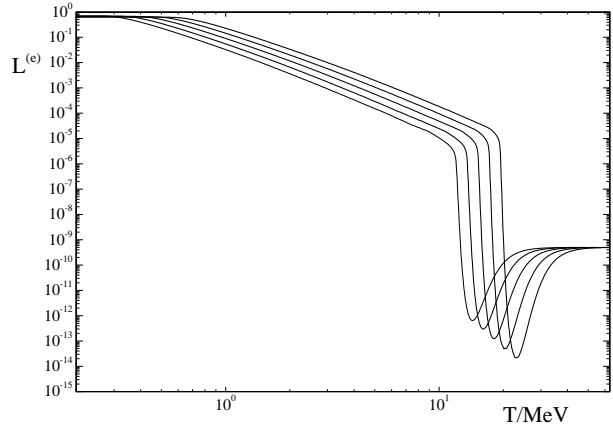}
\end{center}
\bigskip
\caption{Evolution of $L^{(e)} = 2L_{\nu_e} + \eta$
for $\nu_e \to \nu_s$ oscillations with $\sin^2 2\theta_0
= 10^{-8}$ and, from left to right, $\delta m^2/eV^2 =
-0.25, -0.5, -1.0, -2.0, -4.0$ obtained
from numerical solution of the kinetic
equations. The initial $L_{\nu_e} = 0$ is taken and
$\eta = 5\times 10^{-10}$ is assumed.
The low temperature evolution is weakly dependent on these values.
\label{fig:dbfoot}}
\end{figure}

It is more or less evident that asymmetry cannot rise for very
small mixing angles (in the limit of vanishing mixing the asymmetry
would be just zero). Somewhat more surprising is the absence of
asymmetry generation for large mixing. It can be understood in the
following simple way - 
for large $\sin 2\theta$ the resonance is so strong that
sterile states, both $\nu_s$ and $\bar \nu_s$, become quickly and
completely populated and there is no room for asymmetry. A quantitative
estimate can be done as follows.
At a certain moment $q$ the resonance condition is
fulfilled for neutrinos (antineutrinos) with momentum:
\be
y\,\, ({\rm or}\,\,\,\bar y) = \left({ q^2\over y}\right)
\pm d\,q^{2/3} \left( N - \bar N +\eta_o \right),
\label{ybary}
\ee
where $d = 1.65 \cdot 10^{-3} (\cos 2\theta\,\delta m^2)^{-1/3}$,
$\eta_o$ is the charge asymmetry of other particles, except
for neutrinos and
the number densities of neutrinos (antineutrinos) are given by
\be
N(y) = { 10^{10} \over 4\pi^2 } \int^\infty_y
{ dy' y'^2 \over \exp ( y'-\xi_0) +1 }
\label{ny}
\ee
Correspondingly, the integral for $\bar N$ has the lower limit
$\bar y$ and the sign of initial neutrino chemical potential
$\xi_0$ is changed, $\xi_0 \rar -\xi_0$. It is assumed that
$\xi_0\ll 1$, typically it is about (a few)$\cdot 10^{-10}$.
Such expressions for number densities are based on the following
picture: all active neutrinos (antineutrinos) transform into
their sterile partners at the resonance, so that for momentum
$y$ below the resonance one the spectrum is empty. This could
be true only for large mixing angles and in the case that
repopulation of active neutrinos by the inverse annihilation
is slow. Though this picture is very much oversimplified, it
contains some truth that can be helpful for understanding
the phenomenon.

It follows from eq.~(\ref{ybary}) that
\be
y\,\bar y = q^2,
\label{yq2}
\ee
while the difference $(y-\bar y)$ is given by
\be
2(y-\bar y) = d\, q^{2/3} \left[ -{ 10^{10} \over 4\pi^2}
{q^2 \over e^q +1 } (y-\bar y) + \eta_o +
{10^{10} \xi_0 \over 2\pi^2}
\int_q^\infty {dy\, y^2 e^y \over (e^y +1)^2} \right]
\label{y-bary}
\ee
These estimates are true if the difference between the resonance
momenta, $(y-\bar y)$ is small.
Now it is straightforward to find the difference
of the resonance
momenta and to calculate the running lepton asymmetry:
\be
\eta = \eta_{in} + 2\eta_o d\,{  e^{-q} q^{8/3} \over
1 + e^{-q} + Q  e^{-q} q^{8/3} }
\label{etaq}
\ee
where $Q = 10^{10} /8\pi^2$. One can see
that in the case of strong mixing
the asymmetry cannot be large. This result is obtained
in the limit $(y-\bar y)\ll 1 $. It allows to make simple analytic
estimates presented above.
In particular that one cannot formally go to the 
limit $cos 2\theta =0$ or $d\rar \infty$ in the presented expressions.
If the condition of small $(y-\bar y)$ is not 
fulfilled the result about small asymmetry survives but its derivation
is much more complicated technically.

To conclude, we see that there is a good agreement between the numerical
calculations of the asymmetry generation and the semi-analytical
solution of kinetic equations.
The latter is accurate in the limit
of large value of parameter $K_l$ and for sufficiently small mixing,
$(\sv)^2 < 0.01-0.001$; it is difficult to fix more precisely the
limiting value of the vacuum mixing angle. For bigger mixing
the rate of the population of sterile states becomes large
and $\nus$ and $\bar\nus$ states quickly approach equilibrium
and become equally populated, so the asymmetry is not generated. For
small values of the product $K\sv$ the process of asymmetry generation
is inefficient
and the net result is rather low. Numerical
calculations of the effect for very low values of the mass difference
$\dm = 10^{-7} - 10^{-11}$ eV${^2}$ show that the asymmetry could rise
only up to 4 orders of magnitude~\cite{kirilova97,kirilova99b,kirilova00}
producing the net result at the level of $10^{-5}$.
According to the calculations of the work~\cite{dolgov01a}
the asymmetry strongly rises if $\dm > 10^{-3}$ eV$^2$ and possibly for
smaller values depending upon the mixing angle.

In the case of mixing between several (more than 2) neutrinos,
a more complicated picture could emerge. For a specific case
of $m_{\nut} \gg (\mne,\,\, \mnm,\,\, m_{\nu_s})$
the oscillations between $\nut$ and $\nu_s$ could create a
large asymmetry $L_{\nut}$ (about 0.5) and some of this asymmetry
could be converted into $L_{\nue}$ by
$\nut\lrar \nue$-oscillations~\cite{foot00}. The predictions
of models with different values of mixing angles with
light sterile neutrinos
are strongly parameter-dependent and the results are
quantitatively different.

\subsubsection{Chaoticity. \label{sec:chaos}}

As we have already mentioned,
a very important and interesting
development of the theory of
neutrino oscillations in the early universe was stimulated by
the paper~\cite{ftv} where it was argued that a very
large (up to 9 orders of magnitude) rise of primordial lepton
asymmetry could take place because of transformation of active
neutrinos into sterile ones due to an initial
exponential
instability~\cite{barbieri91}, which later transforms into a
power law one~\cite{ftv,dolgov01a} by the back-reaction from the plasma.
Still, the dominate part of the asymmetry was generated during this
later stage. This statement was originally obtained in the frameworks
of thermally averaged kinetic equations, but the approach was
systematically improved in several subsequent
publications~\cite{foot97}-\cite{dibari01},\cite{dolgov01a}.

The investigation of ref.~\cite{ftv} was reconsidered in the
paper~\cite{shi96}. The author also worked with thermally averaged
equations but used a different approximation in the resonance
regime. His result is
that the absolute magnitude of the asymmetry
is indeed very large in agreement with~\cite{ftv} (and later
papers) but its sign is essentially unpredictable. The sign of the
asymmetry is very sensitive to oscillation parameters
and to the input
of numerical calculations. As a result of this feature the sign
of lepton asymmetry might be different in different causally
non-connected domains~\cite{shi99d1}. This could have interesting
implications and, in particular, would lead to
inhomogeneous nucleosynthesis. The existence of chaoticity
was confirmed in several subsequent
papers~\cite{enqvist99}-\cite{buras00} but again in the
framework of simplified
thermally averaged equations.
On the other hand, the analysis of possible chaoticity performed
in ref.~\cite{dibari99c} on the basis of
the numerical solution of kinetic equations with a full momentum
dependence shows a different picture. The results of this work are
presented in fig.~\ref{xF7a}. Most of the parameter space
is not chaotic,
while in the region where chaoticity is observed numerical calculations
are not reliable.
\begin{figure}[ht]
\begin{center}
\hspace{2.0cm}
\vspace{-1.0cm}
\epsfig{file=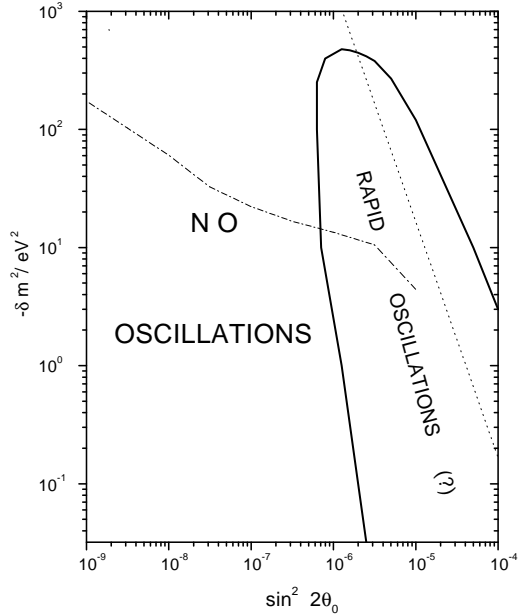,width=7cm}
\end{center}
\caption{\label{xF7a}
Region of parameter space where the final sign
does and does not oscillate for $\nu_\tau \leftrightarrow\nu_s$
oscillations, according to ref.~\protect{\cite{dibari99c}}
}
\end{figure}
This figure should be compared with the calculations based on
thermally averaged equations~\cite{enqvist99} which show that a very
large part of parameter space is chaotic, in contradiction to
ref.~\cite{dibari99c} (see fig.~\ref{plane}).

\begin{figure}[ht]
\centering
\vspace*{-2mm}
\hspace*{-3mm}
\leavevmode\epsfysize=13.0cm \epsfbox{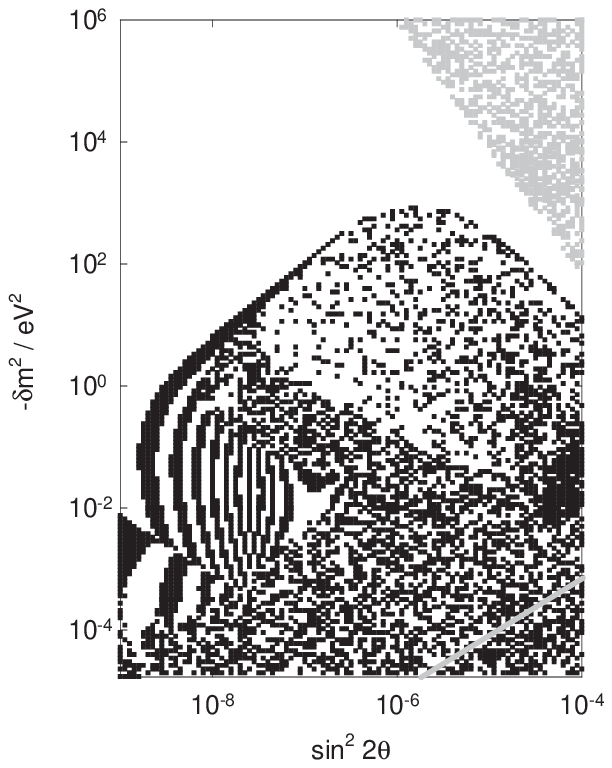}\\[-24mm]
\caption{\label{plane}
 The distribution of the final sign of neutrino asymmetry
 in the mixing parameter space. Negative signs of $L$ are plotted in
  black and positive ones are in white. The initial asymmetry was
  chosen to be \protect{ $L^{\rm in}= 10^{-10}$}.  }
\end{figure}

The analytical 
solution of ref.~\cite{dolgov01a} does not show any
chaoticity. To be more precise, numerical solution of
equation~(\ref{Z'fin}) is chaotic for large values of $K\sv$.
However this chaoticity is related to numerical instability because
with the increasing coefficient in front of the r.h.s. of the equation,
the minimal value of the asymmetry becomes very small and can be smaller
than the accuracy of computation. In this case the calculated value of
the asymmetry may chaotically change sign. However this
regime is well described analytically and it can be seen that the sign
of asymmetry does not change.

The chaotic behavior observed in the 
papers~\cite{shi96}-\cite{buras00} 
has always been present 
in a
simplified approach the kinetic equations were solved for a
fixed ``average'' value of neutrino momentum, $y= 3.15$, so that
the integro-differential kinetic equations are approximated by much
simpler ordinary differential
ones. However, many essential features of the process 
become
obscured in
this approach, in particular
the ``running of the resonance'' over neutrino
spectrum, and it is difficult to judge how reliable the results are.
Moreover, the average value of $1/y$ that enters the refraction
index is $\langle 1/y \rangle \approx 1$, and not
$1/\langle y\rangle \approx 0.3$, though this numerical difference
might not be important for the conclusion.

Possibly fixed-momentum approach is not adequate to the problem, as can
be seen from the following very simple example. Let us
consider oscillations between $\nua$ and $\nus$ in vacuum. Then
the
leptonic charge in active neutrino sector would oscillate with a
very large frequency, $\sim \sin (\dm t /E)$, for a neutrino with a 
fixed energy $E$.
However if one averages
this result with thermal neutrino spectrum, the oscillations
of leptonic charge would be exponentially suppressed. Still this
counter-example is also oversimplified and cannot be considered
a rigorous argument against chaoticity.
It may happen, for example, that for smaller $K$,
when saddle point does not give a good approximation, the differential
asymmetry $\Delta (q,y)$ may be an oscillating function but the
integrated (total) asymmetry is smooth. Another logical option
is that for a smaller $K$ the integrated asymmetry is also chaotic
but not large. However, one cannot exclude a large asymmetry with a
small $K$. It is possible that the fixed momentum approach has an advantage
of simplicity and by using this approach one could discover important
properties that remain obscured in more accurate methods.
Very recently there appeared the paper~\cite{kainulainen01s}
where a numerical solution of complete momentum dependent kinetic 
equations was performed. The authors state that they have found
chaotic behavior of the solution. This statement contradicts analytical 
results of ref.~\cite{dolgov01a}.
It is difficult to make a final judgment at this stage.

\subsection{Active-active neutrino oscillations. \label{ssec:active}}

In the previous sections we considered oscillations between active
and sterile neutrinos. Oscillations between active ones usually do
not lead to modifications of the standard picture
because the oscillating partners are well in thermal equilibrium and
their density matrices do not evolve, remaining of the same
equilibrium form as in the case of no oscillations. However,
initial neutrino states might be out of equilibrium and in that case
oscillations between active partners could produce
interesting effects. We will consider here a
scenario where one of the
active neutrino flavors has a large initial charge asymmetry, which
could be created by processes with leptonic charge non-conservation in
the very early universe. Examples of models leading to such 
state are discussed in secs.~\ref{varabund},\ref{sec:asym}. So the
initial state at the moment when oscillations become active could
be e.g. a state with a large asymmetry in the sector, say, $\num$ or
$\nut$ and zero asymmetry in $\nue$. If there is
mixing between $\nue$ and $\num$ (or $\nut$) then part of muon
asymmetry could be transformed through these oscillations into electronic
asymmetry. The latter could have a noticeable impact on BBN.

Oscillations between active and sterile neutrinos in the presence of a 
large lepton asymmetry would be strongly suppressed because the contribution
of large asymmetry into neutrino effective potential~(\ref{nref}) makes the
mixing angle in vacuum negligibly small. This is not true for
oscillations between active neutrinos. As was noticed in 
ref.~\cite{pantaleone92} and discussed in detail in series of 
papers~\cite{samuel93}-\cite{kostelecky96b} the effective potential of 
oscillating active neutrinos is a non-diagonal matrix in flavor space.
The non-diagonal terms are induced by self-interaction of 
oscillating neutrinos.
The presence of non-diagonal terms strongly alters the oscillation pattern.
In particular, in this case the oscillations are no longer inhibited by
a large charge asymmetry of the plasma and neutrinos with all momenta
oscillate as a single ensemble. Some more surprising results were
presented - in particular, it was argued that under certain
conditions the
neutrino gas that initially consisted only of $\nue$
would go into pure muonic neutrinos and not, as one would expect,
into equilibrium state, equally consisting of $\nue$ and $\num$.
Another surprising statement is that coherence may not be lost
thanks to non-linearity of the system.  Possible CP-odd effects in
neutrino oscillations, as well as neutrino evolution in the presence
of initial non-zero chemical potentials also have been discussed
in the papers cited above. A clear description of some of these phenomena
was presented recently in the paper~\cite{pastor01rs}.

As we have already mentioned oscillations between active neutrinos
would be unobservable if neutrinos are in complete thermal equilibrium.
However neutrino spectrum deviates from equilibrium
due to $e^+e^-$-annihilation following
neutrino decoupling~\cite{df,dt} 
(see sec.~\ref{masslessdistr}). Oscillations could change the 
distorted spectrum and might be in principle observable in 
BBN~\cite{langacker87}.
This effect was considered in ref.~\cite{kostelecky95} and
recently in ref.~\cite{hannestad01sp}. 
According to the calculations of the latter the primordial abundance
of $^4 He$ changes only by $1.5\cdot 10^{-4}$.

The kinetic equations used in the previous section can be easily
modified to apply to this case.
Let us consider for definiteness oscillations
between $\nue$ and $\num$.
One has to take into account the self-interaction processes
$\nue \num \lrar \nue \num$
and $\nue \bar\nue \lrar \num \bar\num$. The refraction index
is determined
by the forward scattering amplitude and since $\nue$ and $\num$ are
considered to be
different states of the same particle one has to include
both processes when there is a $\nue$ with momentum $p_1$ in initial
state and a
$\nue$ or $\num$ with the same momentum in the final state.
The processes of forward
transformation $\nue \lrar \num$ give non-diagonal contributions to
refraction index. Such transformations always exist, even among 
non-oscillating particles, but only in the case of non-vanishing mixing
the non-diagonal terms in the effective potential become observable.

Now instead of expression~(\ref{hint}) we have to use
\be
H_{int}^{(e,\mu)} =
\delta E \left( \begin{array}{cc} h_{ee}  & h_{e \mu} \\
h_{\mu e} & h_{\mu \mu} \end{array} \right) \equiv
{\delta E \over 2}  \left( h_0 + {\bf  \sigma\, h} \right)
\label{hemu}
\ee
where $\delta E = \dm /2E$ and ${\bf \sigma}$ are Pauli matrices.

It is convenient to present the density matrix as:
\be
\rho = {1\over 2} \left( P_0 + {\bf \sigma\, P} \right) =
{1\over 2}  \left( \begin{array}{cc} P_0+P_z  & P_x -iP_y \\
P_x +iP_y & P_0-P_z \end{array} \right).
\label{rho-vec}
\ee
 
The elements of the Hamiltonian matrix (\ref{hemu}) are expressed
through the integrals over momenta of the distribution functions of other
leptons in the plasma and, in particular, of the elements of the density
matrix of oscillating neutrinos themselves. The latter contribute to
non-diagonal matrix elements of the Hamiltonian. The structure of
these terms is essentially the same as the one discussed above for 
mixing between active and scalar neutrinos, see eq.~(\ref{nref}).
The contribution of self-interaction of neutrinos and antineutrinos also
contains two terms. One originates from non-locality of weak interactions 
and is symmetric with respect to charge conjugation:
\be
{\bf h}_+ = {V_{sym} \over 2\pi^2} \int dy y^3 \left( {\bf P} +\bar{\bf P}
\right).
\label{vsym}
\ee
The second is proportional to the charge asymmetry in the plasma and 
equals 
\be
{\bf h}_- = {V_{asym} \over 2\pi^2} \int dy y^2 \left( {\bf P} -\bar{\bf P}
\right)
\label{vasym}
\ee
One can find details e.g. in ref.~\cite{sigl93}. An essential feature,
specific for oscillations between active neutrinos, is the presence
of non-diagonal terms in the Hamiltonian (or in
refraction index). In the
case of large lepton asymmetry in the sector of oscillating neutrinos,
the asymmetric terms in the Hamiltonian strongly dominate and, as a 
result, the suppression 
of mixing angle in the medium, found for $(\nu_a - \nu_s)$-oscillations,
disappears.

If the coherence breaking terms have the simplified form~(\ref{gammarho})
the kinetic equations for density matrix become:
\be
P'_0 &=& -\gamma_e \left( {P_0+P_z \over 2} 
- f_e \right) -
 \gamma_\mu \left( {P_0-P_z \over 2} - f_\mu \right)
\label{dp0}
\\
P'_x &=& -P_y (c_2 + h_z) +P_z h_y -P_x \gamma_{e\mu}
\label{dpx}
\\
P'_y &=& P_x (c_2 + h_z) -
P_z (s_2+h_x) -P_y \gamma_{e\mu}
\label{dpy}
\\
P'_z &=& s_2 P_y +(P_y h_x -P_x h_y) -
 \gamma_e \left( {P_0+P_z \over 2} 
- f_e \right) +
 \gamma_\mu \left( {P_0-P_z \over 2} - f_\mu \right)
\label{dpz}
\ee
where $P'_j = (Hx /\delta E)\partial_x P_j  $ and the ``time'' variable
$x$ is defined according to eq. (\ref{xyi}), $c_2 =\cos 2\theta$, 
$s_2 = \sin 2\theta$, $\theta$ is the vacuum mixing angle,
$ f_{e,\mu}$ are equilibrium distribution functions with 
possibly different chemical potentials for $\nue$ and $\num$, 
$\gamma_a $ is the damping coefficient for $\nue$ or $\num$
normalized to $\delta E$ (see eq. (\ref{gammaj1})), and
$\gamma_{e\mu} = (\gamma_e + \gamma_\mu)/2$. 

If the temperature is sufficiently high and thus $\gamma$ 
is non-negligible, these
equations can be accurately solved in stationary point approximation:
\be
P_x = P_z \,{(s_2+ h_x)(c_2+h_z) + \gamma_{e\mu} h_y 
\over (c_2 +h_z)^2 + \gamma^2_{e\mu} } 
\approx P_z\,{h_x \over h_z} +\epsilon_x
\label{px} \\
P_y = P_z \,{(c_2+h_z)h_y - \gamma_{e\mu} (s_2 +h_x) 
\over (c_2 +h_z)^2 + \gamma^2_{e\mu} }
\approx P_z\,{h_y \over h_z} +\epsilon_y
\label{py}
\ee
where $\epsilon_j$ are small corrections satisfying
$\int dy\,y^2 (\epsilon - \bar \epsilon) =0$.

Substituting these expressions into 
equations~(\ref{dpx}-\ref{dpz}) and integrating over momentum
we find:
\be
h_x = h_z \, { s_2 c_2 \over c_2^2 + 
\langle \gamma_{e\mu}\rangle^2},\,\,\,
h_y = -h_z \, { s_2 \langle \gamma_{e\mu}\rangle
 \over c_2^2 + \langle \gamma_{e\mu}\rangle^2}
\label{hxhy}
\ee
where $\langle ... \rangle$ means thermal averaging.
Substituting these expressions into eq.~(\ref{dpz}) we obtain
\be
P'_z = - P_z\, {s_2^2 \gamma_{e\mu} 
\over c_2^2 + \langle \gamma \rangle^2_{e\mu}}
-\gamma_e \left( {P_0+P_z \over 2} 
- f_e \right) +
 \gamma_\mu \left( {P_0-P_z \over 2} - f_\mu \right)
\label{pz'}
\ee
An analogous equation exists for antiparticles. We can further simplify these
equations if we assume $\gamma_e=\gamma_\mu$. In this approximation
$P_0$ disappears and we arrive at a complete set of equations containing 
only $P_z$ and $\bar P_z$. 

The equilibrium distribution functions $f_{e,\mu}$ depend upon 
the chemical potentials $\xi_{\nue}$ or $\xi_{\num}$. 
They can be expressed through
the lepton asymmetries $(n_{\nu_a} - n_{\bar \nu_a})$. If the asymmetries
are large (compared to the ``natural'' value, 
$\sim 10^{-9} n_\gamma$), then the $z$-component of the Hamiltonian
is proportional to the difference:
\be
(L_{\nue}-L_{\num}) \sim h_z \sim \int d^3 y (P_z -\bar P_z)
\label{hz}
\ee

If the coherence breaking term in eq.~(\ref{pz'}) is sufficiently large,
$P_z$ should be close to its equilibrium value, with running chemical
potentials, $P_z\approx P_z^{(eq)}\equiv (f_e-f_\mu)$. However, one cannot 
neglect the last term in eq.~(\ref{pz'}) because the small difference 
$P_z-P_z^{(eq)}$ is multiplied by a large factor $\gamma$. 
The coherence breaking terms disappear if one subtracts from 
eq.~(\ref{pz'}) the corresponding equation for antiparticles and integrate
the difference over momentum. This follows from separate conservation of 
electronic and muonic charges by the coherence breaking terms.
This conservation permits to impose an evident relation between 
$\gamma$ and $\bar\gamma$. Taking this difference and integrating over
momentum we obtain in the Boltzmann approximation:
\be
h'_z = - {1\over 2}\, h_z\, \int d^3 y\,e^{-y}\,{s_2^2 \gamma_e \over
\langle \gamma_e\rangle^2 + c^2_2}
\label{h'z}
\ee
This equation is rather accurate at sufficiently 
high temperatures,
when coherence breaking terms are non-negligible and $c_2^2$ is not
too small. Otherwise stationary point approximation would be invalid.

The solution of this equation is straightforward. It shows that 
oscillations are not suppressed by matter effects in 
the presence of large lepton asymmetry. 

A detailed numerical investigation of oscillations
between three active neutrinos in the early universe is 
carried out in the
paper~\cite{dolgov02}. An analysis of the impact of oscillating neutrinos
on BBN was performed for the values of oscillation parameters favored by
the solar and atmospheric neutrino anomalies. For the large mixing angle
(LMA) solution flavor equilibrium is established in the early universe and
all chemical potentials $\xi_{e,\mu,\tau}$ acquire equal values. The results
of the calculations for this case are presented in fig.~\ref{f-lma}.
\begin{figure}[t]
\centerline{\epsfig{file=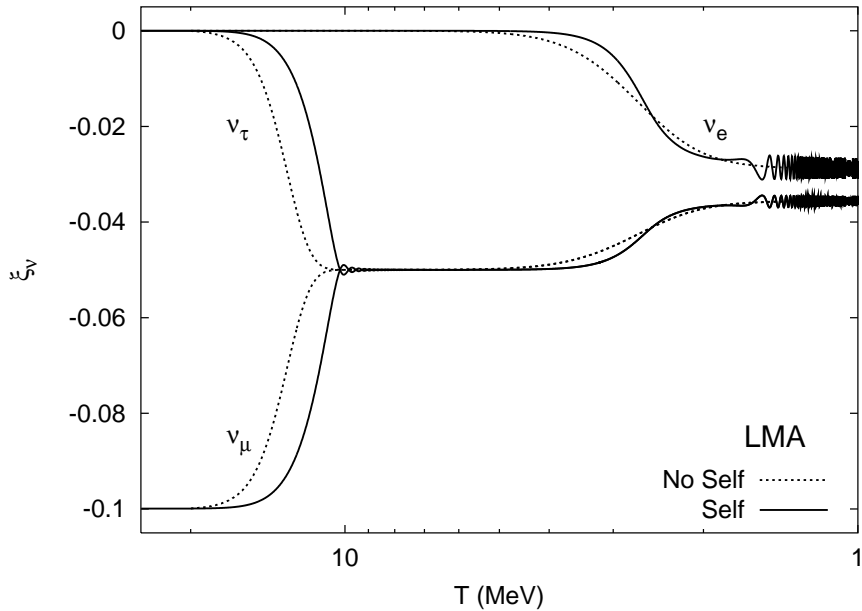,width=12cm}}
\caption{Evolution of neutrino chemical potentials for LMA
case, $\theta_{13}= 0$, and initial values $\xi_e=\xi_\tau=0$ and
$\xi_\mu=-0.1$. Solid and dotted curves are obtained with and
without neutrino self-interactions respectively.
}
\label{f-lma}
\end{figure}
Since for these values of
the parameters, asymmetries in muonic and tauonic sectors are efficiently
transformed into electronic asymmetry, the BBN bounds on chemical potentials
are quite strong, $|\xi_a| < 0.07$ for any flavor $a=e,\mu,\tau$.

For the LOW mixing angle solution,
the efficiency of the transformation of muon or 
tauon asymmetries into electronic one is not so efficient. The
transformation started at $T<1$ MeV below interesting range for BBN.
The results of calculations are presented in fig.~\ref{f-low}.
\begin{figure}[t]
\centerline{\epsfig{file=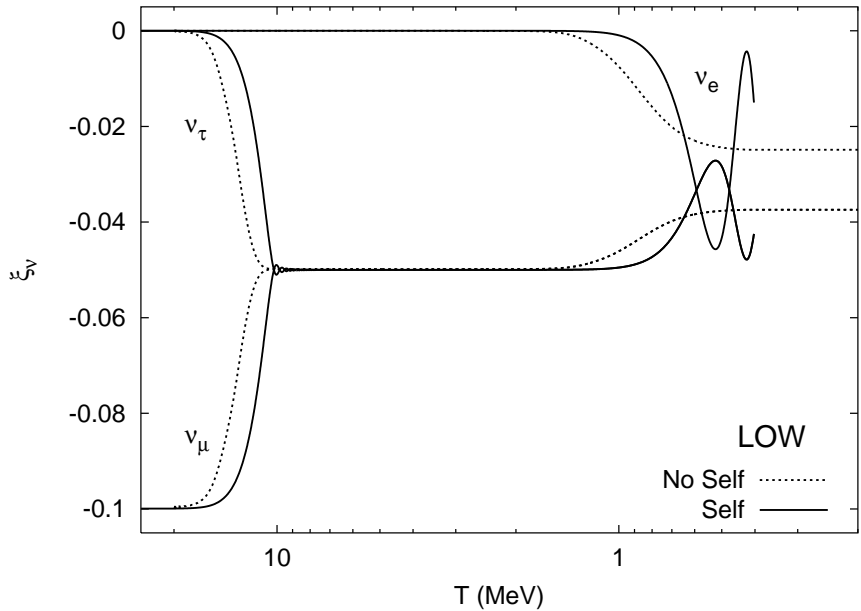,width=12cm}}
\caption{Evolution of the neutrino degeneracy parameters for LOW
case and the initial values $\xi_e=\xi_\tau=0$ and
$\xi_\mu=-0.1$. Notations are the same as in fig.~\ref{f-lma}.
}
\label{f-low}
\end{figure}

The results presented in these figures are valid for 
vanishing mixing angle $\theta_{13}$ (in the standard parameterization of
the $3\times 3$-mixing matrix. An analysis of different non-zero values
of $\theta_{13}$ can be found in the paper~\cite{dolgov02}.

\subsection{Spatial fluctuations of lepton asymmetry. \label{sec:spatfluc}}

A very interesting phenomenon was found in ref.~\cite{dibari99b}.
Neutrino oscillations in the presence of initially small
baryonic inhomogeneities could give rise to domains
with different signs of lepton asymmetry. This effect is different
from the chaotic 
amplification of asymmetry discussed above.
As is shown in section~\ref{sec:br} the initial asymmetry
first drops to an exponentially small value and after that starts
to rise, also exponentially, with a larger integrated exponent.
Since the value of the asymmetry 
at its lowest
could be extremely small,
it is sensitive to small perturbations, 
which could determine the final sign of the asymmetry.
This does not go  
in a trivial way as e.g.
spatial fluctuations of the sign in the minimum but in a somewhat 
trickier manner
as will be explained below.
As is argued in the original paper~\cite{dibari99b}
(see also ref.~\cite{enqvist00}), 
though small spatial fluctuations of the cosmological baryon asymmetry 
do not create a change of sign at the minimum, they would induce 
the formation of a domain with super-horizon sizes (at the moment of 
their creation) with large lepton asymmetries 
of different signs.
The fluctuations of the charge asymmetry
themselves are not directly essential - one can see from the results
of secs.~\ref{sec:br} that the final sign of the
asymmetry is the same as the initial one, if chaoticity
(sec.~\ref{sec:chaos}) is not present. However, initial isocurvature
perturbations in the background asymmetry, which enters the refraction
index of neutrinos~(\ref{etanue},\ref{etanumu}), would induce neutrino
diffusion and the diffusion term might have different signs in different
space points. According to the arguments of ref.~\cite{dibari99b}, it
may dominate the contribution of the background asymmetry in the
minimum and generate 
a sign difference in the final large value of the
lepton asymmetry. We will show below how it works.

This effect would have an important impact on BBN and
on the subsequent neutrino oscillations. This phenomenon was further 
studied in ref.~\cite{enqvist00} in frameworks of 
a one-dimensional model with momentum-averaged kinetic equations.
The authors argued that due to this effect the entire parameter
range where the exponential growth of lepton asymmetry takes place
is excluded by BBN. In particular, the production of sterile neutrinos
through MSW resonance at the domain boundaries (with varying density
of leptonic charge) is so strong that it would invalidate the results
of refs.~\cite{bell98}-\cite{foot00}.

Let us consider the equation, modeling evolution of the lepton
asymmetry in the presence of small spatial inhomogeneities, used in
ref.~\cite{dibari99b}. Following the notations of this paper, let us denote
the asymmetry of active neutrinos of flavor $a$ as $L_{\nu_a}$. The
combination that enters the refraction index of $\nu_a$ is
$2 L_{\nu_a} + \tilde L$, where $\tilde L$ consists of the contributions
of other neutrino species, baryons, and electrons. Due to electric charge
neutrality, the last two are not independent. Preexisting fluctuations
in neutrino asymmetry
would be erased due to a large neutrino mean free
path in cosmic plasma after neutrino decoupling. So background
asymmetry can be written as $\tilde L = \bar L + \delta B (\vec x)$,
where the first term is homogeneous and the last could be
inhomogeneous and related to the fluctuations in the baryon number.
The evolution of the neutrino asymmetry is described by the equation:
\be
\dot{L}_{\nu_{\alpha}}(\vec{x},t)=
a(t) \left[ 2\,L_{\nu_{\alpha}}(\vec{x},t) + \bar L +
\delta B(\vec{x})\right ]
+D(t)\nabla^{2}L_{\nu_{\alpha}}(\vec{x},t)
\label{dotLnu}
\ee
where $D(t)$ is the diffusion coefficient, and the function $a(t)$
is initially negative and generates an exponential decrease of the
asymmetry, but at some critical time $t_c$ it changes sign and 
thus creates a huge rise of the asymmetry. It is essential that, while
$a(t)$ is negative, the asymmetry drops down to a very small value.
In a more accurate formulation $a$ would also
depend on the asymmetry itself, but in what follows we are
interested in rather small values of the asymmetry, where non-linear
effects are not important.

The solution to this equation can be found by the Fourier transform
and we obtain:
\be
L_{\nu_{\alpha}}(x,t) & = &\bar{L}\, \int_{t_{\rm in}}^t dt'\,a(t')
\,e^{2\int_{t'}^t\,dt''\,a(t'')} +\\ \nonumber
& +& e^{2\,\int_{t_{\rm in}}^t dt'\,a(t')}
\int\,d^3k\,e^{i{\vec k}{\vec x}}\,
\hat{L}_{\nu_a}({\vec k},t_{\rm in})\,e^{-k^2\,
\int_{t_{\rm in}}^t dt'\,D(t')}+ \\
\nonumber
& + & \int_{t_{\rm in}}^t dt'\,a(t')\,e^{2\int_{t'}^t\,
dt''\,a(t'')} \,\int \,d^3k\,e^{i{\vec k}{\vec x}}\,
\delta \hat{B}({\vec k},t_{\rm in})\,e^{-k^2\,
\int_{t'}^t dt''\,D(t'')}
\label{solltx}
\ee
Here ``hut'' indicates the Fourier transform of the corresponding function.
The first term in this expression can be explicitly integrated
because the integration measure $dt' a(t')$ is exactly the differential
of the exponential. The integration of this term gives:
\be
(1/2) \bar L \left[ \exp \left( 2\int^t_{t_{in}} dt_2 a(t_2) \right)
-1 \right]
\label{barL}
\ee
So we obtained a rising term (after some initial decrease) plus a constant
initial value of $\bar L$.

The second term can be also integrated
because the initial value $\hat{L}({\vec k},t_{\rm in})$ is
supposed to be homogeneous and so its Fourier transform is just
delta-function, $\delta^3 (k)$. The integral gives
\be
L_{\nu_a}^{(in)}\, \exp \left( 2\int^t_{t_{in}} dt_2 a(t_2) \right)
\label{Lnua}
\ee
where $L_{\nu_a}^{(in)} $ means the initial value, i.e.
taken at $t=t_{in}$. So if we forget about the constant term $\bar L/2$,
we would have $(L_{\nu_a}^{(in)} +\bar L/2)$
multiplied by the rising exponent. One would get exactly this expression
if one solves the equation for $\dot L_{\nu_a}$ in the homogeneous case.

Let us now consider the last term which, according to the
arguments of ref.~\cite{dibari99b}, could change the
sign of the rising asymmetry, i.e. this last term could become the dominant
one. To evaluate the integral let us substitute:
\be
\delta \hat{B}({\vec k},t_{\rm in}) =
\int d^3 x_1 e^{i{\vec k}{\vec x}_1}\, \delta B ({\vec x}_1)
\label{deltabk}
\ee
where $\delta B ({\vec x}_1)$ is the initial value of the inhomogeneous
term. Now we can integrate over $d^3 k$. We have the integral of
the type
\be
\int d^3k \exp [- S^2 k^2 +i {\vec k} (\vec{x}-{\vec x}_1)]
\label{intdk}
\ee
the scalar product of vectors $\vec k$ and $\vec r = \vec x-\vec x_1$
is equal to ${\vec k}({\vec x}-{\vec x_1}) = kr\cos\theta$,
and
\be
S^2 (t) = \int_{t_{in}}^t dt_2 D(t_2)
\label{S}
\ee
Integration over angles in $d^3 k = 2\pi k^2 dk d(\cos\theta)$ is
trivial, it gives $\sin kr /kr$. The remaining integration can be done
as follows:
\be
\int dk k \sin kr \exp [-S^2k^2] =
(d/dr) \int dk \cos kr \exp [-S^2k^2]
\label{intk2}
\ee
and the remaining integration can be performed
if we expand the range of integration
from minus to plus infinity.
Introducing a new variable
$\vec x_1 = \vec x - S(t_1)\vec \rho$ we finally obtain
\be
\int dt_1 a(t_1)\, e^{2\int_{t_{1}}^t dt_2 a(t_2)} \int d^3 \rho \,
\delta B (\vec x -S(t_1) \vec \rho) \, e^{-\rho^2}
\label{fin}
\ee
This is the contribution the lepton asymmetry $L_{\nu_a}$ generated by
the (small) baryonic inhomogeneities. Its asymptotic rise at large
$t$ is similar to the rise of other terms, but its exponential decrease
at intermediate stage could be considerably milder. As a result,
this term could become dominant with the sign determined by the
sign of the fluctuations in the baryon asymmetry. 
We can see this in a simple example 
assuming that the function $a(t)$ has the form
$a(t) = a_1 (t-t_c)$ and that the fluctuations of the asymmetry
are described by one harmonic mode:
$\delta B({\vec x}) = \epsilon_B \cos {\vec k_0}{\vec x}$. This
form of $\delta B$ could be inserted either into eq.~(\ref{fin})
or into initial eq.~(\ref{solltx}) and we find for the oscillating
part of the asymmetry (up to a constant coefficient):
\be
\delta L({\vec x}) &=& a_1 \epsilon_B \cos {\vec k_0}{\vec x} \,\,
e^{a_1 (t-t_c)^2 - S^2 (t)\,k_0^2}
\left[ \int_{t_c-t_{in}}^{t-t_c} dt_1 t_1
e^{-a_1 t_1^2 + S^2(t_1)\, k_0^2}+ \right.
\nonumber \\
&&\left.
\int_0^{t_c-t_{in}} dt_1 t_1 e^{-a_1 t_1^2}
\left( e^{S^2(t_1)\, k_0^2 } - e^{-S^2(t_1)\, k_0^2 }\right)\right]
\label{deltaL}
\ee
Both terms rise as $\exp [a_1 (t-t_c)^2]$, i.e. in the same way as
the other homogeneous terms (we assume that $S(t)$ is finite at large
$t$ and not too large). The first term is exponentially suppressed
as $\exp [-a_1 (t_c-t_{in})^2]$ also at the same level as the homogeneous
terms. The second term, which vanishes in the homogeneous case ( $k_0 =0$
or $ S=0 $) is not exponentially suppressed. In the limit of 
a large
$a_1$ the integral can be evaluated as $\sim S^2(0)\,k_0^2 /a_1$. It
is small but not exponentially small. Thus, 
one can easily imagine a
situation
when the last term dominates and the resonance enhancement of
lepton asymmetry in the background of small fluctuations of baryon
asymmetry could create domains with large and different lepton
asymmetry. The effect is very interesting and deserves more
consideration. 

Chaotic diffusion of neutrinos from these domains would generate 
electric currents by scattering of neutrino flux on electrons or positrons
in primeval plasma. These currents, in turn, would create cosmic magnetic
fields which could serve as seeds of coherent galactic 
magnetic fields \cite{dolgov02dg}.

\subsection{Neutrino oscillations and big bang nucleosynthesis.
\label{sec:nuoscbbn}}

There are several effects 
through which
neutrino oscillations may have influenced
primordial abundances of light elements (we will speak here mostly 
about mixing between active and sterile neutrinos):
\begin{enumerate}
\item{}
If sterile neutrinos are created by 
oscillations before active neutrino decoupling,
then the effective number of neutrino species at nucleosynthesis
would be larger than 3. This effect,
as is well known, results in an increase of mass fraction of helium-4
and deuterium. If the oscillations were efficient after decoupling of
active neutrinos, then the total number density, active + sterile,
would remain
unchanged and the effect on BBN of $\num - \nus$ or
$\nut -\nus$ mixing would be absent.
On the other hand, for the mixing between $\nue$
and $\nu_s$, if excitation of sterile neutrinos took
place after $\nue$ decoupling, the production of $\nu_s$
would be accompanied by the
corresponding decrease in the number/energy density of $\nue$. This
in turn would result in a higher temperature of 
$n/p$-freezing and also in
a larger mass fraction of $^4 He$, though the total energy density
of all neutrinos would remain the same as in the standard model.
\item{}
Oscillations may distort the spectrum of neutrinos and, in particular, of
electronic neutrinos. The sign of the effect differs, depending on
the form of spectral distortion. 
A deficit of electronic neutrinos at
high energy results in a smaller mass fraction of helium-4, while a deficit
of $\nue$ at low energy works in the opposite direction.
A decrease of total number/energy density of $\nue$ (as discussed in
the previous point) would result in an earlier
freezing of neutrino-to-proton ratio and in a larger fraction of helium-4.
\item{}
Oscillations may create an asymmetry between $\nue$ and anti-$\nue$.
If the spectra of $\nue$ and $\bar\nue$ have the equilibrium form with
a non-zero chemical potential then the $n/p$ ratio would change as
$n/p \sim \exp (-\mu_{\nue}/T)$.
Present day data permit the asymmetry in the sector
of electronic neutrinos to be at the level of a few per cent (see
sec.~\ref{degnubbn}), i.e. much larger than the standard $10^{-10}$.
In reality the generation of charge asymmetry by oscillations
may strongly distort the spectrum of active neutrinos, in particular,
of $\nue$, and a more complicated analysis is necessary.
Even if asymmetry was strongly amplified or if it was a primordial one
but still remained below 0.01, its direct influence on BBN would
be negligible. It may, however, have an impact on nucleosynthesis in an
indirect way. Namely, the asymmetry that is larger by several orders
of magnitude than the standard one, could suppress neutrino oscillations
through refraction index so that new neutrino species corresponding to
sterile neutrinos would not be efficiently excited and/or the
spectrum of $\nue$ would not be distorted.
\end{enumerate}
Thus, one can see
that the effects of oscillations 
may result either in a reduction or an increase 
of primordial abundances of $^4He$ and $D$.
This effect is usually described by the effective number of neutrino
species, though the latter is different for $^4He$ and $D$. The impact
on $^7Li$ is more complicated.

Historically, the study of the impact of oscillating neutrinos was 
honed with time, as additional effects were taken into account and 
more precise calculations were performed.
In ref.~\cite{ad1} only excitation of extra neutrino species by
oscillations was considered. It was assumed that neutrinos have both
Dirac and Majorana masses and 
and therefore sterile states could be
produced through oscillations. The condition that only one
extra neutrino species is permitted by BBN 
prompted the conclusion that 
the mixing angle should be smaller than 0.01 and/or mass difference
cannot be larger than $10^{-6}$ eV$^2$. However, the refraction
of neutrinos in the primeval plasma was neglected, and thus the
results of paper~\cite{ad1}
were valid only for a sufficiently large mass
difference, $\delta m^2 > ({\rm keV})^2$. The refraction
index of
neutrinos~\cite{nora} was correctly taken into account in the
paper~\cite{barbieri90}, where the probability of excitation
of sterile states by oscillations was calculated. 
The paper mostly considered the non-resonant case.
Consideration of resonance was postponed for the subsequent
paper~\cite{barbieri91}. Still it was mentioned
in~\cite{barbieri90} that resonance oscillations might have a
strong impact on the lepton asymmetry in the active neutrino sector
and could even change the sign of the neutrino asymmetry.

The probability of non-resonant production of sterile neutrinos
was first estimated in ref.~\cite{barbieri90} where the following
expression was presented:
\be
\Gamma_s = \langle \sin^2 2\theta_{m}
\sin^2 \left( t \omega_{osc} \right) \Gamma_a \rangle
\label{gammas}
\ee
Here $\Gamma_a$ is the production rate of ordinary (active)
neutrinos in the primeval plasma and the averaging is made over the
thermal cosmic background. The mixing angle $\theta_{m}$ and the
frequency of oscillations in the medium $\delta E$ are given
respectively by eqs.~(\ref{sinthetam}) and (\ref{omegaosc}).
This frequency is normally very high so one can substitute
$\sin^2 (\delta E\,t) = 1/2$. 
The rate of active neutrino production
can be parameterized as
\be
{\Gamma_a/ H} = \left({T/ T_a}\right)^3
\label{gammaa}
\ee
where $H$ is the Hubble parameter and $T_a$ is the
freezing temperature of the active neutrinos of flavor 
$a=e,\,\mu,\,\tau$. For
$T<T_a$ the production of $\nu_a$ is effectively switched off.
Using this and other expressions above we conclude
that the
equilibrium of sterile neutrinos is not established~\cite{barbieri90} for:
\be
\sin^4 2\theta |\delta m^2 | \leq 6\cdot 10^{-3} {\rm eV}^2
\left( T_a /3\, {\rm MeV} \right)^6
\label{ster-eq}
\ee
If BBN permits $\Delta N_\nu<1$ additional effective neutrino species
the r.h.s. of this bound would be smaller by the factor
$(\Delta N)^2$. The
values of the freezing temperature taken in ref.~\cite{barbieri90} were
$T_{\nue} = 3$ MeV and $T_{\num,\nut} = 5$ MeV. They
correspond to the freezing of (inverse) annihilation
$l\,\bar l \rar \nu\,\bar\nu$, where $l$ is a light lepton (electron or
any active neutrino). As we saw in sec.~\ref{sec:nonres} this is
not so and the limit is underestimated.

The other groups~\cite{enqvist90a}-\cite{enqvist91}, \cite{enqvist91}-
\cite{cline92}, \cite{shi93} used formally the same result~(\ref{gammas})
but argued that the total rate of reactions with active
neutrinos should be substituted for $\Gamma_a$. The latter is approximately
an order of magnitude larger than the annihilation rate and the
corresponding limit would be much stronger. The argument in favor of
this choice was
that sterile neutrinos were produced in any reaction with
a related active $\nu$ and not only by inverse annihilation.
On the
other hand, it is evident that pure oscillations conserve the total number
$n_{\nu_a} + n_{\nu_s}$, as well as elastic scattering does.
If no new active
neutrinos are produced by some inelastic processes, this conservation
law remains intact and the effective number of neutrino species
at BBN is not changed by the oscillations.
However, if the bulk of sterile neutrinos is produced at sufficiently
high temperatures, when annihilation is in equilibrium, then active
neutrino states are quickly re-populated by inverse annihilation and
the rate of their production is proportional to the total neutrino
reaction rate as is argued in the papers quoted above. According e.g.
to ref.~\cite{enqvist92b} the limits are
$\dm\sin^4 2\theta < 5\cdot 10^{-6} \Delta N^2_\nu$ for $\nue$ and
$\dm\sin^4 2\theta < 3\cdot 10^{-6}\Delta N_\nu^2$ for $\nu_{\mu,\tau}$.

A somewhat more accurate treatment of kinetic
equations ~(\ref{dotrhoaa}-\ref{dotrhosa}) reveals 
that in all
the estimates made in the previous literature the factor 1/2 has been
omitted in the rate of production of $\nus$ (see eq.~(\ref{sprod})).
A correction by
this factor makes the bounds 4 times weaker. The solution of
kinetic equations made in sec~\ref{sec:nonres} under assumption 
of kinetic equilibrium of active neutrinos leads to the bound 
which is weaker than the quoted ones
by another factor 1.5 (so that the total factor is 6).
The last discrepancy is not too large and may be explained by different
approximations made in the solutions.

A surprisingly strong and different in power of $\sin^2\theta$
limit was claimed in ref.~\cite{enqvist95}:
$\dm\sin^2 2\theta < 1.6\cdot 10^{-6}$. The authors argued that the
probability of creation of sterile neutrinos is proportional to
$(\Gamma_W /H)^2$ in contrast to the usually 
obtained
first power of this
factor. This discrepancy is probably related to misinterpretation of
the conversion probability versus total probability of production
found in ref.~\cite{enqvist95}.

There is a continuing activity in the field and more bounds for
different special cases are obtained. The limits on oscillation
parameters that were found in the
papers.~\cite{enqvist92b,cline92,shi93} were reconsidered (and relaxed)
in ref.~\cite{cardall96} for the case of high primordial deuterium.
Nucleosynthesis constraints on the oscillation parameters in a
concrete model of four-neutrino mixing (three active and one sterile)
were considered in refs.~\cite{okada97}-\cite{athar00}.
The values of the parameters were taken in the range indicated by the
direct experimental and solar neutrino data, e.i.
$\delta m^2_{21} \sim 10^{-5} {\rm eV}^2$,\,\,\,
$\delta m^2_{43} \sim 10^{-2} {\rm eV}^2$, and
$\delta m^2_{31} \sim 1 {\rm eV}^2$.
Complexities due to possible resonance transitions
and the related rise of lepton asymmetry were disregarded.
Because of that the results could be applicable only to
non-resonance signs of mass differences.
The effective rate of active neutrino production included both
inverse annihilation and elastic scattering, however,
as we argued above, the results should be reconsidered for the weaker
production rate by the factor 1/2.

It is worth noticing that the bounds are obtained under the assumption
that active neutrinos have standard weak interaction. In the
case of additional stronger interactions, the refraction index of
neutrinos would be larger and the oscillations in the medium
would be more strongly suppressed.
For example, in the case of additional
coupling of neutrinos to majorons the limits
discussed above are
relaxed by several orders of magnitude~\cite{babu92} and sterile
neutrinos would not be dangerous for BBN. Similar arguments were
presented recently in ref.~\cite{bento01}.

If oscillations take place between $\nue$ and $\nu_s$,
another effect may be important~\cite{barbieri90}: if
the equilibrium is not reached,
the number density of $\nue$ would be depleted because of
transformation into $\nu_s$ after $\nue$
decoupled from the cosmic plasma. This is turn would result
in an earlier freezing of neutron-proton transformations and in a
larger $n/p$-ratio. This effect permits to exclude small
values of the mass difference, down to $10^{-7}$ eV$^2$, for large
mixing angles, $\sin^2 2\theta > 0.4$. However the spectrum
distortion of electronic neutrinos caused by oscillations
was neglected in derivation of this bound; this effect is
discussed below.

Resonance oscillations of neutrinos were considered in the early
papers~\cite{enqvist90a}-\cite{shi93}
in adiabatic approximation. In refs.~\cite{enqvist90a,enqvist91}
it was argued that the oscillations drive lepton asymmetry in
active neutrino sector down to zero independently of the
existence of resonance transition. This result disagrees with
with refs.~\cite{barbieri90,barbieri91}. In the second of these
papers it was found that lepton asymmetry
is exponentially
unstable with respect to oscillations and
is not driven to zero but, on the opposite, is enhanced.
Another interesting effect mentioned in ref.~\cite{barbieri90}
is a possible parametric resonance phenomenon in neutrino
oscillations in medium.
In the case of resonance oscillations the parameter space
excluded by BBN is significantly larger and surprisingly the
results of the papers~\cite{enqvist90b},\cite{barbieri91},
and \cite{shi93} are rather close to each other.
For $\sin^2 2\theta > ({\rm a\,\, few\,})\cdot 10^{-2}$
the mass difference above $2\cdot 10^{-7}\,{\rm eV}^2$ is excluded;
for a smaller mixing the limit is roughly
$ \sin^4 2\theta \delta m^2 \geq 5\cdot 10^{-10} {\rm eV}^2$.
However, one should take these results with great caution because
the simplifications made in the calculations may be non-adequate
for the resonance case. A very essential approximation that
strongly simplified the calculations was the use of thermally
averaged kinetic equations. In this case, instead of infinitely many
modes for different neutrino momenta, all matrix elements of
neutrino density matrix were taken at the average momentum
$\langle p \approx 3T\rangle$. Hence
instead of integro-differential
equation containing an integral over $d^3 p$ from neutrino
distributions, which is especially important for the charge asymmetry
term in the refraction index, one got a much simpler set of
ordinary differential equations. However, spectral effects are
of primary importance and the average approach could be misleading.
A detailed derivation of quantum kinetic equations and an
analysis of applicability of momentum-averaged approach can be
found in ref.~\cite{mckellar94}.

Unfortunately in the early papers the oscillations were considered
in one mode approximation, when all relevant quantities were
averaged over neutrino spectrum, as we have already mentioned above.
This permits reducing the problem to 
the functions of one variable - 
time - instead of both time and momentum. Of course in the frameworks
of this formalism
one cannot even pose the question about the real spectrum of
neutrinos. First works 
that did not make this simplifying assumption,
where the elements of density matrix depending of both variables
$p$ and $t$ were considered, were done in 
refs.~\cite{kirilova97}-\cite{kirilova00}. Kinetic equations
governing evolution of density matrix of oscillating neutrinos
were numerically solved for relatively small mass difference,
$\delta m^2 \leq 10^{-7}$ eV$^2$ and arbitrary vacuum mixing
angle. Such small mass difference allows the following
simplifications: 1) the effect of coherence breaking are not
important, so one could use
eqs.~(\ref{dotrhoaa}-\ref{dotrhosa}) without damping terms;
2) a small value of $\delta m^2$
results in a small oscillation frequency and this stabilizes
computations; for a larger $\delta m^2$ a significant numerical
instability appears.

The authors noticed
 an important role
 of the distortion of the
spectrum of electronic neutrinos by the oscillations as well as
the effect of depletion of $\nue$ by transformation into $\nu_s$
on BBN. 
According to their
observation the spectrum distortion cannot be
adequately described by the shifting of the effective neutrino
temperature. The analytical fit to the bound on the oscillation
parameters that follows from the consideration of primordial
$^4 He$ can be written as~\cite{kirilova99b}:
\be
\delta m^2\,\left(\sin^2 2\theta \right)^4 \leq
1.5\cdot 10^{-9}\,\, {\rm eV}^2,\,\,\,
{\rm for}\,\,\, \delta m^2 < 10^{-7}\,{\rm eV}^2.
\label{dmsintheta}
\ee

The evolution of lepton asymmetry due to oscillations was
studied both for resonant and non-resonant cases.
In the non-resonant case the initially small asymmetry, i.e. of the
order of the baryonic one, remains unnoticeable for
nucleosynthesis~\cite{kirilova98a}. In the resonant case the
asymmetry might be considerably amplified by the resonance
transition. An analysis of $^4 He$-production in the presence of
neutrino oscillations with a small mass difference, 
$\dm < 10^{-7}$ eV$^2$ was performed recently in ref.~\cite{kirilova01he}.
It was shown that the mass fraction of $^4 He$ may increase up to
14\% in non-resonance case and up to 32\% in the resonance case.

The impact of the amplification of the asymmetry on BBN 
for a larger mass difference has been studied
in refs.~\cite{foot97b}-\cite{foot00}, \cite{dibari00}.
In the case of $(\nue-\nus)$-mixing the impact of asymmetry generation
on BBN is very strong and the result depends upon the final sign of the
asymmetry in $\nue-\bar\nue$ sector. If the asymmetry is negative
so that $n_{\nue} < n_{\bar\nue}$, the mass fraction of the produced
$^4He$ increases. This corresponds to a positive contribution to the
number of effective neutrino species. In the opposite case the effect
is negative. The magnitude of this contribution into $N^{(eff)}_\nu$,
according to the calculations of ref.~\cite{dibari00}, as a function of
mass difference is presented in fig.~\ref{dbfbbn}.

\begin{figure}[ht]
\begin{center}
\hspace{2.0cm}
\vspace{-1.0cm}
\epsfig{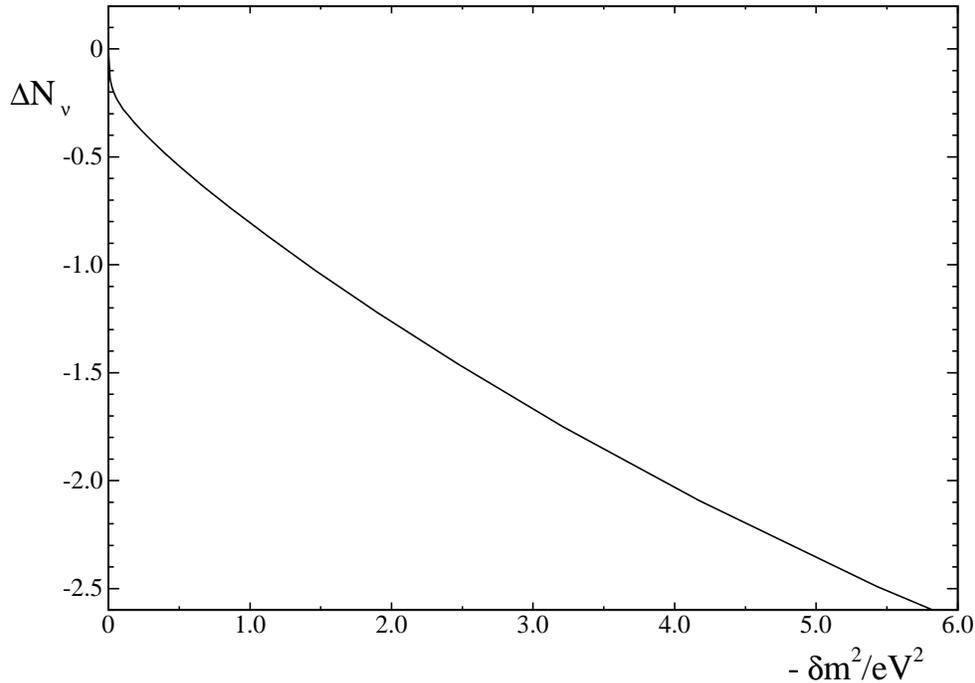}
\end{center}
\caption{\label{dbfbbn}
Change in the effective number of neutrino species for BBN,
$\Delta N_{\nu}$ versus $-\delta m^2$ for the case
$\sin^2 2\theta = 10^{-8}$ and $L_{\nu_e} > 0$,
according to ref.~\protect{\cite{dibari00}}
}
\end{figure}

In the case of more complicated mixing between all active neutrinos and
one sterile $\nu$ (which
is necessary to explain all neutrino anomalies)
the process of asymmetry generation in $\nue$-sector might proceed in
two stages~\cite{foot97b,bell98,foot99c}. First, a large
$\nut$-asymmetry was generated by the 
resonance transition
$\bar\nut-\bar\nus$. At the second stage, the oscillations
$\bar\nut-\bar\nue$ created some asymmetry in $\nue-\bar\nue$ sector.
As was argued in ref.~\cite{foot97b} for the model with
$m_{\nut}\gg m_{\num,\nue,\nus}$ the net result is
either $N^{(eff)}_\nu = 3.4$ or $N^{(eff)}_\nu = 2.5$ in a rather
wide range of mass difference, $\dm = 10-3000$ eV$^2$.
In ref.~\cite{bell98} a different mass pattern was considered:
$m_{\nut} \sim m_{\num} \gg m_{\nue,\nus}$ with the mixing angles favored
by the experimental data. In this case, according to
ref.~\cite{bell98}, $N^{(eff)}_\nu = 3.1$ or $2.7$.
The mirror universe model with three extra mirror neutrinos mixed
with three ``our'' ones is considered in ref.~\cite{foot99c}. A
bigger freedom of the model permits to explain the atmospheric
neutrino anomaly by $(\num-\num')$-mixing (where $\num'$ is the mirror
muon neutrino) without distorting successful prediction of BBN. The
number of extra neutrino species in the model can change from (-1)
to (+1). The results of these works 
disagree with
ref.~\cite{shi99} where it was argued that the $\num-\nus$ solution
to the Super Kamiokande data~\cite{sk98,sk00} could evade the BBN
bounds only for a very large mass difference,
$200 {\rm eV}^2< \dm < 10^4 {\rm eV}^2$. It was also argued
in this paper~\cite{shi99} that the effects of time variation of neutrino
asymmetry and non-instantaneous repopulation would strongly change
the BBN limits found in the papers~\cite{foot97b,bell98}.
On the other hand, the calculations of refs.~\cite{foot97b,bell98}
were re-examined in the work~\cite{foot00},
where it was found that the corrected results are in rough agreement
with the earlier papers~\cite{foot97b,bell98} and the corrections are
not crucial.
Some more discussion of the problem can be found in
refs.~\cite{foot98,shi98a}. All the groups at least agree that
the Super Kamiokande data~\cite{sk98,sk00} on atmospheric neutrino anomaly
in principle could be compatible with BBN in the resonance case
when a large lepton asymmetry in $\nue$-sector are generated,
while it is not so  if asymmetry is
small~\cite{bell98,shi99}.

As we have already mentioned the new data from SNO and 
SuperKamiokande experiments disfavor pure active-sterile neutrino 
mixing (see however~\cite{barger01}).
An analysis of four neutrino mixing of $2+2$ and $3+1$ types has been
performed recently in ref.~\cite{dibari01n}. It has been shown 
there that in
both schemes sterile neutrinos are completely excited in the plasma and
$\Delta N_\nu =1$. According to the same paper, this result in the case
of small lepton asymmetry contradicts the combined 
BBN + CMBR data which permit only $\Delta N_\nu < 0.3$. A possible way
out is to suggest another sterile neutrino with such a mixing that allows 
to generate a noticeable electron neutrino asymmetry.

The effect on BBN of {\it primordial} lepton asymmetry, considerably
larger than $10^{-10}$ in $\nue$ sector, was analyzed in
refs.~\cite{kirilova98b,kirilova99a} for small mass differences,
$|\dm|<10^{-7}$ eV$^2$. While in the absence of
oscillations, BBN is sensitive only to asymmetry at the level above
$10^{-2}$, much smaller values of asymmetry become important if
oscillations take place. The effect is indirect - the asymmetry
changes the light element production through its influence
on neutrino oscillations and they, in turn, change number
density and spectrum of active neutrinos. The asymmetry
in the range $10^{-7} < |L| < 5\cdot 10^{-6}$ has a complicated
impact on the oscillations. It could either enhance or suppress
oscillations depending on the concrete values of the parameters.
Thus the restrictions on the values of $\delta m^2$, $\sin 2\theta$,
and $L$ could be either weaker or stronger.
Smaller $|L|$ have very little influence on BBN, while $|L|$
larger than $5\cdot 10^{-6}$ strongly suppresses the oscillations.
These results can be only obtained if complete (not thermally
averaged) equations are used. As is argued in these works, in the
case of $\delta m^2 > 0 $, when
the ``averaged'' resonance transition
is absent, there could be a resonance, say, in neutrino transition
but only for a certain momentum value. This transition can change
the lepton number and after some time the resonance condition
for neutrinos ceases to be fulfilled, but becomes valid for
antineutrinos. So the process proceeds through ``alternating''
resonances. Of course this phenomenon would be lost after thermal
averaging.

A hot dispute arose between two groups \cite{shi99b,shi99c}
and \cite{kirilova98b,kirilova99a} in connection with
the role of lepton asymmetry in oscillations and BBN. Though one
cannot say that there is direct contradiction between them because
the oscillations are considered in non-overlapping range of
parameters,
$\delta m^2 < 10^{-7}$ eV$^2$ in ref.~\cite{kirilova98b} and
$\delta m^2 \sim 10^{-2} - 10^3 $ eV$^2$ in ref.~\cite{shi99b},
some qualitative difference between the conclusions of the two
groups can be found. An advantage of the
calculations~\cite{kirilova98b} is an exact numerical solution
of kinetic equations for each
momentum mode (see discussion above), in contrast to an approximate
approach of the group~\cite{shi99b}, who works in the parameter
region where numerical calculations are much more complicated.
More work in this direction is definitely in order.

Possibly the case of a large mixing angle, $\sin 2\theta >10^{-3}$
and a large mass differences $\delta m^2 \sim {\rm eV}^2$
can be treated rather reliably even
in approximate resonance approach~\cite{shi99d2,abazajian99} (compare
with sec.~\ref{sec:leptas}). The authors considered the parameter range
favored by the existing experimental indications on neutrino
oscillations. They found a large suppression in the low energy
part of the spectrum of both neutrinos and antineutrinos
relative to the thermal distribution
and showed that this spectral distortion has an 
impact on primordial $^4 He$ at the level of several per cent - 
which is within a range sensitive to observational constraints.

A recent review of some of the problems discussed in this subsection
can be found in ref.~\cite{kirilova01bbn}.

\subsection{ Summary \label{summarynu}}

Although great progress has been made  in understanding the
physics of neutrino
oscillations in the early universe, still a few unsettled problems
remain. The case of small mass difference,
$|\delta m^2 | \leq 10^{-7}$ eV$^2$ is relatively simple. The oscillations
in this case became  effective late enough to allow neglecting the loss of
coherence and the repopulation of active neutrinos. Numerical
solutions~\cite{kirilova97}-\cite{kirilova00} look stable and
sufficiently accurate.

The approximate treatment of 
the non-resonant case and the corresponding
bounds on the oscillation parameters from BBN in principle does not
create any serious problem, though all the papers have missed 
the factor
$1/2$ in the estimate of production rate of sterile
neutrinos~(\ref{sprod},\ref{gammas}). The
disagreement between the works~\cite{barbieri90} and subsequent
ones~\cite{enqvist92b,cline92,shi93} on whether
the rate of
reproduction of active/sterile neutrinos should include only inverse
annihilation or elastic scattering as well (see discussion
in secs.~\ref{sec:nonres},\ref{sec:nuoscbbn}) is resolved in favor
of the later papers but the BBN bounds on the oscillation parameters
are between the results of ref.~\cite{barbieri90} and 
refs.~\cite{enqvist92b,cline92,shi93}, because 
according to 
the calculations of sec.~\ref{sec:nonres} the correct bound is
weaker by factor 6, from which factor 4 is explained by
the missed factor 1/2 mentioned above and the origin of the other 
1.5 is unclear.

The huge rise of lepton asymmetry~\cite{ftv} up to 0.375 in the
resonance case 
is confirmed by several different
methods~\cite{ftv},\cite{foot95}-\cite{dibari01}, \cite{dolgov01a}
and can be very interesting for BBN~\cite{dibari00}.
An important unresolved issue related to the rising asymmetry
is its possible chaotic, oscillating behavior. It
was observed in the
papers~\cite{shi96}-\cite{kainulainen01s}, \cite{shi99}-\cite{abazajian99}
and is not supported by other cited in sec.~\ref{sec:chaos}.
The difficulty of the 
numerical approach is that the integral of the charge asymmetric
diagonal part of the density matrix over momentum~(\ref{Delta'})
contains quickly oscillating functions and enters kinetic equations
with a large coefficient. Hence it is difficult to separate the real
oscillating behavior from numerical artifacts. The effect of chaoticity
was definitely observed in the framework of the simplified thermally
averaged equations but was not seen or proven
within the exact momentum
dependent ones, except for the recent work~\cite{kainulainen01s}.
On the other hand, chaoticity might appear in the range
of parameters where neither analytical methods nor numerical ones are
applicable.

Oscillations among active neutrinos do not lead to noticeable effects in
BBN if initial lepton asymmetries are small. If it is not so, the asymmetries 
would be equilibrated in the case of the LMA solution and the BBN bound
on neutrino degeneracy would be more restrictive than that without 
oscillations. For the LOW solution the impact of neutrino oscillations on 
BBN is much weaker (sec. \ref{ssec:active}).


\section{Neutrino balls. \label{nuball}}

A very interesting phenomenon may take place in left-right symmetric
electro-weak theory. 
Since usually the left-right symmetry is
assumed to be spontaneously broken,
and so right-handed bosons and
right-handed Majorana neutrinos are very heavy over our ``left'' vacuum,
while they are light over ``right'' vacuum,
and, vice versa, our neutrinos that are light over our vacuum are heavy
over ``right'' vacuum state. However, if the symmetry breaking was
spontaneous, but not explicit, these two vacuum states would be
degenerate. As a result of a phase transition in the
early universe mostly the
left vacuum state was formed (at least in the
visible part of the universe. However, small bubbles of another, right,
vacuum state might also remain. Models leading to such an apparent
violation of left-right symmetry can be built, but we will not go into
detail here. Discussion and reference can be found 
in the
papers~\cite{holdom87,dolgov90,dolgov91}. The small bubbles
of wrong (i.e. ``right'') vacuum remaining after hot cosmological epoch
could be stable or, to be more exact, long-lived and
could have survived to the
present day. They form quasi-stable non-topological solitons, or so called
neutrino balls~\cite{holdom87}, supported against collapse by the pressure
of right-handed neutrinos that are light inside the ball and heavy outside.
Thus the wall between two vacuum states is impenetrable for neutrinos,
while it is transparent  for electrons, positrons and photons that are
common for the left-handed and right-handed worlds.

The pressure exerted by the surface tension
\be
p_{st} = 2\sigma /R,
\label{pst}
\ee
where $\sigma $ is the surface energy density and $R$ is the radius of the
ball, should be balanced by the pressure of the neutrino gas inside the
ball. The mass of the ball consists of the mass of the wall separating
the two vacua and the mass of neutrino gas inside:
\be
M_{ball} = 4\pi R^2 \sigma + {4\over 3} {6\sigma \over R} \pi R^3
= 12\pi R^2\sigma
\label{mball}
\ee
When the radius of a ball enters cosmological horizon micro-physical
processes inside the ball determine its evolution. Typically,
cosmological expansion is turned into contraction forced by the
surface tension. The neutrino gas inside becomes degenerate
with the chemical potential (the same for neutrinos and antineutrinos)
equal to~\cite{holdom87,dolgov90}:
\be
\mu \approx 0.15 \, {\rm MeV}\, \sigma_{{\rm TeV}}^{3/8} M_6^{-1/8}
\label{muball}
\ee
where $\sigma_{{\rm TeV}}=\sigma /{\rm TeV}^3$,
$M_6 = M_{ball} /10^6 M_\odot$, and $M_\odot = 1.99\cdot 10^{33}$ g
is the solar mass. The equilibrium radius of such a ball is
\be
R_{eq} = 3\cdot 10^{12}\,{\rm cm}\,M_6^{1/2}\sigma_{{\rm TeV}}^{-1/2}
\label{req}
\ee

If the ball mass is greater than $10^8 M_\odot \sigma_{{\rm TeV}}^{-1}$,
the equilibrium radius is smaller than the gravitational radius
and such balls would form primordial black holes and exist forever.
Neutrino balls with masses below
$10^4 M_\odot \sigma_{{\rm TeV}}^{3}$ would be very short-lived due to
the process $\nu_R \bar\nu_R \rar e^-e^+$. For heavier balls  and
correspondingly for $\mu < m_e$ this process is energetically impossible
and other weaker mechanisms of burning would be essential. The process
$\nu_R \bar \nu_R \rar 2\gamma$ proceeds only in the second order
in weak interaction~\cite{levine67} and is negligible. The reaction
$\nu_R \bar \nu_R \rar 3\gamma$ would burn the balls out
during the time~\cite{holdom87}:
\be
\tau_{3\gamma} = 10^{14}\, \left( m_e /\mu \right)^{13}\,{\rm sec}
\label{tau3gamma}
\ee
Another possible reaction that could destroy the ball is
\be
\nu_R \bar \nu_R e \rar e \gamma.
\label{nunue}
\ee
It gives the life-time~\cite{dolgov90}:
\be
\tau_\gamma = 4\cdot 10^{21} \left( m_e \over \mu \right)^4
\left( 1\, {\rm eV}^3 \over n_e \right) \, {\rm sec}
\label{taugamma}
\ee
where $n_e$ is the electron number density inside the ball. Normally this
process is sub-dominant with respect to $3\gamma$-burning but matter
accretion on neutrino balls could create 
a much larger number density of
electrons, up to~\cite{dolgov91}
$n_e = 10^{9} \sigma_{{\rm TeV}}^{3/2} M_6^{-1/2}\,{\rm eV}^3$.
In fact the number density should be somewhat smaller because the
matter accretion must stop when the ball reaches critical (Eddington)
luminosity. In this scenario neutrino balls remain stable for a long period - 
practically until the present time - and then emit their mass during
10-100 million years emitting about 0.1 solar mass per year.

Some further studies of the properties of neutrino balls,
with gravity effects taken into account, can be found in the
papers~\cite{manka93,holdom94,wang94}. Properties of neutrino balls in
a supersymmetric model are considered in ref.~\cite{manka94}.

Possible cosmological and astrophysical implications of these objects
are rather interesting. Firstly, they could form primordial black holes
if their mass exceeds the value
$10^8 M_\odot /\sigma_{{\rm TeV}}$. The latter could
be seeds for galaxy formation.  For smaller masses they are unstable
and may have luminosity close to that of quasars. This opens a competing
mechanism for 
the central engine of quasars~\cite{dolgov90,dolgov91,manka93}
instead of the canonical one by the accretion of matter on 
a superheavy
black hole. Possibly both mechanisms could be operating and heavy
neutrino balls that formed black holes are active galactic nuclei
(and heavy quasars), while the lighter ones are (or better to say, were)
short-lived quasars that can be observed only in distant parts of the
sky (at relatively large red-shifts).

As argued in ref.~\cite{holdom94a}, neutrino balls could present a
viable model for gamma bursters if a supernova were captured by a ball
and exploded inside. However, more probable is a capture of
an ordinary or neutron star by a neutrino ball. This captured matter
could create a strong outburst of energy by
the reactions similar to (\ref{nunue}).

The term ``neutrino balls'' (or ``neutrino stars'')
was later used in the
literature~\cite{bilic96}-\cite{bilic00} in a very different content.
Namely it was assumed that massive neutrinos could form gravitationally
bound stellar-like structures 
and the works focused on the properties and implications of such objects.
Although it is possible that non-relativistic
self-gravitating neutrinos may form stable stellar mass
(or million stellar mass) objects,
the mechanism of their formation is unclear. It is a difficult problem
to form a gravitationally binded system at stellar scale from
dissipationless particles. This problem is addressed in a recent
paper~\cite{bilic00}, where authors argue that
``dissipationless formation of a heavy neutrino star in gravitational
collapse is numerically demonstrated''.  The value of neutrino mass
assumed in the models
is about 10 keV, so an additional
annihilation mechanism must be introduced to avoid contradiction with
Gerstein-Zeldovich bound. This can be achieved by an anomalous neutrino
interaction with a new light boson. Such new, stronger-than-weak,
interaction 
could also aid the formation of a
gravitationally bound system
of heavy neutrinos, helping with energy dissipation. Massive neutrinos
with 10 keV mass that escaped capture into neutrino stars, 
could form halo of
dark matter around different astronomical objects and, in particular,
around the Sun. Possible observational manifestation of this form of
dark matter are discussed in ref.~\cite{munyaneza99b}.

The massive object in the center of our Galaxy (Sgr A$^*$) may possibly
be identified with a neutrino ball. Normally it is assumed that the Galaxy
hosts a supermassive black hole with the mass about $3\cdot 10^6\,M_\odot $.
This hypothesis implies the luminosity close to $10^{41}$ erg/sec but the
observed luminosity from radio to $\gamma$-ray frequencies is below
$10^{37}$ erg/sec. A possible way to solve this problem is to assume that
the object in the galactic center is a gravitationally bound system of
massive neutrinos discussed in the previous paragraph. This hypothesis
was recently analyzed in ref.~\cite{paolis01}, where it was found that
to satisfy astrophysical constraints, the neutrino mass should be in a
rather narrow interval: $11\,\,{\rm keV} < m_\nu < 24\,\,{\rm keV}$.


\section{Mirror neutrinos. \label{s:mirror}}

The idea that there may exist another world, which communicates with
ours through gravity and possibly, though not necessarily,
through some other very weak interaction, has a long history. 
After it became known that CP-parity is broken~\cite{christenson64}, 
Kobzarev, Okun, and Pomeranchuk~\cite{kobzarev66} suggested that 
invariance with respect to a modified CP
reflection could be restored if the particle content of the theory is 
doubled, i.e. there exist two parallel worlds, ours and
the mirror one, related by a new generalized CP-transformation.
Of course, if these worlds communicate/interact in any way
they should have common gravitational interaction.
An analysis of other possible 
interactions, that could connect the two worlds, was performed in
ref.~\cite{kobzarev66}. It was shown that, 
in addition to gravity, mirror particles might possibly be connected to us 
only through some new very weak,
weaker than normal weak, interaction. In the same
paper~\cite{kobzarev66} a possible existence and detection of 
macroscopic astronomical bodies consisting of mirror matter were
discussed. Later, in 1980's Okun~\cite{okun80} considered a possibility 
that an interaction between the two worlds might be not very weak if it
proceeded through an exchange of new neutral mesons.

Another type of particle doubling was assumed in
ref.~\cite{nishijima65} in an attempt to explain the decays seen by 
Cronin et al~\cite{christenson64} without breaking
CP by introducing shadow K-mesons and other shadow particles. It was
proposed in this paper that the two worlds were
not symmetric, the mirror symmetry was broken and the properties of
ours and mirror, or now better to say, shadow particles were different.

A possibility of some particle doubling was mentioned in the paper 
by Lee and Yang~\cite{lee56}, where parity violation was proposed.
The authors suggested that parity might remain an exact symmetry of 
the theory if, in addition
to left-handed protons of our world, there exist mirror symmetric
protons, so in a broad sense there is left-right symmetry in the
Lagrangian. In this picture the observed right-left asymmetry is prescribed
to a local preponderance of, say, left-handed protons over right-handed
ones. According to ref.~\cite{lee56}, the interactions between $p_R$ and
$p_L$ is not necessarily weak and they might interact with the same
electromagnetic and even the same pion fields, but it was later 
shown~\cite{kobzarev66} that this could not be true.

The hypothesis of mirror (shadow) particles/universe
was later elaborated in many papers~\cite{mirror}. An interesting
implication of the idea of a
mirror world is the possibility of explaining the
smallness of neutrino mass~\cite{zeldovich81,silagadze95}
by an analog of see-saw mechanism (see sec.~\ref{s:bs}).

As was suggested in two pioneering papers, there are two possible types
of scenarios of parallel worlds: the first is the case of exact symmetry
between the worlds, so that the physics there is identical
to ours and the other one is the case of different physics, which could
be due to a different pattern of
symmetry breaking in the two worlds. The first version is often called
the exact parity symmetry or mirror case, while the second one is referred
to as a shadow universe.

Cosmological implications of the existence of another
practically sterile (except for gravity and possibly very weak
interactions) universe, especially in connection with dark matter,
were considered in many papers~\cite{mir-cosm}, but in what follows
we will concentrate only on those which are related to neutrino physics.
In a model with a
broken mirror symmetry~\cite{berezhiani95,berezhiani96}
the scale of electroweak symmetry breaking in the shadow world was
taken to be 30 times larger than our electroweak scale.
Correspondingly shadow neutrinos would be about 1000 (i.e. $30^2$)
times heavier than our neutrinos, and if the latter have the mass in eV
range the former could have keV mass and contribute to warm dark
matter (see sec.~\ref{11dm}). Due to asymmetric inflationary
reheating that could give different temperatures to our and 
the shadow
worlds, the number density of the heavy shadow neutrinos would be
sufficiently small to avoid a contradiction with the Gerstein-Zeldovich
limit. Smaller temperature of the shadow world would make it non-dangerous
at BBN as well.

In the model of refs.~\cite{berezhiani95,berezhiani96}
interactions between the worlds, in addition to gravity,
proceeds through mixing between neutrinos, so the shadow neutrinos
would be perceived in our universe as sterile neutrinos slightly mixed
with the active ones. This idea was proposed earlier in
refs.~\cite{foot92,foot95v} in the framework of 
the exact parity model. In a simultaneous paper~\cite{akhmedov92} it was
argued that even if the original coupling between the worlds was only
due to gravity, quantum gravity effects on 
the Planck scale would induce
mixing between our and mirror neutrinos. Thus, the mixing between
neutrinos in the two worlds is inevitable or, at least, quite natural
and they would be produced in the early universe
through coherent oscillations that are discussed in
sec.~\ref{nuosceu}, and where the BBN bound on the oscillation
parameters are presented.

The implications of mirror neutrinos for the early universe 
cosmology in the exact parity models are considered
in detail in ref.~\cite{foot99c},
see also~\cite{volkas00m}. In the simplest versions of these models one
would expect that the cosmological energy density of mirror particles
and ours should be equal. 
In particular, they should be equal
at BBN. The contribution of mirror particles in this case corresponds to
the effective number of neutrino species slightly over 9,
and is definitely excluded (see sec.~\ref{s:bbn}). However, in the case
of resonance oscillations between ours and mirror neutrinos
a large lepton, especially electron, asymmetry can be generated in our
world (see secs.~\ref{sec:leptas}, \ref{sec:nuoscbbn}). This asymmetry
would have a strong impact on BBN, either enlarging or diminishing the
effective number of neutrino species. The analysis of oscillations
including 3 active and 3 sterile flavors was performed in
ref.~\cite{foot99c} (for the earlier work see ref.~\cite{foot96vm}),
where it was shown that the exact parity model
was consistent with BBN for a large region of the oscillation
parameter space. On the other hand, even in the case of exact mirror
symmetry in the Lagrangian, 
the cosmological evolution of the different
worlds could be different (a kind of spontaneous symmetry breaking)
and the energy density of mirror particles might be much smaller
than ours.

\section{Neutrino and large extra dimensions.\label{s:largeD}}

It was suggested several years ago~\cite{arkani98} that the characteristic
scale of gravity might be a few TeV instead of the usual Planck scale
$10^{19}$ GeV. This could be realized if there are extra dimensions in
addition to the usual $3+1$. The Standard Model (SM) fields live in the
$(3+1)$-dimensional brane, while gravity and possibly some other
fields could propagate in the bulk including the brane
plus additional dimensions. The latter are compact and the compactification
scale could be as large as
a fraction of mm. The standard left-handed neutrinos
localized in the bulk could mix with SM-singlet fermions propagating in
the bulk~\cite{dienes98}-\cite{ioannisian99v} and would be interpreted in
our world as sterile neutrinos. In the model of
refs.~\cite{dvali99,barbieri99} the mass of active neutrinos and the
non-diagonal matrix elements of the mass matrix are determined by one
free parameter,
$m_{i}$, while the diagonal entries for the bulk neutrinos are
equal to $n/R$, where $R$ is the size of the extra dimensional manifold.
According to ref.~\cite{barbieri99} the mass matrix of mixed bulk and brane
neutrinos has the form:
\be
M_i = \pmatrix{m_i & \sqrt{2} m_i & \sqrt{2} m_i & \cdots \cr 0 & 1/R & 0
& \cdots \cr 0 & 0 & 2/R & \cdots \cr
\vdots & \vdots & \vdots & \ddots}
\label{mmm}
\ee
The model describes one active neutrino with mass $m_{\nu_a}$
and the infinite tower of sterile ones with the masses $n/R$ and
mixing angles essentially given by $\sin \theta \sim m_{\nu_a} R$ (if
the latter is small, $m_\nu R \ll 1$). For details see e.g.
ref.~\cite{barbieri99}.

The infinite tower of sterile neutrinos might be dangerous for BBN. This
problem was sketched
in ref.~\cite{barbieri99}, where it
was argued 
that the effect of these $\nu_s$ is not catastrophically large
and even might be compatible with BBN. The essential point on which this
conclusion is based is that the mixing angle diminishes with the rising
number of excitation, $n$, so the effect of all the tower on BBN is
at most as that of one neutrino species.
An accurate treatment demands solving the
infinite system of the coupled kinetic equations for the density 
matrix, and that has not yet been done.

A more extensive analysis was 
carried out
in ref.~\cite{abazajian00p}, where
not only BBN but also data on CMBR, structure formation and diffuse
photon background were taken into consideration. The effective number of
neutrino species at BBN corresponding to $j$-th Kaluza-Klein (KK)
excitation of the bulk neutrinos,
analytically estimated in this work, is
\be
\Delta N_{\nu_j} \sim 10^{-3} \left( {m_\nu \over 1\,\,{\rm eV}}\right)^2
\left( { g_*^f \over g_{*k}^p }\right)
\label{deltaKbulk}
\ee
where $m_\nu$ is the active neutrino mass, and the last factor approximates
the dilution effect caused
by the ratio of relativistic species at the
active neutrino decoupling and at the maximum production of the
Kaluza-Klein mode $j$. The temperature of the latter is approximately:
$T_{max} = 133\,\,{\rm MeV}\,\,(m_j/1\,{\rm keV})^{1/3}$
(see eq.~(\ref{tprodnus})).
The result~(\ref{deltaKbulk}) noticeably differs from that of
ref.~\cite{barbieri99} and more work is desirable to resolve the
controversy.

Taken together, the cosmological constraints discussed in
ref.~\cite{abazajian00p} deal a serious blow to the simple model of
bulk neutrinos mixed with the usual active ones. However, as the
authors noted there could be simple modifications of the models
that might allow circumventing the constraints. First, the mixing angle
and life-time of KK neutrinos are model dependent and it is possible to
modify the particle physics framework to avoid contradictions with
cosmology. Second, the result depends upon the geometry of the bulk. In
particular, the compactification would not necessarily be toroidal one
or there might be additional branes in the bulk that could change the
decay properties of heavier neutrinos. Thus cosmology can definitely
restrict certain models of sterile neutrinos coming from large extra
dimensions, but at the present time it cannot rule out the 
general idea.

The BBN constraints on the models with large extra 
dimensions~\cite{barbieri99,abazajian00p} have been derived for the 
case of $\geq$ 2 extra dimensions. In ref.~\cite{goh01} the models 
with one extra dimension and with the string scale $\sim 10^9$ GeV 
have been studied. It was shown there that such models are compatible 
with BBN and the mixing of bulk to active neutrinos is in the range
interesting for solar neutrino oscillations.

\section{Neutrinos and lepto/baryogenesis. \label{s:bs}}

If cosmological baryon asymmetry originated from previously
created lepton asymmetry~\cite{fukugita90}, one can obtain quite
restrictive bounds on neutrino mass in many realistic models of
particle physics. In short, the idea is as follows. It is believed
that sphaleron processes~\cite{kuzmin85}, operating above the electroweak
phase transition, would destroy all preexisting cosmological
charge density of $(B+L)$.
In principle, the same processes could generate $(B+L)$
at lower 
temperatures when the phase transition was in process. 
This would require deviations from thermal equilibrium, which could be
effective only if the phase transition was first order.

Heavy Higgs makes
this option practically excluded. For the reviews see
refs.~\cite{adbs}, \cite{cohen93}-\cite{riotto99}.
If electroweak 
processes are able only
to destroy preexisting asymmetry
and not to create one, we would need either low temperature baryogenesis
or some mechanisms to create a non-zero $(B-L)$ prior to EW phase 
transition.
This difference, $(B-L)$, is conserved and electroweak processes can
transform it into some nonvanishing and generically non-equal lepton
and baryon asymmetries (see below, eq.~(\ref{BplusL})). This way, 
the observed baryon asymmetry of the universe might be generated.
This scenario is reviewed in 
refs.~\cite{pilaftsis98,buchmuller99,buchmuller01}.

A possible mechanism of creation of lepton asymmetry is a decay of
a heavy Majorana neutrino~\cite{fukugita86}, $\nu_M$.
The existence of such a particle permits 
a natural explanation of the small masses of observed 
left-handed neutrinos by a mixing with the heavy Majorana
companion (the so called see-saw mechanism~\cite{see-saw}).
Lepton asymmetry, created by out-of-equilibrium decays of heavy $\nu_M$,
was generated at a period 
when the rate of lepton charge production:
\be
\Gamma_L =\Gamma_M n_{\nu_M}/T^3 \approx
\alpha_M m_M \exp({-m_M/T})(m_M/T)^{3/2}
\label{}
\ee
was of the order of the
Hubble expansion rate, $H\sim T^2/m_{Pl}$. Here we assume that the
decay rate of $\nu_M$ is equal to $\Gamma_M = \alpha_M\, m_M$.
Since the magnitude of the
lepton asymmetry should not be smaller than the observed baryon
asymmetry we should request that the number density of $\nu_M$ must
be larger than $10^{-9}$ of the entropy density
at the moment of equilibrium breaking, when
$T=T_f$. It means that
$\exp (-m_M /T_f)(m_M/T_f)^{3/2} > 10^{-9}$ or
$m_M > 10^{-9} \alpha_M (m_M /T_f)^2 m_{Pl} \sim (10^8-10^{10})$ GeV.
We have assumed
here that the freezing temperature $T_f \sim (0.1-0.01) m_M$
and $\alpha_M \sim 10^{-2} - 10^{-4}$.
This lower limit on $m_M$ can be translated into an upper limit on the
mass of light neutrinos because the latter, according to 
the see-saw mechanism
is inversely proportional to $m_M$. In this way a very restrictive
upper bound
$m_\nu \sim m_l^2/m_M < 3\cdot 10^{-3}$ eV can be obtained (here $m_l$
is the charged lepton mass). 
One should keep in mind, however, that the mass of the heavy 
Majorana neutrino is strongly model dependent. As argued in
ref.~\cite{pilaftsis97} the isosinglet Majorana mass may be in
the interval from 1 TeV up to grand unification scale depending
upon the mechanism of CP violation and the flavor structure of 
neutrino mass matrix.

Another line of arguments is based on the request that lepton
asymmetry, or $(B-L)$, generated at the earlier stage is not destroyed
by the simultaneous action of $L$-nonconserving processes
and sphaleron interactions. If both types of processes are
efficient and thermal equilibrium is established, then both $B$ and $L$
must vanish. The sphaleron interactions are 
known to be in equilibrium in the temperature interval
\be
(T_{EW} \approx 100\,\, {\rm GeV}) < T < (T_{SPH} \approx 10^{12}\,\,
{\rm GeV}).
\label{TewTsph}
\ee
In addition to the heavy neutrino decays, leptonic charge
non-conservation  with $\Delta L =2$ could originate
e.g. from the effective coupling of lepton of flavor $i$, ($l_i$) and
Higgs ($H$) fields (see e.g.~\cite{buchmuller99}):
\be
{\cal L}_{\Delta L = 2} = g_{ij} H^2 l^T_i Cl_j +\,{\rm h.c.}
\label{deltaL2}
\ee
This interaction could be generated by the exchange of heavy Majorana
neutrino. The Yukawa coupling constants $g_{ij}$ enter the mass matrix
of light neutrinos which appears in the phase where the Higgs field
acquires a
non-vanishing vacuum expectation value, $\langle H \rangle \neq 0$.
Hence the rate of reactions with $\Delta L \neq 0$ is proportional
to the light neutrino mass squared. From the condition that $B$ and
$L$ are not destroyed in the temperature interval~(\ref{TewTsph})
we obtain the limit~\cite{buchmuller99}:
\be
\sum_{i} m_{\nu_i}^2 < \left[ 0.2 \,{\rm eV}\, \left({T_{SPH} \over
T_B }\right)^{1/2} \right]^2
\label{sumimnu}
\ee
where $T_B$ is the baryogenesis temperature, which is usually taken
equal to $T_B=T_{EW} = 100$ GeV. In the first paper~\cite{fukugita90}
a slightly weaker limit, $m_\nu < 50$ keV, was derived.
The more stringent bound~(\ref{sumimnu}) was obtained in
ref.~\cite{harvey90} where the anomalous non-conservation of fermion
number was taken into account at the temperatures above the
electro-weak phase transition.
In a series of papers~\cite{nelson90}-\cite{cline93} much
stronger bounds on neutrino masses than~(\ref{sumimnu})
have been derived from the condition of successful
baryogenesis in the framework of concrete particle physics models.
In particular, if one takes $T_B = 10^{10}$ GeV which is a typical
leptogenesis temperature, the upper limit on the light neutrino
mass would be very strong, $m_\nu < 2$ eV.
A more general, model independent limit, similar, to~(\ref{sumimnu})
can be found from the condition~\cite{sarkar97u} (for a review
see~\cite{sarkar99u}) that the lepton number
non-conserving processes $W^\pm + W^\pm \rar e^\pm + e^\pm$
at $T \sim m_W$ do not wash out lepton charge generated earlier by
$\nu_M$ decays. In other words the reaction rate should be smaller than
the Hubble expansion rate:
\be
\Gamma ( W\,W \rar l_i\,l_j ) =
{\alpha_W^2\,\, (m_\nu)^2_{ij} T^3 \over m_W^4} <
5.44 \sqrt{{g_* \over 10.75}}\,{T^2 \over m_{Pl}}
\label{gammaL}
\ee
where $l_i$ is a charged lepton ($e,\,\mu$ or $\tau$) and
$m_\nu$ is an entry (not necessary diagonal one) of the Majorana
mass matrix of light neutrinos. The reaction rate is estimated at
$T\sim m_W$ where it has the maximum value. From the
condition~(\ref{gammaL}) we find
\be
(m_\nu) < 20\,\,{\rm keV}
\label{mnuij}
\ee

If the above mentioned conditions 
 are fulfilled, then after the
electroweak phase both baryon and lepton asymmetries would be created.
Naively one would expect that, since $(B+L)$ is not conserved by
sphalerons, then in thermal equilibrium $B+L=0$ or, in other words,
$B=-L = (B-L)_{in}/2$. The difference $(B-L)$ is conserved by
sphalerons and, as we mentioned above, should be generated at an earlier
stage by some other
$(B-L)$-non-conserving interactions. However, generally the statement
$(B+L)=0$ is not true. Some combinations of non-conserved charges in thermal
equilibrium must vanish but this depends upon the particle content of the
theory and this vanishing combination is not necessarily $(B+L)$. 
It can be easily seen from the following example.
Let us assume that there
is only one generation of left-handed quarks, $u$ and $d$ (which have
three colors) and leptons, $(\nu, l)$. There are weak interaction
reactions: $u+\bar d \lrar \nu + \bar l$ and all crossed ones, where
$l$ is a negatively charged lepton.
There are also two processes induced
by sphalerons which break lepton and baryon numbers:
$ uudl\lrar {\rm vacuum}$ and $ udd\nu \lrar {\rm vacuum}$. In a realistic
case, the reactions include more particles, but the main features are the
same. In equilibrium the following relations between chemical potentials
hold:
\be
2u +d + l =0,\,\,\, u +2d +\nu =0
\label{udl-eq}
\ee
where for chemical potentials we use the particle symbols, i.e.
$u\equiv \mu_u$, etc.

In the case of small chemical potentials the corresponding charge density
is a linear function of the potentials, and thus the baryonic, leptonic,
and electric charge densities can be written as:
\be
B\sim u+d,\,\,\, L \sim l+\nu,\,\,\, {\rm and}\,\,\,
Q \sim 2u-d -l
\label{BLQ}
\ee
Here we made use of the fact
that the baryonic charge of an individual quark is
$1/3$ but there are three quark colors having equal chemical potential.
Using eqs.~(\ref{udl-eq},\ref{BLQ})
we easily find $(B+L) = -(B-L)/2$. A realistic case includes right-handed
quarks and charged leptons, intermediate bosons, and higgses. It was
considered in ref.~\cite{harvey90} where the following relations between
$B$ and $L$ were obtained:
\be
B+L = -{6N+5m \over 22N + 13m}\, (B-L),\,\,\,
B = {8N + 4m \over 22N + 13m} \, (B-L),
\label{BplusL}
\ee
where $N$ is the number of quark-lepton families and $m$ is the number
of Higgs doublets.

This result is obtained at high temperatures, above
the electroweak phase
transition and under the assumption that all particle species are in thermal
equilibrium. The latter assumption was critically reanalyzed in
refs.~\cite{cline93b,cline93c}. It was found that right-handed electrons
could be out of equilibrium because of their small Yukawa coupling to
Higgs bosons. On the other hand, anomalous sphaleron interactions are
effective
only for left-handed particles. Because of that, lepton asymmetry would be
preserved in the sector of right-handed electrons and not erased as was
suggested in the earlier papers quoted above.
This result leads to a significant
weakening of the previously found bounds. According to the detailed
calculations of ref.~\cite{cline93c} the upper limit on the neutrino
mass is about 20 keV, an order of magnitude weaker than the earlier
estimates by the same authors~\cite{cline93b} and we essentially return
to (\ref{sumimnu}) with $T_B = 100 $ GeV.

There has been
a recent burst of activity in the field, stimulated by the
observed neutrino anomalies and a related 
indication to nonzero neutrino
masses. In ref.~\cite{ellis99ln} the scenario of baryo/leptogenesis based
on the flipped $SU(5)$ model was considered. In ref.~\cite{buchmuller00}
two models with the symmetry groups $SU(5)\times U(1)_F$ and
$SU(3)_c\times SU(3)_L\times SU(3)_R\times U(1)_F$, where $U(1)_F$
is the flavor group, were discussed in detail with an accurate solution
of kinetic equations governing the evolution
of the asymmetry. Relations between
leptogenesis, neutrino masses, and (supersymmetric hybrid) inflation
were discussed in refs.~\cite{lazarides99v,lazarides99}. Reheating after
inflation and related gravitino problem was analyzed in
ref.~\cite{delepine99}. It was shown that in minimal supersymmetric
extension of the standard model, almost all existing scenarios of
leptogenesis and neutrino masses, except for the one
involving 
right-handed sterile neutrinos, are ruled out for a large range
of gravitino mass. The models with Abelian and discrete family
symmetries and their impact on leptogenesis and neutrino masses were
studied in ref.~\cite{berger99}. A scenario of baryo/leptogenesis
with degenerate neutrinos was further considered in ref.~\cite{chun00}.
A modification of see-saw mass equation in left-right symmetric theories
with two Higgs triplets was considered in the papers~\cite{joshipura01}.
Baryon asymmetry would be successfully described if neutrino masses are
not smaller than $10^{-6}-10^{-8}$ eV.

Further development in this area was related to a modification 
of the
leptogenesis scenario, 
which did not necessarily proceed through
non-equilibrium decays of heavy Majorana neutrinos. Such a new
version was proposed in refs.~\cite{ma98s,sarkar98u}. It was
assumed that, instead of right-handed
neutrinos, there exist additional heavy Higgs scalars, and lepton number
was generated through decays of these new heavy Higgs particles, whose
interactions explicitly break lepton number.
The model permits the accommodation of light sterile neutrinos strongly mixed
with the usual active ones. The masses of neutrinos were taken
in eV
down to $10^{-3}$ eV range and all neutrino anomalies were explained.
It is unclear however, if the BBN constraints can be satisfied and
how the new SNO data would change the parameters of the model.

In ref.~\cite{ma99s} a left-right symmetric model
was considered, where both
discussed above possibilities
of leptogenesis could be realized. As is
shown in the paper, successful leptogenesis requires
the mass of the right-handed
neutrinos to be quite high, $m_{N} \geq 10^{16}$ GeV, if $m_N$ exceeds
the mass of right-handed intermediate bosons. However in supergravity models
this option is excluded because of the cosmological gravitino problem.
The case of $m_N < m_{W_R}$ is more realistic and could lead to successful
leptogenesis.

Successful baryogenesis could proceed even without generating a
non-vanishing $(B-L)$, as was necessary for the scenarios discussed above.
A model of this kind is presented in ref.~\cite{dick99}. According to
the model, either a good old GUT baryogenesis operated at the
GUT scale and created a non-zero $(B+L)$ or some other processes
generated lepton asymmetry
in the left-handed lepton sector which might be compensated by the
asymmetry in right-handed neutrino sector. The latter could proceed
e.g. in heavy particle decays
even without lepton number violation and no heavy Majorana lepton 
would be necessary.
By assumption the decays of heavy 
bosons create both left- and right-handed quarks and leptons, in
particular, $\Phi \rar \bar l_L + \nu_R$. Hence the $(B+L)$-asymmetry
or just $L$-asymmetry
produced in these decays is shared between left and right handed
quarks and leptons. At smaller temperatures sphaleron processes become
effective. However right-handed particles do not interact with sphalerons
and the total, left plus right, asymmetry 
could be equilibrated only by the
Yukawa coupling of left- and right-handed fermions to Higgs bosons.
They would wash out all $(B+L)$ if $(B-L)=0$ and if the Yukawa coupling
is sufficiently strong. This is not the case for neutrinos if their mass
is smaller than 10 keV (see sec.~\ref{ssec:nur}). In this case
the lepton asymmetry stored in right-handed neutrinos, with Dirac mass,
does not communicate with the asymmetry in the sector of left-handed
particles. Correspondingly $(B-L)_L$ in this sector is non-vanishing and
as a result the baryon asymmetry processed by sphalerons becomes
non-zero as well, see eq.~(\ref{BplusL}).

A new idea that lepton asymmetry might be produced through CP-violating
oscillations of sterile neutrinos was explored in ref.~\cite{akhmedov98}.
This asymmetry is then transferred to ordinary neutrinos through their
Yukawa coupling with sterile ones. The lepton asymmetry in the active
neutrino sector produces baryon asymmetry at electro-weak scale through
sphaleron processes, as discussed above. In this model the total lepton
number (active plus sterile) is conserved and the redistribution of
lepton asymmetry between active and sterile neutrinos leads
to generation of baryon asymmetry.
An important ingredient is the freezing of sphaleron
transformation. Otherwise all asymmetry in the sterile sector would be
transformed into the active sector
and since $L_{tot} =0$ the net result
would be also zero. From the condition that sterile neutrinos decayed
before the nucleosynthesis epoch, their mass should be larger than
$\sim$GeV. On the other hand, the condition that the Majorana mass does
not wash out baryon and lepton asymmetry leads to the upper bound
on the mass, $M_s \ll 100$ GeV. In the model of ref.~\cite{akhmedov98}
it means that the Yukawa coupling constants are bounded by
$ h^2 \ll 10^{-10}$ and the mass of active neutrinos, which is given by
$ m_a = h^2 v^2 /M_s$, should be in the range $m_a =(10^{-2} - 10^3)$ eV.
Depending upon the version of the scenario this limit could be more
restrictive. Two of the active neutrino species would have masses in
the range $(10^{-6}-10^{-1})$ eV and the third one might have a mass in eV
range and make hot dark matter.

The idea to generate $(B-L)$ through Affleck-Dine mechanism~\cite{afdi}
with subsequent electro-weak reprocessing was suggested in the recent
paper~\cite{berezhiani01} in a supersymmetric hybrid inflation model.
The right-handed neutrino superfield naturally appears in the model
to fine-tune dynamically the necessary initial conditions for inflation.
According to the model the masses of light neutrinos created by the
see-saw mechanism are in the range indicated by the data on neutrino
anomalies, if the latter are interpreted as manifestations of neutrino
oscillations.

In ref.~\cite{bento01b} the idea to generate lepton asymmetry in
scattering processes was explored. Usually light particles that
efficiently participate in scattering are in thermal equilibrium and no
charge asymmetry can be generated. However in the version of
ref.~\cite{bento01b} leptons of our world communicated with leptons in
a hidden sector, e.g. mirror or shadow world (see sec.~\ref{s:mirror}).
There is no equilibrium between these worlds and lepton asymmetries in
both of them could be generated.

In all the papers described above the characteristic scale of generation
of lepton asymmetry was very high, roughly speaking $10^9-10^{10}$ GeV or
even higher. In ref.~\cite{nasri01} the model of low energy leptogenesis
was suggested at the expense of extending the standard model of particle
physics by adding three right-handed neutrinos with the mass about 10 TeV
and two new charged Higgs fields which are singlets under $SU(2)_L$.
This model operates at the TeV-energies accessible to new
accelerators and can be checked in upcoming experiments.

Another possible way of generating small neutrino masses is realized
in the Zee model, where the masses are induced by radiative corrections
and leptonic charge is broken explicitly~\cite{zee80}.
If the non-conservation of leptonic charge in this model is
strong at electroweak scale, then both $B$ and $L$ would be washed
out. However, as is shown in ref.~\cite{haba01}, this is not the
case because of 
the approximate conservation of the difference of lepton
numbers $L_e - L_\mu - L_\tau$. Thus earlier generated $B_L$ would not
be destroyed by the combined action of sphalerons and explicit
$L$-nonconserving interaction in the Zee model.

Leptogenesis in theories with large extra dimensions was considered in
ref.~\cite{pilafsis99}. It was assumed that in addition to the
particles of the standard model localized on the brane, there exists an
isosinglet neutrino field living in the bulk (see sec.~\ref{s:largeD}).
For compactified extra dimensions this field describes an infinite
Kaluza-Klein tower of Dirac neutrinos which, after an introduction of
leptonic charge non-conservation, split into pairs of nearly
degenerate Majorana neutrinos. Each pair of the Majorana neutrinos
presents a strongly mixed two-level system producing a large C- and
CP-violation. Lepton asymmetry generated by decays of these Majorana
neutrinos could be very big and could transform into baryon asymmetry.
For successful implementation of the scenario the universe should
be reheated above 5 GeV.

There is a wealth of literature
on the subject and many papers on the field
may have been
omitted here. In specific models of particle physics the upper
limits on the masses of light neutrinos could be much more restrictive
than those presented in this section. However the bounds are strongly
model dependent and sometimes do not have an important impact on
cosmology. So we confine our discussion to the presented
material. More references can be found in the quoted above papers,
especially in the review
ones~\cite{buchmuller99,buchmuller01,buchmuller01b}.

A few more scenarios relating cosmological baryon asymmetry with neutrino 
masses have appeared recently~\cite{brahmachari01}-\cite{falcone01}.
A supersymmetric model of ref.~\cite{brahmachari01} with additional
supermultiplets and with string scale
unification at $\sim 10^{13}$ GeV may lead to a cosmological lepton 
asymmetry through the decay of this heavy superfield and to a conversion 
of this asymmetry into the observed baryon asymmetry. The masses of light 
neutrinos arising in this model naturally fit the atmospheric and solar
neutrino data. In ref.~\cite{hambye01} a general analysis of models of
lepto/baryo-genesis at the low energy scale 1-10 TeV is presented.
Several known models of leptogenesis are discussed and a new model
based on three-body decays of right-handed neutrinos is proposed. The
latter allows successful lepto/baryo-genesis and neutrino mass generation
at low scale. A relation between the Dirac neutrino mass matrix and the
magnitude of the baryon asymmetry is studied in ref.~\cite{falcone01}. As
is shown in the paper, if the neutrino mass matrix is related to quark or
charged lepton mass matrix the baryon asymmetry would be 2-3 orders of
magnitude smaller than the observed one. For successful baryogenesis 
a less pronounced hierarchy of neutrino mass matrix is necessary.


\section{Cosmological neutrino background and ultra-high energy cosmic
rays. \label{s:cr-cnb}}

The observation of ultra high energy cosmic rays (UHECR)
with $E> 10^{20}$ eV poses a serious problem to the standard theory
of the origin and propagation of energetic cosmic particles. At the
present day more than twenty such events have been observed by
different groups~\cite{uhecr}. It is traditionally assumed that the
primaries that induce energetic atmospheric showers are protons formed
somewhere in violent cosmic sources, e.g. in active galactic nuclei.
However such protons cannot come from a very large distance because of
the Greisen-Zatsepin-Kuzmin cutoff~\cite{gzk}. Protons with the energy
$E>E_{GZK} = 5\cdot 10^{19}$ eV strongly interact with CMBR, producing
pions in the resonance reaction with the excitation of $\Delta$-isobar:
$p + \gamma \rar \Delta \rar N + \pi$, where $N$ is either proton or
neutron. The necessary energy for this process can be roughly found from
the condition $s\equiv (p_p +p_\gamma)^2 \sim 2E_p E_\gamma >2m_pm_\pi$
which gives a result rather close to $E_{GZK}$ presented
above. Due to this inelastic process a proton loses half of its
energy at a distance of approximately 20 Mpc. Since no possible sources
of high energy cosmic rays are seen in the directions of the primaries
of the ultra high energy events up to the distance $\sim 100$ Mpc,
and the interstellar magnetic fields are too weak ($<$ a few micro-gauss)
to bend protons with such a high energy, the
observations~\cite{uhecr} present a serious challenge to the standard
theory of the origin of UHECR, for a review see~\cite{bhatta98}.

The directions of arrival of cosmic rays with $E>E_{GZK}$ 
are not correlated with Galactic or Supergalactic plane~\cite{uhecr}. 
Possible sources of the UHECR are more or less uniformly distributed over 
the sky. However, there are pairs and triplets of UHECR coming from the 
same direction on the sky within the resolutions of 
AGASA \cite{AG1,agasa-web} and Yakutsk (ref. 2 in~\cite{uhecr}).
This small scale clustering component is statistically significant at the
level of 4.5 $\sigma$  and suggests that sources of UHECR are 
point-like \cite{anis1,anis2}.
Another important fact is that events in the clusters have
different energy and uncorrelated arrival times. 
This means that clustered 
UHECR particles are neutral. Natural candidates for such particles 
would be photons, however the 
photons with energies  $E \sim 10^{19} - 10^{20}$~eV lose energy 
within 50 Mpc due to
creation of $e^{\pm}$ pairs on CMBR and cosmic radio background. 

A significant (almost 5 $\sigma$) correlation of UHECR in both AGASA and 
Yakutsk data with BL Lacertae objects (quasars, with jet beamed 
in our direction and no strong emission lines in their spectra)
found recently in ref.~\cite{corr_bllac} creates even more puzzles.
The nearest such objects are located at the distance 150 Mpc 
($z=0.03$), well beyond the GZK volume. The largest part of the known
BL Lacertae are located at moderate $z \sim 0.1$, or unknown redshifts.
If they are indeed the sources of UHECR then they should produce photons
with extremely high energies $E>10^{23}$ eV. Such photons could propagate
several hundred Mpc, constantly losing energy, 
and create secondary photons inside the GZK volume. 
These particles would be the UHECR registered above the GZK 
cutoff~\cite{photons}.
However, this model requires extremely high energies of primary photons, 
a very small magnitude of (unknown) extragalactic radio background,
and extremely small extra galactic magnetic fields (EGMFs),
$B < 10^{-12}$~G. Moreover, if the significant correlation with  
sources at high redshifts, $z > 0.2$, is found, this model will 
be ruled out.

The only known Standard Model particle that is able to 
travel from high redshift sources, $z \sim 1$, is the neutrino. 
The mean free path of very energetic neutrinos with respect
to the resonance production of $Z$-boson by scattering on the
cosmic neutrino background (CNB) is close to the
horizon size~\cite{weiler82,roulet93}. To excite the $Z$-resonance the
energy of the ultra high energy (UHE) neutrinos should be equal to:
\be
E_{UHE} = m_Z^2 /2E_{CNB} \approx 4\cdot 10^{21}\,\,{\rm eV}
(1\,\,{\rm eV} /E_{CNB})
\label{euhe}
\ee
where $E_{CNB}$ is the energy of the cosmic background neutrinos. If they
are massless then
$\langle E_{CNB} \rangle \approx 3T_\nu \approx 5\cdot 10^{-4}$ eV. For
massive neutrinos with $m_\nu > T_\nu$, $E_{CNB} = m_\nu$.

The energy averaged cross-section is 
\be
\bar \sigma_{Z} = \int (ds/M_Z^2) \sigma (\bar\nu \nu \rar Z) =
2\pi\sqrt2 G_F = 4\cdot 10^{-32}\,{\rm cm}^2
\label{sigmanunu}
\ee
Because of the resonant nature of the process the cross-section contains
only first power of $G_F$ and is much larger than typical weak interaction
cross-sections. The mean free path of UHE neutrinos with respect to this
reaction is
\be
l_{free} = 1/(\sigma_Z n_\nu)
\approx 5\cdot 10^{29}\,\,{\rm cm}\, (n_\nu /55 {\rm cm}^{-3})^{-1}
\approx 150\,\,{\rm Gpc}\,\,(n_\nu/55 {\rm cm}^{-3})^{-1}
\label{lfreenu}
\ee
where $n_\nu$ is the neutrino number density; it is normalized to
the standard average cosmological number density of neutrinos
$n_\nu^{(0)}=55/{\rm cm}^3$, see eq.~(\ref{nnungamma}).

Thus a possible source of ultra high energy cosmic ray events could be
the decay of $Z$-boson produced by very energetic neutrinos annihilated
on CNB within 100 Mpc. The primary energetic neutrinos could be produced
by active galactic nuclei at very large distances. This explanation was
suggested in the papers~\cite{fargion97,weiler97}. The $Z$-boson produced
in the reaction $\bar \nu \nu \rar Z$ would have gamma-factor
$\gamma = m_Z /2m_\nu = 4.5\cdot 10^{10}({\rm eV}/m_\nu)$. The average
proton multiplicity in $Z$-decay is 2 and the proton energy in the rest
frame of $Z$-boson is roughly 3 GeV. Hence the
energy of protons from such a source would be about
$1.3 \cdot 10^{20}({\rm eV}/m_\nu) $ eV, which is very close to the
registered signal for $m_\nu \sim 1$ eV.
If this mechanism is indeed operative, the registration of UHECR could
mean that the cosmic neutrino background has been discovered. 
The energetic neutrinos can be considered as messenger fields from distant
violent sources. 

However, the ``Z-burst'' mechanism is severely constrained by at least
two types of observational data. First, there are upper limits on the
UHE neutrino flux, based on non-observation of horizontal
air showers by the old Fly's Eye~\cite{baltrusaitis} 
or by the AGASA~\cite{AgAsA} experiments 
and from non-observation of radio pulses that would
be emitted from the showers initiated by the UHE neutrinos
on the Moon rim~\cite{goldstone}. Second, even if the sources 
exclusively emit neutrinos, the electroweak interactions would also 
produce photons and electrons initiating electromagnetic (EM) cascades 
and transferring the injected energy down below the pair production 
threshold for the energetic photons on CMBR~\cite{bhatta98}. 
The cascades would
give rise to a diffuse photon flux in the GeV range which is 
constrained by the flux observed
by the EGRET instrument on board of the Compton $\gamma-$ray
observatory~\cite{egret}.

The first self-consistent calculation of the Z-burst model was done 
in ref.~\cite{yoshida98}. It was assumed there that the 
distribution of neutrino 
sources evolves with redshift as $(1+z)^3$ or in the way similar 
to the evolution of active galaxies. 
It was shown that in this case the secondary photons with energies 
$E< 100$~GeV would overshoot the measured EGRET flux several times.
This means that the "Z-burst" mechanism 
in its simplest version contradicts the data.

A possible solution to this problem  is to 
increase the local neutrino densities by a factor $N > 20$ 
on scales $l \sim 5$ Mpc \cite{yoshida98}.
The probability of neutrino interactions would locally increase 
and for the same flux of UHECR (normalized to the
experimental data) the secondary EM flux in the EGRET region of energies 
would be below the measured values.     
A development of the local overdensity idea and its different 
applications were discussed 
in refs.~\cite{fargion01} and \cite{fargion01a}.

The necessary flux of primary energetic neutrinos,
which would excite $Z$-resonance and would
agree with the data, depends upon the number density of CNB
and the neutrino mass. The cosmological number density of relic neutrinos
in the case of vanishing lepton asymmetry is $n_\nu^{(0)}=55/{\rm cm}^3$.
However massive neutrinos could cluster around gravitationally bound
astronomical systems and their number density might be much higher.
According to estimates presented in ref.~\cite{weiler99} the enhancement
factors for 1 eV neutrinos are respectively $10^2$, $10^3$, and $10^4$
for galactic supercluster, cluster, and galactic halo. In the subsequent
paper~\cite{pas01} the results are approximately an order of magnitude
weaker. There is an upper limit to the number density of clustered
neutrinos due to their Fermi statistics~\cite{tremaine79} (see
sec.~\ref{normalnu}). The number density of degenerate neutrinos is
$n_\nu = p_F^3/6\pi^2$, where $p_F$ is the Fermi momentum. The average
velocity is equal to $ \langle V \rangle = p_F/ 4m_\nu$. Hence
\be
{n_\nu \over n_\nu^{(0)}} \leq 4\cdot 10^2
\left( {m_\nu \over {\rm eV}}\right)^3\,
\left({ V \over 200\,{\rm km/sec}}\right)^3
\label{nnuclstr}
\ee
where $V$ is the virial velocity, see discussion before eq.~(\ref{mnutg}).
This bound does not permit a too-large enhancement of $n_\nu$.

The UHE neutrino flux, which is necessary for the explanation of the
observed cosmic rays events by the $Z$-burst mechanism, can be estimated
as follows. One can see from the AGASA data~\cite{agasa-web} that
the flux of UHECR with $E\sim 10^{20}$ eV is
\be
F E^2 \approx 5 {\rm eV /sec/sr/cm}^2.
\label{fluxE2}
\ee
It corresponds to the energy density of
the UHECR $\rho_{UHECR} \approx 2\cdot 10^{-9}$ eV/cm$^3$. The rate of
production of such energetic cosmic rays can be estimated assuming that
they were produced throughout all cosmological time, $t_c = 1/H$:
\be
\dot \rho_{UHECR} = H\rho_{UHECR} = 0.5\cdot 10^{-26}
{\rm {eV \over sec\,cm^3}} = 0.7\cdot 10^{43}\,\,
{{\rm erg \over Mpc^3\,\, year}}
\label{dotuhecr}
\ee
Now we have to take into account 
the fact that these UHECR are
secondary, produced by the interaction of $10^{21}$ eV neutrinos with
CNB inside roughly 30 Mpc. It means that the energy production rate of the
primary UHE neutrinos should be $\sim 10 (l_{free}/30\,{\rm Mpc})$ times bigger
than $\dot \rho_{UHECR}$~(\ref{dotuhecr}). It gives
\be
\dot \rho_{prim} = 2\cdot 10^{47} (n_\nu^{(0)} / n_\nu)\,
{{\rm erg \over Mpc^3\,\, year}}
\label{dotrhoprim}
\ee
Such a large rate may require an unusual mechanism of production 
of UHE neutrinos.

The detailed analysis of constraints on the neutrino flux which 
came from non-observation of horizontal air-showers was done in 
ref.~\cite{pillado99}. The authors made a comprehensive study of 
possible observational signatures of energetic neutrinos taking 
their spectral index and the local
neutrino density enhancement as free parameters.
It was shown that the existing data on horizontal showers practically 
exclude clustering of background 
neutrinos with a small halo size for explanation of the UHECR. 
Marginal room is left for models
with low neutrino mass, $m_\nu \sim 0.1$ eV, a very large halo 
size, about 50 Mpc, and rather flat spectrum with spectral 
index $\gamma \sim 1.2$.
The analysis made in ref.~\cite{wibig00} also disfavors the model 
with clustering of background neutrinos. According to this paper 
the annihilation of UHE neutrinos on CNB
could give no more than 20\% of the observed UHECR flux. 

An even  stronger bound comes from the fact that neutrinos with 
masses $m_\nu < 1$~eV are the Hot Dark Matter particles and their 
distribution is less clustered than the distribution of 
the total mass~\cite{ma99,primack00}. The clustering scale for
neutrinos with such a small mass is of the order of the size of 
clusters of galaxies, i.e. several Mpc. 
The local CDM distribution is well known on such scales
from peculiar velocity measurements and does not allow to 
have overdensities more than by the factor 3-4 \cite{daCosta96,dekel99}. 
One could conclude that high overdensities
by factor 20 or larger contradict the data.

A possible way to "reanimate" the Z-burst model was suggested 
in ~\cite{gelmini99}. Instead of an unrealistic local neutrino 
overdensity the model of relic neutrinos with 
a large neutrino chemical potential  was considered.
Degenerate neutrinos could have a much larger
cosmological number density than the usual $55/$cm$^3$ and the 
constraints of ref.~\cite{pillado99} are not applicable to them. 
With the concrete
value of neutrino mass, $m_\nu = 0.07$ eV taken from the
Super-Kamiokande data on atmospheric neutrinos (see sec.~\ref{2prop})
the authors of ref.~\cite{gelmini99} concluded that the necessary value
of neutrino chemical potential should be $\mu_\nu = (4-5)\,T$. 
Unfortunately this value is outside the limits~(\ref{limitsxi}) found 
from CMBR and from BBN bounds for mixed active neutrinos~\cite{dolgov02}
(see sec.~\ref{ssec:active}).

One can assume of course that the Z-burst
mechanism is responsible only for a part of UHECR ~\cite{fodor01}.
In this case both primary neutrino and secondary photon
fluxes can be reduced to obey all existing limits. However, the
origin of the remaining dominant part of UHECR remains mysterious.

Recently a detailed numerical study of the Z-burst model was 
performed in ref.~\cite{kalashev01}. The calculations were based 
on the solution of the Boltzmann transport 
equations for the spectra of nucleons, $\gamma-$rays, electrons,
$\nue$, $\num$, $\nut$, and their antiparticles moving
along straight lines.  Arbitrary injection spectra and
redshift distributions of the sources can be substituted into the
code and all relevant strong, electromagnetic, and weak 
interaction reactions
can be taken into account~\cite{kalashev00}. This code was
compared on the level of individual reactions with the older 
version of such code, used in ref.~\cite{yoshida98}. Contrary to 
the latter, the code of ref.~\cite{kalashev01} allows to use 
arbitrary neutrino masses and 
distribution of the sources. The neutrino injection spectrum
per comoving volume was parametrized as:
\begin{eqnarray}
  \phi_\nu(E,z)&=&f(1+z)^m\,E^{-q_\nu}\Theta(E^\nu_{\rm max}-E)\,
  \nonumber\\
  &&z_{\rm min}\leq z\leq z_{\rm max}\,,\label{para_nu}
\end{eqnarray}
where $f$ is the normalization factor to be fitted from the
data. The free parameters are the spectral index $q_\nu$, the maximal
neutrino energy $E^\nu_{\rm max}$, the minimal and maximal
redshifts $z_{\rm min}$, $z_{\rm max}$, and the redshift
evolution index $m$. 

The possibility to vary the distribution of neutrino sources
suggested in ref.~\cite{kalashev01} permitted to avoid the
disagreement of the Z-burst model with the data.
In order to reduce the photon flux in the EGRET region
one can introduce non-uniformly
distributed sources more abundant at low redshifts,
instead of assuming a local neutrino overdensity.
The fluxes of cosmic rays for different values of $m$ are
presented in fig.~\ref{F1}. 
The value $m=3$ corresponding to the spectrum of ref.~\cite{yoshida98},
predicts an excessive photon flux in the EGRET energy region and 
is excluded by the data.
The uniform source distribution, $m = 0$, is already
in agreement with the EGRET flux, while the negative value,
$m = -3$, leads to GeV photon flux well below it.
The latter corresponds to the sources which 
are more abundant now than at high redshifts. 
For example, the BL Lacertae
objects which are correlated with UHECR according to
ref.~\cite{corr_bllac}, are distributed in such a way.

\begin{figure}[t]
\begin{center}
  \leavevmode
  \hbox{
    \epsfysize=3.in
 \epsffile{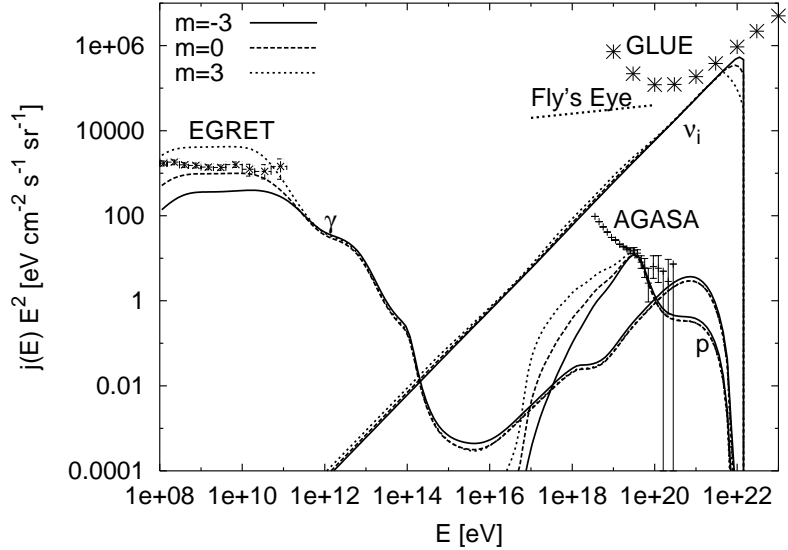}}
\end{center}
\caption{Fluxes of neutrinos, $\gamma-$rays, and nucleons  predicted by the
Z-burst mechanism for $m_\nu = 0.5$~eV, assuming sources
exclusively emitting neutrinos with fluxes equal for all
flavors ~\cite{kalashev01}. Three cases of the source evolution parameter,
$m=-3, 0, 3$ are shown by solid, dashed, and dotted lines, respectively.
Values assumed for the other parameters are:
extragalactic magnetic field strength $B=10^{-9}$ G, minimal radio background strength, 
$z_{\rm min}=0$, $z_{\rm max}=3$, $E^\nu_{\rm max}=2 \times 10^{22}$ eV,
$q_\nu=1$. For each case the neutrino flux amplitude $f$
is obtained from minimizing $\chi^2$ for
$E_{\rm min} = 2.5 \times 10^{19}~{\rm eV}$.
Also shown are experimental upper limits on $\gamma-$ray
and neutrino fluxes (see text and Ref.~\cite{bhatta98} for more
details).}
\label{F1}
\end{figure}

If the Z-burst model works
one could reverse the arguments and ``measure'' neutrino
mass using the observed spectrum of UHECR because 
according to eq.~(\ref{euhe}) the resonance
energy in production of $Z$-bosons depends upon $m_\nu$~\cite{pas01}. 
The main problem in interpretation of the data  
is that the protons and photons, produced in Z-boson decays, interact
with cosmic electromagnetic backgrounds and the observed spectra 
are very far from those produced by the Z-decays, 
see fig.~\ref{F1}. Still an upper bound on neutrino mass about a few eV
can be found because for neutrinos with higher mass the
secondary protons and photons from Z-decay would have 
energies below the observed UHECR values.

Nevertheless, according to the papers~\cite{fodor01,ringwald01} 
the best fit to the data can be obtained with
$m_\nu = 2.34^{+1.29}_{-0.84}$ eV for the production of UHECR in the 
Galactic halo and
$m_\nu = 0.26^{+0.20}_{-0.14}$ eV for the extragalactic origin. 
Later, in ref.~\cite{kalashev01} was shown that it was 
possible to obtain the strong results 
of refs.~\cite{fodor01,ringwald01} 
because many unknown parameters, e.g. neutrino injection spectra 
eq.~(\ref{para_nu}), were fixed at certain definite values. 
Moreover, the authors of refs.~\cite{fodor01,ringwald01}
did not take into account the flux of secondary photons, assuming 
that all of them were cascaded in energy in the EGRET region.  
The conclusion of ref.~\cite{kalashev01} was that the current state
of knowledge does not allow to extract any restrictive
information on neutrino masses from the UHECR data. 
The results of refs.~\cite{fodor01,ringwald01} were reanalyzed in the
recent paper~\cite{ringwald02} where was found 
that $m_\nu = 0.08 ~{\rm eV} - 1.3 ~{\rm eV}$ in agreement with
the conclusion of ref.~\cite{kalashev01}. 

The Z-burst model discussed above is based on the assumption 
that astrophysical sources emit only neutrinos. However, according
to standard scenarios, the UHE neutrinos are secondaries from 
the interactions of primary accelerated protons. Together with 
neutrinos the protons produce gamma-rays
with approximately the same power as neutrinos (because both 
neutrinos and photons are secondaries of pion decays). The photon 
spectrum, typically ends up at the energies 
$E^\gamma_{\rm max}=100\,$TeV~\cite{agn_obs}. 
In this case the Z-burst scenario is difficult to make
consistent with observations. A possible 
solution to this problem is to down-scatter most of the
EM energy into sub-MeV range within the source. Only if
such mechanism is efficient the EGRET bound could be satisfied.

One more problem would appear if the source of high energy 
particles is not completely opaque to the primary nucleons. 
Even if a small fraction of them could
leave the source, then according to ref.~\cite{waxman98},
the nucleon flux between $10^{18}\,$eV
and $10^{19}\,$eV would be much higher than observed. If the
high energy neutrinos are produced in photo-meson or
proton-proton interactions in the sources that are not much larger than
the mean free path with respect to these reactions then the following
upper bound on neutrino flux can be 
obtained~\cite{waxman98b,bahcall99w}:
\be
E^2_\nu F < 2\cdot 10^{-8} {{\rm GeV \over cm^2\,sec\, sr}}
\label{E2F}
\ee
This result depends upon the spectrum of primary protons.
There are two possible ways to escape this bound, either
through production of neutrinos in sources that are optically thick 
to photo-nucleon and nucleon-nucleon interactions, so 
the protons are trapped in the sources, or 
in the processes that do not create simultaneously high
energy cosmic rays. Conventional astrophysical sources of this 
kind are unknown.

Another difficulty of the Z-burst mechanism is the necessity to 
accelerate primary neutrino producing protons up to energies
$E^p_{\rm max}\sim10E^\nu_{\rm max}\sim 4\times10^{22}
\,{\rm eV}\,({\rm eV}/m_\nu)$. 
On the other hand, known mechanisms
are usually limited by $E^p_{\rm max}< 10^{22}\,$eV~\cite{bier-rev}.

Thus, the Z-burst model imposes the following requirements
for the sources \cite{kalashev01}: they should emit energy only in 
neutrinos and, possibly, in sub-MeV $\gamma-$rays and trap most of the
primary protons. These protons should be initially accelerated up to 
very high energies $E^p_{\rm max}> 10^{22}\,$eV, which requires an 
unknown acceleration mechanism.

According to the paper~\cite{kalashev01}
the contribution to the UHECR flux from such hypothetical extragalactic
neutrino sources due to the Z-burst mechanism would exhibit the 
GZK-cutoff for nucleons and would be dominated by $\gamma-$rays at 
higher energies. The required UHE neutrino fluxes are close to the
existing upper limits and should be easily
detectable by future experiments such as Auger~\cite{auger},
Euso~\cite{euso}, RICE~\cite{rice}, or by other radio detection 
techniques~\cite{radhep}. Another possibility to observe energetic 
neutrinos by searching for $\tau$-air showers is discussed in
ref.~\cite{fargion00tau}.

In view of the difficulties of the Z-burst model discussed above, 
non-traditional sources of production of UHE neutrinos have been
considered, namely superheavy particles or topological defects, 
which decay predominantly into neutrinos and other invisible particles. 
In ref.~\cite{gelmini99a} was assumed that there exists a
long-lived superheavy relic particles, $X$, 
with the mass twice larger than the neutrino resonance
energy~(\ref{euhe}). The decays of these particles into neutrino and
a light invisible partner could produce the necessary neutrino flux
determined by~(\ref{dotrhoprim}) if their mass, life-time, and number
density satisfy the condition:
\be
{n_X m_X \over \tau_X} = \dot \rho_{prim} \approx 2\cdot 10^{-22} \left(
{n_\nu^{(0)} \over n_\nu}\right) {\rm{eV \over sec\,\,cm^3}}
\label{nxmxtaux}
\ee
Correspondingly 
$\tau_X \sim 0.5\cdot 10^{26}\,\, {\rm sec} \,{\Omega_X} h^2
\left(n_\nu/ n_\nu^{(0)}\right)$. This simple estimate is reasonably close
to the results presented in refs.~\cite{gelmini99a,crooks00}.

Both the bound~(\ref{E2F}) and Fly's Eye bound (see Fig.~\ref{F1})
are not applicable to the case
of neutrinos coming from heavy particle decays because their spectrum
is peaked at $E_\nu \sim m_X$ and is very much different from the neutrino
spectrum from traditional astrophysical sources. However, the GLUE 
bound Fig.~\ref{F1}
still constrains the neutrino flux near its maximum value. 
The results for proton and photon fluxes presented in the
paper~\cite{kalashev01} are applicable to the model 
of ref.~\cite{gelmini99a}.

Let us also note that if superheavy particles
or topological defects have the branching ratio of the order of 
$0.01$ or larger into
visible channels (quarks, gluons or charged leptons), both photon and 
proton fluxes would be dominated by these channels and the 
contribution from Z-boson decays could be
neglected. Though it is possible to  explain the UHECR spectrum in this 
case \cite{bhatta98}, the distribution of such sources would 
contradict the statistically 
significant clustering in UHECR data \cite{anis1,anis2}.  

In addition to the usual astrophysical sources of high energy cosmic rays,
several particle physics candidates have also been proposed. Among them
are heavy particle decays discussed above and topological defects pioneered
in ref.~\cite{hill87}. A review of different possibilities and the list
of references can be found in~\cite{berezinsky00,domokos01}. These sources
are nor directly related to neutrinos and thus to the subject of the
present review, except for ref.~\cite{berezinsky99v}, where the production
of UHE neutrinos from hidden/mirror sector topological defects
was considered.

The idea to invoke decays of superheavy quasistable particles $X$ for
the explanation of UHECR events was proposed in
refs.~\cite{berezinsky97,kuzmin98}. In this scenario the observed high
energy protons come directly 
from the $X$-decays and not through
the two-step process $X \rar \nu \rar Z \rar p$ discussed above.
However this more
direct explanation seems to be excluded or disfavored by
non-observation of directional correlations of the observed UHE events
with the shape of galactic dark matter halo where these superheavy
particles should be accumulated~\cite{takeda99}. The two-step process
through the $Z$-burst does not suffer from this restriction because in this
case the sources of UHE protons could be at much larger distances, up to
100 Mpc. Long-lived unstable particles producing high energy neutrinos
were discussed in the literature~\cite{henu} independently of the problem
of UHECR. The masses of the particles involved, however, were much
lighter than $\sim 10^{13}$ GeV which are necessary for an
explanation of UHECR events. 
A specific example example of the superheavy
particle $X$ being a right-handed neutrino was recently considered in
ref.~\cite{uehara01} in a multidimensional model.
The mass of the $\nu_R$ was assumed to be about
$10^{14}$ GeV and a large life-time was realized by the separation of
the wave function of $\nu_R$ from other (normal) fermions along the
fifth extra dimension.
The prospect of observation of high energy neutrinos from 
superheavy relics is discussed in ref.~\cite{halzen01}.
However all those models will be ruled out if a significant 
correlation with 
astrophysical sources BL Lacertae ~\cite{corr_bllac} is confirmed by
future observations.

Another possible explanation of the observed UHECR events is that
neutrinos possess a new stronger-than-weak interactions at high energies.
This idea was first suggested in ref.~\cite{beresinsky69}. Earlier
works include also refs.~\cite{domokos87}.
New observations revived the activity in the field and there appeared
several new papers~\cite{bordes97}-\cite{jain00b} where it was assumed
that neutrinos could have anomalously strong
interactions at energies around or above TeV scale, while their
interactions with photons
remain the standard negligibly weak ones. In
this case neutrinos are not subject to GZK cutoff but could interact with
protons in the atmosphere directly with sufficient efficiency inducing
the observed UHECR events. For strong nucleon-neutrino interactions
the flux of the primary neutrinos and their
energies could be much smaller than in the $Z$-burst model and the
restrictions discussed above may be easily satisfied without
contradicting the existing data on cosmic rays. An essential point
is that, despite a large neutrino-nucleon cross-section, the mean free
path of energetic neutrinos in the universe would be much larger than
the present day horizon because the cosmological number density of nucleons
is very small, it is 9-10 orders of magnitude smaller than the number
density of photons in CMBR. Within a galaxy the mean free path would be
about Mpc but it is still much larger than the galactic size.
The idea of strong interactions of neutrinos gained new momentum
following the suggestion  
that gravity may be unified with other interactions
at TeV scale, due to large extra dimensions (see sec.~\ref{s:largeD}).
In this case
an exchange by the ladder of Kaluza-Klein spin-2 excitations of
graviton could give rise to neutrino-nucleon cross-section compatible
with the data~\cite{jain00,anchor00,jain00b}. However, the conclusion
of ref.~\cite{kachel00} disagrees with this optimistic statement. It
is argued there that the neutrino-nucleon cross-section and the
transferred energy per interaction is too small to explain the observed
vertical air showers.
The issue of the high energy behavior of interactions mediated by
spin-2 exchange is rather subtle and deserves more consideration.
If the hypothesis of neutrino strong interactions is confirmed, it
could  be a serious
indication in flavor of 
a modification of physics at TeV scale as was
suggested in recent years.
On the other hand, for $E_\nu =2\cdot 10^{20}\, {\rm eV}$ the corresponding
center of mass energy is only 0.6 TeV and this is somewhat below
theoretical expectations for the new unification scale.

In a recent paper~\cite{wigmans01} the scattering of high energy protons
on cosmic neutrino background, $p+\bar\nu \rar n+e^+$, was suggested as
an explanation of the knee in the spectrum of the cosmic rays with the
energy $10^{16}$ eV. If neutrino mass is 0.4 eV then the threshold of this
reaction is just $10^{16}$ eV. However one needs a very high number density
of neutrinos to make this process noticeable,
$n_\nu > 10^{12}{\rm cm}^{-3}$. On the other hand, the number density
of background protons, $n_p$, in the same regions should be 14 orders of
magnitude smaller than $n_\nu$ because otherwise proton-proton scattering
would dominate. It is difficult to imagine that there
might exist cosmological regions with such a high neutrino density with
simultaneous suppression of $n_p$. The characteristic time of 
neutron-proton
transformation on the background neutrinos is about
$\tau_{np}=10^9 (n_\nu /10^{12}{\rm cm}^{-3}$ years. In other words,
$\dot n_p /n_p = 1/\tau_{np}$, and to obtain an observable effect
on the
spectrum of cosmic rays one should have a too large number
density of protons
in the interaction region in contradiction with the above mentioned
constraints.

A slight modification of the mechanism~\cite{wigmans01} is discussed 
in the paper~\cite{dova01}. The authors explored essentially the same 
idea but suggested that neutrino interaction with protons  
may be stronger 
than is usually supposed
This assumption would allow weakening criticism
aimed at the original version of the scenario discussed above.
However, this model which involves 
the new interaction with the anomalous magnetic moment of
neutrinos, demands the latter to be very big, 
$\mu_\nu \approx 5\cdot 10^{-6} \mu_B$. This value is much larger than
the cosmological
bounds discussed in sec.~\ref{ssec:magnmom} and even than direct
experimental bounds~(\ref{muexsp}).
Possibly the model
may be cured by introducing a
different form of anomalous $\nu \, p$-interaction.

To conclude, the problem of UHECR remains unsolved and it is not clear
if neutrinos play any role in its resolution. If not, then this
section would have nothing to do with the subject of the present review.

\section{Conclusion. \label{s:conclusion}}

As one can see from the material presented above, cosmological implications
of neutrinos as well as implications of cosmology for
neutrino physics are
two vast fields that include diverse physical phenomena that have
different ``raison d'{\^e}tre''.
It is important to distinguish the ``confidence level'' of particular
physical models and assumptions discussed in this review. The existence of
three families of neutrinos is a well established fact, while the 4th
generation, even very heavy (see sec.~\ref{sec:heavynu}), is most probably
excluded. It is quite natural to expect that neutrinos are massive and the
accumulated experimental data present quite strong evidence in favor of
non-vanishing $m_\nu$. Gerstein-Zeldovich upper bound on $m_\nu$
(sec.~\ref{gerzel}) is robust, practically assumptionless, and is
competitive with direct experimental measurements. Together with
Tremaine-Gunn limit (sec.~\ref{normalnu}), it excludes neutrinos as
a dominant component
of cosmological dark matter for any spectrum of primordial
density perturbations. This conclusion can be avoided only if neutrinos
have an unknown new interaction which is much stronger than the usual
weak interaction. If neutrinos are thermally produced in the
early universe and have a non-vanishing mass
they would form hot dark matter and inhibit structure formation at
small scales. A non-thermal production of neutrinos is not excluded.
In particular, neutrino oscillations into sterile partners could strongly
distort the spectrum.

The present day analysis of large scale structure
formation is sensitive to $m_\nu$ of several eV and the future data from
SDSS will measure neutrino mass with the accuracy of a fraction of eV
(sec.~\ref{normalnu}). A distortion of neutrino spectrum which should be
at the per cent level in the standard model (sec.~\ref{masslessdistr})
is not observable in the large scale structure but in the optimistic case
may be observed in the angular variation of CMBR measured by the future
Planck mission (see sec.~\ref{9cmb}).

The number of neutrino species is well measured by BBN (sec~\ref{knu}),
where one may expect the accuracy at the level $\Delta N_\nu \sim 0.1$.
In the coming years similar accuracy may be achieved in CMBR as well
(sec.~\ref{9cmb}) and the existence of the cosmic neutrino background
will be independently confirmed.

A few years ago the value of $\nut$ mass in MeV region, allowed by the
direct experimental limit~(\ref{mnt}), was discussed in connection with
possible cosmological effects: cold dark matter, BBN, etc. Nowadays, the
interpretation of neutrino anomalies in terms of oscillations demands
the mass difference 
squared between $\nut$ and $\nue$ or $\num$ smaller or
about one eV. This practically excludes $\nut$ with MeV mass. If this is
indeed the case, the limits on $\mnt$ and $\nut$ life-time obtained from
consideration of primordial nucleosynthesis
(secs.~\ref{massstabnu} and \ref{massunstbl}), taken literally, become
not interesting. On the other hand, a twist that would allow a MeV mass
of $\nut$ is not 100\% excluded and the material of these sections might
become relevant again. Moreover, the physics and arguments presented
here
are applicable to any other hypothetical particles that might be
present during BBN. Anyhow, the upper limit 0.2-0.3 MeV for $\mnt$, if
the latter is stable on BBN time scale, is sufficiently well founded.
Another interesting implication for
BBN of massive (unstable) $\nut$ is
that its effect is non-monotonic and could both enlarge or diminish
primordial abundances of light elements.

Massive unstable neutrinos could escape the Gerstein-Zeldovich mass
limit if their life-time is sufficiently short but they would have
a noticeable impact on the structure formation
(sec.~\ref{cosden},\ref{anomalous}) and on the angular fluctuations
of CMBR (sec.~\ref{9cmb}). If the decay goes into electromagnetic
channel it may be registered by cosmic electromagnetic backgrounds,
either by the CMBR spectrum or by other forms
of radiation
(sec.~\ref{cmbr},\ref{cer}).

Experimental data in favor of neutrino oscillations require studying
the role of oscillations in cosmology (sec.~\ref{nuosceu}). Unfortunately,
in the case of mixing between the three known neutrinos 
only, there are no
observable effects in the standard case of thermal equilibrium. If
the initial state 
is not the equilibrium one, for example, if there is a
non-negligible cosmological lepton asymmetry, neutrino oscillations
would lead to interesting effects in BBN (sec.~\ref{ssec:active})
and, in particular, to quite strong bounds on neutrino degeneracy. 

If there is a new sterile neutrino (or several sterile species)
the mixing between active and sterile ones would lead to striking
consequences. In particular, in the resonance case
a large lepton asymmetry in the sector of
active neutrinos could be generated (sec.~\ref{sec:leptas}). Moreover,
this asymmetry could strongly fluctuate as a function of space point
(sec.~\ref{sec:spatfluc}). Neutrino oscillations in the early universe
would lead to the excitation
of new neutrino species, to the distortion 
of the
spectrum of active neutrinos, to large (and possibly inhomogeneous)
lepton asymmetry. All that would have a strong impact on BBN and, as
a result, restrictive bounds on the oscillation parameters could be
obtained (sec.~\ref{sec:nuoscbbn}). In a sense these results are at the
second level of plausibility, because they invoke an additional hypothesis
of the existence of sterile neutrinos. However, theoretically
such case is quite
natural. Sterile neutrinos would appear if Dirac mass is non-zero, or if
there exists a mirror or shadow world coupled to ours through very weak
interactions (sec.~\ref{s:mirror}), or if large
extra dimensions generate the
Kaluza-Klein tower of sterile neutrinos in our space.

Right-handed sterile neutrinos could be created in the early universe
either by their coupling to weak currents, proportional to their Dirac
mass, $(m_\nu / E_\nu)$, or due to new interactions with right-handed
intermediate bosons (secs.~\ref{ssec:nur}). In the latter case,
non-zero neutrino mass is not necessary. In the first case, BBN
considerations lead to the upper limit on neutrino mass of the order
of 100 keV. In the second case one can obtain the lower limit on the
mass of the right-handed intermediate bosons of about 1 TeV
or larger.
If neutrinos have a non-zero magnetic moment, the spin-flip in
magnetic fields in the early universe would produce additional neutrino
species and BBN permits to derive restrictive upper bounds on neutrino
magnetic moment at the safe level $(10^{-10}-10^{-11}) \mu_B$. With
rather conservative hypotheses about the magnitude of primordial
magnetic fields the limit could be considerably stronger, though less
reliable (sec.~\ref{ssec:magnmom}).

Very heavy sterile neutrinos with masses
in the range 10 - 200 MeV
are practically excluded by the combined experimental data, BBN,
cosmic electromagnetic background, and supernova 1987A
(sec.~\ref{ss:nush}). Lighter sterile neutrinos with masses
in keV range could contribute to the cosmological warm dark matter. The
latter may be an important component of the total dark matter permitting
to solve some problems present in CDM scenario of large scale structure
formation (sec.~\ref{sterilenu}).

A simple deviation from the standard cosmological framework can be
realized if there is a large lepton asymmetry of the universe, i.e.
a large excess of neutrinos over antineutrinos or vice versa. This
looks rather exotic but there are several possible models that
might generate a large lepton asymmetry, while a small baryon
asymmetry remains undisturbed. Neutrino degeneracy would influence
BBN, large scale structure formation, and CMBR
(sec.~\ref{sec:asym}, \ref{leptaslss}).
A combined analysis of the data permits putting the
limits~(\ref{limitsxi}) on the values of chemical potentials of $\nu_e$
and $\nu_{\mu,\tau}$ neutrinos. They are not very restrictive and a large
asymmetry is still allowed but better bounds may appear in the near
future with new more precise observational data. Moreover, if
lepton asymmetry varies on cosmologically large scales the universe
could be strongly chemically inhomogeneous, while energetically very
smooth (sec.~\ref{varabund}).

Some other cosmological implications of neutrinos demands more ``ifs''.
For example astronomically large objects consisting of neutrinos
(sec.~\ref{nuball}) may in principle exist, but 
but this would require
either spontaneously broken left-right symmetric theories or new
anomalous neutrino interactions. Such exotic objects could help 
to solve some astrophysical mysteries.

If leptonic charge is non-conserved and if baryogenesis proceeds through
leptogenesis (sec.~\ref{s:bs}), then from the condition that the
processes with $L$-nonconservation did not destroy charge asymmetry
of the universe (both leptonic and baryonic), one can derive quite
restrictive limits on neutrino masses. However, one should keep in mind
the both ``ifs'' mentioned above.

The explanation of the spectral features of high energy cosmic rays by
the scattering 
of energetic particles on cosmic neutrino background
(sec.~\ref{s:cr-cnb}) might be a promising way to observe the latter.

To summarize, we see that cosmology definitely confirms neutrino
existence, moreover, the existence of three neutrino species. In the near
future astronomers will be able to measure or to constraint neutrino
mass more accurately than direct experiments do.
Allowing minor modifications
of the standard cosmological model or the minimal standard model
of particle physics new exciting phenomena related to neutrinos
could be observed in the sky in the near future.

\bigskip
{\bf Acknowledgment}

I am deeply grateful to L. Okun, F. Villante, and my collaborators on
the subject on cosmological implications of neutrinos S. Hansen,
S. Pastor, and D. Semikoz for discussions and many helpful and stimulating
comments on the manuscript.

\newpage


\def\prd{{Phys.~Rev.~D}}
\def\prl{{Phys.~Rev.~Lett.}}
\def\apj{{Ap.~J.}}
\def\apjl{{Ap.~J.~Lett.}}
\def\apjsuppl{{Ap.~J.~Supp.}}
\def\mnras{{M.N.R.A.S.}}


\end{document}